\documentclass[rmp,aps,amsfonts,twocolumn,superscriptaddress,preprintnumbers]
{revtex4}
\input epsf
\usepackage{graphics,epsfig}

\begin{document}
\def\rem{$\clubsuit$}
\def\EFT{EFT}
\def\MP{M_P}
\def\MPd#1{M_{P,#1}}
\newcommand{\etal}{{et~al.}}
\def\cite#1{\citep{#1}}
\def\citeasnoun#1{\citet{#1}}
\newcommand\be{\begin{equation}}
\newcommand\ee{\end{equation}}
\def\bea{\begin{eqnarray}}
\def\eea{\end{eqnarray}}
\def\crt{\nonumber \\}
\def\eqn#1#2{\be \label{eq:#1} #2  \ee}
\def\CPbar{\hbox{{\rmCP}\hskip-1.80em{/}}}
\def\ctn#1{{{\cite{#1}}}}
\def\reffig#1{Figure \ref{fig:#1}}
\def\rfn#1{Eq. {({\ref{eq:#1}})}}
\def\rfs#1{Sec.~{\ref{sec:#1}}}
\def\rfss#1{Sec.~{\ref{ss:#1}}}
\def\rfsss#1{Sec.~{\ref{sss:#1}}}
\def\lref#1#2{{\bibitem{#1}#2}}
%
\def\np#1#2#3{Nucl. Phys. {\bf B#1} (#2) #3}
\def\pl#1#2#3{Phys. Lett. {\bf #1B} (#2) #3}
\def\prl#1#2#3{Phys. Rev. Lett. {\bf #1} (#2) #3}
\def\pr#1#2#3{Phys. Rev. {\bf #1} (#2) #3}
\def\aph#1#2#3{Ann. Phys. {\bf #1} (#2) #3}
\def\prep#1#2#3{Phys. Rep. {\bf #1} (#2) #3}
\def\rmp#1#2#3{Rev. Mod. Phys. {\bf #1}}
\def\cmp#1#2#3{Comm. Math. Phys. {\bf #1} (#2) #3}
\def\mpl#1#2#3{Mod. Phys. Lett. {\bf #1} (#2) #3}
\def\ptp#1#2#3{Prog. Theor. Phys. {\bf #1} (#2) #3}
\def\jhep#1#2#3{JHEP {\bf#1}(#2) #3}
\def\jmp#1#2#3{J. Math Phys. {\bf #1} (#2) #3}
\def\cqg#1#2#3{Class.~Quantum Grav. {\bf #1} (#2) #3}
\def\ijmp#1#2#3{Int.~J.~Mod.~Phys. {\bf #1} (#2) #3}
\def\atmp#1#2#3{Adv.~Theor.~Math.~Phys.{\bf #1} (#2) #3}
\def\ap#1#2#3{Ann.~Phys. {\bf #1} (#2) #3}
%
%
%
%
\def\IB{\relax\hbox{$\inbar\kern-.3em{\rm B}$}}
\def\IC{\relax\hbox{$\inbar\kern-.3em{\rm C}$}}
\def\ID{\relax\hbox{$\inbar\kern-.3em{\rm D}$}}
\def\IE{\relax\hbox{$\inbar\kern-.3em{\rm E}$}}
\def\IF{\relax\hbox{$\inbar\kern-.3em{\rm F}$}}
\def\IG{\relax\hbox{$\inbar\kern-.3em{\rm G}$}}
\def\IGa{\relax\hbox{${\rm I}\kern-.18em\Gamma$}}
\def\IH{\relax{\rm I\kern-.18em H}}
\def\IK{\relax{\rm I\kern-.18em K}}
\def\IL{\relax{\rm I\kern-.18em L}}
\def\IP{\relax{\rm I\kern-.18em P}}
\def\IR{\relax{\rm I\kern-.18em R}}
\def\IT{{\bf T}}
\def\IZ{\relax\ifmmode\mathchoice{\hbox{\cmss Z\kern-.4em Z}}{\hbox{\cmss Z\kern-.4em Z}}
{\lower.9pt\hbox{\cmsss Z\kern-.4em Z}} {\lower1.2pt\hbox{\cmsss
Z\kern-.4em Z}} \else{\cmss Z\kern-.4em Z}\fi}
\def\II{\relax{\rm I\kern-.18em I}}
\def\IIa{{\II a}}
\def\IIb{{\II b}}
\def\IX{{\bf X}}
\def\ttb{Type $\II$B string theory}
\def\ndt{{\noindent}}
\def\bx{{\bf G}}
\def\sssec#1{\ndt$\underline{#1}$}
\def\CA{{\cal A}}
\def\CB{{\cal B}}
\def\CC{{\cal C}}
\def\CD{{\cal D}}
\def\CE{{\cal E}}
\def\CF{{\cal F}}
\def\CG{{\cal G}}
\def\CH{{\cal H}}
\def\CI{{\cal I}}
\def\CJ{{\cal J}}
\def\CK{{\cal K}}
\def\CL{{\cal L}}
\def\CM{{\cal M}}
\def\CN{{\cal N}}
\def\CO{{\cal O}}
\def\CP{{\cal P}}
\def\CQ{{\cal Q}}
\def\CR{{\cal R}}
\def\CS{{\cal S}}
\def\CT{{\cal T}}
\def\CU{{\cal U}}
\def\CV{{\cal V}}
\def\CW{{\cal W}}
\def\CX{{\cal X}}
\def\CY{{\cal Y}}
\def\CZ{{\cal Z}}
\def\B#1{{\bf #1}}
\def\BH{\B{H}}
\def\BM{\B{M}}
\def\BN{\B{N}}
\def\Ba{\B{a}}
\def\Bb{\B{b}}
\def\Bk{\B{k}}
\def\Br{\B{r}}
\def\p{\partial}
\def\pb{\bar{\partial}}
\def\eV{{\rm eV}}
\def\TeV{{\rm TeV}}
\def\GeV{{\rm GeV}}
\def\cm{{\rm cm}}
\def\dir{{\CD}\hskip -6pt \slash \hskip 5pt}
\def\dd{{\rm d}}
\def\Dslash{\rlap{\hskip0.2em/}D}
\def\cb{\bar{c}}
\def\ib{\bar{i}}
\def\jb{\bar{j}}
\def\kb{\bar{k}}
\def\lb{\bar{l}}
\def\mb{\bar{m}}
\def\nb{\bar{n}}
\def\ub{\bar{u}}
\def\wb{\bar{w}}
\def\sb{\bar{s}}
\def\tb{\bar{t}}
\def\vb{\bar{v}}
\def\xb{\bar{x}}
\def\zb{\bar{z}}
\def\Cb{\bar{C}}
\def\Db{\bar{D}}
\def\Tb{\bar{T}}
\def\Zb{\bar{Z}}
\def\half{{1\over 2}}
\def\bra#1{{\langle}#1|}
\def\ket#1{|#1\rangle}
\def\bbra#1{( #1|}
\def\kket#1{|#1 )}
\def\vev#1{\langle{#1}\rangle}
\def\codim{{\mathop{\rm codim}}}
\def\Hol{{\rm Hol}}
\def\cok{{\rm cok}}
\def\rank{{\rm rank}}
\def\coker{{\mathop {\rm coker}}}
\def\diff{{\rm diff}}
\def\Diff{{\rm Diff}}
\def\Tr{{\rm Tr~}}
\def\tr{{\rm tr~}}
\def\Id{{\rm Id}}
\def\vol{{\rm vol}}
\def\Vol{{\rm Vol}}
\def\c{\cdot}
\def\sdtimes{\mathbin{\hbox{\hskip2pt\vrule height 4.1pt depth -.3pt
width.25pt\hskip-2pt$\times$}}}
\def\dim{{\rm dim~}}
\def\Re{{\rm Re~}}
\def\Im{{\rm Im~}}
\def\imp{$\Rightarrow$}
\def\danger{{NB:}}
\def\Lie{{\rm Lie}}
\def\mod{{\rm mod}}
\def\lieg{{\underline{\bf g}}}
\def\liet{{\underline{\bf t}}}
\def\liek{{\underline{\bf k}}}
\def\lies{{\underline{\bf s}}}
\def\lieh{{\underline{\bf h}}}
\def\clieg{{\underline{\bf g}}_{\scriptscriptstyle{\IC}}}
\def\cliet{{\underline{\bf t}}_{\scriptstyle{\IC}}}
\def\cliek{{\underline{\bf k}}_{\scriptscriptstyle{\IC}}}
\def\clies{{\underline{\bf s}}_{\scriptstyle{\IC}}}
\def\CCK{K_{\scriptscriptstyle{\IC}}}
\def\inbar{\,\vrule height1.5ex width.4pt depth0pt}
\def\IIa{IIa}
\font\cmss=cmss10 \font\cmsss=cmss10 at 7pt
\def\sdtimes{\mathbin{\hbox{\hskip2pt\vrule height 4.1pt
depth -.3pt width .25pt\hskip-2pt$\times$}}}
\def\ao{{\bf a}}
\def\co{{\bf a}^{\dagger}}

\def\a{{\alpha}}
\def\ap{{l_s^2}}
\def\b{{\beta}}
\def\d{{\delta}}
\def\g{{\gamma}}
\def\e{{\epsilon}}
\def\z{{\zeta}}
\def\ve{{\varepsilon}}
\def\vf{{\varphi}}
\def\kk{{\kappa}}
\def\m{{\mu}}
\def\n{{\nu}}
\def\u{{\Upsilon}}
\def\l{{\lambda}}
\def\s{{\sigma}}
\def\t{{\theta}}
\def\o{{\omega}}
\def\seealso{{See also }}
\def\hepth#1{}
\def\hepph#1{}
%
\title{Flux Compactification}

\author{Michael R. Douglas}
\email{mrd@physics.rutgers.edu}
\affiliation{Department of Physics and Astronomy, Rutgers University,
Piscataway NJ 08855 U.S.A.}
\affiliation{Institut des Hautes Etudes Scientifiques, 35 route des
Chartres, Bures-sur-Yvette France 91440}
\author{Shamit Kachru}
\email{skachru@stanford.edu}
\affiliation{Department of Physics and SLAC, Stanford University, Stanford,
CA 94305}
\affiliation{Kavli Institute for Theoretical Physics, University of
California, Santa Barbara, CA 93106}
\begin{abstract}  
We review recent work in which compactifications of string and M
theory are constructed
in which all scalar fields (moduli) are massive, and supersymmetry is broken
with a small positive cosmological constant, features needed
to reproduce real world physics.  We explain how this work implies that
there is a ``landscape'' of string/M theory vacua, perhaps containing
many candidates for describing real world physics, and present the arguments
for and against this idea.  We discuss statistical surveys of the
landscape, and the prospects for testable consequences of this picture,
such as observable effects of moduli, constraints on early cosmology, and 
predictions for the scale of supersymmetry breaking.
\end{abstract}                                                                 

\maketitle
\tableofcontents


\section{INTRODUCTION}
\label{sec:intro}

It is an old idea that unification of the fundamental forces may be
related to the existence of extra dimensions of space-time.  Its first
successful realization appears in the works of \citeasnoun{Kaluza:1921tu}\ %
and \citeasnoun{Klein:1926tv}, which postulated a fifth dimension of
space-time, invisible to everyday experience.  

In this picture, all physics is described at a fundamental level by a
straightforward generalization of general relativity to five
dimensions, obtained by taking the metric tensor $g_{\mu\nu}$ to
depend on five-dimensional indices $\mu=0,1,2,3,4$, and imposing
general covariance in five dimensions.
Such a theory allows five-dimensional Minkowski space-time as a
solution, a possibility in evident contradiction with experience.
However, it also allows many other solutions with different or less
symmetry.  As a solution which could describe our universe, consider a
direct product of four-dimensional Minkowski space-time, with a circle
of constant periodicity, which we denote $2\pi R$.  It is easy to check that
at distances $r>>R$, the gravitational force law reduces to the familar
inverse square law.  Furthermore, at energies $E << \hbar/Rc$, all
quantum mechanical wave functions will be independent of position on
the circle, and thus if $R$ is sufficiently small (in 1926,
subatomic), the circle will be invisible.

The point of saying this is that the five dimensional metric
$g_{\mu\nu}$, regarded as a field in four dimensions, contains
additional, non-metric degrees of freedom.  In particular, the
components $g_{\mu 5}$ transform as a vector field, which turns out
to obey the Maxwell equations in a curved background.  Thus, one
has a unified theory of gravitation and electromagnetism. 

The theory contains one more degree of freedom, the metric component
$g_{55}$, which parameterizes the radius $R$ of the extra-dimensional
circle.  Since the classical Einstein equations are scale invariant,
in the construction as described, there is no preferred value for this
radius $R$.  Thus, Kaluza and Klein simply postulated a value for it
consistent with experimental bounds.

Just like the other metric components, the $g_{55}$ component is 
a field, which can vary in four-dimensional space-time in any way
consistent with the equations of motion.  We will discuss these
equations of motion in detail later, but their main salient feature is
that they describe a (non-minimally coupled) massless scalar field.
We might expect such a field to lead to physical effects just as
important as those of the Maxwell field we were trying to explain.
Further analysis bears this out, predicting effects such as
new long range forces, or time dependence of parameters, in direct
conflict with observation.

All this would be a historical footnote were it not for the discovery,
which emerged over the period 1975--1985, that superstring theory
provides a consistent quantum theory of gravity coupled to matter in
ten space-time dimensions \cite{Green:1987sp,Green:1987mn}.  At
energies low compared to its fundamental scale (the string tension),
this theory is well described by ten-dimensional supergravity, a
supersymmetric extension of general relativity coupled to Yang-Mills
theory.  But the nonrenormalizability of that theory is cured
by the extended nature of the string.

Clearly such a theory is a strong candidate for a higher dimensional
unified theory of the type postulated by Kaluza and Klein.  Around
1985, detailed arguments were made, most notably by
\citeasnoun{Candelas:1985en}, 
that starting from the heterotic superstring theory, 
one could derive supersymmetric grand
unified theories (GUTs) of the general class which, for completely
independent reasons, had already been postulated as plausible
extensions of the Standard Model up to very high energies.
This construction, the first quasi-realistic string compactification,
took ten-dimensional
space-time to be a direct product of four-dimensional Minkowski
space-time, with a six dimensional Ricci flat
manifold, one of the so-called Calabi-Yau manifolds.
Performing a Kaluza-Klein type analysis, one obtains a four-dimensional
theory unifying gravity with a natural extension of the Standard Model,
from a single unified theory with no free parameters.

However, at this point, the problem we encountered above rears
its ugly head.  Just like the classical Einstein-Maxwell equations,
the classical supergravity equations are scale invariant.  Thus, if we
can find any solution of the type we just described, by rescaling 
the size $R$ of the compactification manifold, we
can obtain a one-parameter family of solutions, differing only in the
value of $R$.  Similarly, by making a rescaling
of $R$ with a weak dependence on four-dimensional position,
one obtains approximate solutions.  Thus, again $R$ corresponds to
a massless field in four dimensions, which is again
in fatal conflict with observation.

In fact, the situation is even worse.  Considerations we will discuss
show that typical solutions of this type have not just one but
hundreds of parameters, called moduli.  Each will lead to a massless
scalar field, and its own potential conflict with observation.  In
addition, the interaction strength between strings is controlled by
another massless scalar field, which by a long-standing quirk of
terminology is called the dilaton.  Since this field is present in all
string theories and enters directly into the formulas for observable
couplings, many proposals for dealing with the other moduli problems,
such as looking for special solutions without parameters, founder
here.

On further consideration, the moduli are tied up with many other
interesting physical questions.  The simplest of these is just the
following: given the claim that all of known physics can arise from a
fundamental theory with no free parameters, how do the particular
values we observe for the fundamental parameters of physics, such as
the electron mass or the fine structure constant, actually emerge from
within the theory?  This question has always seemed to lie near the
heart of the matter and has inspired all sorts of speculations and
numerological observations, some verging on the bizarre.

This question has a clear answer within superstring theory, and the
moduli are central to this answer.  The answer may not be to every
reader's liking, but let us come back to this in due course.

To recap, we now have a problem, a proliferation of massless scalar
fields; and a question, the origin of fundamental parameters.
Suppose we ignore the dynamics of the massless scalar fields for a
moment, and simply freeze the moduli to particular values, in other
words restrict attention to one of the multi-parameter family of
possible solutions in an {\it ad hoc} way.  Now, if we carry out the
Kaluza-Klein procedure on this definite solution, we will be able to
compute physical predictions, including the fundamental parameters.
Of course, the results depend on the details of the assumed solution
for the extra dimensions of space, and the particular
values of the moduli.
		
Returning to the problem of the massless scalar fields, a possible
solution begins with the observation that the equations of motion of
general relativity and supergravity are scale invariant only at the
classical level.  Defining a quantum theory of gravity (in more than
two space-time dimensions) requires introducing a preferred scale, the
Planck scale, and thus there is no reason that the quantum theory
cannot prefer a particular value of $R$, or of the other moduli.
Indeed, this can be demonstrated by simple considerations in quantum
field theory.  For example, given a conducting cavity, even one
containing vacuum, one can measure an associated Casimir energy, which 
depends on its size and shape.  This agrees with the theoretically
predicted vacuum energy of the zero-point fluctuations of the quantum
electromagnetic field.  Very similar computations show that a quantum
field in a compactified extra dimensional theory will have a Casimir
energy which depends on the size and other moduli parameters of the
extra dimensions, and which contributes to the four-dimensional
stress-energy tensor.

In a more complete treatment, this Casimir energy would be the first
term in a systematic expansion of the quantum vacuum energy, to be
supplemented by higher order perturbative and nonperturbative
contributions.  In higher dimensional theories, it is also possible to
turn on background field strengths in the extra dimensions without
breaking Lorentz invariance, and these contribute to the vacuum energy as
well.  All of these effects can be summarized in an effective
potential, defined as the total vacuum energy, considered as a function
of assumed constant values for the moduli fields.

We now work on the assumption that this effective potential, defined
in precise analogy to the effective potentials of conventional quantum
field theory and many-body physics, can be used in a very similar way:
to determine the possible (metastable)
vacuum states of the theory, as the local
minima of the effective potential.  Any configuration not at such a
minimum will roll down to one, converting its excess potential energy
into other entropically favored forms, such as radiation.  This
argument is very general and applies to all known physical systems
with many degrees of freedom; it is widely accepted in cosmology as
well, so there is no evident reason not to accept it in the present
context.

Almost all effective potentials for systems in the real world have
more than one local minimum.  The consequences of this fact depend on
the time scales of transitions between minima (quantum or thermally
induced) compared to the time scales under study.  If transitions
proceed rapidly, the system will find the global minimum of the
potential, and if this changes upon varying parameters the system
undergoes a phase transition.  On the other hand, if transitions
between vacua proceed slowly, local minima are effectively stable, and
one speaks of a system with multiple configurations.  Both phenomena
are ubiquitous; examples of extremely long-lived metastable
configurations include most organic molecules (which ``decay'' to
hydrocarbons and carbon dioxide), and all nuclei except $^{62}$Ni, the
nucleus which minimizes the binding energy per nucleon.

The structure of effective potentials responsible for multiple minima,
metastability and transitions is central to a good deal of real world
physics and chemistry.  Although details are always essential, there
are also principles which apply with some generality, which make up the theory
of energy landscapes \cite{Wales:2003}.
The picturesque term ``landscape'' actually originated in
evolutionary biology \cite{Wright:1932}.

For reasons we will discuss in \rfs{qual}, string vacua with small
positive cosmological constant, as would fit present astronomical
observations, are believed to be metastable and extremely long-lived
even compared to cosmological time scales.  Thus, if we find multiple
local minima of the effective potentials derived from string/M theory
compactification, the appropriate interpretation is that string/M
theory has multiple configurations, the vacua.

Now, ever since the first studies of string compactification, it has
appeared that choices were involved, at the very least the choice of 
compactification manifold, and other discrete choices, leading to
multiple vacua.  However, it was long thought that this might be an 
artifact of perturbation theory, or else not very interesting, as 
the constraints of fitting the data would pick out a unique
candidate solution.  While occasional suggestions to the contrary
were made, as in 
\citeasnoun{Linde:2005ht,Banks:1995uh,Smolin:1997,Schellekens:2006}, 
these were not supported by enough evidence to attract serious attention.

This has changed in recent years, as increasingly detailed
arguments have been developed for the existence of a large number of
candidate vacua within string/M theory.  
(The bulk of our review will be devoted to these arguments, so we defer
the references to there.)
These vacua realize different values
of the cosmological term,
enabling an ``anthropic'' solution of the cosmological constant problem,
along the lines set out by 
\citeasnoun{Banks:1984tw,Linde:2005ht,Weinberg:1987dv,
Bousso:2000xa},
which can naturally accomodate the growing evidence for dark energy 
(see \citeasnoun{Copeland:2006wr} for a recent overview).

Does anything pick out one or a subset of these vacua as the preferred
candidates to describe our universe?  At this point, we do not know.
But, within the considerations we discuss in this review, there is no
sign that any of the vacua are preferred.  So far as we know, any
sufficiently long-lived vacuum which fits all the data, including
cosmological observations, is an equally good candidate to describe
our universe.  This is certainly how we proceed in analogous
situations in other areas of physics.  The analogy leads to the term
``landscape of string vacua'' and a point of view in
which we are willing to consider a wide range of possibilities for
what selected ``our vacuum.''  Indeed, an extreme point of view might
hold that, despite the evident centrality of this choice to all that
we will ever observe, nevertheless it might turn out to be an
undetermined, even ``random'' choice among many equally consistent
alternatives.

Of course, such a claim would be highly controversial.  And, while in
our opinion the idea must be taken very seriously, it is far
outrunning the present evidence.  String/M theory is a theory of
quantum gravity, and given our present limited understanding both of
general principles of quantum gravity and of its microscopic
definition, it is too early to take any definite position about such
claims.  Rather, in this review, we will try to state the evidence
from various sides.  To start with, since there is as yet no precise
definition of the effective potential in string theory, we need to
state our working definition, and justify it within our present
understanding of the theory.  Then, there are important differences
between other physical theories and quantum gravity, which suggest
various speculations about why some of the vacua which appear
consistent at the level of our discussion, actually should not be
considered.  Another point in which quantum gravity plays an essential
role is the idea that early cosmology leads to a ``measure factor,''
an {\it a priori} probability distribution on the vacua which must be
taken into account in making predictions.

We discuss all of these points in \rfs{eff}.  While pointing out many
incomplete aspects of the theory, whose development might
significantly change our thinking, we conclude that at present there
is no clear evidence against, or well formulated alternative to, the
``null hypothesis'' which states that each of these vacua is {\it a
priori} a valid candidate to describe our universe.  In fact, many of
the suggested alternatives, at least within the general framework of
string theory, would themselves require a significant revision of
current thinking about effective field theory, quantum mechanics, or
inflationary cosmology.  Compared to these, the landscape hypothesis
appears to us to be a fairly conservative option.  We will argue as
well that it can lead to testable predictions, perhaps by finding
better selection principles, or perhaps by thinking carefully about
the situation as it now appears.

To summarize the situation, while we have a criterion that determines
preferred values for the size and other moduli, namely that our vacuum
is a long-lived local minimum of the effective potential, this
criterion does not determine the moduli uniquely, but instead gives us
a set of possibilities, the vacua.  Let us make an {\it ad hoc} choice
of vacuum, and ask what physics it would predict.

To first address the fate of the moduli, while these would still be
scalar fields, they would be ``lifted,'' gaining masses $M_{moduli}^2$
proportional to the second derivatives of the effective potential.
In general, we would expect to see their effects only in
experiments at energies $E \sim M_{moduli}$ and above.
The remaining effect of this physics, referred to as ``moduli stabilization,''
is to set the parameters in the solution, which enter into
physical predictions.

What values do we expect for $M_{moduli}$?  Although detailed
computations may not be easy, the energy scales which enter
include the Planck scale, the string
tension, and the inverse size of the extra dimensions
$\hbar c/ R$ (often referred to as the ``Kaluza-Klein scale'' or
$M_{KK}$).  There is no obvious need for the lower energy scales of
present-day physics to enter, and thus it seems plausible that a
detailed analysis would lead to all moduli gaining masses comparable
to the new scales of string theory.  In this case, the prospects for
direct observation of physical effects of the moduli would be similar
to those for direct observation of excited string modes or of the
extra dimensions, in other words a real possibility but not a
particularly favored one.

It is possible that some moduli might gain lower masses and thus have
more direct experimental consequences.  One class of observational
bounds on the masses of moduli arise from fifth-force
experiments; these are important for masses less than about $10^{-3} \eV$.  
A stronger bound comes from cosmology; masses up to $10~ \TeV$ or
more are constrained by the requirement that energy trapped in
oscillations of the moduli fields should relax before primordial
nucleosynthesis \cite{Banks:1993en, deCarlos:1993jw}.  Both bounds
admit loopholes, and thus this possibility is of
interest for phenomenology.

How does one compute the effective potential in string theory?  For a
long time, progress in this direction was slow, due to the belief
that in the compactifications of most interest, the
effective potential would arise entirely from nonperturbative effects.
This brought in the attractive possibility of using asymptotic freedom
and dimensional transmutation to solve the hierarchy problem 
\cite{Witten:1981nf},
but also the difficulty that such effects could only be computed in
the simplest of theories.

Other possibilities were occasionally explored.  A particularly simple
one is to turn on background magnetic fields (or generalized $p$-form
magnetic fields) in the extra dimensions.  These contribute the usual
$B^2$ term to the energy, but since they transform as scalars in the
observable four dimensions this preserves Poincar\'e symmetry, and
thus such configurations still count as ``vacua.''  Furthermore,
writing out the $B^2$ term in a curved background, one sees that it
depends non-trivially on the metric and thus on the moduli, and thus
it is an interesting contribution to the effective potential for
moduli stabilization.  However, while this particular construction,
usually called the flux potential, is simple, the lack of
understanding of other terms in the effective potential and of any
overall picture inhibited work in this direction.

Over the last few years, this problem has been solved, by combining
this simple idea with many others: the concepts of superstring
duality, other techniques for computing nonperturbative effects such
as brane instantons, and mathematical techniques developed in the
study of mirror symmetry, to compute a controlled approximation to the
effective potential in a variety of string and M theory vacua.  The
basic result is that these effective potentials can stabilize moduli
and lead to supersymmetry breaking with positive cosmological
constant, just as is required to get a vacuum which could describe our
universe.  One can go on to get more detailed results, with
applications in particle physics and cosmology which we will discuss.

We have now finished the non-technical summary of the basic material
we will cover in this review, and turn to an outline.  In \rfs{qual},
we assume a general familiarity with particle physics concepts, but
not necessarily with string theory.  Thus, we begin with an overview
of the basic ingredients of the different 10d string theories,
and the known types of compactification.  We then discuss some of the
data needed to specify a vacuum, such as a choice of Calabi-Yau
manifold, and a choice of moduli.  We then explain in general terms
how the fluxes can be expected to induce potentials for moduli of the
extra dimensions.  Finally, we describe some applications of flux
vacua: to the cosmological constant problem, to particle physics, and
to early cosmology.

In \rfs{eff}, we begin to assume more familiarity with string
theory, and critically examine the general framework we will use in
the rest of the paper: that of 10d and 4d effective field theory (EFT).
While our present day understanding of physics fits squarely into this
framework, there are conceptual reasons to worry about its validity in
a theory of quantum gravity.

In \rfs{con}, we turn to detailed constructions of flux
vacua.  These include the simplest constructions which seem to fix all
moduli, in both the IIb and the IIa theories.  We also comment on
recent progress, which suggests that there are many extensions of these
stories to unearth. 

It will become clear, from both the general arguments in \rfs{qual},
and the concrete examples in \rfs{con}, that the number of apparently
consistent quasi-realistic flux vacua is very large, perhaps
greater than $10^{500}$.  Therefore, we need to use statistical
reasoning to survey broad classes of vacua.  In \rfs{stat}, we describe
a general framework for doing this, and give an overview of the results.

We conclude with a discussion of promising directions for further research
in \rfs{conclu}.

\section{A QUALITATIVE PICTURE}
\label{sec:qual}

We begin by briefly outlining the various known classes of
quasi-realistic compactifications, to introduce terminology, give the
reader a basic picture of their physics, and explain how observed
physics (the Standard Model) is supposed to sit in each.  A more
detailed discussion of each class will be given in \rfs{con}, while
far more complete discussions can be found in
\cite{Green:1987sp,Green:1987mn,Polchinski:1998rq,Polchinski:1998rr,
Johnson:2003gi,Zwiebach:2004tj}. 

We then introduce the mathematics of compactification
manifolds, particularly the Calabi-Yau manifolds, to explain why
moduli are more or less inevitable in these constructions.  Even more
strikingly, this mathematics suggests that the number of types of
matter in a typical string/M theory compactification is of order
hundreds or thousands, far more than the 15 or so (counting the
quarks, leptons and forces) observed to date.  Thus, a
central problem in string compactification is to explain why most of
this matter is either very massive or hidden (so far), and give us a
good reason to believe in this seemingly drastic exception to Occam's
razor.

In the next subsection, we 
explain flux compactification, and how it solves the problem
of moduli stabilization.  In particular, it becomes natural that almost
all moduli fields should be very massive, explaining why they are not seen.

We then explain, following \citeasnoun{Bousso:2000xa}, why flux
compactifications in string theory lead to large numbers of similar
vacua with different values of the cosmological constant, leading to
an ``anthropic'' solution of the cosmological constant problem.
This solution depends
crucially on having the many extra types of unobserved matter we just
mentioned and might be regarded as the ``justification'' of this
generic feature of string compactification.

Finally, we 
outline some of the testable consequences this picture might lead to.
These include not just observable effects of the moduli, but also calculable
models of inflation, and new mechanisms for solving the hierarchy problem
of particle physics.

\subsection{Overview of string and M theory compactification}
\label{ss:compact}

String/M theory is a theory of quantum gravity, which can
at present be precisely formulated in several weakly coupled limits.
There are six such limits; five of these are the
superstring theories in ten space-time dimensions
\cite{Polchinski:1998rq,Polchinski:1998rr}, 
called type IIa, type IIb, heterotic $E_8
\times E_8$, heterotic $SO(32)$ and type I.  
In addition, there is an eleven dimensional limit, usually called M
theory \cite{Duff:1996aw}.  
All of these limits are described at low energies by
effective higher dimensional supergravity theories.  Arguments
involving duality \cite{Polchinski:1996nb,Bachas:2002mi} as well as
various partial nonperturbative definitions 
\cite{Banks:1999az,Aharony:1999ti} strongly
suggest that this list is complete,
and that all are limits of a single unified theory.

While there is a rich theory of string/M theory compactification to diverse
dimensions, we will focus on quasi-realistic compactifications.
These are solutions of
the theory which look to a low energy observer like a four
dimensional approximately Minkowski space-time, with physics roughly
similar to that of the Standard Model coupled to general relativity.
The meaning of ``roughly similar'' will become apparent as we proceed,
but requires obtaining the correct gauge group, charged
matter content and symmetries, as well as arguments that the observed
coupling constants and masses can arise.

Now, of the six weakly coupled limits, the type II theories and
M theory have 32 supercharges and (at least at first sight) do not
include Yang-Mills sectors, a problem which must be solved to get
quasi-realistic compactifications.  The other three theories have 16
supercharges and include Yang-Mills sectors, $SO(32)$ from the open
strings in type I, and either $E_8 \times E_8$ or $SO(32)$ in the
heterotic strings.  On the other hand, by ten-dimensional
supersymmetry, the only fermions with Yang-Mills quantum numbers are
the gauginos, transforming in the adjoint of the gauge group.  Thus,
we must explain how such matter can give rise to the observed quarks
and leptons, to claim we have a quasi-realistic compactification.

Let us briefly discuss the important physical scales in a compactification.
Of course, one of the main goals is to explain the observed four dimensional
Planck scale, which we denote $\MPd4$ or simply $\MP$.  
By elementary Kaluza-Klein reduction of $D$-dimensional supergravity,
$$ 
\MPd{D}^{D-2} 
\int
 d^Dx\ \sqrt{g}R^{(D)} \rightarrow 
(\Vol{M}) \MPd{D}^{D-2}
\int
 d^4x\ \sqrt{g}R^{(4)} + \cdots, 
$$ 
this will be related to the $D$-dimensional Planck scale $\MPd{D}$, and the
volume of the compactification manifold $\Vol(M)$.
Instead of the volume, let us define the Kaluza-Klein scale
$
M_{KK}=1/Vol(M)^{1/(D-4)} ,
$
at which we expect to see Kaluza-Klein excitations; the relation then
becomes
\begin{equation} \label{eq:kkscale}
\MPd{4}^2 = \frac{\MPd{D}^{D-2}}{M_{KK}^{D-4}} .
\end{equation}
In the simplest (or ``small extra dimension'')
picture, used in the original work on string
compactification, all of these scales are assumed to be roughly equal.
If the Yang-Mills sector is also $D$-dimensional, this is 
forced upon us, to obtain an order one four-dimensional gauge coupling;
there are other possibilities as well.

\subsubsection{Supersymmetry}
There are many reasons to focus on compactifications with low
energy ${\cal N}=1$ 
supersymmetry. 
From a bottom-up perspective,
SUSY suggests natural extensions of the
Standard Model such as the minimal supersymmetric Standard Model
(MSSM) \cite{Dimopoulos:1981zb}, or non-minimal SSM's with additional fields.
These models can help solve the hierarchy problem, can explain coupling
unification, can contain a dark matter candidate, and have other
attractive features.  But so far, all this is only suggestive, and
these models tend to have other problems, such as reproducing
precision electroweak measurements and a (presumed) Higgs mass $M_H\ge
113 \GeV$.  Thus, many alternative models which can explain the hierarchy,
and even the original ``desert'' scenario which postulates no new
matter below the GUT scale, are at this writing still in play.

Since collider experiments with a good chance of detecting TeV scale
supersymmetry are in progress at Fermilab and scheduled to begin soon
at Cern, the question of what one can expect from theory has become
very timely.  We have just given the standard bottom-up arguments for
low energy SUSY, and these were the original motivation for the large
effort devoted to studying such compactifications of string/M theory
over the past twenty years.  From this study, other top-down reasons
to focus on SUSY have emerged, having more to do with the
calculational power it provides.  Let us summarize some of these
motivations.

First, there are fairly simple scenarios in which an assumed high
scale ${\cal N}=1$ supersymmetry, is broken by dynamical effects at low
energy.  In such compactifications, supersymmetry greatly simplifies
the computation of the four dimensional effective Lagrangian, as
powerful physical and mathematical tools can be brought to bear.  Now
this may be more a question of theoretical convenience than principle,
as in many models (such as the original GUT's) perturbation theory
works quite well at high energy.  But, within our present 
understanding of string/M theory, it is quite important.

Second, as we will discuss in \rfsss{moduliproblem}, supersymmetry
makes it far easier to prove metastability, in other words that a
given vacuum is a local minimum of the effective potential.  In
particle physics terms, metastability is the condition that the scalar
field mass terms satisfy $M_i^2\ge 0$.  Now in supersymmetric
theories, there is a bose-fermi mass relation, $M_{Bose}^2 =
M_{Fermi}(M_{Fermi} - X)$, where $X$ is a mass scale related to the
scale of supersymmetry breaking.  Thus, all one needs is
$|M_{Fermi}|>>X$, to ensure metastability.

At first, this argument may not seem very useful, as in many realistic
models the observed fermions all have $M_{Fermi}\le X$.  But, of course,
this is why these fermions have already been observed.  
Typical string compactifications have many more particles,
and this type of genericity argument will become very powerful.

Note that neither of these arguments refers directly to the
electroweak scale and the solution of the hierarchy problem.  As we
formulate them more carefully, we will find that their requirements
can be met even if supersymmetry is broken so far above the
electroweak scale that it is irrelevant to the hierarchy problem.

This will lead to one of the main conclusions of the line of work
we are reviewing, which is that {\it TeV scale supersymmetry is not 
inevitable} in string/M theory compactification. Rather, it is an
assumption with good theoretical motivations, which should
hold in some string/M theory compactifications.
In others, supersymmetry is broken at scales which are
well described by 4d effective field theory, allowing us to
use these tools to control the analysis, but SUSY is not directly
relevant to solving the hierarchy problem.
A third class of models is known in which SUSY is broken at the KK scale,
and there are even arguments 
for models in which supersymmetry is broken at the string scale,
as in \citeasnoun{Silverstein:2001xn}.

In any case, 
we will proceed with the 
assumption that our compactification
preserves $d=4$, ${\cal N}=1$ supersymmetry at the KK scale.
A standard argument \cite{Green:1987mn} shows that
this is related to the existence of
covariantly constant spinors on $M$, which is determined by its
holonomy group $\Hol(M)$.  More precisely,
the number of supersymmetries in $d=4$, is equal to the
number of supercharges in the higher dimensional theory, divided by
$16$, times the number of singlets in the decomposition of a
spinor ${\bf 4}$ 
of $SO(6)$ under $\Hol(M)$.  In the generic case of
$\Hol(M)\cong SO(6)$ this is zero, so to get low energy supersymmetry
we require $\Hol(M)\subset SO(6)$, a condition on the manifold and
metric referred to as {\it special holonomy}.

All possible special holonomy groups were classified by
\citeasnoun{Berger:1955}, and the results relevant for supersymmetry in
$d=4$ are the following.  For $\dim M=6$, as is needed in
string compactification, the special holonomy groups are
$U(3)$ and $SU(3)$, and subgroups thereof.  The only choice of $\Hol(M)$
for which the spinor
of $SO(6)$ contains a unique singlet is $SU(3)$.
Spaces which admit a metric with this special holonomy are known
as Calabi-Yau manifolds (or threefolds,
from their complex dimension).  
Their existence was proven in
\citeasnoun{Yau:1977ms}, and we will discuss some of their properties
later.  One can show that the special holonomy metric is Ricci flat,
so this choice takes us a good part of the
way towards solving the 10d
supergravity equations of motion.

For $\dim M=7$, as would be used in compactifying M theory, the only
choice leading to a unique singlet is $\Hol(M)\cong G_2$.
Manifolds of $G_2$ holonomy also carry Ricci flat metrics, 
and this leads to a second class of geometric compactifications.  
A third class,  F theory compactification, is based on
elliptically fibered Calabi-Yau fourfolds.
We will defer discussion of these to 
\rfs{con}; physically they are closely related to certain type IIb
compactifications.

The outcome of the discussion so far, reflecting the state of the
field in the late 1980's, is that the three theories with $16$
supercharges and Yang-Mills sectors, all admit compactification on
Calabi-Yau manifolds to $d=4$ vacua with ${\cal N}=1$ supersymmetry, and
gauge groups of roughly the right size to produce GUT's.  On the other
hand, the theories with $32$ supercharges have various problems; the
type II theories seem to lead to ${\cal N}=2$ supersymmetry and too small
gauge groups, while M theory on a smooth seven dimensional manifold
cannot lead to chiral fermions \cite{Witten:1981me}.  In fact all
of these problems were later solved, but let us here follow the
historical development.

\subsubsection{Heterotic string}

The starting point for \citeasnoun{Candelas:1985en} (CHSW)
was the observation that the grand
unified groups are too large to obtain from the Kaluza-Klein
construction in ten dimensions, forcing one to start with a theory
containing 10d Yang-Mills theory; furthermore the matter
representations $5+\bar{10}$, $16$ and $27$ can be easily obtained by
decomposing the $E_8$ adjoint (and not from $SO(32)$), forcing the
choice of the $E_8 \times E_8$ heterotic string.

General considerations of \EFT\ make it natural for
the two $E_8$'s to decouple at low energy, so in the simplest models,
the Standard Model is embedded in a single $E_8$, leaving the other as
a ``hidden sector.''  But what leads to spontaneous symmetry breaking
from $E_8$ to $E_6$ or another low energy gauge group?  This comes
because we can choose a non-trivial background Yang-Mills connection
on $M$, let us denote this $V$.
Such a connection is not invariant under $E_8$ gauge transformations
and thus will spontaneously break some gauge symmetry, at the natural
scale of the compactification $M_{KK}$.  The remaining unbroken group
at low energies is the commmutant in $E_8$ of the holonomy group of $V$.
Simple group theory, which we will see in an example below, implies
that to realize the GUT groups $E_6$, $SO(10)$ and $SU(5)$, the holonomy
of $V$ must be $SU(3)$, $SU(4)$ or $SU(5)$ respectively.

Not only is $E_8$ gauge symmetry breaking possible, it is actually
required for consistency.   As part of the Green-Schwarz anomaly
cancellation mechanism, the heterotic string has a three-form field
strength $\tilde H_3$ with a modified Bianchi identity,
\begin{equation}
\label{hetbianchi}
d\tilde H_3 = {\alpha^\prime\over 4} \left( Tr(R\wedge R) - Tr(F_2 \wedge
F_2) \right)~.
\end{equation} 
In the simplest solutions, $\tilde H_3=0$, and then consistency
requires the right hand side of (\ref{hetbianchi}) to vanish identically.
The solutions of \citeasnoun{Candelas:1985en} accomplish this by taking
the ``standard embedding,'' in which one
${\it equates}$ the $E_8$ gauge connection on $M$ (in one of the two
$E_8$'s) with the spin connection $\omega$, i.e. considers an $E_8$ vector
bundle $V \to M$ which is $V=TM$.  
In this case, since $F=R$ for one of the $E_8$'s, and vanishes
for the other, (\ref{hetbianchi}) is trivially satisfied,
and (by considerations we give in \rfs{con})
so are the Yang-Mills equations.

Thus, any Calabi-Yau threefold $M$, gives rise to at least one
class of heterotic string compactifications, the CHSW compactifications.
The holonomy of $V$ is the same as that of $M$, namely $SU(3)$,
and thus this construction leads to an $E_6$ GUT.
Below the scale $M_{KK}$,
there is a 4d ${\cal N}=1$ supersymmetric \EFT\ governing 
the light fields.  In the CHSW models,
these include
\begin{itemize}
\item A pure $E_8$ ${\cal N}=1$ SYM theory, the hidden sector.
\item An ``observable'' $E_6$ gauge group.  One can also make
simple modifications to $V$ (tensoring with Wilson lines) to 
accomplish the further breaking to $SU(3)\times SU(2)\times U(1)$
at $M_{KK}$, so typically in these models  $M_{GUT}\sim M_{KK}$.
\item Charged matter fields.
The reduction of the $E_8$ gauginos will give rise to chiral fermions
in various 4d matter (chiral) multiplets.
The adjoint of 
$E_8$ decomposes under $E_6 \times SU(3)$ as 
\begin{equation}
\label{groupt}
{\bf 248} = {\bf(27,3)} \oplus {\bf (\overline{27},\bar 3)} \oplus {\bf 
(78,1)} \oplus {\bf(1,8)}~.
\end{equation}
Thus we need the spectrum of massless modes arising from charged matter
on $M$ in various $SU(3)$ representations.
As explained in \citeasnoun{Green:1987mn}, this is
determined by the Dolbeault cohomology groups of $M$; thus
\begin{equation}
n_{27}=h^{2,1}(M),~~n_{\overline{27}} = h^{1,1}(M)
\end{equation}
are the numbers of chiral multiplets in the $27$ and $\overline{27}$ 
representations of $E_6$.  Since for a Calabi-Yau manifold
the Euler character $\chi=2(h^{1,1}-h^{2,1})$, we see that
the search for three-generation $E_6$ GUTs in this framework will
be transformed into a question in topology: the existence of Calabi-Yau
threefolds with $\vert \chi \vert = 6$.
This problem was quickly addressed, and quasi-realistic models were
constructed, beginning with \citeasnoun{Tian:1986ic,Greene:1986bm}.

\item Numerous gauge neutral ${\bf moduli}$ fields.
The Ricci-flat metric on the Calabi-Yau space $M$ 
is far from unique.  By Yau's theorem \cite{Yau:1977ms},
it comes in a family of dimension $2 h^{2,1}(M) + h^{1,1}(M)$.
As we will describe at much greater length below, the parameters
for this family, along with $h^{1,1}(M)$ axionic partners,
are moduli
corresponding to infinitesimal deformations of the complex structure  
and the K\"ahler class of $M$.  
In addition, there is also the dilaton chiral multiplet, containing
the field which controls the string coupling, and an axion partner.

\item More model dependent modes arising from the ${\bf(1,8)}$ in
(\ref{groupt}).  These correspond to infinitesimal
deformations of the solutions to the Yang-Mills field equations, and
thus are also moduli, in this case moduli for the gauge connection
$V$.  By giving vacuum expectation values to these scalars, one 
moves out into a larger space of compactifications with $V\ne TM$.
\end{itemize}

It should be emphasized that the CHSW models,  based on the standard
embedding $V\cong TM$, are a tiny fraction of the heterotic Calabi-Yau 
compactifications.
More generally, a theorem of Donaldson, Uhlenbeck and Yau
relates supersymmetric solutions of the Yang-Mills equations
to stable holomorphic vector bundles $V$ over $M$.
Many such bundles exist which are not in any way related to $TM$.
In this case, the Bianchi identity (\ref{hetbianchi}) becomes non-trivial.
Instead of solving it exactly, one can argue that if one solves
(\ref{hetbianchi}) in cohomology, then one can extend
the solution to all orders in an expansion in the inverse radius of $M$
\cite{Witten:1986kg}.

These more general models are of great interest because they allow for
more general phenomenology.  Instead of GUTs based on $E_6$, which
contain many unobserved particles per generation, one can construct
$SO(10)$ and $SU(5)$ models.  The technology involved in constructing
such bundles is quite sophisticated; some state of the art
constructions appear in \citeasnoun{Donagi:2004ub} and references therein.

One can then go on to compute couplings in the \EFT\ %
at the compactification scale.  Perhaps the most characteristic feature
of weakly coupled
heterotic models is a universal relation between the four dimensional
Planck scale, the string scale, and the gauge coupling, which
follows because all interactions are derived from
the same closed string diagram.  At tree level, this relation is
\begin{equation}\label{eq:hetcouple}
M_{Planck,4}^2 \sim \frac{M_{KK}^2}{g_{YM}^{8/3}} .
\end{equation}
Since the observed gauge couplings are order one, this clearly requires
the extra dimensions to be very small.  Actually, if we put in the constants, 
this relation leads to
a well known problem, as discussed in \citeasnoun{Witten:1996mz} and references
there: if we take a plausible grand unified coupling
$g_{YM}^2\sim 1/25$, one finds $M_{KK}\sim 10^{18}\GeV$ which is far too
large for the GUT scale.  Various solutions to this problem have been 
suggested, such as large one-loop corrections to \rfn{hetcouple}
\cite{Kaplunovsky:1987rp}.

Perhaps the most interesting of these proposed solutions is in the
so-called ``heterotic M theory''
\cite{Horava:1995qa,Witten:1996mz}.
Arguments from superstring duality 
suggest that the strong coupling limit of the ten-dimensional 
$E_8\times E_8$
heterotic string is eleven-dimensional M theory compactified on
an interval; the two ends of the interval provide ten-dimensional
``end of the world branes'' each carrying an $E_8$ super Yang-Mills
theory.  In this theory,  while much of the previous discussion
still applies, the relation \rfn{hetcouple} is drastically
modified.

Finally, there are ``non-geometric'' heterotic string constructions,
based on world-sheet conformal field theory, such as \cite{Kawai:1986va,
Kawai:1986ah,Antoniadis:1986rn,Narain:1986qm}.  In some
cases these can be argued to be equivalent to geometric constructions
\cite{Gepner:1987vz}.  

\subsubsection{Brane models}
\label{sss:branemodels}

Following the same logic for type II theories leads to ${\cal N}=2$
supersymmetric theories.  Until the mid 1990s, the only known way to
obtain ${\cal N}=1$ supersymmetry from type II models was through
``stringy'' compactifications on asymmetric orbifolds
\cite{Narain:1986qm}.  A powerful theorem of \citeasnoun{Dixon:1987yp}
demonstrated that this would never yield the Standard Model, and
effectively ended the subject of type II phenomenology for 8 years.

After the discovery of Dirichlet branes \cite{Polchinski:1995mt} this
lore was significantly revised, and quasi-realistic compactifications
can also arise in both type IIa and type IIb theories.  A recent
review with many references appears in
\citeasnoun{Blumenhagen:2005mu}.  Since flux compactifications are
presently most developed in this case, we need to discuss it in some
detail.

Dirichlet branes provide a new origin for non-abelian gauge symmetries 
\cite{Witten:1995im}.  Furthermore, intersecting branes (or branes
with world-volume fluxes) can localize chiral
matter representations at their intersection locus \cite{Berkooz:1996km}.
And finally, an appropriate choice of D-branes can preserve some but
not all of the supersymmetry present in a type II compactification.
Thus, type II strings on Calabi-Yau manifolds, with appropriate
intersecting brane configurations, can give rise to chiral $\CN=1$
supersymmetric low energy \EFT's.

There are three general classes of type II $\CN=1$ brane
compactifications on Calabi-Yau manifolds:
\begin{itemize}
\item IIa orientifold compactifications with D6 branes.
\item IIb orientifold compactifications with D7 and D3 branes.
\item Generalized type I compactifications; in other words
IIb orientifold compactifications with D9 and D5 branes.
\end{itemize}
After the choice of Calabi-Yau, a particular compactification is
specified by a choice of orientifold projection
\cite{Gimon:1996rq}, and a choice of how
the various Dirichlet branes are embedded in space-time.
Each of the branes involved is ``space-filling,'' meaning that
they fill all four Minkowski dimensions; the remaining spatial dimensions
($p-3$ for a D$p$-brane) must embed in a supersymmetric cycle of the
compactification manifold (to be
further discussed in \rfs{con}).  Finally, since Dirichlet branes 
carry Yang-Mills connections, just as in the heterotic construction
one must postulate the background values for these fields.  The
nature of this last choice depends on the class of model; it is almost
trivial for IIa (where supersymmetry conditions require a flat
connection on the internal space, in the simplest cases), and 
for the generalized type I model with D9 branes
one uses essentially the same vector bundles as in the heterotic
construction, while in the IIb model with D7 and D3 branes, the number
of choices here are intermediate between these extremes.

In a full analysis, a central role is played by the so-called tadpole
conditions.  We will go into more detail about one of these (the D$3$
tadpole condition) later.  These conditions have more than one
physical interpretation.  In a closed string language, they express the
condition that the total charge on the compactification manifold,
including Dirichlet branes, orientifold planes and all other sources,
must vanish, generalizing the Gauss' law constraint that the total
electric charge in a closed universe must vanish.  In an open string
language, they are related to anomalies, and generalize the condition
(\ref{hetbianchi}) related to anomaly cancellation in heterotic strings.
In any case, a large part of the general problem of finding and
classifying brane models, is to list the possible supersymmetric
branes, and then to find all combinations of such branes which solve
the tadpole conditions.

The collection of all of these choices (orientifold, Dirichlet branes
and vector bundles on branes) is directly analogous to and generalizes
the choice of vector bundle in heterotic string compactification.  In
some cases, such as the relation between heterotic $SO(32)$ and type I
compactification, there is a precise relation between the two sets of
constructions, using superstring duality.  There are also clear
relations between the IIa and both IIb brane constructions, based
on T-duality and mirror symmetry between Calabi-Yau manifolds.  

The predictions of the generic brane model are rather different from
the heterotic models.    Much of this is because the relation between
the fundamental scale and the gauge coupling, analogous to
\rfn{hetcouple}, takes the form
$$
g_{YM}^2 = \frac{g_s l_s^{p-3}}{\Vol X} ,
$$
where $l_s$ is the string length and 
$\Vol X$ is the volume of the cycle wrapped by the
particular branes under consideration.  Since
{\it a priori,} volumes of cycles
have no direct relation to the total volume of the compactification
manifold, one can have many more possible
scenarios for the fundamental scales in these theories,
including large extra dimension models.  
A related idea is that the branes responsible for the 
observed (standard model) degrees of freedom can be localized to
a small subregion of the compactification manifold, allowing its
energy scales to be influenced by ``warping''
\cite{Randall:1999ee}.

Even if one has small extra dimensions, coupling unification is
generally not expected in brane models.  This is because the different
gauge groups typically arise from stacks of branes wrapping different
cycles, with different volumes, so the couplings have no reason to be
equal.

To conclude this overview, let us mention two more classes of
compactification which can be thought of as strong coupling limits of
the brane constructions, and share many of their general properties.
First, there are compactifications of M theory on manifolds of $G_2$
holonomy; these are related to IIa compactification with D6-branes by
following the general rules of IIa--M theory duality.  Second, there
are F theory compactifications on elliptically fibered fourfolds; these
can be obtained as small volume limits of M theory on  
Calabi-Yau fourfolds (where one shrinks the volume of the elliptic fiber),
and are simply an alternative description of 
IIb compactification with D7 and D3 branes (and a dilaton which varies
over the internal dimensions).
Both of these more general classes have duality relations with the
heterotic string constructions, so that (in a still
only partially understood sense) all of the $\CN=1$ compactifications
are connected via superstring dualities, supporting the general idea
that all are describing vacuum states of a single all-encompassing theory.

\subsection{Moduli fields} 
\label{ss:modulifields}

In making any of the string compactifications we just described, in
order to solve the Einstein equations, we must choose a Ricci-flat
metric $g_{ij}$ on the compactification manifold $M$.  Now, given such
a metric, it will always be the case that the metric $\lambda g_{ij}$
obtained by an overall constant rescaling is also Ricci-flat, because
the Ricci tensor transforms homogeneously under a scale
transformation.  Thus, Ricci-flat metrics are never unique, but always
come in families with at least one parameter.  

Now, the ``universal'' parameter space describing all such metrics, is
by definition the moduli space of Ricci-flat metrics.  This is a
manifold, possibly with singularities, which we denote as $\CM$.
It can be described in terms of coordinate patches; let us denote a
set of local coordinates as $t^\alpha$ with $1\le \alpha\le n$.

What is the physics of this?  In general treatments of Kaluza-Klein
reduction, one decomposes the $D$-dimensional equations of motion
as a sum of terms, say for a massless scalar field as $\phi$
$$
0 = (\eta^{\mu\nu} \partial_\mu \partial_\nu + \Delta_M) \phi ,
$$
where $\Delta_M$ is the scalar Laplacian on $M$.
One then writes the higher dimensional field $\phi$ as a sum over
eigenfunctions $f_k$ of $\Delta_M$,
\begin{equation}\label{eq:KKdecomp}
\phi(x,y) = \sum_k \phi_k(x) f_k(y) .
\end{equation}
Substituting, one finds that the eigenvalues $\lambda_k$ become
the masses squared of an infinite set of fields,
the ``Kaluza-Klein modes.''

Doing the same in the presence of moduli, we
might consider the parameters $t^i$ as undetermined, and write
$$
\int_{M\times\IR^{3,1}} d^Dx\ \sqrt{G}R^{(D)}
 \left[~G({\vec t}) + \delta G~\right] ,
$$
where $G(\vec t)$ is an explicit parameterized family of Ricci flat
metrics on $M$, and $\delta G$ are the small fluctuations around it.
We then expand $\delta G$ in eigenfunctions, to find the
spectrum of the resulting \EFT.

Within the space of all variations $\delta G$, there is a finite
subspace which corresponds to varying the moduli,
$$
\delta G_\alpha \equiv \frac{\partial G(\vec t)}{\partial t^\alpha} .
$$
One can check that these variations form a basis for the space of
solutions of the linearized Einstein equation on $M$.
Decomposing $\delta G(x,y)$ in analogy to \rfn{KKdecomp}, one
finds that each independent variation gives rise to a 
massless field in the four-dimensional theory.
Conversely, the parameters $t^\alpha$ can be regarded as
the expectation values of these massless fields.

Now, since we get a valid compactification for any particular
choice of Ricci-flat metric, locality demands that we should be able
to vary this choice at different points in four dimensional space-time.
By general principles, such a local variation must be described by
a field.  The situation is analogous to that of a spontaneously
broken symmetry.  By locality, we can choose the symmetry breaking
parameter to vary in space-time, and if the parameter was continuous
it will lead to a massless field, a Goldstone mode.

However, 
there is a crucial difference between
the two situations.  The origin of the Goldstone mode in symmetry breaking
implies that the physics of any constant configuration of this
field must be the same (since all are related by a symmetry).  On the
other hand, moduli can exist {\it without a symmetry}.  In this case,
physics can and usually will depend on their values.  Thus, one finds a
parameterized family of {\it physically distinct vacua}, the moduli
space ${\CM}$, 
connected by simply varying massless fields.

While this situation is almost never encountered in real world physics, this
is not because it is logically inconsistent.  Rather, it is because in
the absence of symmetry, there is no reason the effective potential
should not depend on all of the fields.  Thus, even if we were to find a
family of vacua at some early stage of our analysis, in practice the
vacuum degeneracy would always be broken by corrections at some later
stage.

A well known loophole to this statement is provided by supersymmetric
quantum field theories.  Due to nonrenormalization theorems, such
moduli spaces often persist to all orders in perturbation theory or
even beyond.  These theories manifest different low energy physics at
distinct points in ${\CM}$, and thus provide a theoretical
example of the phenomenon we are discussing here.

Conversely, one might argue that, given that supersymmetry is broken in
the real world, any moduli we find at this early stage will be lifted
after supersymmetry breaking.  We will come back to this idea later,
once we have more of the picture.  We will eventually argue
that while this is true, in models with low energy supersymmetry breaking,
it is more promising to consider stabilization of many of the moduli 
${\it above}~$ the scale of supersymmetry breaking.
However, this is a good illustration of the idea that it is
acceptable to have moduli at an early stage in the analysis, if they are
lifted by corrections to the potential at some lower energy scale.
A useful term for these is ``pseudo-moduli'' \cite{Intriligator:2006dd}.

Finally, 
whether or not the moduli play an
important role in observable physics, they are very important in
understanding the configuration space of string theory.  In
particular, in many of the explicit constructions we discussed above,
as well as in the explicit non-geometric constructions we briefly
mentioned, one finds that apparently different constructions in fact
lead to vacua which differ only in the values of moduli, and thus one
can be turned into another by varying moduli.  In this situation,
there need be no direct relation between the number of constructions,
and the final number of vacua after moduli stabilization.

In early exploratory work, this point was not fully appreciated.  As a
relevant example, in \citeasnoun{Lerche:1986cx}, the number of lattice
compactifications was estimated to be $10^{1500}$.  Thus already this work
raised the possibility that the number of string vacua might be very
large.  However, these were very simple vacua, either with unbroken
supersymmetry, or else with uncontrolled supersymmetry breaking.
At the time, it was generally thought that the number of
quasi-realistic vacua would be much smaller.  An argument to this
effect was that since moduli were not stabilized in these models, it
might be (as is now thought to be the case) that this large number of
compactifications were simply special points contained in a far
smaller number of connected moduli spaces of vacua.  Then, in similar
quasi-realistic models with broken supersymmetry and an effective
potential, the number of actual vacua would be expected to be
comparable to the (perhaps small) number of these connected moduli
spaces.

Such debates could not be resolved at that time.  To make convincing
statements about the number and distribution of vacua, one needs to
understand the effective potential and moduli stabilization.

\subsection{Calabi-Yau manifolds and moduli spaces}

While our main concern is moduli spaces of Ricci flat metrics,
we should first give the reader some examples of
Calabi-Yau threefolds, so the discussion is not completely abstract.
Let us describe the simplest known constructions, as discussed in
\citeasnoun{Green:1987mn}.

\subsubsection{Examples}
The simplest example to picture mentally
is the ``blown-up $T^6/\IZ_3$ orbifold.''
We start with a six-torus with a flat metric and volume $V$, chosen 
to respect a discrete $\IZ_3$ symmetry.  To be more precise, we
take three complex coordinates $z^1$, $z^2$, and $z^3$, and define
the torus by the identifications 
$$
z^i \cong z^i+1 \cong z^i + e^{2\pi i/3} .
$$
We then identify all sets of points related by the symmetry
$$
z^i \rightarrow e^{2\pi i/3} z^i \ \ {\rm for\ all}\ i.
$$

While this is not a manifold (it has singularities at 
the $27$ fixed points of
the group action), one can still it for string
compactification.  One can also modify it to get a smooth Calabi-Yau,
the ``blown-up orbifold.''
This process introduces topology at each of the fixed
points; as it turns out, a two-cycle and a four-cycle.  Thus, the
final result is a smooth Calabi-Yau with second Betti number $b^2=\dim
H^2(M,\IR)=27$.  One can also show that the third Betti number 
$b^{3}=2$.\\[0.1in]

A second simple example is the ``quintic hypersurface'' in $\IP^4$.
This is the space of solutions of a complex equation of degree five
in five variables $z_i$, such as
\begin{equation}\label{eq:fermatquintic}
z_1^5 + z_2^5 + z_3^5 + z_4^5 + z_5^5  = 0 ,
\end{equation}
where the variables are interpreted as coordinates on complex
projective space, {\it i.e.} we count the vectors
 $(z_1,z_2,z_3,z_4,z_5)$
and
 $(\lambda z_1,\lambda z_2,\lambda z_3,\lambda z_4,\lambda z_5)$
as representing the same point, for any complex $\lambda\ne 0$.
One can show that the Euler character
$\chi=-200$ for this manifold by elementary topological arguments
(\cite{Green:1987mn}, vol. II, 15.8).  With a bit more work, one finds all
the Betti numbers, $b^0=b^2=b^4=b^6=1$, and $b^3=204$.
We omit this here, instead computing $b^3$ by other means in the next
subsection.

The main point we take from these examples
is that it is easy to find Calabi-Yau threefolds with Betti
numbers in the range 20--300; indeed, 
as we will see later in \rfsss{cystat}, 
this is true of most known Calabi-Yau threefolds.

\subsubsection{Moduli space - general properties}

The geometry of a moduli space of Calabi-Yau manifolds as they appear
in string theory has been nicely described
in \citeasnoun{Candelas:1989js} 
(see also \citeasnoun{Seiberg:1988pf, Strominger:1990pd}).
Locally, it takes a product form
\begin{equation}
{\cal M} = {\cal M}_{C} \times {\cal M}_K
\end{equation}
where the first factor is associated with the complex structure
deformations of $M$ and the second is associated with the K\"ahler
deformations of $M$, complexified by the B-field moduli.  

These two factors enter into physical string compactifications
in rather different ways.  At the final level of the effective ${\cal N}=1$
theory, the most direct sign of this is that the tree
level gauge couplings are
controlled by a subset of the moduli:
\begin{itemize}
\item ${\cal M}_K$ for IIb compactifications;
\item ${\cal M}_{C}$ for IIa compactifications.
\end{itemize}

The main results 
we need for the discussion in this section, are the relations
between the Betti numbers of the Calabi-Yau manifold $M$, and
the dimensions of these moduli spaces:
\begin{equation}\label{eq:dimcymod}
b_2 = \dim {\cal M}_K; \qquad
b_3 = 2\dim {\cal M}_C + 2 .
\end{equation}
The first relation follows from Yau's theorem, and is not hard to explain
intuitively.  Since the Ricci flatness condition is a second order PDE,
at a linearized level, it reduces to the condition that a deformation of
a Ricci flat metric must be a harmonic form.  The K\"ahler moduli space
parameterizes deformations which come from deforming the K\"ahler form,
and thus its dimension is the same as that of the space of harmonic
two-forms, which by Hodge's theorem is $b_2$.
The second relation can be understood similarly by relating the
remaining metric deformations to harmonic three-forms, 
given a bit more complex geometry.

Mathematically, one can understand these moduli spaces in great
detail, and in principle exactly compute many of the quantities which
enter into the flux potential we will discuss shortly.  Without going
into the details of this, let us at least look at an example,
the complex structure moduli space of the quintic hypersurfaces we just
discussed.  

The starting point is the observation that we do not need to take
the precise equation \rfn{fermatquintic}, to get a Calabi-Yau manifold.
In fact, a generic equation of degree five in the variables,
\begin{equation}\label{eq:generalquintic}
f(z) \equiv \sum_{1\le i,j,k,l,m\le 5} c^{ijklm} z_i z_j z_k z_l z_m = 0 ,
\end{equation}
can be used.  This equation contains
$5\cdot 6\cdot 7\cdot 8\cdot 9/5!= 126$ 
adjustable coefficients, denoted $c^{ijklm}$, and varying these
produces Calabi-Yau manifolds with different complex structures.
To be precise, there is some redundancy at this point.  
One can make linear redefinitions $z_i \rightarrow g_i^j z_j$
using an arbitrary $5\times 5$ matrix $g_i^j$, to absorb $25$ of the
coefficients.
This leaves $101$ undetermined coefficients, so $\dim {\cal M}_C=101$.
By \rfn{dimcymod}, this implies that the Betti number $b^3=2\cdot 101+2=204$.

One can continue along these lines, defining the meaning of a
``generic equation,'' and taking into account the redundancy we just
mentioned, to get a precise definition of the $101$-dimensional moduli
space ${\cal M}_C$ for the quintic, and results for the moduli space
metric, periods and other data we will call on in \rfs{con} and \rfs{stat}.
Similar results can be obtained for more or less any Calabi-Yau moduli space,
and many examples can be found in the literature on mirror symmetry.
While the techniques are rather intricate, it seems fair to say that
at present this is one of the better understood elements of the theory.

\subsection{Flux compactification: Qualitative overview}
\label{ss:fluxcompact}

Each of the weakly coupled limits of string/M theory has certain
preferred ``generalized gauge fields,'' which are sourced by the elementary
branes.  For example, all closed string theories (type II and heterotic)
contain the ``universal Neveu-Schwarz (NS) 2-form potential'' $B_{ij}$
or $B_2$ (we will use numerical subscripts to indicate the degree
of a form).
Just as a one-form Maxwell potential can minimally couple to a point
particle, this two-form potential minimally couples to the 
fundamental string world sheet.  At least in a quadratic (free field)
approximation, the space-time action for the $B$ field is a direct
generalization of the Maxwell action; we define a field strength
$$
H_3 = dB_2,
$$
in terms of which the action is
$$
S = \int d^{10}x \sqrt{g}\left(R - (H_3)_{ijk} (H_3)^{ijk}
 + \ldots \right) ,
$$
leading to an equation of motion
$$
\partial^i (H_3)_{ijk} = \delta_{jk} + \ldots
$$
where $\delta_{jk}$ is a source term localized on the world-sheets
of the fundamental strings.

The analogy with Maxwell theory goes very far.  For example, recall
that some microscopic definitions of Maxwell theory contain magnetic
monopoles, particles which are surrounded by a two-sphere on which the
total magnetic flux $g=\int F_2$ is non-vanishing.  This magnetic
monopole charge must satisfy the Dirac quantization condition, $e\cdot
g=2\pi$ (in units $\hbar=1$).  So too, closed string theories contain
five-branes, which are magnetically charged.  A five-brane in ten
space-time dimensions can be surrounded by a three-sphere, on which
the total generalized magnetic flux $\int H_3$ is non-vanishing.
Again, it must be quantized, in units of the inverse electric charge
\cite{Nepomechie:1984wu,Teitelboim:1985yc}.

Besides the NS two-form, the type II string theories also contain
generalized gauge fields which are sourced by the Dirichlet branes,
denoted $C_{p+1}$ with $p=0,2,4,6$ for the IIa theory, and
$C_{p+1}$ with $p=1,3,5$ for the IIb.  We denote their
respective field strengths $F^{(p+2)}$; these are not all independent
but satisfy the general ``self-duality'' condition
$$
*F_{p+2} = F_{10-p-2} + {\rm non-linear\ terms}.
$$
To complete the catalog, 
the type I theory has $C_{2}$ (as it has a D-string),
while M theory has a three-form potential $C_{3}$, which
couples to the supermembrane.

We are now ready to discuss flux compactification.  The general idea
makes sense for any higher dimensional theory containing a $p+1$ form
gauge field for any $p$.  Let us denote its field strength as
$F_{p+2}$.  

Now, suppose we compactify on a manifold with a non-trivial
$p+2$ cycle $\Sigma$; more precisely the homology group $H_{p+2}(M)$ should 
be non-trivial, and $\Sigma$ should be a non-trivial element of homology.
In this case, we can consider a configuration with a non-zero
${\it flux}$ of the field strength, defined by the condition
\begin{equation}
\int_{\Sigma} F_{p+2} = n \ne 0
\end{equation} 
As a simple illustration -- not directly realized in string theory --
we might imagine starting with Maxwell's theory in six dimensions, and
compactifying on $M=S^2$.  In this case, $H_2(S^2,\IZ)\cong\IZ$, and
we can take a generator $\sigma$ to be the $S^2$ itself.  Thus, we
are claiming that there exists a field configuration whose
magnetic flux integrated over $S^2$ is non-zero.  Indeed there is;
we can see this by considering the field of an ordinary magnetic
monopole at the origin of $\IR^3$, and restricting attention to an
$S^2$ at constant radius $R$, to obtain the field strength
$$
B_{\theta\phi} = g \sin\theta~d\theta~ d\phi .
$$
While this solves Maxwell's equations in three dimensions by construction,
one can easily check that
such a magnetic field actually solves Maxwell's equations restricted to the 
$S^2$.
Thus, this is a candidate background field configuration for compactification
on $S^2$.

Note that we have defined a flux which threads a nontrivial cycle in
the extra dimensions, with no charged source on the $S^2$.
The monopole is just a pictorial device with which to construct it.
Appealing to the monopole also allows us to call on Dirac's argument,
to see that quantum mechanical consistency requires the flux $n$ to
be integrally quantized (in suitable units).

The same construction applies for any $p$.  Furthermore, if we have
a larger homology group, we can turn on an independent flux for each
basis element $\sigma_i$ of $H_{p+2}(M,\IR)$, 
\begin{equation}
\int_{\sigma_i} F_{p+2} = n_i ,
\end{equation} 
where $1\le i\le \dim H_{p+2}(M,\IR)\equiv b_{p+2}$, the $p+2$'th 
Betti number of the manifold $M$.  In the case $p=0$ of Maxwell theory,
one can see that any vector of integers $n_i$ is a possible field
configuration, by appealing to the mathematics of vector bundles
(these numbers define the first Chern class of the $U(1)$ bundle).
Equally precise statements for $p>0$, or for the case in which the
homology includes torsion, are in the process of being formulated
\cite{Moore:2003vf}.

Now, in Maxwell's theory and its generalizations, 
turning on a field strength results in a potential energy
proportional to ${\bf B}^2$, the square of the magnetic field.  
Of course, the presence of nontrivial ${\bf E}$ or ${\bf B}$ in our
observed four dimensions would imply
spontaneous breaking of Lorentz symmetry.  By contrast, in our case,
we can turn on magnetic fluxes in the extra dimensions ${\it without}$
directly breaking 4d Lorentz invariance.  
However, there will still be an energetic cost, now proportional
to $F^2$, the square of the flux.

Now, the key point is that because the fluxes are threading cycles in
the compact geometry, this energetic cost will depend on the precise
choice of metric on $M$.  In other words, it will generate a potential
on the moduli space ${\cal M}$.  If this potential is sufficiently
generic, then minimizing it will fix the metric moduli.

In principle, this potential can be computed by
starting from the standard Maxwell lagrangian
coupled to a curved metric.  One finds for the potential energy
\begin{eqnarray}
\label{eq:emenergy}
V &=& \int_M d^Dy\ \sqrt{G} G^{ij} G^{kl} (F_2)_{ik} (F_2)_{jl} \\
 &=& \int_M F_2 \wedge (*F_2) \nonumber
\end{eqnarray}
where $G$ is the metric on $M$.  The second version, 
in differential form notation and where $*$ denotes the $D$-dimensional
Hodge star, applies for any $F_{p+2}$ with the replacement
$2 \to p+2$; here the metric enters in the 
definition of $*$.

Now, if we substitute for $G$ the family of Ricci flat metrics
$G(\vec t)$ introduced in \rfss{modulifields}, and do the integrals,
we will get an explicit expression for $V(t)$, which we can minimize.
This is the definition of the flux potential; we now have the technical
problem of computing it.  

At first, it is not clear that this can be done at all; indeed we
cannot even get started as no closed form expression is known for any
Ricci flat metric on a compact Calabi-Yau manifold.  
In principle the computations could be done numerically, but working with
solutions of six dimensional nonlinear PDE's is not very easy either,
and this approach is in its infancy 
\cite{Headrick:2005ch,Douglas:2006hz}.
Fortunately, by building
on many mathematical and physical works, we now have an approach which
leads to a complete analytical solution of this problem, as we will
discuss in \rfs{con}.

\subsubsection{Freund-Rubin compactification}
\label{sss:freundrubin}
There are other Kaluza-Klein theories in which the technical problem
of computing \rfn{emenergy} is far simpler, and was solved well before
string theory became a popular candidate for a unified theory.  While
these theories are too simple to be quasi-realistic, they serve as
good illustrations.  Let us consider one here, leaving 
more detailed discussion to \rfs{con}.

After it was realized that Nature employs non-abelian gauge fields, the
earliest idea of 5d unification was augmented.  Instead, theorists
considered $4+D$ dimensional theories, with $D$ of the dimensions
compactified on a space with a non-abelian isometry group.
This leads to a gauge group which contains
the isometry group.  One can even find seven dimensional manifolds
for which this is the Standard Model gauge group, although chiral fermions
remain a problem for this idea.

In any case, the problem of explaining how and why the extra $D$
dimensions were stabilized in whatever configuration was required to
obtain 4d physics was first studied in this context.  A collection of
historically significant articles on Kaluza-Klein theory, with modern
commentary, can be found in \citeasnoun{Appelquist:1987nr}.

The first serious attempt we know of to explain the ``spontaneous
compactification'' of extra dimensions appeared in the work 
\citeasnoun{Cremmer:1976ir}.
This work was extended by \citeasnoun{Luciani:1977zv} and reached
more or less modern form with the seminal paper of
\citeasnoun{Freund:1980xh}.

Let us see how the Freund-Rubin mechanism works by again
considering six dimensions, now in Einstein-Maxwell theory. 
Compactifying to 4d on an $S^2$, they found that
inclusion of a magnetic flux piercing the $S^2$ allows one to stabilize
the sphere.
One can understand this result by a
scaling argument; such arguments are discussed in modern contexts 
in \citeasnoun{Giddings:2003zw,Silverstein:2004id,Kachru:2006em}. 
We start with a 6d Einstein/Yang-Mills Lagrangian
\begin{equation}
S = \int d^6x \sqrt{-G_6} \left({\cal R}_6 - |F_2|^2 \right) ,
\end{equation}
where all dimensions are made up with powers of the fundamental scale $M_6$.
We then consider reduction to 4d on a sphere of radius $R$:
\begin{equation}
ds^2 = \eta_{\mu\nu}dx^{\mu}dx^{\nu} + R^2 g_{mn}(y) dy^m dy^n
\end{equation}
where $m,n$ run over the two extra dimensions, and $g$ is the 
metric on a two-sphere of unit radius.
Let us then thread the $S^2$ with $N$ units of
$F_{2}$ flux
\begin{equation}
\int_{S^2} F_2 = N~.
\end{equation}

In the 4d description, $R(x)$ should be viewed as a field.  
Naively reducing, we will find a Lagrangian where $R^2(x)$ multiplies
the curvature tensor ${\cal R}_{4}$.  To disentangle the graviton
kinetic term from the kinetic term for the modulus $R(x)$, we should
perform a Weyl rescaling.  After this rescaling, we find an
effective potential with two sources.

First, before Weyl rescaling, the 6d Einstein term would contribute
to the action a term
proportional to the integrated curvature of the $S^2$, {\it i.e.}
the Euler character.
In particular, positive curvature makes a negative contribution
to the potential.  After the rescaling, this term is no longer constant;
instead it scales like $- R^{-4}$.  

In addition,
the $N$ units of magnetic flux through the $S^2$ contribute
the positive energy described in \rfn{emenergy}.
By flux quantization, $F_2 \sim {N\over R^2}$, while the integral
over the internal space contributes a factor of $R^2$.  Therefore,
the flux potential scales like $N^2/R^6$.  The dimensions are
made up by powers of the fundamental scale, in terms of which
the flux quantum is defined. 

Thus, the total potential as a function of $R(x)$ takes
the form
\begin{equation}
\label{FRpot}
V(R) \sim {N^2\over R^6} - {1\over R^4}~.
\end{equation}
It is not hard to see that this function has minima where $R \sim N$.
So with moderately large flux, one can achieve radii which are large
in fundamental units, and curvatures which are small, justifying 
the use of supergravity.

Strictly speaking, the original Freund-Rubin vacua are ${\it not}$
compactifications which yield lower-dimensional \EFT's.
The vacuum energy following from
(\ref{FRpot}) is negative, and gives rise to a 4d curvature scale
comparable to the curvature of the $S^2$.  Therefore, 4d effective
field theory is not obviously a valid approximation scheme in these
vacua.  
It is plausible, however, that by using more complicated manifolds and
tuning parameters to decrease the
4d vacuum energy, one could use the Freund-Rubin idea to obtain quasi-realistic
vacua \cite{Acharya:2003ii}.

\subsection{A solution of the cosmological constant problem}
\label{ss:fluxsol}

Einstein's equations, relating the curvature of space-time to
the stress-energy of matter, admit an additional term on the
right hand side,
$$
g_{ij} = 8\pi G_N \left( T_{ij} + \Lambda g_{ij} \right) .
$$
The additional ``cosmological constant'' term $\Lambda$ is a
Lorentz-invariant vacuum energy and is believed to be generically
present in any theory of quantum gravity; it receives corrections
from known quantum effects (somewhat analogous to the Casimir effect)
at least of order $(100 \GeV)^4$.  On the
other hand, elementary considerations in cosmology show that any
value $|\Lambda| > 1 (\eV)^4$ or so is in violent contradiction with
observation.  More recently, there is observational evidence of
various types (the acceleration of the expansion of the universe;
and detailed properties of the cosmic microwave background spectrum)
which can be well fit by assuming $\Lambda \sim 10^{-10} (\eV)^4 > 0$.

This is by now a very long-standing question with which most readers
will have some familiarity; we refer to
\cite{Weinberg:1988cp,Carroll:2000fy,Padmanabhan:2002ji,Nobbenhuis:2004wn}
for introductory overviews, and the history of the problem.  A very
recent discussion from the same point of view we take here is in
\citeasnoun{Polchinski:2006gy}, along with general arguments against many
of the other approaches which have been taken towards the problem.

One approach which cannot be ruled out on general grounds is
to simply assert that the fundamental theory contains the small
observed parameter $\Lambda$.  More precisely, the large quantum
contributions $\Lambda_q$ from all types of virtual particles (known
and unknown), are almost precisely compensated by an adjustable ``bare
cosmological constant'' $\Lambda_{bare}\sim -\Lambda_q$.  However,
besides being unesthetic, this idea {\it cannot} be directly realized
in string/M theory, which is formulated without free parameters.
Rather, to address this problem, we must find out how to compute the
vacuum energy, and argue that the energy of the vacuum we observe
takes this small value.

Taken purely as a problem in microscopic physics, the
prospects for accurately computing such a small vacuum energy seem
very distant; furthermore it seems very unlikely that any vacuum would
exhibit the remarkable cancellations between the large known
contributions to the vacuum energy, and unknown contributions, required
to make such an argument.  But here is precisely the loophole; what is
indeed very unlikely for a single vacuum, can be a likely property
for {\it one} out of a large set of vacua.

Simple toy models in which this is the case were proposed in
\citeasnoun{Abbott:1984qf,Banks:1991mb}.  
The general idea is to postulate a potential
with a large number of roughly equally spaced minima, for example
$$
V(\phi) = a \phi - b \sin \phi + \Lambda_q,
$$
(with $b>0$) whose minima $\phi=(2n+1/2)\pi$ 
have energies $\Lambda_n=\Lambda_q + 2\pi a n - b$.
Thus, if $a$ is very small, then no matter what value $\Lambda_q$ takes,
at least one minimum will realize the small observed $\Lambda$.
By postulating more terms, one can even avoid having to postulate a
small number $a$ \cite{Banks:1991mb}.  For example, consider
$$
V(\phi) = E_1 \sin (a_1\phi+b_1) +  E_2 \sin (a_2\phi+b_2) + \Lambda_q .
$$
The reader may enjoy checking that if the ratio $a_1/a_2$ is
irrational, any $\Lambda$ (within the range $\Lambda_q\pm E_1\pm E_2$)
can be approximated to any desired accuracy.

While in \EFT\ terms these models might be
reasonable, the actual potentials arising from string/M theory
compactification appear not to take this form.  Besides verifying this in
explicit expressions, there
is a conceptual problem.  This is that these models assume that the
field $\phi$ can take extremely large values, of order $1/\Lambda$.
However, taking a modulus $\phi$ to be so large, implies that the
Calabi-Yau manifold is decompactifying, or undergoing some similar
limit.  In such a limit, the potential can be computed more
directly and does not take the required form.

However, there is another mechanism for producing potentials with
large numbers of minima, introduced by \citeasnoun{Bousso:2000xa},
which relies on having a very large number of degrees of freedom.\footnote{
Amazingly, this idea was anticipated in \citeasnoun{Sakharov:1984ir}.
}
Let us consider a toy model
of flux compactification, where there are $N$ different p-cycles in the
compact geometry that may be threaded by the flux of some
p-form field $F$ 
\begin{equation}
\int_{\Sigma^i} F = n^i,~i=1,\cdots,N 
\end{equation}

Let us also assume a simple {\it ad hoc} cutoff on the allowed
values of the fluxes, of the form
\begin{equation}
\label{eq:cutoff}
\sum_i n_i^2 \leq L
\end{equation}
where $L$ is some maximal amount of flux.  One can view \rfn{cutoff} as
a toy model of the more complicated tadpole conditions that arise in real
string models.  Finally, let us assume that for each value of the fluxes
$F$, the resulting potential function for moduli admits a minimum with
energy
\begin{equation}
\label{eq:eform}
V \sim -V_0+ c_i n_i^2
\end{equation}
Here we take the $c_i$ to be distinct order one constants, while
$-V_0$ is assumed to be a large fixed negative energy density,
for example representing the quantum contribution to the cosmological
constant $\Lambda_q$ we discussed earlier.

A striking fact follows from these simple assumptions and
known facts about compactification topologies --
the number of vacua will be huge.
As we discussed, typical 
Calabi-Yau threefolds have betti numbers of order $100$.  
For a space with $N=100$ and $L=100$, 
\rfn{cutoff} indicates that the number of vacua can be
approximated by the volume of a sphere of radius 
${\sqrt L}\sim 10$ in a 100-dimensional space.
This is roughly ${\pi^{50}\over 50!} \times 10^{100} \sim 10^{60}$.
Here it was important that $\sqrt{L}$ is much larger than the unit of
flux quantization, so that one can approximate the number of possible
flux choices by computing the volume in flux space of the region defined
by \rfn{cutoff}.

We will justify this toy model in \rfs{con} and
\rfs{stat}, by showing that the real
counting of flux vacua -- while differing in details -- is similar,
that classes of vacua with significantly larger $N$ and $L$ exist,
and that for a sufficiently
large fraction of the flux choices, one has
controlled approximations in which the vacua exist.

Now, let us consider the cosmological constant in this model.
In a vacuum with flux vector $n_i$, this will be given by \rfn{eform}.
Thinking of the quadratic term in \rfn{eform} as defining a 
squared distance from
the origin in $N$-dimensional space, we see that
to have a small vacuum energy of order $\epsilon$, 
a flux vacuum must sit within a shell bounded by two ellipsoids,
of radius $\sqrt{V_0}$ and $\sqrt{V_0 + \epsilon}$.
(These are ellipsoids because the $c_i$ are not all equal, though we
assume them all to be ${\cal O}(1)$).)

As argued in \citeasnoun{Bousso:2000xa}, if the number of vacua
exceeds $\sim 10^{120}$, quite plausibly this shell is populated by
some choices of flux.  The simplest argument for this is that, given
that the fluxes $n_i$ and the postulated coefficients $c_i$ are independent, 
we can expect the values of the vacuum energy  attained by \rfn{eform}
to be roughly uniformly distributed over scales much smaller than the 
coefficients $c_i$.  Thus, in a set of $N_{vac}$ vacua, we might expect
the typical ``level spacing'' to be $1/N_{vac}$, and that a vacuum energy
of order $1/N_{vac}$ will be realized by at least one vacuum.  We will
make more precise arguments of this type in \rfs{stat}.

Thus, this toy model can explain why
at least some vacua exist with the very small 
cosmological constant consistent with observation.
Furthermore, the essential features of the toy model, namely
a very large number of vacua with widely distributed vacuum energies,
distinguished by the values of hundreds of microscopic parameters,
seem to be shared by more realistic stringy flux compactifications.
This energy landscape of potential string vacua has been called
the ``string landscape'' \cite{Susskind:2003kw}; a detailed
and very clear qualitative discussion can be found in 
\citeasnoun{Susskind:2005bd}. 

\subsubsection{Anthropic selection}
\label{sss:anthropic}

Suppose we grant that a few, rare vacua will have small $\Lambda$. 
How do we go on to explain why we find ourselves in such a vacuum?

There have been many attempts to find dynamical mechanisms which strongly
prefer the small $\Lambda$ solutions (including relaxation mechanisms
\cite{Brown:1987dd,Brown:1988kg,Feng:2000if,Steinhardt:2006bf}, peaking of the
wave-function of the Universe \cite{Hawking:1984hk,Coleman:1988tj},
and many others \cite{Rubakov:1999aq,Itzhaki:2006re}). 
Each seems in some sense problematic: for instance the relaxation
proposals typically suffer
from an ``empty universe problem,'' whereby they favor completely
empty vacuum solutions with small $\Lambda$, incompatible with our
cosmological history.
For a much more detailed
discussion of the problems that bedevil dynamical selection
mechanisms, and possible loopholes, see \citeasnoun{Polchinski:2006gy}.

Absent a dynamical selection mechanism, one can try to use so-called
``anthropic'' criteria to explain why we inhabit a vacuum with small
$\Lambda$.  A better term for the generally accepted criteria
of this type is {\it selection effect}; in other words we take the
evident fact that the circumstances of a particular experiment or
observation might skew the distribution of observed outcomes, and
apply it to the problem at hand of why we observe ``our universe''
instead of another.

In practice, what is meant by this, is an argument which focuses on
some macroscopic property of our universe, and derives constraints on
microphysics by requiring the microphysics to be consistent with the
macroscopic phenomenon.  The most famous example, and the one we will
cite, is the argument of \citeasnoun{Weinberg:1987dv} that the
existence of structure (i.e. galaxies) puts stringent bounds on the
magnitude of the cosmological term.\footnote{While we cannot fully
review the history here, important earlier works along these lines include
\cite{Banks:1984tw,Banks:1984cw,Linde:1984ir,Barrow}.}
For positive cosmological
constant, the bound arises due to two competing effects.  On the one
hand, primordial density perturbations gravitate and attract each
other; in a universe with vanishing $\Lambda$, the Jeans instability
will eventually lead to the formation of large scale structure.
On the other hand, a large $\Lambda$ and the consequent accelerated
expansion, lead to such rapid dilution of matter that structure
can never form.  The requirement that structure has time to form
before the accelerated expansion takes over, leads to a bound on
$\Lambda$ within an order of magnitude or two of the observed value.

Weinberg's logic suggests that if structure is required for observers,
and if there are many possible vacua with different values of $\Lambda$,
then selection effects will explain why any given observer sees an
atypically small value of $\Lambda$.  It is also important that
since the scales of microphysics differ so drastically from the scale
of the required $\Lambda$, one can expect the distribution of vacua
in $\Lambda$-space to be reasonably flat over the anthropically
acceptable range.  Hence, all else being equal, one should expect to
find a value of $\Lambda$ close to the upper bound compatible with
structure formation.

This seems to be true in our universe.  It is notable that
Weinberg's bound was published well before the detection of dark
energy, and the amount of dark energy is very close to his estimate of
the maximal value compatible with the formation of structure.  There
is some controversy about how close our universe is to the bound;
see {\it e.g.} \citeasnoun{Loeb:2006en}.

An important question which must be asked before accepting this logic
is whether early cosmology actually can give rise to all of these
possible vacua.  According to the currently favored picture, based on
the theory of eternal inflation, this is so: our universe sits in a
``multiverse'' with many different inflating regions, corresponding to
the different de Sitter critical points in the set of vacua.  We
discuss this further, and the arguments that different flux vacua can
be connected by physical processes in string theory, in
\rfss{measure}.

Anthropic arguments are typically met with suspicion for the simple
reason that it seems hard to convincingly and quantitatively verify a
physical theory based on such arguments.  There are many reasons
(discussed in e.g. \citeasnoun{Banks:2003es,Arkani-Hamed:2005yv,
Wilczek:2005aj}) to believe that more traditional, dynamical
explanations will be required to resolve some of the outstanding
mysteries of physics.  But unless another convincing solution to the
cosmological constant problem is found, this argument is likely to stay
with us.

\subsection{Other physical consequences}
\label{ss:fluxphys}

While explaining the cosmological constant would be an important achievement,
the resolution provided by the landscape of flux vacua does not suggest
immediate tests.
 
Happily, the study of flux vacua also leads to new and testable
string models of particle physics and cosmology; indeed this has driven
much of the interest in the subject.
Over the past few years, these studies 
have motivated new models of TeV scale particle physics
\cite{Arkani-Hamed:2004fb,Giudice:2004tc,%
Arkani-Hamed:2005yv,Giudice:2006sn}, new models
of inflation \cite{Kachru:2003sx,Silverstein:2003hf,Chen:2004gc,
Chen:2005ad}, which can have testable signatures
via cosmic strings \cite{Sarangi:2002yt,Jones:2003da,Copeland:2003bj,
Dvali:2003zj}
or non-gaussianities of the spectrum of density perturbations
\cite{Alishahiha:2004eh,Babich:2004gb,Chen:2006nt}, and new 
testable proposals
for the mediation of supersymmetry breaking \cite{Choi:2005ge}.  
Some of these models have large or warped extra dimensions and manifest
very low scale quantum gravity, raising the exciting possibility of
producing black holes at future colliders (for a recent review, see
\cite{Landsberg:2006mm}).

As things stand, none of these models appear as inevitable top-down
consequences of string theory; rather they are special choices made
out of a wide range of possibilities in the fundamental theory,
proposed in part because they have clearly identifiable or at least
unusual characteristic signatures.  The hope is that an influx of new
data on TeV scale particle physics and inflationary cosmology over
the next decade will help select between these ideas, or else
suggest new, testable proposals.

Here, we briefly describe, at a very qualitative level, three areas
where studies of flux vacua may be directly relevant to phenomenological
questions in string theory.  We will need to call upon some basic
results from the theory of supersymmetry breaking, so we review this
first.

\subsubsection{Overview of spontaneous supersymmetry breaking}
\label{sss:susybr}

By spontaneous supersymmetry breaking, we mean that although the
vacuum breaks supersymmetry, at some high energy scale dynamics is
described by an $\CN=1$ supergravity \EFT.
As discussed in \citeasnoun{Wess:1992cp}, 
the effective potential in such a theory is determined by the
superpotential $W$, a holomorphic ``function'' of the chiral fields,\footnote{
To be more precise, the superpotential in supergravity 
is a section of a holomorphic line bundle.}
and the K\"ahler potential $K$, a real-valued function of these fields.
Let us denote the chiral fields as $\phi^i$, then the effective
potential takes the form
\begin{equation}\label{eq:sugraV}
V =  e^{K/M_{Pl,4}^2} \left(\sum_i |F_i|^2 - 3\frac{|W|^2}{M_{Pl,4}^2}\right) 
 + \half \sum_\alpha D_\alpha^2 
\end{equation}
where 
$F_i=DW/D\phi^i\equiv\partial W/\partial\phi^i +
{1\over M_{Pl,4}^2} (\partial K/\partial\phi^i)W$ 
are the so-called F terms, associated to
chiral fields, while the D terms 
$D_\alpha \sim \sum \phi^\dagger t_\alpha\phi$ 
are associated to generators of the gauge group.

While any solution of $\partial V/\partial\phi^i=0$ with $\partial^2
V/\partial\phi^i\partial\phi^j$ positive definite is a metastable
vacuum, spontaneous supersymmetry breaking is characterized by
non-zero values for some of $F_i$ and $D_\alpha$.  The most basic
consequence of this is that
the gravitino gains a mass $m_{3/2} = e^{K/2 M_{Pl,4}^2} |W|/M_{Pl,4}^{2}$ 
by
a super-Higgs mechanism.
If we assume that the cosmological constant $V\sim 0$, $|W|$
and thus $m_{3/2}$ are determined by \rfn{sugraV} in terms of $|F|^2$
and $|D|^2$.

One should be careful to distinguish the various energy scales which
appear in supersymmetry breaking; we will define 
\begin{equation} \label{eq:defMsusy}
M_{susy}^4= \sum_i |F_i|^2  + \half \sum_\alpha D_\alpha^2 ,
\end{equation}
the energy scale associated to supersymmetry breaking in the microscopic
theory.  Note that many authors use a different definition in which
$M_{susy} \sim m_{3/2}$.

A third set of energy scales are set by the MSSM soft supersymmetry
breaking terms, such as masses for the gauginos and scalars.  These
are more model dependent, but usually fall into two general classes.
The first class are effects which lead to masses proportional to $F/M_P$.
One fairly generic source of scalar masses is coupling through irrelevant
terms in the K\"ahler potential, with the general structure
\begin{equation}
\int d^2\theta d^2\bar\theta {c^2\over M_P^2}
 X^\dagger X (\phi^i)^\dagger \phi^i~. 
\end{equation}
where $X$ should be thought of as a chiral field containing the
largest F-term.
Such terms are not forbidden by any symmetry (unless
$\phi_i$ is a Goldstone boson, but compactification moduli in
general are not, with the notable exception of axions). 
If they are present and $F_X\ne 0$, the field $\phi^i$ obtains a mass
\begin{equation}\label{eq:gravmed}
m_i \sim \frac{c~F_X}{M_P} .
\end{equation}
Similarly, if $X$ appears in the gauge-coupling function $f$ for some
gauge group $G$, i.e. in the term 
\begin{equation}
\int d^4x ~d^2\theta f(X) Tr(W_{\alpha}W^{\alpha}), 
\end{equation}
then $F_X \neq 0$ gives rise to a gaugino mass as well.
Another generic source of masses for charged particles
is anomaly mediation \cite{Randall:1998uk,Giudice:1998xp}; 
in particular this produces gaugino masses
$
m_{1/2} \sim b~g_{YM}^2~m_{3/2}
$
where $b$ is a beta function coefficient.

Once one has a soft supersymmetry breaking mass term for charged fields $X$, 
one can get further supersymmetry breaking effects suppressed not by $M_P$
but by $M_X$, the mass of the $X$ fields.  One loop diagrams 
of $X$ particles will produce soft mass terms for charged gauginos, and
at higher loop order soft masses for all charged particles.  
This is known as gauge mediation; for references and
a review see \citeasnoun{Giudice:1998bp}.  

Let us now consider a quasi-realistic model which solves the hierarchy
problem by spontaneous supersymmetry breaking, in the sense that 
the small number $M_{EW}/M_P$ comes out of some dynamics.  In general, one
expects the EFT to be a sum of several parts; a supersymmetric
Standard Model (SSM); a sector responsible for supersymmetry breaking;
possibly a messenger sector which couples supersymmetry breaking to
the SSM; and finally sectors which are irrelevant for this discussion.
After integrating out all non-SSM fields, one obtains an SSM with
soft supersymmetry breaking terms, such as masses for the gauginos and
scalars.  The first test of the model is that
the resulting potential leads to electroweak symmetry breaking.
This depends on two general
features of the supersymmetric extension.  Recall that an SSM must
have at least two Higgs doublets; let us suppose there are two, $H_u$
and $H_d$.  First, the Higgs doublets can get a supersymmetric mass term
$$
W = \ldots + \mu H_u H_d ,
$$
the so-called $\mu$ term.  This must be small, $\mu\sim M_{EW}$.
In addition, one must get soft supersymmetry breaking masses coupling
the two Higgs doublets (the $b$ term), also of order $M_{EW}$.
Of course, there are many, many more constraints to be satisfied by
a realistic model, most notably on flavor changing processes.

Now, one can distinguish two broad classes of supersymmetry breaking
models.  In the first class, generally known as 
``gravity mediated'' models,
supersymmetry breaking is mediated only by
effects which are suppressed by powers of $M_P$.  In this case,
to obtain soft masses at $M_{EW}$,
the natural expectation is $F \sim (10^{11} \GeV)^2$, the so-called
intermediate scale, and $m_{3/2} \sim M_{EW}$.

On the other hand, if the SSM soft masses come from gauge mediation,
the sparticle masses are suppressed by powers of $M_X$, not $M_P$.
Therefore, depending on $M_X$, 
one can get by with a much smaller F breaking, perhaps as low
as $F \sim (100 ~\TeV)^2$.  Such a gauge mediated model will have 
$m_{3/2} << M_{EW}$ as well as many other differences from the first
class.

This more or less covers the basic facts we will need for this review;
further discussion can be found in many good reviews such as 
\citeasnoun{Martin:1997ns,Giudice:1998ic,Luty:2005sn}

\subsubsection{The moduli problem}
\label{sss:moduliproblem}

As we discussed, string compactifications preserving 4d ${\cal N}=1$
supersymmetry typically come with dozens or hundreds of moduli fields.
These are chiral multiplets $\phi_i$ which have gravitational strength
couplings and a flat potential to all orders in perturbation theory.

In general, all scalar fields, including the moduli, will receive mass
after supersymmetry breaking.  In a few cases, namely the moduli which
control the Standard Model (or grand unified) gauge couplings, we can
put a lower bound on this mass, around $100 \GeV$, just by considering
quantum effects in the Standard Model.  As pointed out in
\citeasnoun{Banks:2001qc}, this precludes any observable variation of
the fine structure constant (and the other SM gauge couplings), even
on cosmological time scales.  Thus, while the underlying theory
allowed for such time variation in principle, it is inconsistent with
known properties of our vacuum combined with the effective potential
hypothesis.  This is perhaps the simplest testable prediction of 
string/M theory for which contrary evidence has ever been reported 
\cite{Murphy:2003hw}; the present status is discussed in 
\citeasnoun{Uzan:2002vq,Uzan:2004qr}.\footnote{
String/M theory also leads to many testable predictions for which we have no
reason at present to expect contrary evidence, for example CPT conservation, 
unitarity bounds in high energy scattering, and so forth.}

More generally, one can estimate moduli masses in particular models
of supersymmetry breaking.  Using \rfn{gravmed}, and assuming
a gravity mediated model with $F \sim (10^{11} \GeV)^2$, we find
a rough upper bound
$$
m_{moduli} \sim 1 ~\TeV.
$$
As for gauge mediated models, since
moduli which do not couple directly to the Standard Model
also get their leading masses from \rfn{gravmed}, their masses
will be far lower, even down to the eV range.

In general,
such particles would not be subject to direct detection, because
of their very weak (nonrenormalizable) coupling to the Standard Model.
One can construct optimistic scenarios (including the large-extra
dimensions scenario \cite{Arkani-Hamed:1998rs} and models of gauge
mediation with very low SUSY breaking scale) in which
the moduli 
masses come down to $10^{-3}$ eV, so that one could hope to detect
such fields in fifth-force experiments studying the strength of gravity
at short distances \cite{Dimopoulos:1996kp}.  

However, granting the usual discussion of inflationary cosmology,
scalar fields masses less than about $100 \TeV$ will cause significant
phenomenological problems.  In particular, they cause a Polonyi
problem -- the oscillations of such scalars about the minima of their
potential, in a cosmological setting, will overclose the universe
\cite{Banks:1993en, deCarlos:1993jw}.
One way of understanding this is as follows.  The equation of motion
for the moduli $\phi_i$ in the early universe is
\begin{equation} \label{eq:cosmod}
\label{gradflow}
\ddot \phi_i + 3H \dot \phi_i = -{\partial V \over \partial \phi_i}
\end{equation}
Taking a single modulus $\phi$ and
Taylor expanding $V(\phi) \sim m^2\phi^2 + \cdots$, we see that the
``Hubble friction'' (the second term on the LHS) dominates over
the restoring force from the potential energy if $H >> m$. 
Via the relation $H^2 M_P^2 \sim V_{tot}$ (where $V_{tot}$ is the total
energy density of the Universe), we see that in the early Universe,
Hubble friction will dominate for light fields.  This means that until
$H$ decreases to $H < m$, such fields will ${\it not}$ reach the minima
of their potential; they will be trapped by Hubble friction at some
random point.\footnote{This discussion is oversimplified, since $V$ itself
may receive significant thermal corrections.  The point then is that
for a modulus field, the  
true minimum only appears, typically very far away ($\sim M_P$ in 
field space) from the
finite-temperature minimum, after $H$ drops below the typical scales
of the zero-temperature potential.} 
After the Hubble constant drops below $m$, the energy
density in these fields can dominate the Universe, 
leading to a variety of possible problems (overclosure, 
modifications of the successful predictions of BBN, etc.).

There are scenarios with moduli in this mass range where the cosmological
problems are avoided, say by a stage of low-scale inflation
\cite{Randall:1994fr,Dvali:1995mj}.
In general, however, this 
suggests that the idea that string moduli get their mass
through radiative corrections after SUSY breaking is disfavored.
Rather, we should look for the physics of moduli
stabilization at higher energy scales.

As we discussed, we can expect the flux potential to produce moduli
masses.  A first naive estimate for the energy scale of this potential
would be $M_{KK}$, since this is an effect of compactification.
However, this neglects the fact that the unit of quantization of the
fluxes is set by the fundamental scales, in string theory the string
scale.  This discussion is somewhat model dependent \cite{Kachru:2006em}; 
in \rfs{con} we
will discuss the case of IIb flux vacua.
In general in such vacua, the complex structure moduli which get a mass
from fluxes end up with a typical mass $M_F$ 
\begin{equation} \label{eq:defMF}
M_F \sim {\alpha^\prime \over R^3}
\end{equation}
which satisfies $M_F << M_{KK}$ at 
moderately large radius, but is still well above the supersymmetry
breaking scale $M_{SUSY}$ (and far above the even smaller gravitino mass
$M_{3/2} \sim M_{SUSY}^2/M_P$) 
for low energy supersymmetric models with moderate
$R$.

In a top-down discussion, one must check that these masses squared
are positive, {\it i.e.} metastability.  Actually, one can argue that
this is generic in supersymmetric theories, in the following sense.
The mass matrix $V''$ following from \rfn{sugraV} takes the form
\begin{equation}\label{eq:susymassrel}
M_{boson}^2 = M_{fermion}(M_{fermion} - \alpha M_{3/2})
\end{equation}
for some order one $\alpha$.\footnote{The reader should not confuse this
with formulae governing the sparticle partners of standard model
excitations, for which the soft-breaking terms give the dominant effects,
and can lead to splittings much larger than this estimate in various 
scenarios.}  
Thus, any bosonic partner to a fermion with $|M_{fermion}| >> M_{3/2}$
will automatically have positive mass squared.  Since for moduli,
$M_{fermion}\sim M_F >> M_{3/2}$, this entire subsector will be stable.

\paragraph{Quintessence}
There is one cosmological situation in which the existence of
an extremely light, weakly coupled scalar field has been proposed as a
feature instead of a bug.  One of the standard alternatives to a cosmological
constant in explaining the observed dark energy is ``quintessence'' 
\cite{Peebles:1987ek}.  In this picture, a slowly rolling scalar field dominates
the potential energy of the Universe, in a sort of late-time analogue
of early Universe inflation (though perhaps lasting only for ${\cal O}(1)$
e-foldings).  In light of our discussion, it is natural to ask, can
string theory give rise to natural candidates for quintessence?

The observational constraints on time variation of coupling constants
make it necessary to keep the relevant scalar very weakly coupled to
observable physics.  The necessary mass scale of the scalar, comparable
to the Hubble constant today, means also that this scalar must ${\it not}$
receive the standard $\sim M_{SUSY}^2/M_P$ mass from SUSY breaking.
The most natural candidate is therefore a pseudo Nambu-Goldstone boson,
and in string theory, these arise plentifully as axions.  An axion with
weak enough couplings and whose shift symmetry is broken by dynamics
at very low energies, could conceivably serve as quintessence; it has been
Hubble damped on the side of its potential until the present epoch, and may 
just be beginning its descent.  

The prospects for this scenario are described in the recent paper
\citeasnoun{Svrcek:2006hf}.  While it is plausible,
the scenario suffers from all of the tuning problems of the
cosmological constant scenario, and an additional ``why now'' problem
-- there is no good reason for the field to become undamped only in
the recent past.  

\subsubsection{The scale of supersymmetry breaking}
\label{sss:susyscale}

Perhaps the most fundamental question in string phenomenology is the
scale of supersymmetry breaking.  As we discussed, there are many hints
in the present data which point towards TeV scale supersymmetry.
It has long been thought that low energy supersymmetry would also
follow from a top-down point of view.  One of the simplest arguments to
this effect uses the concept of ``naturalness,'' according to which an
\EFT\ can contain a small dimensionful parameter,
only if it gains additional symmetry upon taking the parameter to
zero.  This is not true of the Higgs mass in the Standard Model,
but can be true for supersymmetric theories.

On the other hand, the solution we just described for the cosmological
constant problem seems to have little to do with this sense of
naturalness; indeed it may seem in violent conflict with
it.\footnote{Actually, in \rfs{stat}, we will show that
the c.c. is uniformly distributed in some classes of vacua, 
in a way consistent with traditional 
naturalness.  In our opinion, anthropic arguments are {\it not}
in contradiction with naturalness, rather they presuppose some idea of
naturalness.}  Should this not give us pause?  How do we know that
the small ratio $M_{EW}^2/M_{Pl,4}^2 \sim 10^{-33}$ might not have a
similar explanation?  Following this line of thought, one might seek
an anthropic explanation for the hierarchy, as has been done in
several works \cite{Arkani-Hamed:2004fb,Giudice:2004tc,Arkani-Hamed:2004yi,
Arkani-Hamed:2005yv,Giudice:2006sn}.  
While interesting, the possibility of
such an explanation would not bear directly on whether the underlying
theory has low energy supersymmetry, unless we could argue that our
existence required this property (or was incompatible with it), which
seems implausible.

However, there is a different set of arguments, which we will now
describe, that low energy supersymmetry, and the naturalness
principle which suggested it, may not be the prediction of string/M
theory.  Rather, one should define a concept of {\it stringy naturalness,}
based on the actual distribution of vacua of string/M theory, which
leads to a rather different intuition about fine-tuning problems. 

The starting point is the growing evidence that there are many classes
of string vacua with SUSY breaking at such high scales that it does
not solve the hierarchy problem, starting with early works such as
\cite{Scherk:1979zr,Alvarez-Gaume:1986jb,Seiberg:1986by,
Dixon:1986iz}, and more recently models with stabilized moduli such as
\cite{Silverstein:2001xn,Saltman:2004jh}.  Despite their disadvantage
in not solving the hierarchy problem, might such vacua
``entropically'' overwhelm the vacua with low-scale breaking?
Let us illustrate how one can study this question
with the following top-down approach to deriving
the expected scale of supersymmetry breaking, 
along the lines advocated in \citeasnoun{Susskind:2004uv,Douglas:2004qg}.

Suppose for a moment that one has classified the full set of
superstring vacua, obtaining some set with elements labelled by $i$.
Suppose, for sake of argument, that we had a complete model of how
early cosmology produces these vacua, which leads to the claim that
``the probability to observe vacuum $i$ is $P(i)$.''  Finally, suppose
that the SUSY breaking scale in the $i$'th vacuum is $F_i$.  Then, we could
use this data to define a probability distribution over SUSY breaking
scales.  Similarly, if we have more observables for each vacuum
we could define a joint distribution over all of them.

To make a simple discussion, let us focus on two parameters, the
supersymmetry breaking scale $F$, and the scale of electroweak 
symmetry breaking
$M_{EW}\sim 100 \GeV$.
Now, imagine that we are about
to do an experiment which will detect superpartners if 
$F < F_{exp} = 1 \TeV$.  Then,
the probability with which we expect to discover supersymmetry would be
\begin{equation}\label{eq:probsusy}
P_{susy} = \sum_{F_i \le F_{exp},M_{EW,i}=100 \GeV} P(i).
\end{equation}
If this probability were high, we would have derived
a top-down prediction of TeV scale supersymmetry.

But, from what we know about string theory, do we know it will be
high?  Might it instead be low, so that the discovery of TeV scale
supersymmetry would in some sense be evidence against string theory?

Before continuing, we hasten to say that any top-down ``prediction''
of this sort would only be as good as the assumptions which went in,
and furthermore would probably rely on drastic simplifications of the
full problem.  We fully expect that the problem of testing string
theory, like any other theory, will involve the same sort of
interaction between theory and experiment which characterizes all
successful science.  Our goal here is to make an idealization of this
complex problem, in order to gain understanding.  We will discuss the
assumptions and simplifications which would go into any such
prediction in \rfs{stat}, here let us continue in order to make the
point that {\it given what we know now, TeV scale supersymmetry is
not an inevitable prediction of string theory.}

First, given our ignorance of the correct probabilities $P(i)$, a
simple hypothesis to get a feel for the problem is to set the
probability $P(i) = 1/N$ for each of the $N$ vacua in the landscape.
In other words, we assume that the more string/M theory vacua realize a
certain property, the more likely we are to observe it.  In
\rfs{stat}, we will critically examine this hypothesis, and see how
far one can go without making any appeal to probabilities at this
point, but let us grant it for the moment.

Now, let us rephrase the usual argument from naturalness in this
language.  We focus attention on the subset of string/M theory vacua
which, while realizing all the other properties of the Standard model,
may have a different value for the electroweak scale $M_{EW}$.  Since this is
quadratically renormalized, in the absence of any other mechanism,
we expect that the fraction of theories with $M_{EW}<M_{EW,max}$ should
be roughly
$$\frac{M_{EW,max}^2}{M_{cutoff}^2} \sim 10^{-30}
$$
taking $M_{cutoff} \sim M_{GUT}$ for definiteness.  While small,
of course given enough vacua, we will find vacua in which the
hierarchy is a result of fine tuning.\footnote{
See \citeasnoun{Silverstein:2004sh} for a toy model of how fluxes
can do this.}

Let us now grant that we have some subset of the string theory vacua
in which the Higgs mass is determined by supersymmetry
breaking in the general way we discussed in \rfsss{susybr}.
More specifically, let us grant that the Higgs mass satisfies a 
relation like \rfn{gravmed}, with $F_X \sim 10^{11} \GeV$, 
the intermediate scale, so that we can expect to see supersymmetry
at the TeV scale.  Then, while there are further conditions to check,
one might expect an order one fraction of these models to work.

Now, the naturalness argument is the claim that, since most of the TeV
scale supersymmetry vacua work (fit the data), while only $10^{-30}$
of the fine tuned vacua work, we should expect to live in a universe
with TeV scale supersymmetry, or at least prefer this
alternative to the fine-tuned models.

Of course, we arranged our discussion in order to make the essential
gap in this argument completely evident.  It is that, even though the
fraction of fine-tuned vacua which work is relatively small, if their
number is large, we might find in the end that far more of these vacua
work than the supersymmetric vacua.  Given our hypothesis, string
theory would then predict that we should {\it not} see supersymmetry at
the TeV scale.

Is this what we expect or not?  Before taking a position, one should
realize that the additional structures being postulated in the
supersymmetric models -- the scale of susy breaking, a solution to the
$\mu$ problem, a mediation mechanism in which FCNC and the other
problems of generic supersymmetric models are solved, and so on --
each come with a definite cost, not in terms of some subjective
measure of the complexity or beauty of the theory, but in terms of
what fraction of the actual string/M theory vacua contain these
features.  Is this cost greater than $10^{-30}$ or not?

We will describe results bearing on this question in \rfs{stat},
but we are still far from having sufficient knowledge of
the set of string vacua to make convincing statements.  But given toy
models which incorporate some of the detailed structure of flux vacua
in computable limits, there are already interesting suggestions
about how the computation might turn out
\cite{Susskind:2004uv,Douglas:2004kp,Dine:2004is, Silverstein:2004sh}

What is already clear, is that claims that string theory
naturally `prefers' low energy supersymmetry are, as yet, far from
being justified.  Indeed, the simplest toy models suggest the opposite.
It would be very important to improve our understanding of this point.

\subsubsection{Early universe cosmology}
\label{sss:cosmo}

There is substantial and growing evidence for a period of early
universe inflation to explain the homogeneity, isotropy, and large-scale
structure of our Hubble volume \cite{Linde:2005ht,Spergel:2006hy}. 
However, obtaining a reasonable model of inflation in string theory
requires a detailed understanding of moduli stabilization
\cite{Kachru:2003sx}.  The reason is as follows.

Recall the equations \rfn{cosmod}, which govern
the dynamics of scalar fields $\phi_i$ evolving in a scalar potential $V$
in an FRW cosmology with Hubble constant $H$.
To obtain slow-roll inflation, one requires that the
gradient in the potential is not very steep, so that this
dynamics reduces to gradient flow.
This leads to the standard slow-roll conditions, that
\begin{equation}
\label{eq:slowroll}
\epsilon = {M_P^2\over 2}({V^\prime\over V})^2 << 1,~~
\eta = M_P^2 {V^{''} \over V} << 1~.
\end{equation}
Primes denote derivatives with respect to the inflaton $\phi$; the first
condition provides for a period of accelerated
expansion, while the second guarantees that this period of accelerated
expansion will last sufficiently long to solve the horizon and
flatness problems (for reviews of basic facts about inflationary 
cosmology, see \citeasnoun{Lyth:1998xn,Linde:2005ht}).  

Now,
in string models at moderately weak coupling $g_s \to 0$ and/or large volume 
$R >> l_s$ for the internal dimensions
(so that 10d supergravity can be used),
one ${\it knows}$ that \rfn{slowroll} is not true.
${\it All}$ known sources of potential energy fall rapidly to zero
as $R^{-n}$ with $n\geq 6$
\cite{Giddings:2003zw,Silverstein:2004id,Kachru:2006em}. 
Similarly all known sources vanish as a positive power
of $g_s$.  These power laws are far too fast to allow slow-roll inflation,
or late-time acceleration for that matter \cite{Hellerman:2001yi,
Fischler:2001yj}. 

Thus, to achieve slow-roll inflation, one must either work in a regime
of strong coupling/small radius where it is difficult at present to
compute \cite{Brustein:2002mp}, or else one must find models where the
radii/dilaton and other rapidly rolling moduli have been stabilized by
a computable potential, as we have for the flux vacua.

Let us consider a concrete proposal in this light, that of
\citeasnoun{Dvali:1998pa}.  In these models,
branes and anti-branes (or more generally,
branes which do not preserve the same supersymmetry) are both present,
in different parts of the compact space $M$.
The candidate inflaton is the distance between the branes and 
anti-branes on $M$, the inflationary potential is generated by
interbrane RR and gravitational forces,
while the exit from inflation can occur when the brane
and anti-brane reach a distance $\sim l_s$ from one another, where
the lightest stretched string becomes tachyonic.  This picture is
reminiscent of hybrid inflation \cite{Linde:1993cn}, 
with the tachyon playing
the role of the ``waterfall field'' that causes the exit from
inflation. 
Such brane inflation models were generalized and explored in
\citeasnoun{Burgess:2001fx,Burgess:2001vr,Alexander:2001ks,Dvali:2001fw,
Shiu:2001sy,Garcia-Bellido:2001ky, 
Herdeiro:2001zb,Jones:2002cv,Gomez-Reino:2002fs,Dasgupta:2002ew} 
without addressing
the issue of moduli stabilization; a review appears in
\citeasnoun{Quevedo:2002xw}. 

It was argued in \citeasnoun{Kachru:2003sx} that considering brane inflation
in the absence of moduli stabilization does not make sense; 
the predictions derived from considerations of the open string
potential ignoring the closed string modes, are corrected 
very significantly by inclusion of the closed strings.  The interbrane
potential, for D3 and anti-D3 branes separated by a distance $d$
in $M$, is given by
\begin{equation}
\label{branepot}
V(d) = 2 T_3 \left(1 - {1\over (2\pi)^3} {T_3 \over M_{10}^8 d^4}
\right) 
\end{equation}
where $T_3$ is the brane tension.
Note that $d$ is related to a canonically normalized scalar field via
the relation $\phi = \sqrt{T_3} d$.

Can (\ref{branepot}) plausibly satisfy the slow-roll conditions
\rfn{slowroll}?  There is a 
well known problem.  On a space of radius $R$, using the relation
between $M_{10}$ and the 4d Planck scale $M_P$,  
one can quickly see that
$\eta \sim (R/d)^6 \times {\cal O}(1)$. 
Since one expects $d \leq R$, such models will have trouble
giving rise to slow-roll inflation.  Many clever model building
tricks were postulated to surmount this kind of difficulty in 
the papers cited previously; arguments presented in 
\citeasnoun{Kachru:2003sx} show that ${\it generically}$ the problem persists.

Even having fixed this, perhaps by some fine tuning, there is
a more basic problem.
The correct 4d Einstein frame
potential is not quite (\ref{branepot});
it must undergo a Weyl rescaling to reach 4d Einstein frame, and
this multiplies (\ref{branepot}) by an overall factor of ${1\over R^{12}}$.
Thus regardless of the interbrane potential, 
the system of equations (\ref{gradflow}) will lead to rapid
decompactification!
A similar argument shows that one must prevent
relaxation to $g_s \to 0$.  So achieving
slow-roll inflation requires
stabilizing the radion and the dilaton, a problem which can be solved
in flux compactification.

One still must engineer a flat enough
interbrane potential to satisfy 
\rfn{slowroll}, as generic moduli stabilization  mechanisms 
do not yield sufficiently flat interbrane potentials.  
The state of the art 
is described in \citeasnoun{Baumann:2006th};
while some tuning is involved, construction of quasi-realistic models
seems well within reach in many scenarios.

While we have focused on brane inflation,
similar issues arise in other inflationary models using moduli fields
\cite{Binetruy:1986ss,Banks:1995sw,Blanco-Pillado:2004ns,
Blanco-Pillado:2006he,Greene:2005rn} 
or axions \cite{Freese:1990rb,Adams:1992bn,
Arkani-Hamed:2003wu,Banks:2003sx,Dimopoulos:2005ac,Easther:2005zr}.

\section{QUANTUM GRAVITY, THE EFFECTIVE POTENTIAL AND STABILITY}
\label{sec:eff}

As the subsequent discussion will be quite technical, before going more
deeply into details we should ask more basic questions, such as
\begin{itemize}
\item What are our implicit assumptions?
Can we trust them, and the formalism which they lead to?
\item Might there be {\it a priori} arguments that the type of vacuum
we seek (with stabilized moduli and positive cosmological constant)
does not exist, or is extremely rare?
\item 
Related to this, might there be unknown additional consistency
conditions, which are satisfied by only a few of the vacua?
\end{itemize}

Since as yet we have no fully satisfactory nonperturbative definition
of any string theory or M theory, clearly our discussion cannot start
from first principles; we need to make assumptions about how the
theory works and what constitutes a ``solution'' to proceed.
Thus, our arguments will not be conclusive, but rather are meant
to summarize existing work and suggest new approaches to
addressing these questions.

\subsection{The effective potential}
\label{ss:effpot}

Our point of view, as we explained in the introduction and implicitly
assumed throughout \rfs{qual}, is that the vacuum structure of
string/M theory is determined by an effective potential $V_{eff}$.
This is a function of the many scalar fields which parameterize the
local choices (moduli) determining a particular solution, and whose
value is the exact vacuum energy of that solution.  Granting this, our
problem is to define $V_{eff}$ incorporating all classical and quantum
contributions to the energy, compute it in a controlled way, and find
its local minima.

While this is how all known physical theories work, there are good
reasons not to accept this uncritically in a quantum theory of
gravity, as has been particularly emphasized in
\citeasnoun{Banks:2003es,Dine:2004fw,Banks:2004xh}.
Let us cite a
few of these reasons, and then consider the various candidates we have
for complete definitions of the theory, to try to evaluate them.

We begin by asking whether the concepts which enter into the effective
potential are well defined.  First, there is
no universal way to define energy in a generally covariant theory.
The standard formal definition of energy is the dynamical variable
conjugate to time translation, in the sense of Hamiltonian mechanics,
or in quantum commutation relations.  However, in a generally covariant
theory, time translation invariance is simply an arbitrariness of the
choice of global time coordinate, on which no observable can depend.
The logical conclusion is therefore that the energy, and thus the
effective potential, in any such theory must be identically zero.

Of course, this conclusion is not acceptable in a theory which
can describe conventional non-gravitational physics, as clearly the
concept of energy is sensible and useful in that context.  Formally,
one simple way around it is to consider only asymptotically flat
solutions, which at large distances (in any space-like direction)
approach Minkowski space-time (or its product with the internal
dimensions).  In such a solution, one can define the generators of the
Poincar\'e group purely in terms of the asymptotic fields; in
particular the energy $E$ is related to the term in the metric $g_{00}
\sim -2E/r$ which expresses the Newtonian potential of a source with
mass $E$.  The vacuum solutions we will mostly be interested in
are maximally symmetric dS or AdS spacetimes with cosmological
constant which is very small compared to the other
scales of microscopic physics,
and thus are extremely close to asymptotically flat.
For such solutions, 
the flat space definition of energy should be operationally
adequate (and is indeed the one we use in everyday physics).

Actually, there are major loopholes in the argument we just made,
coming from caveats such as ``in the present epoch,'' and ``extremely
close to asymptotically flat.''  We will discuss these below in
\rfss{semigrav}, with the conclusion that they all rely on some sort
of non-locality in the theory.  While this does not make them
unthinkable, let us postpone this discussion and proceed to discuss the
definition of the effective potential in Minkowski space-time.

We recall the standard definition of the effective potential in a
quantum field theory, for definiteness a theory of a single scalar
field $\phi$.  We first couple $\phi$ to a source $J$, and compute
(say using the functional integral) the partition function $Z(J)$, to
define the generating function of connected Green's functions
$F(J)=\log Z(J)$.  We then set the expectation value $\phi_0$
of $\phi$ by solving the equation
$$
\frac{\partial F}{\partial J} = \phi_0 ,
$$
which formally amounts to a Legendre transform.  The resulting
functional $\Gamma(\phi_0)$, specialized to constant $\phi_0$, is
the effective potential.

In trying to repeat this definition in string theory, we face the
problem that it is not possible to couple a string theory to a local
source, nor to a local current; this was one of the main problems with
the early proposals for using strings to describe hadronic physics.
This led to the general observation that the theory tends not to
provide natural definitions of ``off-shell'' quantities, meaning
quantities defined in terms of space-time histories which are not
solutions.  For example, computations of scattering amplitudes using
the string world-sheet formalism are unambiguous only if all of the
external states are on mass shell.  
This is not
considered a flaw in the theory, as the S-matrix is defined purely in
terms of scattering of on-shell external states.  However, 
the effective potential is an off-shell quantity.

Two general ways around this problem are known.  
The first approach is to do without the coupling to a local source,
instead manipulating the value of $\phi_0$ by adjusting the boundary
conditions.  This is not completely general, but can be satisfactory
in some situations.  For example, if the effective potential is zero,
any constant $\phi_0$ will be a solution, and we can pick a particular
solution by choice of boundary conditions.  More generally, if we know
the effective potential in advance, we can find the solutions of the
\EFT, and pick one by choice of boundary conditions.

This is implicitly what is done in most work on string
compactifications with extended supersymmetry.  For example, in a
family of compactifications to Minkowski space, supersymmetry
guarantees that the effective potential is zero, so there is no
difficulty in adjusting moduli by varying boundary conditions.  
Another class of examples is flux compactifications with ${\cal N}\ge 4$
supersymmetry in anti-de Sitter space.  Again, supersymmetry
determines the effective potential uniquely, so that one can study
solutions with prescribed boundary conditions without detailed string
theoretic computation.  This is used implicitly in many works on the
AdS/CFT correspondence.

While at first this definition seems inadequate for the problem at
hand, in which we want to compute a non-trivial effective potential
which we do not know in advance, one can still try to follow this
route.  One would start with a known extended supersymmetry
background, and then postulate boundary conditions (probably
time-dependent) which, if it were the case that the effective potential 
described a second non-trivial metastable vacuum, would lead to a
solution matching on to this solution in the interior.  
We will say more about this approach in \rfss{semigrav}, and explain
the obstacles to it there.

This brings us to the second approach, which is simply to couple the
string theory to a non-local source.  For example, one can do this
in string field theory, the framework which is most directly analogous
to quantum field theory \cite{Zwiebach:1993cs}.  
Just as QFT can be defined in terms of an operator $\phi(x)$
which creates or destroys a particle at a point in space-time $x$,
here one introduces a string
field operator, call it $\Phi[L]$, which creates or destroys a
string on a one-dimensional loop $L$ in space-time.
One can then introduce a source $J[L]$ for the string field into the
action in the standard way, say as
$$
S = S_0 + \int dL\ \Phi[L] J[L] ,
$$
where the definition of the integral over loops is taken from the
string field theory framework.  One then follows the logic
which led to the field theory definition, to get a string field
theoretic effective potential $\Gamma[\Phi_0]$.

While such a definition might be difficult to use in practical
computations, the
point is to have a precise definition which 
could be used to justify our approximate considerations.  To do this,
the next step would be to identify the light modes in
$\Gamma[\Phi_0]$, and solve for all of the others, to obtain an
effective potential which is a function of a finite number of fields.
To the extent that we could do this, we would have made precise the
intuition that string theory reduces to field theory at long
distances, where the effective potential is a valid concept.

However, there are formidable obstacles to making such a definition
precise.  At present there is very little understanding of string
field theory beyond its perturbative expansion, 
and just as for quantum
field theory, this expansion is only asymptotic \cite{Shenker:1990uf}.
It is also not obvious that all of the nonperturbative effects we
will call upon below are contained in string field theory,
see \citeasnoun{Schnabl:2005gv} for relevant progress on this.
In any case, verifying or refuting the approximate discussion we
will make below would be an important application of a
nonperturbative definition.

For four-dimensional quantum field theory, such a definition is made
by appealing to the renormalization group.  One must find some
asymptotically free UV completion of the theory of interest, and then
find some approximate finite description of the weakly coupled short
distance theory, such as lattice field theory.

While we have no comparable theoretical understanding of string
theory, there is a widely shared intuition that, at least in
considering low energy processes and vacuum structure, string theory
is weakly coupled at short distances (the string scale and below).
This intuition has several sources: first, the extended nature of the
string cuts off interactions at these distances.  Second, asymptotic
supersymmetry makes the leading contributions of massive states to the
effective action cancel.  Finally, other effects of massive states are
suppressed by inverse powers of the fundamental scales.  
Presumably, this intuition justifies matching on to a field theoretic 
description at distances around the string scale, and then following the 
standard RG paradigm.

\subsection{Approximate effective potential}

Let us now grant that the problems we just discussed are only
technical, and consider how we would make a precise definition of the
effective potential we use in this review, namely in weakly coupled
string field theory, taking into account nonperturbative effects in a
semiclassical expansion.

We start with ten-dimensional \EFT, {\it i.e. supergravity}
with $\alpha'$ and $g_s$ corrections.  We then compactify to get a
four-dimensional effective action with massive KK modes, string modes
and the like.  At this level, the discussion is precise.
Even in the presence of fluxes, in many models, the leading results
can be inferred from supersymmetry and considerations of 4d ${\cal N}=2$
supergravity.

We then need to add in semiclassical nonperturbative effects, such as
instantons and wrapped branes.  The basic features of these can already
be seen in supersymmetric field theory.  After much development,
originating in the study of 4d 
supersymmetric QCD \cite{Affleck:1983mk,Shifman:1991dz}, and using
holomorphy and duality arguments,
quite a lot is known about exact superpotentials
in ${\cal N}=1$ field theories \cite{Intriligator:1994jr},
and exact prepotentials
in ${\cal N}=2$ field theories, following
\cite{Seiberg:1994rs}.

In string theory, mirror symmetry relates
exact prepotentials in type IIa Calabi-Yau models and type IIb Calabi-Yau
models, where an infinite (worldsheet) instanton sum in the 
prepotential on one side maps to a completely
classical geometric computation on the other \cite{Candelas:1990rm}.
String duality maps these worldsheet 
instanton sums to spacetime instanton sums, allowing one to recover
nonperturbative effects
from string duality \cite{Kachru:1995wm,Kachru:1995fv}.  This grew
into the realization that one could design stringy configurations of branes
or singularities to give rise to a given low energy field theory,
and compute the instanton sums via string techniques
\cite{Katz:1996fh,Katz:1996th}. 

More generally, holomorphy arguments allow one to classify which 
Euclidean branes, wrapping which topologies, can
contribute to a holomorphic superpotential.  The theory
of D-branes \cite{Polchinski:1995mt} allowed finding
the full list of possible BPS instantons relevant for a variety of ${\cal N} 
\geq 1$ vacua.  A prototypical example of a macroscopic
argument classifying the
branes and topologies which are relevant for instanton effects in
F-theory is \citeasnoun{Witten:1996bn}.  While exact computation of
the superpotential in a general compactification is
still beyond our reach, this does allow for principled estimates
of the leading instanton contributions in many backgrounds.  In the
particular cases where the instanton effect can be re-interpreted in
the low energy effective theory as a dynamical effect in quantum field
theory, even the coefficient can be estimated with some confidence, by
matching to exact field theory results.  In many examples, even this
crude level of understanding suffices to exhibit vacua in
the reliable regime of weak coupling and large volume.

The main issue we now have to address, is that we want to take the
sum of various terms, some inferred directly from supergravity or
world-sheet physics, and others computed (or even inferred) from
nonperturbative effects.  Typically, a solution of $\partial
V_{eff}/\partial\phi^i= 0$ for the full effective potential will not
be a critical point of the various terms which enter into $V_{eff}$,
so these terms will be ambiguous.  But if there are ambiguities, how
can we be sure that we have fixed them for every term in a consistent
way?

Our eventual answer to this question in this review will be to exhibit
examples of solutions in which contributions to the
effective potential with different origins have parametrically different
scales.  Thus, although individual terms may have some ambiguity, a very
weak control over this ambiguity will suffice to prove that the full
effective potential has minima.

We see no reason that such a separation of scales should be needed for
consistency, so this type of argument is not completely satisfactory;
it does not apply to large numbers (perhaps the vast majority) of
solutions.  
However, already
within the limits of this argument, we will find sufficiently many
stabilized vacua to justify the basic claims of \rfs{qual}.

\medskip
Even restricting attention to these solutions, we are still not done.
Another pitfall to guard against in extrapolating results for the
effective potential is the possibility of phase transitions.  This is
especially worrisome for first order transitions, which unlike second
order transitions have no clear signal such as a field or order
parameter becoming massless.  Such transitions are not possible in
global supersymmetry, in which the energy of a supersymmetric vacuum
is always zero; however this is not true after supersymmetry breaking
and in supergravity.  Should we worry about this possibility?

Actually, the rules here are somewhat different from equilibrium
statistical mechanics and field theory, in that sufficiently
long-lived metastable configurations will count as vacua.  However, a
possibility which needs to be considered is that additional
fields, perhaps arising from the Kaluza-Klein modes of dimensional
reduction, or composite fields expressing quantum correlations, might
destabilize our candidate vacua.

The first possibility, that Kaluza-Klein modes destabilize vacua, will
be considered in \rfs{con}.  The basic argument that this generically
does not happen was given in \rfsss{moduliproblem}.

The second possibility is handled by a combination of arguments.  In
most of the \EFT, quantum fluctuations are
controlled by the string coupling, which we have assumed to be small.
Thus, mass shifts for composite fields will be small, so given that
the moduli are all massive, we do not expect phase transitions.  This
argument has the flaw that some subsectors of the theory must be
strongly coupled at low energy (after all we know this is the case for
QCD).  For these sectors, we appeal to existing field theory analyses,
and the assumption that the supersymmetry breaking scale is smaller
than the fundamental scale, so that supergravity effects are a small
correction.

\subsection{Subtleties in semiclassical gravity}
\label{ss:semigrav}

In our discussion so far, we assumed that our local region of the
universe can be well modelled as Minkowski space-time.  
Of course, no matter how slow the
time evolution of the universe, or how small the cosmological
constant, if these are non-zero, at sufficiently large scales the
nature of the solution will be radically different from Minkowski
space-time.  Thus, we might wonder whether even if a solution looks
consistent on cosmological scales, it could be inconsistent as a 
full solution of the theory.

At first this might sound like it could only happen if the underlying
framework were non-local.  However, while string/M theory is believed
to be in some sense non-local at the fundamental (Planck and string)
length scales, in all known formalisms and computations these effects are
either exponentially small at longer distances, or appear to be gauge
artifacts, analogous to the apparent instantaneous force at a distance
one finds in Coulomb gauge.  Thus it is hard to see how they could be
relevant.  Still, some feel that paradoxes involving black hole
evaporation and entropy point to non-locality \cite{Giddings:2006sj}.

Even in a local theory, a solution which is consistent on short time
scales, can in principle be inconsistent on longer time scales, by
developing a singularity with no consistent physical interpretation
or ``resolution.''  Although one often hears the slogan that ``string
theory resolves space-time singularities,'' there are examples of
space-like singularities with no known resolution, nor any
proof that this cannot be done, making this an active field of research.

Now in an ordinary physical theory, one would say that the possibility
of developing a singularity with no consistent interpretation shows
that the theory is not fundamental; rather it should be derived from a
more fundamental theory in which the corresponding solution is not
singular.  Familiar examples include Navier-Stokes and other
phenomenological many-body theories, and of course classical general
relativity.  

In the present context, one might attempt a different interpretation.
If it turned out here that some subset of vacua generically led to
singularities, while another subset did not, it might be
reasonable to exclude the first set of vacua as inconsistent.  
Now it seems strange to us, indeed acausal, to throw out a
solution because of an inconsistency which appears (say) $10^{10^{10}}$ years
in the future.  Still, if such an approach led to interesting claims,
it might be worth pursuing.

Another idea along these lines is that there might be approximately
Minkowski solutions which, while themselves consistent, cannot be
embedded in a solution with a sensible cosmological origin.  This test
seems better as it is consistent with causality.  It could be further
refined by asking not just that the cosmology be theoretically
consistent, but that it agree with observation.  Of course, we will
eventually need to address this issue in the course of testing any given
solution, but we might ask if there are simple arguments that some
solutions cannot be realized cosmologically, or cannot satisfy
the constraints discussed in \rfsss{cosmo}, before going into details.
We know of no results in this direction however.

Let us now come back to a point raised in \rfss{effpot}, and explain the
obstacles to performing thought experiments which prove the existence of
multiple (isolated) vacua of an effective potential 
\cite{Farhi:1989yr,Banks:2000zy}.  
For instance, suppose
the effective potential for a single scalar $\phi$ has two vacua at
$\phi_{\pm}$.  One can make a vacuum bubble interpolating between the two
vacua, whose surface tension we can call $\sigma$.  
Starting from the $\phi_+$ vacuum, suppose one nucleates a bubble of radius $R$
in the $\phi_{-}$ phase.  
The Schwarzschild radius of the bubble is $\sigma (R/M_P)^2$.
So the bubble will be smaller than its Schwarzschild radius unless
$ R > \sigma (R/M_P)^2$, i.e. unless
\begin{equation}
\label{critrad}
R < {M_P^2 \over \sigma}~.
\end{equation}
This is interesting for the following reason.  
A $\phi_+$ experimentalist can only use the bubble to infer the existence
of the $\phi_-$ vacuum and study its properties, if (\ref{critrad}) is
satisfied.
We would expect that the
potential barrier between two typical vacua in a quantum
gravity theory should be
$\sim M_P$, as there is no small parameter to change the
scaling in typical solutions.  Then, one would also find $\sigma \sim M_P^3$, 
and only bubbles smaller than the Planck length would be outside
their Schwarzschild radius!  Of course such bubbles are not a priori
meaningful solutions, and could not be used by an experimentalist to verify the
existence of other vacua.  

This argument is a bit quick, for example because the vacua we will discuss in
\rfs{con} do have small parameters, but the conclusion is largely correct, as
explained further in \cite{Farhi:1989yr,Banks:2000zy}.

\subsection{Tunneling instabilities}
\label{ss:tunnel}

We have argued that in string theory, the effective potentials one infers
from direct computation typically have many minima.  For such a minimum
to be considered a metastable vacuum, its lifetime
$\tau$ should be parametrically long compared to the string time.
We now explain why there are large
numbers of vacua that satisfy this criterion, and even
the more stringent criterion 
$\tau >> \tau_{today} \sim 15$ billion years, using
the theory of the decay of false vacua in field theory
developed in
\citeasnoun{Coleman:1977py,Callan:1977pt,Coleman:1980aw}.

Let us consider a toy model consisting of a single scalar field $\phi$, with
a metastable de Sitter vacuum of height $V_0$ at $\phi_0$, and a 
second Minkowski vacuum at infinity in field space.  This 
can be thought of as modeling the potential
for a volume modulus in a string
compactification, where the second vacuum represents the 
decompactification limit \cite{Kachru:2003aw}.  
Suppose the barrier height separating
the dS vacuum from infinity is $V_1$.

The tension of the bubble wall for the bubble of false vacuum decay is
easily computed to be
\begin{equation}
\label{bubble}
T = \int_{\phi_0}^\infty d\phi \sqrt{2V(\phi)}
\end{equation}
The dominant tunneling process differs depending on whether
$V_0 M_P^2  >> T^2$ or $V_0 M_P^2 << T^2$.

Since we see that $T \sim \sqrt{V_1} \Delta\phi$, this translates into
the question of whether $\Delta \phi << \sqrt{{V_0\over V_1}}$ or
$\Delta \phi >> \sqrt{{V_0 \over V_1}}$.  The former regime is
called the ``thin wall limit'' for obvious reasons.  In this limit,
the analysis of Coleman et al applies.  The tunneling probability,
is given by
\begin{equation}
P = {\rm exp} \left( -{27\pi^2 T^4 \over 2 V_0^3} \right)
\end{equation} 
For dS vacua with small $V_0 << V_1$, the rate is clearly highly
suppressed, easily yielding a lifetime in excess of $10^{10}$ years. 

In the opposite regime of a low but thick potential barrier, 
$V_0 M_P^2 << T^2$, the dominant instanton governing
vacuum decay would instead be the
more enigmatic Hawking-Moss instanton \cite{Hawking:1981fz}.
The physical interpretation of this instanton is unclear; a description
in terms of thermal fluctuations of the $\phi$ field which yields
the same estimate for the rate can be found in \citeasnoun{Linde:2005ht} and
references therein.
The action of this instanton is the difference between the dS entropies
of dS vacua with vacuum energies $V_0$ and $V_1$, resulting in a tunneling rate
\begin{equation}
\label{eq:hm}
P \simeq {\rm exp} \left({-S(\phi_0)}\right) = {\rm exp} \left(- {24\pi^2
\over V_0} \right)~.
\end{equation}
For small $V_0$, this again is completely negligible.  
The formula \rfn{hm} neglects a small multiplicative correction factor
of ${\rm exp}({24\pi^2\over V_1})$ which accounts for the entropy at
the ``top of the hill."  

For $V_0 << V_1$, this factor is not numerically
important, but its presence serves to prove a conceptual point.  Because
of the existence of the Hawking-Moss instanton, ${\it any}$ dS vacuum
which is accessible in the \EFT\ approximation to string
theory, will have a lifetime which is parametrically
short compared to the Poincare recurrence time
of de Sitter space (considered as a thermal system with a number of
degrees of freedom measured by the de Sitter entropy) \cite{Kachru:2003aw}.  

This discussion illustrates how, within the regime of \EFT,
one can find long-lived vacua.  However, a questionwhich already
appeared in \citeasnoun{Bousso:2000xa}, and has not been settled in more
realistic models, is that besides the approximate Minkowski vacua at
infinity we just discussed, there are many other possible endpoints
for the decay of a vacuum, both dS and AdS vacua.  Some of these
tunneling rates have been computed in
\citeasnoun{Kachru:2002gs,Kachru:2003aw,Frey:2003dm, Ceresole:2006iq}, and
generically they are also very small.  However, one might wonder whether
the large degeneracy of possible targets could lead to enough
``accidentally'' low barriers to substantially increase the overall
decay rates.  This might be addressed using the statistical techniques
of \rfs{stat}.

\subsection{Early cosmology and measure factors}
\label{ss:measure}

In any theory with many vacua, one could ask: are some vacua preferred
over others?  A natural answer in the present context is that if so,
it will be for cosmological reasons: perhaps the ``big bang'' provides
a preferred initial condition, or perhaps the subsequent dynamics favors
the production of certain vacua.

This type of question has been studied by cosmologists for many years;
some recent reviews include
\citeasnoun{Guth:2000ka,Linde:2005ht,Tegmark:2004qd,Vilenkin:2006xv}.
At present the subject is highly controversial and thus we are
only going to sketch a few of the basic ideas here.

One general idea is that a theory of quantum gravity will have a
preferred initial condition.  The most famous example is the wave
function of \citeasnoun{Hartle:1983ai}, 
which is defined in terms of
the Euclidean functional integral.  Presumably, time evolving this
wave function and squaring it would lead to a probability distribution
on vacua.  In the present context, this suggests looking for a
natural wave function on moduli space, or on some larger configuration
space of string/M theory.  An idea in this direction appears in
\citeasnoun{Ooguri:2005vr}.

Another idea, more popular in recent times, is that the distribution
of vacua is largely determined by the dynamics of inflation.
Inflation involves an exponential expansion of spatial volume, which
tends to wash away any dependence on initial conditions.  In particular,
many of the standard arguments for inflation in our universe, such as
the explanations of homogeneity, flatness, and the non-observation of
topological defects, rely on this property.
While these standard arguments
do not in themselves bear on the selection of a
particular vacuum, it is widely believed that inflation also
washes away all dependence on the initial conditions relevant for 
vacuum selection (say the choice of compactification manifold,
moduli and fluxes), because of the phenomenon of {\it eternal inflation}
\cite{Vilenkin:1983xq, Linde:1986fc,Linde:1986fd}.  

Without going into details, eternal inflation leads to a picture in which
any initial vacuum, will eventually nucleate bubbles containing all the
other possible vacua, sometimes called ``pocket universes.''
Because of the exponential volume growth, the number distribution 
of these pocket universes will ``very quickly'' lose memory of the
initial conditions and, one hopes, converge on some universal
distribution.

For this to happen, the microscopic theory must satisfy certain
conditions.  First, the effective potential must either contain
multiple de Sitter vacua, or contain regions in
which inflation leads to large quantum fluctuations (essentially, one
needs $\delta\rho/\rho \sim 1$).  Then, to populate all vacua from any
starting point, and thus have any hope to get a universal
distribution, all vacua must be connected by transitions.  These
conditions are fairly weak and very likely to be true in string/M
theory.  The first can already be satisfied by models of the type we
discussed in \rfsss{cosmo}.  One can find much evidence for the second
condition, that all vacua are connected through transitions, from the
theory of string/M theory duality.  For example, it is true for a wide
variety of models with extended supersymmetry (for example, ${\cal N}=2$ type
II compactifications on Calabi-Yau \cite{Greene:1995hu,Avram:1995pu}),
and thus will be true for flux compactifications built from these. 

Either way, the result of such considerations would be a probability
distribution on vacua, usually referred to as a ``measure factor.''  
This probability distribution would then be used to make probabilistic
predictions, along the lines we suggested in an example in \rfsss{susyscale}.

At this point, many difficult conceptual questions arise.  After all,
our (the observable) universe is a unique event, and most statisticians and
philosophers would agree that the standard ``frequentist'' concept of
probability, which assumes that an experiment can be repeated an
indefinite number of times, is meaningless when applied to unique
events.  While this may at first seem to be only a philosophical
difficulty, it will become practical at the moment that our
theoretical framework produces a claim such as ``the probability with
which our universe appears within our theory is $.01$'', or perhaps
$10^{-10}$, or perhaps $10^{-1000}$.  How should we interpret such
results?

Many cosmologists have argued that the interpretation of a measure
factor requires taking into account the selection effects we discussed
in \rfsss{anthropic} in a {\it quantitative} way, estimating the
``expected number of observers'' contained in each pocket universe, to
judge whether a ``typical observer'' should expect to make a certain
observation.  Such an analysis would be very complex, involving a good
deal of astrophysics, and perhaps even input from other disciplines
such as chemistry and biology, leading many to wonder whether generally
accepted conclusions could be obtained this way.

Other interpretations have been suggested.  One is
to consider a probability as having a clear
interpretation only if it
is extremely small, and consider such unlikely vacua as
``impossible.'' In other words, we choose some $\epsilon$, and if our
observations can only be reproduced by vacua with probability less
than $\epsilon$, we consider the theory with this choice of measure
factor as falsified.  While one might debate the appropriate choice of
$\epsilon$, since some ideas for measure factors lead to extremely
small probabilities for some vacua (say proportional to tunneling
rates, which as we saw in \rfss{tunnel} are extremely small), this
might be interesting even with an $\epsilon$ so small as to meet
general acceptance.

Another approach, which is probably the most sound philosophically, is
not to try to interpret absolute probabilities defined by individual
theories, but only compare probabilities between different theories,
considering the theory which gives the largest probability as preferred.
Even without a competitor to string/M theory, this might be useful in
judging among proposed measure factors, or dealing with other theoretical
uncertainties.
Following up this line of thought would lead us into Bayesian statistics;
see \citeasnoun{MacKay:2003} for an entertaining and down-to-earth introduction
to this topic.

Anyhow, these questions are somewhat academic at this point, as
general agreement has not yet been reached about how to define a
measure factor, or what structure the result might have.  In
particular, doing this within eternal inflation is notoriously
controversial, though recent progress is reported in
\citeasnoun{Bousso:2006ev,Vilenkin:2006xv} and references there, and 
perhaps generally
accepted candidate definitions will soon appear.

Thus let us conclude this subsection by simply listing a few of the
claims for measure factors which appear in existing literature.
One such is the entropy $\exp 24\pi^2 M_P^4/E$ of de Sitter
space with vacuum energy $E$.  This is the leading approximation to
the Hartle-Hawking wave function, where $E$ is usually interpreted as
the vacuum energy in some initial stage of inflation.  Since such a
factor is extremely sharply peaked at small $E$, its presence is more
or less incompatible with observed inflation, ruling this wave
function out.  There are some ideas for how corrections in higher
powers in $E$ could fix this, see
\citeasnoun{Firouzjahi:2004mx,Sarangi:2006eb}.

Another common result is $\exp 24\pi^2 M_P^4/\Lambda$, formally the same
entropy factor, but now as a function of the cosmological constant at
the present epoch.  This arose in the early attempts to derive a
measure from eternal inflation, and has a simple interpretation there:
the probability that a randomly chosen point sits in some vacuum, includes
a factor of the average lifetime of that vacuum, as predicted by \rfn{hm}.
This interpretation suggests that this measure factor is also
incorrect, as during almost all of this lifetime the universe is cold
and empty, so this factor has no direct bearing on the expected number
of observers.  More technical arguments have also been made against it.

If we ignore this problem, since this measure is heavily peaked on
small $\Lambda$, we might claim to have a dynamical solution to the
c.c. problem.  From the point of view we are taking, this proposal has
the amusing feature that it predicts that the total number of
$\Lambda>0$ vacua is roughly $10^{120}$, which presumably could be
checked independently.  If so, this would seem superficially
attractive, as in principle it predicts a unique overwhelmingly
preferred vacuum, the one with minimum positive $\Lambda$.  On the
other hand, the prospects for computing $\Lambda$ accurately enough to
find this vacuum seem very dim.  Even if we could get exact results
for $\Lambda$, there are arguments from computational complexity
theory that the problem of finding its minimum is inherently
intractable \cite{Denef:2006ad}, making this measure factor nearly
useless in practical terms.  It is also far from obvious at present
that the number of $\Lambda > 0$ vacua is in the neighborhood of
$10^{120}$.

Perhaps a better response to the problem is to define away the entropy
factor.  There are various closely related ways to do this
\cite{Vilenkin:2006xv}; for example
in \citeasnoun{Vanchurin:2006qp} it is argued that it can be done by
restricting attention to the world-line of a single ``eternal
observer,'' and counting the number of bubbles it enters.  This leads
to a prescription in terms of the stationary distribution of a Markov
process constructed from intervacuum tunneling rates; its detailed
properties are being explored, but at first sight this appears to lead
to a wildly varying probability factor $P(i)$ which, since it is
determined by the structure of high energy potential barriers, would have
little correlation to most observable properties of the vacua
themselves.  As we discuss in \rfss{interp}, this might still allow making
probabilistic predictions.

Other factors which have appeared in such proposals, and while 
probably subleading to the ones we covered might be important, include
a volume expansion factor (the overall growth in volume during slow-roll
inflation), the volume in configuration space
of the basin of attraction leading to the local minimum \cite{Horne:1994mi},
a canonical measure on phase space \cite{Gibbons:2006pa},
dynamical symmetry enhancement factors \cite{Kofman:2004yc}, and
the volume of the extra dimensions \cite{Firouzjahi:2004mx}.

\subsection{Holographic and dual formulations}

The advent of string/M theory duality in the mid-90's led to an
entirely novel perspective on many questions, and several new
candidate nonperturbative frameworks, such as matrix models, Matrix
theory, and the AdS/CFT correspondence.  While at present it is not
known how to use any of them to directly address the problem at hand,
perhaps the most general of these
is the AdS/CFT correspondence, which bears on the definition of
solutions with negative cosmological constant.

Consider a maximally
symmetric four-dimensional solution of string theory with negative
cosmological constant, in other words a product of $3+1$ dimensional
anti-de Sitter space-time with a $6$ dimensional internal space.
According to AdS/CFT, there will exist a dual $2+1$-dimensional
conformal field theory (without gravity), which is precisely equivalent
to the quantum string theory in this space-time.  This can be made more
concrete for questions which only involve observables on the boundary
of AdS; for example a scattering amplitude in AdS maps into a correlation
function in the CFT, and boundary conditions of the fields in AdS map
into the values of couplings in the CFT.

This dictionary has been much studied.  The most important
entries for present purposes are the relation between the $3+1$
dimensional AdS c.c. and the number of degrees of freedom of the CFT,
and the relation between masses in AdS and operator dimensions in the
CFT.  For example, in Freund-Rubin compactification
of IIb string theory on $AdS_5 \times S^5$, 
the curvature radius ${R^4 / (\alpha^\prime)^2} \sim g_sN$,
so the number of degrees of freedom $N^2$ scales
to very large values for weakly curved, weakly coupled vacua.
Similarly, the map between operators and gravity modes shows
that operators with dimension $\Delta \sim {\cal O}(1)$ map
to KK modes with masses $\sim {1 / R}$.  

For the Freund-Rubin examples, the AdS curvature radius and the radius
of the internal sphere are equal.  For the $AdS_4$ vacua which arise
in discussions of the landscape, one is usually interested instead
in theories with compact dimensions having $R_{KK} << R_{AdS}$, so
there is an effective 4d description.  Such theories will have dual
CFTs that differ ${\it qualitatively}$ from those appearing in 
standard examples of AdS/CFT.  By the mapping from gravity modes
to field theory operators, we see for instance that the number of
operators with $\Delta \sim {\cal O}(1)$ will be much smaller in 
these theories.  Instead of an infinite tower of operators with
regularly spaced conformal dimension (dual to the KK tower in 
Freund-Rubin models), these dual CFTs will have a sporadic set of
low dimension operators (dual to the compactification moduli),
and then a much larger spacing between the operators dual to KK modes. 

So, given a class of AdS vacua in the landscape, it seems reasonable
to search for candidate dual CFTs that could provide their exact
definition.
Further thought leads to difficulties with this idea.
First, the AdS vacua whose existence is established
using effective potential techniques, by definition lie in the regime 
in which the
gravity description is weakly coupled.  Since they have no exact moduli
(the compactification moduli are given small but nonzero masses by fluxes
and other effects, and are
dual to irrelevant operators whose dimension is however of ${\cal O}(1)$),
they do not extrapolate (along lines of fixed points) to a dual regime
where the field theory would be weakly coupled.  So trying to find the
dual field theory, involves working on the wrong (strongly coupled) side
of the duality, a difficult procedure at best.

Second, we are not primarily interested in typical landscape vacua.
Rather, we are most interested in those highly atypical vacua in which
fortuitous cancellations give rise to small $\Lambda$, as in
the Bousso-Polchinski argument.  Such vacua rely on complicated
cancellations between many terms, and there are reasons to think they
are exceedingly hard to find explicitly even in the more computable
gravity description.  This is the familiar problem, that one would need
to include Standard Model and other loop corrections to very high orders in
perturbation theory, to claim that one had found a specific vacuum with
small $\Lambda$. 
Even worse, a small
variation to one of these complicated solutions (such as changing a
flux by one unit) will spoil the cancellation and give a large cosmological
term. 
This suggests that the CFT's we would be most interested in finding
(which are dual to the AdS vacua with atypically small $\Lambda$) 
are also complicated, and
furthermore that we might need to compute very precisely to see the
cancellations which single out the few solutions with small cosmological
term.

Nevertheless, in principle we should be able to get the general
features of the problem to agree on both sides.  The basic picture
would seem to be that we start with QFT's with many, many degrees of
freedom, perhaps the dual theory of
\citeasnoun{Silverstein:2003jp}, and then flow down to CFT.
To recover agreement with our effective potential analyses, we would
need to find that a generic RG flow either loses almost all the
degrees of freedom, and thus is dual to large $\Lambda$, or else has no
weakly coupled space-time interpretation at all.  On the other
hand, given appropriate tunings in the bare theory, more degrees of
freedom would survive the flow, leading to a theory whose dual had a
tuned small $\Lambda$.  

\section{EXPLICIT CONSTRUCTIONS}
\label{sec:con}

We now 
describe how flux vacua can be constructed in type II string theories.
The most studied case involves IIb/F-theory vacua, so we will begin there.
We then present more recent results about IIa flux vacua, discuss
mirror symmetry in this setting, and 
provide some definite indications that many new classes of vacua
are waiting to be explored.
We cannot exhaustively review all approaches
to the subject; rather, we hope that this review, together
with the excellent reviews \cite{Frey:2003tf,Silverstein:2004id,Grana:2005jc},
will provide a good overview of various approaches and
classes of models.  For discussions of models without low energy
supersymmetry, see \citeasnoun{Silverstein:2004id}.

Before we begin, we should describe the approach we will use here to
compute the effective potential.  
The dimensional reduction of classical 10d supergravity (supplemented
by branes and fluxes) to 4d, is a well defined and straightforward (if
technically challenging) procedure.  In various flux compactifications,
it yields a tree approximation to the effective potential, as a function
of the moduli fields.  This effective potential should
be the leading term in a systematic expansion if $g_s$ 
and $l_s^2/R^2$ (in the leading order solution) are sufficiently small.
In some cases, we will supplement this classical potential with
quantum corrections which have been computed in the $g_s$ or $\alpha^\prime$
expansion, or with known non-perturbative contributions to the
4d effective superpotential.  The general form of the latter can in many
cases be inferred from holomorphy and symmetry considerations,
or from known field theory results (applied to the low-energy limit of
the string compactification).
While this general strategy is likely to apply only to a small fraction
of all vacua (those which are self-consistently stabilized at weak coupling
and large volume), we will see that this fraction alone is enough to indicate
the existence of a landscape, and to suggest concrete problems for further
work.  While this leaves open the conceptual questions of section
III.A, it
is also fair to say that the success of \EFT\ in describing
particle physics to date gives us good reason to hope that it will
be applicable to the study of pseudo-realistic string vacua.

\subsection{Type IIb D3/D7 vacua}
\label{sec:IIbclass}

In this subsection, we consider type IIb / F-theory vacua whose 
4d ${\cal N}=1$ supersymmetry is of the type preserved by D3/D7 branes
in a Calabi-Yau orientifold.  

\subsubsection{10d solutions}

Here, we describe the 10d picture of flux compactifications in the
supergravity limit.  We follow the treatment in \citeasnoun{Giddings:2001yu}. 
Closely related solutions (related to the IIb solutions via the
F-theory lift of M-theory) were first found in M-theory compactifications
on Calabi-Yau fourfolds in \citeasnoun{Becker:1996gj}, and some
aspects of their F-theory
lift were described in \citeasnoun{Gukov:1999ya,Dasgupta:1999ss}.

The type IIb string in 10 dimensions has a string frame action
\begin{eqnarray}
\label{IIblag}
L &=& {1\over 2\kappa_{10}^2} \int ~d^{10}x~(-G)^{1/2}
e^{-2S} \bigg( R + 4 \partial_{\mu}S \partial^{\mu}S  \nonumber \\
& & - {1\over 2} \vert F_1 \vert^2 - 
{1\over 12} G_3 \cdot \overline{G_3} - {1\over 4 \cdot 5!}\tilde F_5^2
\bigg) \nonumber \\
& &+{1\over 8i\kappa_{10}^2} \int e^{S} C_4 \wedge G_3 \wedge \overline{G_3}
+ S_{loc}~.
\end{eqnarray}
The theory has an NS field strength $H_3$ (with potential $B_2$) 
and RR field strengths
$F_{1,3,5}$ (with corresponding potentials $C_{0,2,4}$).  
The field strength
\begin{equation}
G_3 = F_3 - \phi H_3
\end{equation}
is a combination of the RR and NS three-form field strengths,
\begin{equation}
\phi = C_0 + ie^{-S}
\end{equation}
is the axio-dilaton, and
\begin{equation}
\tilde F_5 \equiv F_5 - {1\over 2} C_2 \wedge H_3 + {1\over 2}B_2 \wedge F_3~.
\end{equation}
The 5-form field is actually self-dual; one must impose the constraint
\begin{equation}
\tilde F_5 = * \tilde F_5
\end{equation}
by hand when solving the equations of motion.
Finally, $S_{loc}$ in
(\ref{IIblag}) allows for the possibility that we include the action of
any localized thin sources in our background; possible sources which could
appear in string theory include D-branes and orientifold planes.

We will start by looking for solutions with 4d Poincare symmetry.
The Einstein frame metric should take the form 
\begin{equation}
ds_{10}^2 = e^{2A(y)} \eta_{\mu\nu} dx^{\mu} dx^{\nu} +
e^{-2A(y)} \tilde g_{mn}(y) dy^m dy^n
\end{equation}
$\mu, \nu$ run over $0,\cdots,3$ while $m,n$ take values $4,\cdots,9$ and
$\tilde g_{mn}$ is a metric on the compactification manifold $M$.
We have allowed for the possibility of a warp factor $A(y)$. 
In addition one should impose
\begin{equation}
\phi = \phi(y),~~\tilde F_5 = (1+*) [ d\alpha(y) \wedge dx^0 \cdots \wedge
dx^3]
\end{equation}
and allow only ${\it compact}$ components of the $G_3$ flux
\begin{equation}
F_3, H_3 \in H^3 (M,\IZ)~.
\end{equation}
The $G_3$ equation of motion then tells one to choose a harmonic representative
in the given cohomology class.

One can show by using the trace-reversed Einstein equations for the $\IR^4$
components of the metric, that
\begin{eqnarray}
\label{4deinstein}
\tilde \nabla^2 e^{4A} &=& e^{2A} {G_{mnp} \overline G^{mnp} \over 12 Im(\tau)}
+ e^{-6A}[\partial_m \alpha \partial^m \alpha + \partial_m e^{4A} \partial^m
e^{4A}] \nonumber \\ 
& &+ \kappa_{10}^2 e^{2A}(T^{m}_{m} - T^{\mu}_{\mu})_{loc}~.
\end{eqnarray}
We have denoted the stress-energy tensor of any localized objects (whose action
appears in $S_{loc}$) by $T_{loc}$.

This equation already tells us something quite interesting.  The first
two terms on the right hand side are $\geq 0$, but on a compact manifold,
the left hand side integrates to zero (being a total derivative).  Therefore,
in compact models, and in the {\it absence} of localized sources, there is a
no-go theorem: the only solutions have $G_3=0$ and $e^A = {\rm constant}$,
and IIb supergravity does not allow nontrivial warped compactifications.
This is basically the no-go theorem proved in various ways in
\citeasnoun{Gibbons:1984kp,deWit:1986xg,Maldacena:2000mw}.

This does not mean that one cannot find warped solutions in the full
string theory.  String theory does allow localized sources.  It was 
emphasized already in \citeasnoun{Verlinde:1999fy} that one can make warped
models by considering compactifications with $N$ D3 branes, and stacking
the D3 branes at a point on the compact space; then as is familiar
from the derivation of the AdS/CFT correspondence 
\cite{Maldacena:1997re}, the geometry near the branes can become highly
warped.  

For this loophole to be operative, one needs 
\begin{equation}
(T^{m}_{m} - T^{\mu}_{\mu})_{loc} < 0
\end{equation}
to evade the global obstruction to solving Eq. (\ref{4deinstein}).
Before finding nontrivial warped solutions with flux, we will also need
one more fact.  The Bianchi identity for $\tilde F_5$ gives rise to
a constraint
\begin{equation}
\label{fbianchi}
d \tilde F_5 = H_3 \wedge F_3 + 2\kappa_{10}^2 T_3 \rho_3^{loc}
\end{equation}
where $T_3$ is the D3-brane tension, and $\rho_3^{loc}$ is the 
local D3 charge density on the compact space.  The integrated Bianchi
identity then requires, for tadpole cancellation,
\begin{equation}
\label{d3tadpole}
{1\over 2\kappa_{10}^2 T_3} \int_M H_3 \wedge F_3 + Q_3^{loc} = 0
\end{equation}
where $Q_3^{loc}$ is the sum of all D3 charges arising from localized
objects.

Now, one can rewrite the equation (\ref{fbianchi}) more explicitly in
terms of the function $\alpha(y)$ as
\begin{equation}
\tilde\nabla^2 \alpha = i e^{2A} {G_{mnp} *_{6} \overline{G}^{mnp} \over
12 Im(\tau)} + 2 e^{-6A} \partial_m \alpha \partial^m \alpha
+ 2\kappa_{10}^2 e^{2A}T_3 \rho_3^{loc} .
\end{equation}
Subtracting this from the Einstein equation (\ref{4deinstein}), one finds
\begin{eqnarray}
\label{theeqn}
\tilde\nabla^2(e^{4A}-\alpha) &=& {e^{2A}\over 24 Im(\tau)}\vert iG_3 -
*_6 G_3\vert^2 + e^{-6A} \vert \partial(e^{4A}-\alpha)\vert^2 \nonumber \\
& & + 2\kappa_{10}^2 e^{2A}[{1\over 4}(T^{m}_{m} - T^{\mu}_{\mu})^{loc} -
T_{3}\rho_3^{loc}] .
\end{eqnarray}
Let us make the {\it assumption}
\begin{equation}
\label{sourceas}
{1\over 4}(T^{m}_m - T^{\mu}_{\mu})^{loc} \geq T_{3} \rho_3^{loc}~.
\end{equation}
This serves as a constraint on the kind of localized sources we will want
to consider in finding solutions.
The inequality is saturated by D3 branes and O3 planes, as well as
by D7 branes wrapping holomorphic cycles; it is satisfied
by $\overline{D3}$ branes; and it is violated by O5 and $\overline{O3}$
planes.

Assuming we restrict our sources as above, it follows from
(\ref{theeqn}) that $G_3$ must be imaginary self-dual
\begin{equation}
\label{ISDcond}
*_6 G_3 = i G_3 ,
\end{equation}
that the warp factor and $C_4$ are related
\begin{equation}
e^{4A} = \alpha ,
\end{equation}
and that the inequality (\ref{sourceas}) is actually {\it saturated}.
So solutions to the tree-level equations should include only D3, O3 and
D7 sources.  In the quantum theory, one can obtain solutions
on compact $M$ with
$\overline{D3}$ sources as well; we will describe this when we discuss
supersymmetry breaking.

We did not write out the extra-dimensional Einstein equation and the
axio-dilaton equation of motion yet; their detailed form will not be
important for us.  Imposing them, we find that this class of solutions
describes F-theory models \cite{Vafa:1996xn} in the supergravity
approximation, including the possibility of background flux. 
As noted earlier, these solutions are closely related to those of
\citeasnoun{Becker:1996gj}, whose F-theory interpretation has also 
been described in \citeasnoun{Gukov:1999ya,Dasgupta:1999ss}.

The simplest examples of such solutions are perturbative IIb orientifolds.
An argument of \citeasnoun{Sen:1997gv}
shows that every compactification of F-theory on a Calabi-Yau fourfold
has, in an appropriate limit, an interpretation as a IIb orientifold
of a Calabi-Yau threefold.  We will therefore develop the story in
the language of IIb orientifolds, but the formulae generalize in a
straightforward way to the more general case.
In this special case of perturbative orientifolds, at leading order,
the metric on the internal space is ${\it conformally}$ Calabi-Yau;
it differs by the warp factor $e^{2A}$. 

\subsubsection{4d effective description}

In this section we describe the construction of the 4d effective action
for IIb orientifolds with RR and NS flux, 
following \citeasnoun{Giddings:2001yu}.  
The main result will be an explicit and computable result for the 4d
effective potential, which can be analyzed using analytical, numerical or 
statistical techniques.
Earlier work in this direction appeared in 
\citeasnoun{Gukov:1999ya,Dasgupta:1999ss,Taylor:1999ii,
Mayr:2000hh}), while related results in gauged supergravity were
presented in \citeasnoun{Polchinski:1995sm,Michelson:1996pn,
Dall'Agata:2001zh,Andrianopoli:2001gm,
Andrianopoli:2002rm,Andrianopoli:2002mf,D'Auria:2002tc,Ferrara:2002bt,
D'Auria:2002th,Ferrara:2002hb,Andrianopoli:2003jf,D'Auria:2003jk,
Angelantonj:2003rq,Andrianopoli:2003sa,Angelantonj:2003zx,
Dall'Agata:2004dk,Dall'Agata:2004nw}.   
Generalizations of this formalism to include effects of the warp
factor appear in \citeasnoun{DeWolfe:2002nn,Giddings:2005ff,Frey:2006wv}.

We consider a Calabi-Yau threefold $M$ with $h^{2,1}$ complex structure
deformations, and choose a symplectic basis $\{ A^a, B_b \}$ for the
$b_3 = 2h^{2,1} + 2$ 
three-cycles $a,b = 1,\cdots,h_{2,1}+1$, with dual cohomology elements
$\alpha_a, \beta^b$ such that:
\begin{equation}
\int_{A^a} \alpha_b = \delta^a_b,~~\int_{B_b} \beta^a = - \delta^a_b,~~
\int_M \alpha_a \wedge \beta^b = \delta^b_a~.
\end{equation}
Fixing a normalization for the holomorphic three-form $\Omega$, we
then define the periods
\begin{equation}
z^a = \int_{A^a} \Omega,~~{\cal G}_b = \int_{B_b} \Omega
\end{equation}
and the period vector $\Pi(z) = ({\cal G}_b,z^a)$.  The $z^a$ are
projective coordinates on the complex structure moduli space of the
Calabi-Yau threefold, with ${\cal G}_b = \partial_{b}G(z)$
($G(z)$ is commonly known as the ``prepotential'').   The
K\"ahler potential ${\cal K}$ for the $z^a$ as well as the
IIb axio-dilaton $\phi = C_0 + {i\over g_s}$ is given by
\begin{equation}
\label{Kis}
{\cal K} = - {\rm log} \left(i \int_M \Omega \wedge \bar \Omega \right) -
{\rm log} (-i(\phi - \bar\phi))~. 
\end{equation}
Note that given the period vector, one can re-write 
\begin{equation}
\int_M \Omega \wedge \bar \Omega = - \Pi^{\dagger} \Sigma \Pi
\end{equation}
where $\Sigma$ is the symplectic matrix.   
This structure on the complex structure moduli space follows from
so-called special geometry, as derived in 
\citeasnoun{Dixon:1989fj,Candelas:1990pi, 
Strominger:1990pd}.  
The special geometry governs the moduli space of vector multiplets in
${\cal N}=2$ supersymmetric compactifications.  However, to
leading approximation (i.e. tree level), it also governs
the complex structure moduli space of ${\cal N}=1$ orientifolds 
of these models, which is the application of interest here.
In general, some of the complex structure moduli could be projected out
in any given orientifold construction; in this circumstance, one should
appropriately restrict the various quantities to the surviving submanifold
of the moduli space.

Now, we consider turning on fluxes of the RR and NS-NS 3-form field
strengths $F_3$ and $H_3$.  In a self-explanatory notation, we define
these via integer-valued $b_3$-vectors $f, h$:  
\begin{equation}
F_{3} = - (2\pi)^2 \alpha^\prime (f_a \alpha^a
+ f_{a+h_{2,1} + 1}\beta_a),
\end{equation}
\begin{equation}
H_3 = - (2\pi)^2 \alpha^\prime
(h_a \alpha^a + h_{a+h_{2,1}+1}\beta_a)~.
\end{equation}
These fluxes generate a superpotential for the complex structure moduli
as well as the axio-dilaton \cite{Gukov:1999ya}
\begin{equation}
\label{fluxsup}
W = \int_{M} G_3 \wedge \Omega(z) = (2\pi)^2 \alpha^\prime (f-\phi h)
\cdot \Pi(z)
\end{equation}
where $G_3 = F_3 - \phi H_3$. 

To write down a general expression for the potential, we need to introduce
one more ingredient.
Thus far, we have described only a K\"ahler potential on the complex
structure moduli space.  In general models, there are also K\"ahler moduli
(up to $h^{1,1}(M)$ of them, depending on how many survive the orientifold
projection).  However, they will cancel out of the tree-level effective
potential 
in the IIb supergravity, in the following way \cite{Giddings:2001yu,
Grimm:2004uq}. 
The K\"ahler potential for these moduli is
\begin{equation}
\label{Kk}
{\cal K}_k = -2 {\rm log}(V)
\end{equation}
Given a basis of divisors $\{ S_\alpha \}$, $\alpha=1,\cdots,h_{1,1}$,
the volume $V$ is determined in terms of the K\"ahler form
$J = t^{\alpha} S_{\alpha}$ by 
\begin{equation}
V = {1\over 6} S_{\alpha \beta \gamma} t^\alpha t^\beta t^\gamma ~. 
\end{equation}
Here $S_{\alpha\beta\gamma}$ is the triple 
intersection form, and counts the
intersections of the divisors $S_{\alpha}$, $S_{\beta}$ and $S_{\gamma}$.
Note that for this class of vacua, the flux superpotential
(\ref{fluxsup}) does ${\it not}$ depend explicitly on the K\"ahler moduli.

Now, on general grounds, the expression for the potential in ${\cal N}=1$
supergravity takes the form \cite{Freedman:1976xh}
\begin{equation}
\label{sugrapot}
V = e^{{\cal K} + {\cal K}_k} \left( \sum_{i,j} g^{i \bar j} D_i W \overline{
D_j W} - 3 |W|^2 \right)
\end{equation}
where $i,j$ run over indices labeling 
the complex structure and K\"ahler moduli
as well as the dilaton.  
$D_i W$ is the K\"ahler covariantized derivative $D_i W = \partial_i W +
K_{,i}W$.
At this point, 
because of the special structure where $W$ is independent of the $t$s at
tree level, as well as the tree-level form of the K\"ahler potential, 
in the expression (\ref{sugrapot}) the $-3|W|^2$ term precisely
cancels the terms where $i,j$ run over $\alpha,\beta$ (the
K\"ahler moduli). Therefore, one can
express the full tree-level flux potential as \cite{Giddings:2001yu}
\begin{equation}
\label{noscale}
V = e^{{\cal K}_{tot}} \left(\sum_{a,b} g^{a \bar b} D_a W \overline
{D_b W} \right)
\end{equation} 
where here the sum over $a$ also includes $\phi$.

So surprisingly, despite the fact that we are working in an ${\cal N}=1$
supergravity, the potential is positive semi-definite with vacua precisely
when $V=0$!  
Furthermore, one sees immediately that generic vacua are not
supersymmetric; supersymmetric vacua have $D_a W = D_{\phi} W = 
D_\alpha W = 0$,
while non-supersymmetric vacua have $D_{\alpha} W \neq 0$ for some
$\alpha$.  
This is precisely a realization of the cancellation that occurs in a
general class of supergravities known as no-scale supergravities
\cite{Cremmer:1983bf,Ellis:1983sf}. 
Unfortunately, the miracle of vanishing cosmological constant
for the non-supersymmetric vacua depended on the tree-level structure
of the K\"ahler potential
(\ref{Kk}) which is not radiatively stable.  
Therefore this miracle, while suggestive, does not lead to any mechanism
of attacking the cosmological constant problem.
The potential (\ref{noscale}) receives important corrections both in
perturbation theory and nonperturbatively. 

A simple characterization of the points in moduli space which give solutions
to $V=0$ for a given flux arises as follows.  We need to solve the
equations
\begin{equation}
D_\phi W = D_a W = 0
\end{equation}
which, more explicitly, means
\begin{equation}
(f-\bar\phi h) \cdot \Pi(z) = (f-\phi h) \cdot (\partial_a \Pi +
\Pi \partial_a {\cal K}) = 0~. 
\end{equation}
In fact, these equations have a simple geometric interpretation: for a
given choice of the integral fluxes $f, h$, they require the metric
to adjust itself (by motion in complex structure moduli space) so that
the (3,0) and (1,2) parts of $G_3$ vanish, leaving a solution where
$G_3$ is ``imaginary self-dual'' (ISD), as in (\ref{ISDcond}).

At this stage, since we are solving $h_{2,1}+1$ equations in $h_{2,1}+1$
variables for each choice of integral flux, 
it seems clear that generic fluxes will fix all of the complex structure
moduli as well as the axio-dilaton.  Furthermore,
one might suspect that the
number of vacua will diverge, since we have not yet constrained the fluxes
in any way.  

However, the fluxes also induce a contribution
to the total D3-brane charge, arising from the term in the 10d IIb 
supergravity Lagrangian
\begin{equation}
{\cal L} = \cdots + {1\over 8i \kappa_{10}^2} \int {C_{(4)} \wedge G_3 \wedge
\overline{G}_3 \over {Im\phi}} + \cdots
\end{equation}
where $C_4$ is the RR four-form potential which couples to D3 branes.
This results in a tadpole for D3-brane charge, in the presence of the
fluxes:
\begin{equation}
N_{\rm flux} = {1\over (2\pi)^4 (\alpha^\prime)^2} \int_M 
F_3 \wedge H_3 = f \cdot \Sigma \cdot h~. 
\end{equation} 
This is important because: i) one can easily check that for ISD fluxes
$N_{\rm flux} \geq 0$, and ii) in a given orientifold of $M$, there is a
tadpole cancellation condition
(\ref{d3tadpole}), which we can write in the form
\begin{equation}
\label{tadpole}
N_{\rm flux} + N_{D3} ~=~L
\end{equation}
where $L$ is some total negative D3 charge which needs to be cancelled, 
arising by induced charge on D7 and O7 planes \cite{Giddings:2001yu}, 
and/or explicit O3 planes.  In practice, for an orientifold which
arises in the Sen limit 
\cite{Sen:1997gv} of an F-theory compactification on elliptic fourfold
$Y$, one finds \cite{Sethi:1996es}
\begin{equation}
L = {\chi(Y)\over 24}~. 
\end{equation} 
What this means is that the allowed flux choices in an orientifold
compactification on $M$, and hence the numbers of flux vacua, are 
stringently constrained by the requirement $N_{\rm flux} \leq L$.
This will be important later in the review, when we discuss vacuum
statistics for this class of models.

We note here that in describing this classical story, we have simplified
matters by turning on only the background closed string fluxes. In general
orientifold or F-theory models, D7 branes with various gauge groups are
also present, and one can turn on background field strengths of the D7
gauge fields, generating additional contributions to the tadpole condition
(\ref{tadpole}) and the space-time potential energy.  Because our story
is rich enough without considering these additional ingredients, we proceed
with the development without activating them, 
but discussions which incorporate them in this class of vacua can be
found in {\it e.g.} 
\citeasnoun{Burgess:2003ic,Jockers:2004yj,%
Jockers:2005zy, GarciadelMoral:2005js, Haack:2006cy}.

\paragraph{Example: The conifold}
\label{para:conifold}

We now exemplify our previous considerations by finding flux vacua in
one of the simplest non-compact Calabi-Yau spaces, the deformed conifold.
The metric of this space is known explicitly \cite{Candelas:1989js}.
The vacua we discuss below have played an important role in 
gauge/gravity duality \cite{Klebanov:2000hb}, the study of geometric
transitions \cite{Vafa:2000wi}, and warped compactifications of string
theory \cite{Giddings:2001yu}, including models of supersymmetry
breaking \cite{Kachru:2002gs}.  We will encounter some of these
applications as we proceed.

The deformed conifold is a noncompact Calabi-Yau space, defined by
the equation
\begin{equation}
\label{conifold}
P(x,y,v,w) = x^2 + y^2 + v^2 + w^2 = \epsilon^2
\end{equation}
in $\IC^4$.  
As $\epsilon \to 0$, the geometry becomes singular: the origin is
non-transverse, since one can solve $P = dP = 0$ there.  It is not
difficult to see that an $S^3$ collapses to zero size at this point
in moduli space; e.g. for real $\epsilon^2$, the real slice of
(\ref{conifold}) defines such an $S^3$.  
In this limit, the geometry can be viewed as a cone over $S^3 \times S^2$.
There are two topologically nontrivial three-cycles; the A-cycle $S^3$
we have already discussed, which vanishes when $\epsilon \to 0$, and a
dual B-cycle swept out by the $S^2$ times the radial direction of
the cone.

The singularity (\ref{conifold}) arises locally in many compact
Calabi-Yau spaces (at codimension one in the complex structure moduli space).
In such manifolds, the B-cycle is also compact; the behavior of the periods
of $\Omega$ is partially universal, being given by
\begin{equation}
\label{conper}
\int_{A} \Omega = z,~~\int_{B} \Omega = {z\over 2\pi i} {\rm log}(z) + {\rm
regular} = {\cal G}(z)~.
\end{equation}
Here $z \to 0$ is the singular point in moduli space where A collapses,
and the regular part of the B-period is non-universal.

We can now study flux vacua using the periods (\ref{conper})
and the explicit formulae (\ref{fluxsup}), (\ref{noscale}) for the
superpotential and potential energy function.
Choosing
\begin{equation}
\label{KSflux}
\int_A F_3 \sim M, ~~\int_B H_3 \sim -K
\end{equation}
we find that the superpotential takes the form
\begin{equation}
\label{VYsup}
W(z) = - K \phi z + M{\cal G}(z)
\end{equation}

Given the logarithmic singularity in ${\cal G}$, this 
superpotential bears a striking resemblance to the Veneziano-Yankielowicz
superpotential of pure ${\cal N}=1$ supersymmetric $SU(M)$ gauge theory,
conjectured many years ago \cite{Veneziano:1982ah}.  We'll see that this
is no accident.

The K\"ahler potential can be determined using the equation
(\ref{Kis}).  We will be interested in vacua which arise close to the
conifold point where $z$ is exponentially small; to obtain such vacua
we will consider $K/g_s$ to be large.  In this limit, the dominant terms
in the equation for classical vacua are
\begin{equation}
D_z W = {M\over 2\pi i}{\rm log}(z) - i{K\over g_s} + \cdots 
\end{equation}
where $\cdots$ are ${\cal O}(1)$ terms that
will be negligible in a self-consistent manner.  
For $K/g_s$ large, one finds that
\begin{equation}
z \sim {\rm exp} (-2\pi K/g_sM)
\end{equation}
and these flux vacua are exponentially close to the conifold point in
moduli space.
Due to the ambiguity arising from the logarithm,
there are $M$ vacua, distributed in phase but with the
magnitude of $z$ as given above.

For the noncompact Calabi-Yau, these are good flux vacua.  In fact, 
the conifold with fluxes
(\ref{KSflux}) is dual, via gauge/gravity duality, to a certain
${\cal N}=1$ supersymmetric
$SU(N+M) \times SU(N)$ gauge theory, with $N=KM$ \cite{Klebanov:2000hb}.
While it is beyond the scope of our review 
to discuss this duality in detail, the IR physics of the gauge theory
involves gluino condensation in pure $SU(M)$ ${\cal N}=1$ SYM.
This fact, together with the duality, explains the appearance of the
Veneziano-Yankielowicz superpotential in (\ref{VYsup}).
The $M$ vacua we found in the $z$-plane, are the $M$ vacua which saturate
the Witten index of pure $SU(M)$ SYM.

In a compact Calabi-Yau, the dilaton $\phi$ is also dynamical and we would
need to solve the equation $D_{\phi}W =0$ as well.  Naively, one would
find an obstruction to doing this in the limit described above (large
$K/g_s$ and exponentially small $z$).  In fact, one ${\it can}$ do this
even in compact situations, as described in \citeasnoun{Giddings:2001yu}.  

While this example is quite simple, we will use it to illustrate many
points in our review.
In the literature, one can find many other
examples of explicit vacua, both in toroidal orientifolds
\cite{Dasgupta:1999ss,Greene:2000gh,Kachru:2002he,Frey:2002hf,Cascales:2003zp,
Cascales:2003pt,Blumenhagen:2003vr}
and in more nontrivial Calabi-Yau threefolds \cite{Curio:2000sc,
Curio:2001ae,Tripathy:2002qw,
Giryavets:2003vd,
Giryavets:2004zr,Conlon:2004ds,DeWolfe:2004ns,DeWolfe:2005gy,Aspinwall:2005ad}.

\subsubsection{Quantum IIb flux vacua}
\label{sss:quantumIIb}

At the classical level, the K\"ahler moduli of IIb orientifolds with flux
remain as exactly flat directions of the no-scale potential.  However,
quantum corrections will generally generate a potential for these moduli.
This potential will have at least two different sources:

\begin{enumerate}
\item
In every model, there will be corrections to the K\"ahler potential which
depend on K\"ahler moduli.  The leading such corrections have been computed
in {\it e.g.} \citeasnoun{Becker:2002nn,Berg:2005ja,Berg:2005yu}.   
As soon as ${\cal K}_k$ takes a more general form than (\ref{Kk}), the
no-scale cancellation disappears and the scalar potential will develop
dependence on the K\"ahler moduli.  

\item
The superpotential in these models enjoys a non-renormalization theorem
to all orders in perturbation theory \cite{Burgess:2005jx}. 
Nonperturbatively, it can be violated by Euclidean D3-brane instantons.
The conditions for such instantons to contribute in the absence of
$G_3$ flux, and assuming they have smooth worldvolumes, with vanishing
intersection with other branes in the background, 
are described in \citeasnoun{Witten:1996bn}.  The basic condition is familiar
also from supersymmetric gauge theory: there should be precisely two
fermion zero modes in the instanton background.  
These zero modes can be counted as follows.  One can lift the Euclidean
D3 brane to an M5 brane wrapping a divisor $D$ in the M-theory dual
compactification on a Calabi-Yau fourfold.  Then, the number of fermion
zero modes can be related to the holomorphic Euler character $\chi$ of
the divisor: 
\begin{equation}
\label{cond}
{\rm number ~of~ zero~ modes} = 2 \chi(D) = 2 \sum_{p=0}^{3} h^{0,p}(D) ~. 
\end{equation}
In the simplest case of an isolated divisor with $h^{0,0}=1$ and
other $h^{0,p}$ vanishing, the contribution is definitely nonzero.
For more elaborate cases where $\chi = 1$ but the divisor has a moduli
space, it is conceivable that the integral over the instanton moduli
could vanish.
\end{enumerate}

The conditions under
which such instantons contribute in the presence of various fluxes and/or
space-filling D-branes (whose worldvolumes they may intersect) remain
a subject of active investigation 
\cite{Gorlich:2004qm,Saulina:2005ve,%
Kallosh:2005gs,Lust:2005cu,Blumenhagen:2006xt,Ibanez:2006da,%
Haack:2006cy,Florea:2006si}.
The condition (\ref{cond}) is certainly modified.
More generally, there can be contributions from nonperturbative dynamics
in field theories arising on D7-brane worldvolumes, whose gauge coupling
is K\"ahler-modulus dependent \cite{Kachru:2003aw,Gorlich:2004qm}.

It was argued in \citeasnoun{Kachru:2003aw} (KKLT)
that such corrections will
allow one to find flux compactifications of the IIb theory that manifest
landscapes of vacua with all moduli stabilized.
As a simple toy model for how such corrections may be important let us
consider a model with a single K\"ahler modulus $\rho$, with
\begin{equation}
{\cal K}_k = - 3 ~{\rm log} (-i (\rho - \bar\rho))
\end{equation}
Here one should think of
${\rm Im}(\rho) \sim {R^4\over (\alpha^\prime)^2}$ where $R$ is the radius
of $M$, while $Re(\rho)$ is related to the period of an axion arising from
$C_4$ \cite{Giddings:2001yu}.  If there is a D7 stack which
gives rise to a pure SYM sector, whose gauge coupling
depends on $\rho$, one finds a superpotential of the general form
\begin{equation} 
\label{kkltsup}
W = W_0 + Ae^{ia\rho}~. 
\end{equation}
One should view $W_0$ as being the constant arising from evaluating
the flux superpotential (\ref{fluxsup}) at its minimum in complex structure
moduli space.  $A$ is a determinant which a priori depends on complex
structure moduli, and $a$ is a constant depending on the rank of the D7
gauge group.
We noted above that $A$ would generally 
depend on complex structure moduli.  However,
the scales in the flux superpotential make it clear that complex structure
moduli receive a mass at order ${\alpha^\prime \over R^3}$, while any
K\"ahler moduli masses arising from the correction  
in (\ref{kkltsup}) will be significantly smaller.  Therefore, one can view
the supergravity functions above as summarizing the \EFT\ %
of the light mode $\rho$, having integrated out the heavy complex structure
modes and dilaton.  For a detailed discussion of possible issues with
such a procedure, see {\it e.g.} \citeasnoun{Choi:2004sx,deAlwis:2005tg}. 

It is then straightforward to show that one can solve the equation
$D_{\rho}W = 0$, yielding a vacuum with all moduli stabilized and with
unbroken supersymmetry \cite{Kachru:2003aw}.  
For small $W_0$, this vacuum moves into the
regime of control (large ${\rm Im} (\rho)$) with logarithmic speed.
(Small $a$ arising from large rank gauge groups also helps).  
This provides a loose proof-of-principle that
one can find models with all moduli stabilized.  This picture
has been substantially 
fleshed out and extended in further work; the most explicit examples
to date appear in \citeasnoun{Denef:2004dm,%
Denef:2005mm,Lust:2005dy,Lust:2006zg}.

Before moving on to summarize further detailed considerations, we discuss
here two important questions which may concern the reader.  Firstly,
under the assumptions above, one requires an exponentially small value of
$W_0$ to obtain a vacuum which is in the regime of computational control,
where further corrections are expected to be small.  Is it reasonable to
expect such a small value?  Actually,
in all string models of SUSY GUTs in which $M_s \sim M_P$,
such a tuning of $W_0$ is inevitable.
Recall from \rfsss{susybr} that
the largest $F$ term which is allowed in a model where SUSY explains the
gauge hierarchy is roughly $(10^{11} {\rm GeV})^2$.  
Taking into account \rfn{sugraV}, a small cosmological constant
requires 
\begin{equation}
|W|^2 \leq M_P^2 (10^{11} {\rm GeV})^4 \rightarrow ({W\over M_P^3}) \leq 
10^{-14}~. 
\end{equation}
For models of gauge mediation with low-scale breaking, the tune becomes
even larger.  This tune is absolutely necessary in the standard supergravity
picture of unification, and enters directly into cosmology via the gravitino
mass.  It is therefore an ${\it inevitable}$ problem in standard SUSY
scenarios with high string scale, that one will be required to tune $W$ to
be small at any minimum.    

This does not answer the question of whether such small values of $W_0$
are in fact attainable in actual flux vacua.  We will answer this in
the affirmative in \rfs{stat}.  Indeed, we will make far more detailed
claims; for example that the (naive) expectation that this could be
easily attained by an approximate R symmetry, is not true for known flux 
vacua.

Another interesting question is: when is one justified in using the
tree-level K\"ahler potential while including the nonperturbative
correction to $W$?  Clearly, at very large volume, corrections to
${\cal K }_k$ (which are power-law suppressed) are more important than
instanton effects.  However, in the spirit of self-consistent
perturbation theory, this is not the relevant question.  The relevant
question is, given the estimates above, if one then includes a first
correction to ${\cal K}_k$ and then re-expands around the solution one
has obtained with the tree level ${\cal K}_k$, how much does the
solution shift?  It is easy to verify that for large ${\rm Im}(\rho)$,
the perturbative corrections to ${\cal K}_k$ (expanded around the
minimum of the potential) shift the solution by a small amount, which
can be tuned by tuning $W_0$.

Naturally, however, this suggests that the corrections to 
${\cal K}$ themselves
can cause interesting new features at large volume, giving rise to further
critical points in the potential distinct from the KKLT minima.
Such critical points were observed in \citeasnoun{Balasubramanian:2005zx,
Balasubramanian:2004uy,vonGersdorff:2005bf,Berg:2005yu}, using 
estimates for the first few quantum corrections to ${\cal K}$. 
These can yield vacua with very large volume, even realizing the
large extra dimensions scenario of \citeasnoun{Arkani-Hamed:1998rs}. 
The phenomenology of such models has been described in 
\citeasnoun{Conlon:2005ki,Conlon:2006us}.

\subsubsection{Supersymmetry breaking}

The vacua we have discussed so far are supersymmetric.  One would hope to
learn also about vacua which have supersymmetry breaking
at or above the TeV scale, and have positive cosmological constant.
Here we discuss three ideas in this direction: one in some detail (largely
because it is novel and uses stringy ingredients), and two more
standard ideas quite
briefly.
We will focus on theories with low energy
breaking (i.e. breaking far below the KK scale).
There are also known solutions with supersymmetry
breaking
at the KK scale \cite{Saltman:2004jh} or even higher scales
\cite{Silverstein:2001xn}.   
Examples of this type are discussed in a pedagogical way in 
\cite{Silverstein:2004id}.

We will also only discuss the mechanisms of SUSY-breaking that have been
explored in the IIb landscape.  One of the most important consequences
of SUSY-breaking is of course the generation of soft terms.  For flux-induced
breaking, these terms have been investigated in 
\cite{Camara:2003ku,Lawrence:2004zk,Lawrence:2004kj,%
Camara:2004jj,Ibanez:2004iv,Lust:2004fi,Lust:2004dn,
Font:2004cx,Marchesano:2004yn,Lust:2005bd,Allanach:2005yq}.
More generally, constructions of models incorporating a standard-like model
together with flux stabilization have appeared in \citeasnoun{Cvetic:2004xx,
Marchesano:2004yq,Marchesano:2004xz,Cvetic:2005bn}.

\paragraph{Warped Supersymmetry Breaking}

The idea presented in \citeasnoun{Kachru:2003aw}, 
is as follows.
Calabi-Yau compactification, at leading order in $\alpha^\prime$, gives
rise to a compactification metric of the form
\begin{equation}
ds^2 = \eta_{\mu\nu} dx^{\mu} dx^{\nu} + g_{mn} dy^m dy^{n}
\end{equation}
with $\mu, \nu$ running over coordinates in our $\IR^4$, and 
$m, n = 1,\cdots,6$ parametrizing 
the coordinates on the ``extra'' six dimensions. 

However, in the presence of fluxes, one finds a more general metric of
the form
\begin{equation}
ds^2 = e^{2A(y)} \eta_{\mu\nu}dx^{\mu} dx^{\nu} + e^{-2A(y)} g_{mn}
dy^m dy^n~. 
\end{equation} 
$A(y)$ is a warp factor, which allows the ``scale'' in the 4d Minkowski
space to vary as one moves along the compact dimensions $y^m$.
\footnote{
The fact that fluxes generate warping was described in
\citeasnoun{Strominger:1986uh,Becker:1996gj}, this was discussed in the
IIb context in \citeasnoun{Dasgupta:1999ss,Greene:2000gh}, and concrete 
ideas about Randall-Sundrum scenarios in string theory were first
developed in 
\citeasnoun{Verlinde:1999fy,Chan:2000ms}.} 
The equation determining $A(y)$ in terms of the flux compactification
data 
can be found in \citeasnoun{Giddings:2001yu}. 
Compactifications where $A(y)$ varies significantly as one moves over the
compact six-manifold $M$, are often called ``warped compactifications."

An important toy model of warped compactification is the Randall-Sundrum
model \cite{Randall:1999ee}.  This is a 5d model where a warp factor
which varies by an exponential amount over the 5th dimension (which is
compactified on an interval), can be used to explain exponential hierarchies
in physics.  The basic idea is that scales at the end of the 5th dimension
where $e^A$ has a minimum, are exponentially smaller than those at the 
UV end where $e^A$ is maximized.

The simplest realization of this idea in string theory uses precisely the same
kinds of (deformed) conical throats 
that arise in describing string duals of confining
gauge theories \cite{Klebanov:2000hb}. 
We found for instance that in the conifold geometry, one can stabilize
moduli exponentially close to a conifold point in moduli space without
tuning
\begin{equation}
\label{conrem}
\int_A F_3 = M,~~\int_B H_3 = -K,~~z = {\rm exp}(-2\pi K/g_sM)~.
\end{equation}
But the fluxes here are precisely those of the warped deformed conifold
solution which appears in gauge/gravity duality; hence the warp factor
$e^{2A(y)}$ at the ``tip'' of the deformed conifold, will take the
same value it does there.  This gives rise to an exponential
warping
\begin{equation} 
e^A \sim e^{-2\pi K/3g_sM}~. 
\end{equation}
As a result,  
compactifications of the conifold with flux, can give rise to string
theory models which accomodate the exponential warping of scales
used in Randall-Sundrum scenarios \cite{Giddings:2001yu}.
The possibilities for making realistic R-S models in this general
context have been
investigated in much more detail in 
\cite{Cascales:2003wn,Franco:2005fd,Cascales:2005rj,Gherghetta:2006yq}
and references therein.

Instead of using the redshifting of scales to explain the Higgs
mass directly, this warping can also be used in another way.  Imagine that
instead of engineering the Standard Model in the region of minimal warp
factor, one arranges for SUSY breaking to occur there.  
The Standard Model can be
localized in the bulk of the
Calabi-Yau space, where $e^A \sim {\cal O}(1)$.  In this situation,
the exponentially small scale of supersymmetry breaking can be explained
by warping, instead of by instanton effects.  
It can be transmitted via gravity mediation or other mechanisms to
the observable sector.  Precisely this scenario, combined with other
assumptions, has been explored in a phenomenological context in {\it e.g.}
\citeasnoun{Choi:2005ge,Choi:2005uz,Choi:2005hd,Kitano:2005wc,Brummer:2006dg}.

To flesh out such scenarios, one should provide explicit microscopic
models of such SUSY breaking.  Such 
a model was proposed already in \citeasnoun{Kachru:2002gs}.  The idea is to
consider the conifold with flux, in the presence of a small number 
$p<<M$ of
anti-D3 branes.  While the throat carries $\int H \wedge F = KM$ units
of D3 brane charge, this is not obviously 
available to perturbatively annihilate
with the anti-branes.  It then becomes interesting to work out the
dynamics of this non-supersymmetric but controlled system.  

For $p << M$, we can consider the $p$ anti-branes as probes of the
exact solution given in \citeasnoun{Klebanov:2000hb}.  Their dynamics
will be governed by their worldvolume action in the fixed
supergravity background.  
This action is a function of the six matrix-valued fields
$\Phi^i$, which are adjoints of $SU(p)$ and parametrize the
brane positions on $M$.
In an appropriate
duality frame, it is given by the sum of two terms: a
Born-Infeld term
\begin{equation}
\label{BI}
S_{BI} = -{\mu_3\over g_s} \int d^4x Tr \sqrt{det(G_{||}) det(Q)}
\end{equation} 
and a Chern-Simons term
\begin{equation}
\label{CS}
S_{CS} = -\mu_3 \int d^4x Tr(2\pi i_{\Phi}i_{\Phi}B_6 + C_4)~.
\end{equation}
Here ${\mu_3\over g_s}=T_3$, $G_{||}$ is 
the pullback of the induced metric along the anti-branes,
$i_{\Phi}$ is the interior derivative so
\begin{equation}
i_{\Phi}i_{\Phi}B_6 = \Phi^n \Phi^m B_{mnpqrs} {1\over 4!} 
dy^p \wedge \cdots \wedge dy^s ,
\end{equation}
$Q$ is the matrix
\begin{equation}
Q^i_j = \delta^i_j + {2\pi i \over g_s} [\Phi^i,\Phi^k](G_{kj} + g_s C_{kj})
\end{equation}
and $B_6$ is given in an ISD flux background by
\begin{equation}
\label{bsixis}
dB_6 = {1\over g_s^2} *_{10} H_3 = -{1\over g_s} dV_4 \wedge F_3
\end{equation}
where $dV_4$ is the volume form on $\IR^4$ at the brane location in
the compact dimensions.

It is best to summarize the dynamics in three steps \cite{Kachru:2002gs,
DeWolfe:2004qx}.

\subparagraph{Weight loss}
The non-commutator terms in the ISD flux background yield the action
\begin{equation}
-p {\mu_3\over g_s} \int d^4x \sqrt{g_4} e^{4A} Tr \left(2 + {1\over 2}e^{-2A}
\partial_{\mu} \Phi^i \partial^{\mu} \Phi^j g_{ij}\right)~.
\end{equation}
Therefore, the leading potential is
\begin{equation}
\label{antipot}
V(y) = 2 e^{4A(y)}~.
\end{equation}
It arises by adding the BI and CS terms; for a D3-brane these would
instead cancel, as D3-branes in the ISD flux backgrounds feel no force.

In the Klebanov-Strassler solution \cite{Klebanov:2000hb}, 
the warp factor depends only on the ``radial'' direction in the cone,
$A(y) = A(r)$ for some radial direction $r$.  
Then the potential
(\ref{antipot}) simply yields a force in the radial direction
\begin{equation}
F_{r}(r) = -2{\mu_3 \over g_s} \partial_{r} e^{4A}(r)~.
\end{equation} 
The warp factor monotonically decreases as one goes towards the
smooth (deformed) tip of the cone, so in the first step of evolution,
the $p$ anti-branes are drawn quickly to the region of minimal warp factor,
the tip of the deformed conifold.
This result is intuitively clear: the branes wish to minimize their energy,
and the minimal energy can be obtained by going to the region where
$e^{A} << 1$.

\subparagraph{Embiggening}
Now, let us analyze the dynamics of the $p$ anti-D3 branes at the tip.
The metric at the tip of the warped deformed conifold is given by
\begin{equation}
ds^2 \simeq (e^{-2\pi K/3 Mg_s})^2 dx_{\mu}dx^{\mu} + R^2 d\Omega_3^2
+ (dr^2 + r^2 d\tilde \Omega_2^2) \times b_0^2~.
\end{equation}
Here, $b_0$ is a number of order 1, and $R^2 \sim g_s M$.  
In particular at the tip $r=0$, the geometry is well approximated by
an $S^3$ of radius $\sqrt{g_sM}$. 

The flux is also easy to determine; the $H_3$ flux is spread over the radial
direction, while the $F_3$ flux threads the $S^3$ at the tip.  In the
supergravity regime where $g_s M >> 1$, we can solve $\int_A F_3 = M$ by
just setting $F$ proportional to the warped volume form $\epsilon$
on the $S^3$:
\begin{equation}
F_{mnp} = f \epsilon_{mnp},~~f \sim {1\over \sqrt{g_s^3 M}}~.
\end{equation}
So the system we are studying consists of p anti-D3 branes transverse to
a diffuse magnetic 3-form flux (that, is a flux whose flux density is
small in the supergravity regime of large $g_sM$)), or 
equivalently, p anti-D3 branes in
an electric 7-form flux.

This system is T-dual to D0-branes in an electric 4-form flux.   
These D0-branes undergo the famous Myers effect \cite{Myers:1999ps};
p D0-branes in a background flux expand into a fuzzy D2-brane carrying
p units of worldvolume gauge flux (to encode the D0 charge).  Similarly here,
the anti-D3 branes should be expected to expand into 5-branes, carrying
p units of worldvolume flux.  Because we are working in a duality frame
where $S_{CS}$ contains a coupling to $B_6$, in fact the anti-D3s will
expand into an NS 5 brane.

We can see this in equations as follows.  On the large $S^3$, one
can approximate
\begin{equation}
C_{kj} \sim {2\pi \over 3} F_{kjl}\Phi^l,~~G_{kj} \sim \delta_{kj}~.
\end{equation}
Therefore
\begin{equation}
Q^i_j = \delta^i_j + {2\pi i \over g_s} [ \Phi^i, \Phi_j ] 
+ i {4\pi^2 \over 3} F_{kjl} [ \Phi^i, \Phi^k ] \Phi^l~.
\end{equation}
Then
\begin{equation}
Tr(\sqrt{det Q}) \simeq p - i {2\pi^2\over 3} F_{kjl} Tr\left( [ \Phi^k,
\Phi^j ] \Phi^l \right) - {\pi^2\over g_s^2} Tr [\Phi^i, \Phi^j]^2~.
\end{equation}

Now the $B_6$ term in $S_{CS}$ would ${\it cancel}$ the cubic term in
the potential if we were considering D3 branes; they do not undergo a
Myers effect in this background.  On the other hand, for anti-D3 branes,
the $B_6$ term ${\it adds}$ and we find an effective potential
\begin{eqnarray}
\label{vanti}
V_{eff}(\Phi) &=& e^{-8\pi K/3 Mg_s} {\mu_3 \over g_s} \left(
p - i {4\pi^2 f \over 3} \epsilon_{kjl} Tr [\Phi^k, \Phi^j ] \Phi^l
\right.  \nonumber \\ & & \left. - {\pi^2\over g_s^2} Tr [\Phi^i, \Phi^j]^2 + \cdots 
\right)~.
\end{eqnarray}
It is important to emphasize that this potential is exponentially small,
due to the warp factor at the tip of the cone.

Now, demanding ${\partial V_{eff} \over \partial \Phi} = 0$, we find
the equation
\begin{equation}
\label{fuzzy}
[ [\Phi^i, \Phi^j ], \Phi^j ] - ig_s^2 f \epsilon_{ijk} [\Phi^j,\Phi^k] = 0~.
\end{equation}
We can solve this equation by choosing constant matrices $\Phi^i$ that
satisfy
\begin{equation}
\label{sutwo}
[\Phi^i, \Phi^j ] = - i g_s^2 f \epsilon_{ijk} \Phi^k~.
\end{equation}

This is a very familiar equation.
Up to a rescaling of fields, (\ref{sutwo}) is just the commutation relation
satisfied by $p \times p$ matrix representations of the $SU(2)$ generators!
Therefore, we can find extrema of the anti-brane potential, by simply
choosing (generally reducible) $p \times p$ matrix representations of
$SU(2)$, i.e. there is an extremum for each partition of $p$.
The full ``landscape'' of these extrema is somewhat complicated
(see {\it e.g.} \citeasnoun{Jatkar:2001uh,DeWolfe:2004qx} for some remarks about
its structure, and \citeasnoun{Gomis:2005wc} for a more general discussion of
open string landscapes).  What is clear is that the energetically preferred 
solution is the $p$ dimensional irreducible representation, for which
\begin{equation}
V_{eff} \simeq e^{-8\pi K/3 M g_s} \times p {\mu_3 \over g_s} \left( 1- {8\pi^2\over 3} {(p^2-1) \over M}
{1\over b_0^{12}} \right) .
\end{equation}
The ${\it radius}$ of the fuzzy $S^2$ the branes
unfurl into, is given by
\begin{equation}
\label{radis}
\tilde R^2 = {4\pi^2 \over b_0^8} {(p^2-1)\over M^2} \times R^2
\end{equation}
where $R^2 \sim g_sM$ controls the size of the $S^3$ at the tip of
the geometry.

It is clear from (\ref{radis}) that we can only trust this solution for
$p << M$; for larger $p$, the radius $\tilde R$ approaches the radius
of the $S^3$, and global features of the geometry may become important.

\subparagraph{Deflation}
We now comment on the ultimate fate of these non-supersymmetric anti-D3
states in the Klebanov-Strassler throat.  The throat is characterized
by $\int_A F_3 = M,~\int_B H_3 = -K$.  At very large values of the
radial coordinate $r$ (the UV of the dual quantum field theory), the
charge $Q_{tot}$ characterizing the throat with the $p$ probe antibranes
is then:
\begin{eqnarray}
\label{before}
&&\int_A F_3 = M, ~\int_B H_3 = -K, N_{\overline{D3}} = p
\nonumber \\  &&\to Q_{tot} = KM - p~.
\end{eqnarray} 

But there are also supersymmetric states carrying this same total charge;
for instance, one could consider
\begin{eqnarray}
\label{after}
&&\int_A F_3 = M,~\int_B H_3 = -(K-1),~N_{D3} = M-p
\nonumber \\&& \to Q_{tot} = KM - p~.
\end{eqnarray}
Since the two charge configurations (\ref{before}) and
(\ref{after}) have the same behavior at infinity
in the radial coordinate, they should be considered as two 
distinct states in the same theory.  In fact, one can explicitly write down
a vacuum bubble interpolating between them; it consists of an NS 5-brane
wrapping the $A$-cycle, and was studied in detail in \citeasnoun{Kachru:2002gs,
DeWolfe:2004qx}.  This bubble can be interpreted as a bubble of
false vacuum decay, carrying the metastable non-supersymmetric vacuum
(\ref{before}) to a stable supersymmetric vacuum. 
Because the scale of supersymmetry breaking in the initial vacuum is
exponentially small, one can control these states quite well for
$1 << p << M$.  A detailed study shows that as $p$ approaches $M$,
the metastable vacuum disappears; the critical value of $p/M$ is
${\cal O}(1/10)$.

This situation is reminiscent of some recent examples where
direct study of 4d supersymmetric field
theories has uncovered metastable non-supersymmetric vacua 
\cite{Intriligator:2006dd}.
Extending our knowledge of such states (using either gauge/gravity duality
or 4d field theory techniques), and the interrelations between them,
remains a very active area of research.

In addition to their interest as an example of the intricate dynamics
that can occur with branes in flux backgrounds, these states have also
been used in the KKLT proposal to obtain de Sitter vacua in string
theory \cite{Kachru:2003aw}, and play an important role in some
models of string inflation \cite{Kachru:2003sx}.  
\footnote{The argument of section IV.A.1 that 
one cannot solve the IIb equations of motion in 
warped Calabi-Yau flux compactifications if one 
includes anti-D3 sources, is true
only at tree level.  The same effects which allow one to stabilize
the K\"ahler moduli, also allow the incorporation of anti-branes with
sufficiently small (warped) tension, as shown in \citeasnoun{Kachru:2003aw}.} 
Of course, for the
former role, other mechanisms of supersymmetry breaking could serve
as well.  We now discuss two less stringy, but very well motivated,
ideas.

\paragraph{Dynamical Supersymmetry Breaking}

An alternative to using warped compactification to obtain
an exponentially small scale of supersymmetry breaking, is to use 
dimensional transmutation and instanton effects \cite{Witten:1981nf}. 
Many examples of field theories which dynamically break supersymmetry
have been discovered over the years, starting with the work 
of Affleck, Dine and Seiberg summarized
in \citeasnoun{Affleck:1984xz}.  
More recent reviews include \cite{Poppitz:1998vd,
Shadmi:1999jy}. 

It is clear that one can incorporate these dynamical breaking sectors
as part of the low energy physics of a string compactification.
The extra-dimensional picture then does not a priori add much to
the 4d discussion, although to some extent it can be useful in 
``geometrizing'' criteria for different mediation mechanisms to 
dominate \cite{Diaconescu:2005pc}.
Discussions of DSB with gauge or gravity mediation of SUSY breaking to
the Standard Model, in fairly concrete pseudo-realistic 
string compactifications,
appear in \citeasnoun{Diaconescu:2005pc,Franco:2006es,Garcia-Etxebarria:2006rw,
Braun:2006em}.

\paragraph{Breaking by fluxes}
\label{sp:fluxbreak}

Perhaps the most direct analog of the original
Bousso-Polchinski proposal, in the IIb flux landscape, is
the following.  We saw in the
previous subsection that one can supersymmetrically stabilize all moduli
after including nonperturbative corrections to the superpotential which
depend on K\"ahler moduli.  Previous to stabilizing K\"ahler moduli, it
would have seemed that one must solve the no-scale equation $V=0$ to find
a IIb flux vacuum.  However, given that one will stabilize K\"ahler moduli
anyway, it is no longer necessary to do this.  Instead, 
consider the potential $V_{\rm flux}(z_a,\phi)$ arising from the three-form
fluxes.  If one finds a critical point of this potential in the
$a, \phi$ directions, with
\begin{equation}
\partial_{a} V = \partial_{\phi} V = 0,~ \partial^2 V \geq 0
\end{equation}
then the vacuum would be stable in the $a,\phi$ directions (despite the
tree-level instability in the K\"ahler modulus directions).  Now, including
the instanton contributions to $V$,
it becomes clear that one may stabilize
the K\"ahler moduli and complex/dilaton moduli 
while using a ``flux vacuum'' for the complex/dilaton
moduli which is ${\it not}$ of the ISD type, as long as the departure from
the ISD condition is not too severe.   
Naively, any violation of the ISD equations yields, via (\ref{fluxsup}), a
nonzero F-term for some complex/dilaton modulus.  
This means that the resulting vacuum will yield spontaneous supersymmetry
breaking.  
This is too quick: one should really go back and solve the full equations
of the \EFT, including any dependence of the (prefactor to the)
non-perturbative effects on the complex structure moduli.  However, in
many cases, this intuition should be borne out.
A toy model vacuum of this type has been exhibited in
\citeasnoun{Saltman:2004sn}.

Because the effects being used to stabilize K\"ahler moduli are
exponentially small, this mechanism is only viable if one ``tunes'' in 
flux space to find proto-vacua with a very small violation of the
ISD condition.  This was shown to be generically possible in IIb vacua
in \citeasnoun{Denef:2004cf}, as we will discuss in \rfss{susyscaletwo}.

\subsection{Type IIa flux vacua} 
\label{sec:IIaclass}

In this section, we briefly discuss the construction of
Calabi-Yau flux vacua in type IIa string theory.  Our exposition follows
the notation and strategy of \citeasnoun{DeWolfe:2005uu}, using the
${\cal N}=1$ supersymmetric formalism developed in
\citeasnoun{Grimm:2004ua}.  Closely related
work developing the basic formalism for IIa flux compactification
and presenting explicit examples also 
appears in \citeasnoun{Derendinger:2004jn,Kachru:2004jr,Villadoro:2005cu,
Derendinger:2005ph,House:2005yc,Camara:2005dc,Bovy:2005qq,Saueressig:2005es,
Aldazabal:2006up,Benmachiche:2006df,Ihl:2006pp,Acharya:2006ne}.  
Candidate M-theory vacua 
which are in many ways similar to these IIa models were first
described in \citeasnoun{Acharya:2002kv}; see also
\citeasnoun{Behrndt:2005im}.

\subsubsection{Qualitative considerations}
\label{sec:qualiia}

Before we launch into a detailed study, it is worth contrasting the present
case with the class of IIb vacua we just described.  
In the IIa string compactified on a Calabi-Yau space $M$, one can 
imagine turning on background fluxes of both the NS three-form field
$H_3$ and the RR $2p$ form fields $F_{0,2,4,6}$. 
The basic intuition that 3-form fluxes should yield complex structure
dependent potentials, while even-form fluxes should yield K\"ahler
structure dependent potentials, then suggests that the IIa flux
superpotential will depend on all geometric moduli already at tree level.

In fact, if we focus for a moment
on just the dilaton and 
the volume modulus, which are normally two of
the more vexing moduli in string constructions, we can see 
by a simple scaling argument that the flux potential will suffice
to stabilize them in a regime of control.

To find the potentials due to fluxes, one should reduce the flux
kinetic and potential terms from 10d to 4d, remembering to perform
the necessary Weyl rescalings to move to 4d Einstein frame.
These are discussed in a pedagogical way in \citeasnoun{Silverstein:2004sh}.
The results are as follows.  If the compactification manifold has radius 
$R$ and the string coupling is $g_s=e^{\phi}$, then: 

\noindent
$\bullet$
N units of RR p-form flux contributes to the scalar potential with the scaling
\begin{equation}
V_{RR} = N^2 {e^{4\phi}\over R^{6+2p}}
\end{equation}

\noindent
$\bullet$
N units of NS 3-form flux contribute
\begin{equation}
V_{NS} = N^2 {e^{2\phi}\over R^{12}}
\end{equation}

\noindent
$\bullet$
N orientifold p+3 planes wrapping a p-cycle in the compact manifold and filling
spacetime, contribute
\begin{equation}
V_{O(p+3)} = - N {e^{3\phi}\over R^{12-p}}
\end{equation}
(while of course N D-branes would, up to an overall coefficient, make the same
contribution with a positive sign).

The simplest class of ${\cal N}=1$ supersymmetric IIa orientifolds arise
by acting with an anti-holomorphic involution ${\cal I}$
on a Calabi-Yau space $M$.
The fixed locus of  
${\cal I}$ is some collection of special Lagrangian cycles, which are
wrapped by O6 planes.  Let us assume for a moment that there are 
${\cal O}(1)$ O6 planes in our construction.  The tadpole condition for
D6-brane charge takes the schematic form
\begin{equation}
\label{sixtad}
N_{D6} + \int_{\Sigma} F_0 \wedge H_3 = 2 N_{06}~. 
\end{equation}
where $\Sigma$ is the three-cycle pierced by $H_3$ flux.
We can therefore cancel the tadpole by introducing ${\cal O}(1)$
units of $F_0$ and $H_3$ flux,
without adding D6 branes.  The other fluxes are unconstrained by tadpole
conditions; so we can, for instance, also turn on $N$ units of $F_4$ 
flux.  The overall result is a potential that takes the schematic form
\begin{equation}
\label{schempot}
V = {e^{4\phi}\over R^6} - {e^{3\phi}\over R^9} + {e^{2\phi}\over R^{12}} 
+ N^2 {e^{4\phi} \over R^{14}}~.
\end{equation}
This potential has minima with $R \sim N^{1/4}$ and $g_s \sim N^{-3/4}$.
Hence, as emphasized in \citeasnoun{DeWolfe:2005uu}, the IIa theory
can be expected to admit 
flux vacua with parametrically large values of the compactification
volume and parametrically weak string coupling, in a $1/N$ expansion.
Unlike standard Freund-Rubin vacua \cite{Freund:1980xh}, 
these theories are effectively
four-dimensional; the 4d curvature scale is parametrically less than
the compactification radius.

It is still important to verify that the qualitative considerations here are
born out in detail in real Calabi-Yau models.  We now describe the
relevant formalism.

\subsubsection{4d multiplets and K\"ahler potential}

To find the chiral multiplets in a 4d ${\cal N}=1$ supersymmetric
orientifold of $M$, we proceed as follows. 
The ${\cal N}=2$ compactification on $M$ gives rise to $h^{1,1}$
${\cal N}=2$ vector multiplets and $h^{2,1}+1$ hypermultiplets
(including the universal hyper).  The projection will choose an
${\cal N}=1$ vector or chiral multiplet from each ${\cal N}=2$
vector, and an ${\cal N}=1$ chiral multiplet from each hyper.

Let us first analyze the projected vector multiplet moduli space.
If in a basis of (1,1) forms on $M$ there are $h^{1,1}_-$ that are
odd under the involution, then the surviving moduli space of 
K\"ahler forms is $h^{1,1}_-$ dimensional.  
(The even basis elements give rise to ${\cal N}=1$ vector multiplets,
which contain no moduli and will not enter in our discussion).
We can write the complexified
K\"ahler form on the quotient as
\begin{equation}
J_c = B_2 + iJ = \sum_{a=1}^{h^{1,1}_-} t_a \omega_a 
\end{equation} 
where $t_a = b_a + i v_a$ are complex numbers, and $\omega_a$ form a basis for
$H^{1,1}_-$.
The rather surprising fact that elements of $H^{1,1}_-$ correspond to moduli
comes about because the supersymmetric IIa 
orientifolds are based on orientation reversing
involutions, which reverse the sign of $J$ and $B_2$.
Then in the dimensional reduction, these should be expanded in a basis of
two-forms which flip sign under the involution \cite{Grimm:2004ua}.

The K\"ahler potential for the reduced moduli space is inherited from
the ${\cal N}=2$ parent Calabi-Yau theory, and is given by
\begin{equation}
\label{iiakahpot}
K^K(t_a) = -{\rm log}\left( {4\over 3} \int_M J \wedge J \wedge J \right)
= - {\rm log} \left({4\over 3}\kappa_{abc}v^a v^b v^c \right)
\end{equation}
where $\kappa_{abc}$ is the triple intersection form
\begin{equation}
\kappa_{abc} = \int_M \omega_a \wedge \omega_b \wedge \omega_c~.
\end{equation}

Now, we turn to the projected hypermultiplet moduli space.  
Here, the formalism is more intricate \cite{Grimm:2004ua}.
Choose a basis for the harmonic three-forms $\{ \alpha_A, \beta_B \}$
where $A,B = 0,\cdots,h^{2,1}$ and
\begin{equation}
\int_M \alpha_A \wedge \beta_B = \delta_{AB}~. 
\end{equation}
Without loss of generality, one can expand $\Omega$ as
\begin{equation}
\Omega = \sum_A Z_A \alpha_A - g_B \beta_B~.
\end{equation}
The $Z_A$ are homogeneous coordinates on complex structure moduli
space; we will denote by $z_C$ ($C=1,\cdots,h^{2,1}$)
the {\it inhomogeneous} coordinates on this
same space.

The complex structure moduli are promoted to quaternionic multiplets
in the ${\cal N}=2$ parent theory by adjoining RR axions.  If we expand
the $C_3$ gauge potential whose field strength is $F_4$
\begin{equation}
C_3 = \xi_A \alpha_A - \tilde \xi_B \beta_B
\end{equation}
then we get $h^{2,1}+1$ axions.  The axions from $\xi_0, \tilde \xi_0$
join the axio-dilaton to yield the universal hypermultiplet, while the
other $h^{2,1}$ axions quaternionize the $z_C$.

The orientifold involution splits $H^3 = H^3_+ \oplus H^3_-$.  Each
of these eigenspaces is of (real) dimension $h^{2,1}+1$.  Let us split
the basis for $H^3$ so $\{ \alpha_k,\beta_\lambda \}$ span the even
subspace, while $\{ \alpha_\lambda, \beta_k \}$ span the odd subspace.
Here $k=0,\cdots,\tilde h$ while $\lambda = \tilde h+1,\cdots, h^{2,1}$. 
Then the orientifold restricts one to the subspace of moduli space
\cite{Grimm:2004ua} 
\begin{equation}
Im Z_k = Re g_k = Re Z_{\lambda} = Im g_{\lambda} = 0~.
\end{equation}
$C_3$ is also even under the orientifold action; hence, one keeps 
the axions $\xi_k$ and $\tilde \xi_\lambda$ while projecting
out the others.
In addition, the dilaton $\phi$ and
one of $\xi_0, \tilde \xi_0$ are kept in the spectrum of the orientifold.  
So as expected, from each hypermultiplet, we get a single chiral
multiplet,
whose scalar components are the real or imaginary part of the complex
structure modulus, and an RR axion.

We can summarize the surviving hypermultiplet moduli in terms of
the object  
\begin{equation}
\Omega_c = C_3 + 2i {\rm Re} \left(C \Omega \right)~. 
\end{equation}
Here, $C$ is a ``compensator'' which incorporates the dilaton
dependence via
\begin{equation}
C = e^{-D + K^{cs}/2},~~e^D = \sqrt{8} e^{\phi + K^K/2}~. 
\end{equation} 
One should think of $e^D$ as the four-dimensional dilaton; $K^{cs}$
is the K\"ahler potential for complex structure moduli
\begin{eqnarray}
K^{cs} &=& -{\rm log} \left( i \int_M \Omega \wedge \bar \Omega \right)
\nonumber \\ &=& -{\rm log}~ 2(Im Z_{\lambda} Re g_{\lambda} - Re Z_k Im g_k )~.
\end{eqnarray}

The surviving chiral multiplet moduli are then the expansion of 
$\Omega_c$ in a basis for $H^3_+$:
\begin{equation}
N_k = {1\over 2} \int_M \Omega_c \wedge \beta_k = {1\over 2}\xi_k + 
i {\rm Re}(CZ_k)
\end{equation}
and 
\begin{equation}
T_{\lambda} = i \int_M \Omega_c \wedge \alpha_{\lambda} =
i \tilde\xi_{\lambda} - 2 {\rm Re}(Cg_{\lambda})~.
\end{equation}

The K\"ahler potential which governs the metric on this moduli space is
\begin{equation}
K^Q = -2 {\rm log}\left( 2 \int_M {\rm Re}(C\Omega) \wedge *
{\rm Re}(C\Omega) \right)~.
\end{equation} 

\subsubsection{Fluxes and superpotential}

Now, we can contemplate turning on the fluxes which are projected in by
the anti-holomorphic involution.  It turns out that $H_3$ and $F_2$
must be odd, while $F_4$ should be even.  So we can write
\begin{equation}
H_3 = q_\lambda \alpha_\lambda - p_k \beta_k, ~~
F_2 = -m_a \omega_a,~~
F_4 = e_a \tilde \omega^a
\end{equation}
where $\tilde \omega^a$ are the 4-form duals of the $H^{1,1}_-$ basis
$\omega_a$.  
To see that they are even under the orientifold involution (as they must
be to enter in the expansion of $F_4$), one can simply recall that
when nonzero, $\tilde \omega_a \wedge \omega_b \sim {\rm Vol}$ and
the volume form changes sign under the involution.
There are in addition two parameters $m_0$ and $e_0$,
parametrizing the $F_0$ and $F_6$ flux on $M$.  

In the presence of these fluxes, one can write the 4d potential
after dimensional reduction as \cite{Grimm:2004ua,DeWolfe:2005uu}
\begin{equation}
V = e^K \left( \sum_{t_a, N_k, T_{\lambda}} g^{i\bar j} D_i W {\overline
{D_j W}} - 3 |W|^2 \right)~. 
\end{equation}
Here the total K\"ahler potential is 
\begin{equation}
\label{IIak}
K = K^K + K^Q 
\end{equation}
and $D_i W = \partial_i W + W \partial_i K$ is the K\"ahler covariantized
derivative.

The superpotential $W$ is defined as follows.  Let 
\begin{equation}
W^Q(N_k, T_{\lambda}) = \int_M \Omega_c \wedge H_3 =
-2p_k N_k - iq_\lambda T_{\lambda}
\end{equation}
and
\begin{equation}
W^K(t_a) = e_0 + \int_M J_c \wedge F_4 - {1\over 2}\int_M J_c \wedge J_c 
\wedge F_2 - {m_0\over 6} \int_M J_c \wedge J_c \wedge J_c~. 
\end{equation} 
The full superpotential is then 
\begin{equation}
\label{IIaW}
W(t_a, N_k, T_{\lambda}) = W^Q(N_k, T_{\lambda}) + W^K(t_a)~. 
\end{equation}

Our first qualitative point is now clear: the potential depends,
in general, on all geometric moduli at tree level.  Detailed examination
of the system of equations governing supersymmetric vacua 
\begin{equation}
D_{t_a}W = D_{N_k}W = D_{T_{\lambda}}W = 0
\end{equation}
shows that 
under reasonable assumptions of genericity, one can stabilize all 
geometric moduli
in these constructions \cite{DeWolfe:2005uu}.  
These same considerations show that in the leading
approximation, $h^{2,1}_+$ {\it axions} will remain unfixed. 
An orientifold of a rigid Calabi-Yau model (i.e., one with $h^{2,1}=0$) was
studied in detail in \citeasnoun{DeWolfe:2005uu}, where it was shown that
this flux potential gives rise to an infinite number of 4d vacua with
all moduli stabilized.
Furthermore, as suggested by the scaling argument in 
(\ref{sec:qualiia}), these solutions can be brought into a regime
where $g_s$ is arbitrarily weak and the volume is arbitrarily large. 
We note that to get a semi-realistic vacuum, one must impose physical 
criteria (a lower bound on KK masses, or a lower bound on the gauge
couplings) which will regulate the infinite number of models found here.
At present, it seems only a finite (though perhaps very large) number
of such models can satisfy even such basic physical cuts.

\subsubsection{Comments on 10d description}

The 10d description of the IIa solutions is less well understood than 
the description of their IIb counterparts.  That is because in the IIb
case, one special class of solutions is conformally Calabi-Yau, at
leading order \cite{Giddings:2001yu}.  In
the IIa case, on the other hand, the metrics of the supersymmetric
compactifications are those of ${\it half-flat}$ manifolds with
$SU(3)$ structure.  The definition of such spaces 
can be found in
\citeasnoun{Chiossi:2002tw}, 
and their relation to supersymmetric IIa compactification
is described in 
\citeasnoun{Behrndt:2004km,Behrndt:2004mj,Lust:2004ig,%
Behrndt:2005bv,House:2005yc}.  

It is natural to wonder what relation these half-flat solutions bear to 
the Calabi-Yau flux vacua we have been discussing, where the fluxes are
viewed as a perturbation of a IIa Calabi-Yau compactification.  This issue
has been clarified in \cite{Acharya:2006ne}.
The description in terms of a Calabi-Yau metric perturbed by backreaction
from the flux (and inclusion of thin-wall brane sources) 
is valid at asymptotically large volume.  
Finite (but large) volume analysis
of the supergravity solution with localized O6-planes, indicates that
the backreaction deforms the metric to a half-flat, non Calabi-Yau
metric with $SU(3)$
structure, outside a small neighborhood of the O-planes.  
The detailed formulae for the stabilization of moduli
derived from the considerations of the previous subsection, can be
recovered precisely from the supergravity solution
in the approximation that the O6 charge is smeared.

\subsection{Mirror symmetry and new classes of vacua}
\label{ss:mirrorsym}

The constructions we have reviewed in detail here, are based at the start
on type II Calabi-Yau models.  
Such models enjoy mirror symmetry, a duality exchanging the
IIb string on $M$ with a IIa string on a ``mirror manifold'' $W$.
Dual theories must give rise to the same 4d physics, though in different
regimes of parameter space one or the other may be a better description.
Therefore, we see immediately that simply to match dimensions of hyper
and vector multiplet moduli spaces, one must have $h^{1,1}(M) = h^{2,1}(W)$
and $h^{2,1}(M) = h^{1,1}(W)$.  This can be viewed as a mirror reflection
on the Hodge diamond of a Calabi-Yau space, which explains the name
of the duality.  

It is natural to wonder whether, since the parent Calabi-Yau theories
enjoy mirror symmetry, the classes of flux vacua we have constructed
above also come in mirror pairs.  Are they dual to one another in some way?
This seems unlikely, given the qualitative differences between the two
classes of 4d \EFT's.  We will see that it isn't the case,
but that nevertheless exploring analogues of mirror symmetry 
for these vacua will lead us to interesting conclusions, 
suggesting that the known portion of the
landscape 
is a small piece of a much larger structure.

While mirror symmetry was used to great effect already in 
\citeasnoun{Candelas:1990rm}, methods of constructing 
$W$ given $M$ were known then only for very special classes of models
\cite{Greene:1990ud}.  An important advance came during the duality
revolution, when it was realized that mirror symmetry is
a generalization of T-duality \cite{Strominger:1996it}.
The known examples of Calabi-Yau spaces admit, in some limit, a
fibration structure where $T^3$ fibers vary over an $S^3$ base.
By matching of BPS states, it
was shown that
mirror symmetry can thought of as the operation
of T-dualizing the $T^3$ fibers of $M$ to obtain $W$, and vice-versa.

Since the study of $T^3$ fibrations is in its infancy, it may
seem surprising that this construction will be of use to us.  However,
it is a conceptually simple relation of a space and its mirror, and it
will allow us to check whether the mirrors of IIb Calabi-Yau flux vacua are 
IIa Calabi-Yau flux vacua.  This general subject has been explored in
\citeasnoun{Gurrieri:2002wz,Kachru:2002sk,Fidanza:2003zi,%
Bouwknegt:2003wp,Fidanza:2004wa,Chiantese:2004tx,Tomasiello:2005bp,%
Grana:2005ny}. 

\subsubsection{A warm-up: The twisted torus}

We first provide an illustrative
example that should make our conclusions intuitively clear.  We follow
the discussion in \citeasnoun{Kachru:2002sk}.  
Imagine string compactification on a square $T^3$, $M$, with metric
\begin{equation}
ds^2 = dx^2 + dy^2 + dz^2 
\end{equation}
and a nonzero NS three-form flux
\begin{equation}
\int_M H_3 = N~.
\end{equation}
Since $H=dB$, we are free to choose a gauge in which
\begin{equation}
B_{yz}=Nx
\end{equation}
with other components vanishing.
This configuration is not
a static solution of the equations of motion; the $T^3$ is flat so there is
no curvature contribution to the lower-dimensional effective potential,
while the $H_3$ flux energy can be diluted by expanding the volume of
the $T^3$.  We ignore this for now; we will use this setup as a module
in a more complicated configuration that provides a static solution of
the full equations of motion momentarily.

With the data at hand, we can proceed to T-dualize in the $z$ direction.
Applying Buscher's T-duality rules \cite{Buscher:1987sk,Buscher:1987qj} 
(their generalizations to include RR fields 
\cite{Bergshoeff:1995as,Hassan:1999bv}
will also play a role momentarily), 
we find a new background with:
\begin{equation}
B = 0, ~~ds^2 = dx^2 + dy^2 + (dz + Nx dy)^2
\end{equation}
The coordinate identifications to be made in interpreting the metric
are
\begin{equation}
(x,y,z)\simeq (x,y+1,z)\simeq (x,y,z+1) \simeq (x+1,y,z-Ny)
\end{equation}
This space is an example of a Nilmanifold -- it has $h^{1}=2$, and
in particular is topologically distinct from $T^3$, which would have
been the expected T-dual target space in the absence of $H_3$ flux. 
So, we see that T-dualizing along a leg of an $H_3$ flux, one can
exchange the NS flux for other NS data -- namely, topology, as
encoded by the metric.  
Here, 
the nontrivial topology arises because as one winds around the $x$ circle,
one performs an $SL(2,\IZ)$ transformation mixing the $y,z$ directions.

If we T-dualize again, now along the $y$ direction,
straightforward application of the rules leads us to the metric
\begin{equation}
ds^2 = {1\over {1+ N^2 x^2}} (dz^2 + dy^2) + dx^2
\end{equation}
and the B-field
\begin{equation}
B_{yz} = {Nx \over {1 + N^2 x^2}}~. 
\end{equation} 
Making sense of this data is not as simple as interpreting the Nilmanifold
metric above.  In particular, as you wind around the circle coordinatized
by $x$, the metric $g$ and $B$ are not periodic in any obvious sense.
There is a ``stringy'' sense in which they ${\it are}$ periodic;
there is an $O(2,2;\IZ)$ transformation that relates the
values at $x=1$ to the values at $x=0$.
However, this $O(2,2;\IZ)$ transformation is not an element of 
$SL(2,\IZ)$, and so this data can at best make sense as the target
space of a ``stringy'' sigma model.  Discussions of such non-geometric
backgrounds (including and generalizing asymmetric orbifolds)
are a subject of current interest; see for instance
\citeasnoun{Silverstein:2001xn,Hellerman:2002ax,%
Dabholkar:2002sy,Hull:2004in,Flournoy:2004vn,Hull:2005hk,%
Dabholkar:2005ve,Flournoy:2005xe,Shelton:2005cf,Lawrence:2006ma,%
Hull:2006qs,Hull:2006va}.  

In the following, we will focus our discussion on open questions about
the geometric vacua.  However, these considerations suggest
that once one considers general vacua, novel 
``stringy'' geometric structures will play
an important role in obtaining a thorough understanding.

\subsubsection{A full example}

We will now provide full string solutions which incorporate the
previous phenomena.  We follow \citeasnoun{Kachru:2002sk}; see also
\cite{Schulz:2004ub,Schulz:2004tt,Grana:2006kf} for further discussion of
these models. 

Consider IIb string theory on 
the $T^6/\IZ_2$ orientifold, where the $\IZ_2$ inverts
all six circles (and is composed with the operation of worldsheet
parity reversal).  
For simplicity, focus attention on a $(T^2)^3$, with complex moduli
$\tau_{1,2,3}$:
\begin{equation}
dz^i = dx^i + \tau_i dy^i,~~\Omega = \Pi_i dz^i
\end{equation}
Flux vacua in this model were studied in {\it e.g.} \citeasnoun{Dasgupta:1999ss,
Kachru:2002he,Frey:2002hf}.  One example from \citeasnoun{Kachru:2002sk}
will suffice for us.  Let
\begin{equation}
F_3 = 2 \left( dx^1 \wedge dx^2 \wedge dy^3 + dy^1 \wedge dy^2 \wedge dy^3
\right)
\end{equation}
and
\begin{equation}
H_3 = 2 \left (dx^1 \wedge dx^2 \wedge dx^3 + dy^1 \wedge dy^2 \wedge dx^3
\right).
\end{equation}
The factor of 2 is inserted in order to avoid subtleties with flux
quantization of the sort described in \citeasnoun{Frey:2002hf}.
We can easily read off the flux superpotential
\begin{equation}
W = 2 (\tau_1 \tau_2 \tau_3 + 1) + 2\phi (\tau_1 \tau_2 \tau_3 + \tau_3)~. 
\end{equation}
It is easy to see that along the locus
\begin{equation}
\label{Mis}
\phi\tau_3 = -1,~~\tau_1 \tau_2 = -1
\end{equation}
the equations ${\partial W \over {\partial \tau_i}} = {\partial W \over \partial
\phi} = 0$ are satisfied, as is $W=0$.  Therefore, there is a moduli
space ${\cal M}$ of supersymmetric vacua.  In 
fact, these vacua preserve ${\cal N}=2$
supersymmetry -- this is a special feature which arises because the torus
is a non-generic Calabi-Yau space. 
The non-genericity of the torus also implies that one should impose
primitivity conditions $J \wedge G_3 = 0$ on $G_3$; for a given choice
of the integral fluxes, this becomes a constraint on K\"ahler moduli. 

We have chosen our fluxes so that in appropriate regions of ${\cal M}$,
the ``best" description (i.e. the description which seems to be most
weakly coupled, among known duality frames)
is either the model above, or its
T-dual on one, two or three circles.  In 
the gauge
\begin{equation}
B_{x^1 x^3} = 2x^2,~~B_{y^1 x^3} = 2y^2
\end{equation}
the relevant T-dual descriptions are the following.

\medskip
\noindent{\it One T-duality along $x^1$:}
\medskip

This gives rise to a IIa model with metric
\begin{equation}
ds^2 = {1\over R_{x^1}^2} (dx^1 + 2x^2 dx^3)^2 + 
R_{x^2}^2 (dx^2)^2 + R_{x^3}^2 (dx^3)^2 + \cdots 
\end{equation}
Here, $x^{1,2,3}$ sweep out a Nilmanifold over the $T^3$ spanned
by the $y^i$.  There are also nonzero fluxes remaining:
\begin{equation}
B_{y^1 x^3} = 2y^2
\end{equation}
in the NS sector and
\begin{equation}
F_2 = 2 dx^2 \wedge dy^3,~~F_4 = 2 (dx^1 + 2x^2 dx^3) \wedge dy^1
\wedge dy^2 \wedge dy^3
\end{equation}
in the RR sector.

This manifold has $h^{1}=5$ and is non-K\"ahler.  In particular, it
isn't just that one does not use a Calabi-Yau metric in describing
the physical theory (that is true even for Calabi-Yau compactification,
where $\alpha'$ corrections deform the metric even in the absence of
flux).  There is a topological obstruction to putting such a metric
on this space. 

\medskip
\noindent{\it Second T-duality along $y^1$:}
\medskip

Now we find the IIb theory with metric
\begin{eqnarray}
\label{twometric}
ds^2 &=& \tilde R_{x^1}^2 (dx^1 + 2x^2 dx^3)^2 + R_{x^2}^2 (dx^2)^2
\nonumber\\& &+ R_{x^3}^2 (dx^3)^2 
+ {1\over R_{y^1}^2} (dy^1 + 2y^2 dx^3)^2 \nonumber \\ & &
+ R_{y^2}^2 (dy^2)^2 + R_{y^3}^2 (dy^3)^2
\end{eqnarray}
and with fluxes
\begin{eqnarray}
B = 0, ~~F_3 &=& 2 (dx^1 + 2 x^2 dx^3) \wedge dy^2 \wedge dy^3 \nonumber \\& &+
2 (dy^1 + 2y^2 dx^3) \wedge dx^2 \wedge dy^3~.
\end{eqnarray} 
This space is also non-K\"ahler.

\medskip
\noindent{\it Third T-duality along $y^3$:} 
\medskip

This just flips the radius of the $y^3$ circle in
(\ref{twometric}) and changes the flux to
\begin{equation}
F_2 = 2 (dx^1 + 2x^2 dx^3) \wedge dy^2 + 2 (dy^1 + 2 y^2 dx^3) \wedge dx^2~. 
\end{equation}

At this point, we have T-dualized on some $T^3$ in the original starting
model, and so we can consider this an analog of mirror
symmetry in the spirit of \citeasnoun{Strominger:1996it}.  
This example suggests that 
the IIb Calabi-Yau flux vacua of the general class studied in
(\ref{sec:IIbclass}) are ${\it not}$ mirror to the IIa  
Calabi-Yau flux vacua described in (\ref{sec:IIaclass}). 
In less simple examples, we expect that dualizing flux vacua with
one leg of the $H_3$ flux along the $T^3$ fiber could lead to a geometric
but non Calabi-Yau dual, while duals of theories with more legs 
of $H_3$ on the $T^3$ fiber
will
in general be ``non-geometric'' vacua.  
 
One might wonder whether every geometric flux vacuum admits
${\it some}$ dual description that brings it
into one of the two large classes we've explored in the IIb and IIa
theories in the previous subsections.
A study of examples strongly suggests that this is false.
For instance, the class of vacua described in \citeasnoun{Chuang:2005qd}
does not admit a dual description involving IIa, IIb or heterotic
Calabi-Yau compactification. 
Furthermore, \citeasnoun{Grana:2006kf} exhibit an explicit vacuum based on
Nilmanifold compactification that is not dual 
to a Calabi-Yau with flux.

\section{STATISTICS OF VACUA}
\label{sec:stat}

Given a systematic construction of a set of string vacua, besides
working out individual examples, one can try to get some understanding
of the possibilities from statistical studies.  As we mentioned
earlier, such studies date back to the late 1980's, and while at that time
moduli stabilization was not understood, still interesting results were
obtained.  Perhaps the most influential of these came out of the
related study of the set of Calabi-Yau threefolds, which provided the
first evidence that mirror symmetry was a general phenomenon
\cite{Candelas:1994bu,Kreuzer:1991gf}.  We will briefly review
some of these results in \ref{sss:cystat}.

The systematic constructions we have discussed of flux superpotentials
and other effective potentials enable us for the first time to find
statistics of large, natural classes of stabilized vacua.  In this
section, we describe a general framework for doing this
\cite{Douglas:2003um}, and some of these results.
See \citeasnoun{Kumar:2006tn,Douglas:2004zg} 
for other recent reviews.

The large number of flux vacua suggests looking for commonalities with
other areas of physics involving large numbers, such as statistical
mechanics.  As it turns out, there are very close analogies with the
theory of disordered systems, in which one constructs idealized models
of crystals with impurities, spin glasses and other disordered
systems, by taking a ``random potential.''  In other words, one
chooses the potential randomly from an ensemble of potentials chosen
to reflect general features of the microscopic physics, and does
statistical studies.  Given a simple choice of ensemble, one can even
get analytic results, which besides adding understanding, are
particularly important in studying rare phenomena.  As we will
explain, by treating the ensemble of flux superpotentials as a random
potential, one can get good analytical results for the distribution of
flux vacua, which bear on questions of phenomenological interest.

We begin our treatment with a careful explanation of the definitions,
as while they are simple, they are different from those commonly used
in statistical mechanics and quantum cosmology.  This is so that we
can avoid ever having to postulate that a given vacuum ``exists'' or
``is created'' with a definite probability, an aspect of the theory
which, as we discussed in \rfss{measure},
is not well understood at present.
Rather than a probability distribution, we will discuss vacuum
counting distributions, which can be unambiguously defined.

One reason to be careful about these definitions, is that the need
for making theoretical approximations will lead us to introduce
approximate vacuum counting distributions, which are also
interpreted in probabilistic terms.  However, the underlying definition of
probability in this case is clear; it expresses our confidence in the
particular theoretical arguments being used, and in this sense is
subjective.  The payoff for this methodological interlude will
be a clear understanding of how statistics of string theory vacua
can lead to a precise definition of {\it stringy naturalness}, as
introduced in \rfsss{susyscale}.

We proceed to describe counting of flux vacua, and some of the exact
results.  This will enable us to continue the discussion of
\rfsss{susyscale} on the scale of supersymmetry breaking.
We continue with a brief survey of what is known about other
distributions, such as of Calabi-Yau manifolds, and distributions
governing the matter content.  We then survey various simpler
distributions which have been suggested as models for the actual
distributions coming from string theory.  
Finally, we discuss the general interpretation of statistical results,
and the prospects for making arguments such as those in \rfsss{susyscale}
precise.

\subsection{Methodology and basic definitions}
\label{ss:method}

Suppose we have a large class of vacua, constructed along the lines
of \rfs{con} or otherwise.  As we discussed, we have no {\it a priori}
reason to prefer one over the other.  While we have many {\it a posteriori}
ways to rule out vacua, by fitting data, computing measure factors, or
otherwise, this requires detailed analysis to do.  In this situation,
we may need to know the distributions of vacua, or of their observable
properties, to make theoretical progress.

As in \rfsss{susyscale}, it is useful to motivate the subject as an
idealization of the problem of testing string theory.  If we had a list
of all possible vacua, call these $V_i$, then all
we would need to do this, is to compute a list of the observables for each,
and check whether the actual observables appear on this list.

To be a bit more concrete, let us grant that physics is well described
by the Standard Model, a 4d EFT $T_{SM}$.
The problem would then reduce to finding the list
of all the \EFT's $T_i$ which are low energy limits of the vacua
$V_i$, and checking whether $T_{SM}$ appears on this list.

To be even more concrete, let us consider the data which goes into
explicitly specifying a particle physics \EFT.  This will include both
discrete and continuous choices.  Discrete choices include the gauge
group $G$ and matter representation $R$ of fermions and bosons.
Choices involving parameters include the effective potential, Yukawa
couplings, kinetic terms and so on; let us denote the vector
containing these parameters as $\vec g$.  While we will not do it
here, in a complete discussion, we would need to specify the cutoff
prescription used in defining the \EFT\ as well.  In any case,
we can regard the sum total of these choices
as defining a point $T_i=(G,R,\vec g)$ in a ``theory space'' ${\cal T}$.

Now, given a set of vacua $\{T_i\}$, the corresponding
vacuum counting distribution is a density on ${\cal T}$,
\begin{equation}\label{eq:dNdef}
dN_{vac}(G,R,\vec g) = \sum_i \delta(G,G_i) 
 \delta(R,R_i) \delta^{(n)}(\vec g - \vec g_i) ,
\end{equation}
or, for conciseness,
\begin{equation}\label{eq:dNdefT}
dN_{vac}(T) = \sum_i \delta(T-T_i) .
\end{equation}
Its integral over a subset of theory space ${\cal R}\subset {\cal T}$
is the number of vacua contained in this subset,
\begin{equation}\label{eq:NvacR}
N({\cal R}) = \int_{\cal R} dN_{vac}~. 
\end{equation}

It should be clear that \rfn{dNdef} contains the same information as
the set of vacua $\{T_i\}$.  What may be less obvious, but will emerge
from the discussion below, is that one can find useful approximations
to such distributions, which are far easier to compute than the actual
vacuum counting distribution.  This is because these distributions
show a great deal of structure, which is {\it not} apparent if one
restricts attention to quasi-realistic models from the start.  This
observation is the primary formal motivation for introducing the
definition.

At this point, if the definition \rfn{dNdef} is clear, one can proceed
to the next subsection.  However, since many similar but different
definitions can be made, and the issue of interpretation may confuse
some readers, let us briefly expand on these points.

To eliminate one possible source of confusion at the start, the list
we are constructing is of ``possible universes'' within string
theory. Our own universe at the present epoch is supposed to
correspond to one of these universes, not some sort of superposition
or dynamical system which explores multiple vacua.  The point of the
list, or the equivalent distribution \rfn{dNdef}, is simply to have a
precise way to think about the totality of possibilities.

Another possible confusion is between \rfn{dNdef}, and the definition
of a measure factor used in quantum cosmology.
As we discussed in \rfss{measure}, to 
define a measure factor, we need to assign a ``probability factor'' to each
vacuum, call this $P(i)$.
The measure factor corresponding to a given list $\{T_i\}$ and
probability factor $P(i)$ is then
\begin{equation}\label{eq:dmudef}
d\mu_P(T) = \sum_i P(i)~ \delta(T-T_i) .
\end{equation}
Its integral over a region ${\cal R}$, gives the probability with which 
a vacuum in ${\cal R}$ will be produced in the 
associated cosmological model.

We already discussed some aspects of the interpretation of such
distributions in \rfss{measure}, and we will continue this in
\rfss{interp}.  The main point we want to
make here is simply that, unlike a measure factor, a vacuum
counting distribution is {\bf not} a probability distribution, and
does not require any concept of a ``probability that a universe of
type $T$ exists'' for its definition.  Rather, it summarizes
information about the set of consistent vacua of the theory.

\subsubsection{Approximate distributions and tuning factors}

A reason to be careful about the difference between a vacuum counting
distribution and a measure factor, is so that we can properly
introduce the idea of an approximate vacuum counting distribution.  To
motivate this, suppose that we know how to construct a set of vacua
$V_i$, but that our theoretical technique is not adequate to compute
the exact value of a coupling $g$ in each vacuum, only some
approximation to it.  In practice this will always be true, but it
gains particular significance for parameters which we must fit to an
accuracy far better than our computational abilities, with the prime
example being the cosmological constant as we discussed earlier.

Suppose for sake of discussion that we are interested in the
cosmological constant $\Lambda$, but can compute it only to an
accuracy roughly $\Delta\Lambda$.  We might model our relative ignorance by
modifying our definition \rfn{dNdef} to
\begin{equation}\label{eq:dNapprox}
dN_{vac}(\Lambda) = \sum_i \frac{1}{\sqrt{\pi}\Delta\Lambda} \exp
 -\frac{(\Lambda-\Lambda_i)^2}{(\Delta\Lambda)^2} ,
\end{equation}
a sum of Gaussian distributions of unit weight.  The choice of
the Gaussian, while not inevitable, would follow if the total error
was the sum of many independent terms, which is reasonable as the
cosmological constant receives corrections from many sectors in the theory.

If we use the resulting approximate vacuum counting distribution
to compute integrals like \rfn{NvacR}, we will get results like
``we expect region ${\cal R}$ to contain half a vacuum,'' or perhaps $10^{-10}$
vacua.  What could this mean?

Of course, given that string theory  and the effective potential
have a precise definition, any
particular vacuum has some definite cosmological constant
$\Lambda_{i,true}$.  The problem is just that we don't know it.  In
modelling our ignorance with a Gaussian (or any other distribution),
we have again introduced probabilities into the discussion -- but note
that this is a different and less problematic sense of probability
than the $P(i)$ we introduced in discussing the measure factor.  It is
not intrinsic to string theory or cosmology, but rather it expresses
our judgement of how accurate we believe our theoretical computations
to be, and nothing else.  As such, it is a technical device, but a
useful one as we shall see.

Having understood this, the meaning of results like ``we expect region
${\cal R}$ to contain $10^{-10}$ vacua'' in this context
becomes clear.  In actual fact, the region
must contain zero, one or some other definite number of vacua.  While
given the theoretical information to hand, we do not know the actual
number, we now have good reasons to think ${\cal R}$ probably does not contain
any vacua.  However, this conclusion is not ironclad; numerical
coincidences in the computations might put one or more vacua into ${\cal R}$.
If our model for the errors is correct, the probability of this happening
is $10^{-10}$, in the usual ``frequentist'' sense: if we have $10^{10}$ 
similar regions to consider, we expect one of them to actually
contain a vacuum.

The reader will probably have already realized that what we have just
discussed, gives a precise sense within string theory to the usual
discussion of fine tuning made in \EFT.  Although
in principle every coupling constant in every string vacuum has some
definite value, and in this sense is ``tuned'' to arbitrary precision,
in practice we cannot compute to this precision, and need to work with
approximations.  The preceding discussion gives us a way to do this and
to combine the results of various approximations.  This
could be used to justify the style of discussion we made in 
\ref{sss:susyscale}, where we compared hypothetical
numbers of vacua with and without low energy supersymmetry.  
In combining the ingredients of
an approximate vacuum counting distribution, small
tuning factors can be compensated by multiplicity factors, to produce 
seemingly counterintuitive results.  We will come back to this idea
after discussing some concrete results.

Of course, the specific ansatz \rfn{dNapprox} was a way to feed in
explicit knowledge about computational accuracy and tuning.  As we
will see, there are many other approximations one might make in
computing a vacuum counting distribution, sometimes with explicit
control parameters and sometimes not, but with the same general
interpretation.  We will discuss the ``continuous flux approximation''
in some detail below.

Finally, let us cite the standard statistical concept of a {\it
representative} sample.  This is a sample from a larger population,
in which the distribution of properties of interest well approximates 
the distribution in the larger population.  Given a representative 
sample of $N_{rep}$ vacua, their distribution $dN_{rep}$,
and the total number of vacua $N_{vac}$, 
we could infer an approximation to the total vacuum counting distribution,
$$
dN_{vac}(T) \sim \frac{N_{vac}}{N_{rep}} dN_{rep}(T) .
$$
While elementary, this idea is probably our main hope of ever characterizing
the true $dN_{vac}$ of string/M theory in practice, 
so making careful use of it is likely to become an increasingly
important element of the discussion.

\subsection{Counting flux vacua}
\label{ss:count}

The simplest example of the general framework we are about to describe
is the counting of supersymmetric IIb flux vacua for a Calabi-Yau with
no complex structure moduli.  This leads to flux vacua with stabilized
dilaton-axion, and a one-parameter distribution which can be worked
out using elementary arguments.  One can explicitly see the nature of
the continuous flux approximation.

We then discuss general results for supersymmetric vacua, an
explicit two parameter distribution, and
some of the general conclusions
from this analysis.  Finally, we discuss the
formalism, which combines elements of random potential theory with the
mathematical theory of random sections of holomorphic line bundles.

\subsubsection{IIb vacua on a rigid CY}
\label{sss:rigidcount}

This problem was studied in \citeasnoun{Ashok:2003gk,Denef:2004ze}.  
We write $\tau$ for the
dilaton-axion; by definition it must satisfy $\Im\tau >0$, in other
words it takes values in the upper half plane.  A rigid CY, for
example the resolved $T^6/\IZ_3$ orbifold,  has
$b^{2,1}=0$ and thus $b^3=2$; thus there are two NS fluxes $a_i$
and two RR fluxes $b_i$, which we take to be integrally quantized.

The flux superpotential Eq. (\ref{fluxsup}) is
$$
W = (f_1+ \Pi f_2)\tau + g_1+ \Pi g_2 \equiv F \tau + G ,
$$
where we group the NS and RR fluxes into two complex combinations $F$
and $G$.  Here 
$$
\Pi\equiv \frac{\int_B \Omega}{\int_A \Omega}
$$
is a complex number which is determined by the geometry of the CY;
let us take $\Pi=i$ for simplicity.

The K\"ahler potential on this moduli space is $K=-\log\Im\tau$, and
it is very easy to solve $DW=0$ for the location of the supersymmetric
vacuum as a function of the fluxes; it is
\begin{equation}\label{eq:solvetau}
DW=0 \leftrightarrow \bar\tau = -{G\over F} ,
\end{equation}
so there will be a unique vacuum if $\Im G/F > 0$, and otherwise none.

The tadpole condition Eq. (\ref{tadpole}) becomes
\begin{equation}\label{eq:simptad}
\Im F^* G \leq L .
\end{equation}
Finally, the $SL(2,\IZ)$ duality symmetry of IIb superstring theory
acts on the dilaton and fluxes as
\begin{equation}\label{eq:sltwoZ}
\left(\matrix{
a& b\cr
c& d }\right) : \ \ %
\tau \rightarrow \tau' = \frac{a\tau+b}{c\tau+d}; \ \ %
\left(\matrix{ F\cr G }\right) 
 \rightarrow
\left(\matrix{ a F + b G\cr c F + d G }\right).
\end{equation}
Two flux vacua which are related by an $SL(2,\IZ)$
transformation are physically equivalent, and should only be counted once.
Since the duality group is infinite, gauge fixing this symmetry is
essential to getting a finite result.

A direct way to classify these flux vacua, is to first enumerate all
choices of $F$ and $G$ satisfying the bound \rfn{simptad}, taking one
representative of each orbit of \rfn{sltwoZ}, and then to list the flux
vacua for each.  Now it is not hard to see 
that this can be
done by taking $f_2=0$, $0\le g_1<f_1$ and $f_1 g_2 \leq L$.  By
\rfn{solvetau}, each choice of flux stabilizes a unique vacuum, and
thus the total number of vacua is finite,
\begin{equation}\label{eq:totalnum}
N_{vac}(L) = 2 \sigma(L) = 2 \sum_{k|L} k ,
\end{equation}
where $\sigma(L)$ is a standard function discussed in textbooks
on number theory, with the asymptotics
$$
\sum_{L\le N} \sigma(L) =  {\pi^2\over 12}N^2 + \CO(N log N) ~. 
$$

Finally, we can use \rfn{solvetau} to get the distribution of vacua
in configuration space.  Let us suppose that in the resulting low energy
theory, $\tau$ controls a gauge coupling, but there is no direct dependence
of the low energy theory on the values $F, G$ of the fluxes apart
from the dependence encoded in $\tau$.  In this case, it is useful to
use $SL(2,\IZ)$ transformations to bring all of the vacua into the
fundamental region $|\tau|\ge 1$ and $|\Re\tau| \le 1/2$, as this
is the moduli space of physically distinct theories, ignoring the
flux.

\begin{figure}
\label{fig:ttwovacua}
\epsfbox{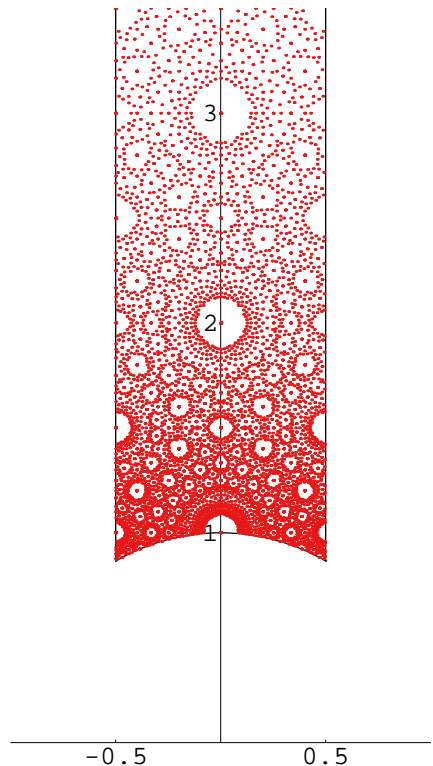} 
\caption{Values of $\tau$ for rigid CY flux vacua with $L_{max} = 150$.
From \citeasnoun{Denef:2004ze}.}
\end{figure}

We plot the results for $L=150$ in \reffig{ttwovacua}.
Each point on this graph is a possible value of $\tau$ in some
flux vacuum; many of the points correspond to multiple vacua.

While the figure clearly displays a great deal of structure, one might
worry about its intricacy and ask: if this is what comes out of the
simplest class of models, what hope is there for understanding the
general distribution of vacua in string theory?  Fortunately, there
is a very simple approximate description, which captures much of the
structure of this distribution.  It is a uniform distribution,
modified by a sort of ``symmetry enhancement'' phenomenon.

We first discuss the uniform distribution.  A very naive first guess
might be $d^2\tau$, but of course this is not invariant under field
redefinitions; rather we must look at the geometry of the
configuration space to decide what is a natural ``uniform''
distribution.  Now the configuration space of an \EFT\ %
always carries a metric, the ``sigma model metric,''
defined by the kinetic terms in the Lagrangian,
\begin{equation}\label{eq:metmod}
\CL = G_{ij} \partial\phi^i\partial\phi^j + \ldots .
\end{equation}
Thus, the natural definition of a ``uniform measure'' on
configuration space, is just the volume form associated to the
sigma model metric,
\begin{equation}\label{eq:volform}
d\mu = d^n\phi \sqrt{\det G(\phi)}.
\end{equation}
In the problem at hand, this is
\begin{equation}\label{eq:uhpmetric}
d\mu = \frac{d^2\tau}{4(\Im\tau)^2} .
\end{equation}

Of course, this is a continuous distribution, unlike the actual vacuum
counting distributions which are sums of delta functions.  However, if
we take a limit in which the number of vacua becomes arbitrarily
large, the limiting distribution of vacua
might become continuous.  Since the
discreteness of the allowed moduli values was due to flux
quantization, and intuitively the effects of this
should become less important as $L$ increases, it is reasonable 
to conjecture that in the limit $L\rightarrow\infty$, the distribution of
flux vacua in moduli space approaches \rfn{uhpmetric}.

If we are a bit more precise and keep track of the total number of vacua,
we can make a similar conjecture for the vacuum counting distribution
itself.  Normalizing \rfn{uhpmetric} so that its integral over a fundamental
region is \rfn{totalnum}, we find
\begin{equation}\label{eq:ttwolimdist}
\lim_{L\rightarrow\infty} dN_{vac} = \pi~L~
\frac{d^2\tau}{(\Im\tau)^2} .
\end{equation}
For example, a disc of area $A$ should contain $4 \pi A L$ vacua in
the large $L$ limit.

While this is true, as can be deduced from the formalism we will
describe shortly, at first glance the finite $L$ distribution may not
look very uniform.  Comparing with the $L=150$ figure, we see that
around points such as $\tau = n i$ with $n \in \IZ$, there are holes
of various sizes containing no vacua.  Where do these holes come from,
and how can they be consistent with the claim?

In fact, at the center of each of these holes, there is a large
degeneracy of vacua, which after averaging over a sufficiently large
region recovers the uniform distribution.  For example, there are 240
vacua at $\tau = 2i$, which compensate for the lack of vacua in the
hole.  As discussed in \citeasnoun{Denef:2004ze}, while this leads to
a local enhancement, just beyond the radius of the hole the uniform
approximation becomes good.

This behavior can be understood as
coming from alignments between the lattice
of quantized fluxes and the constraints following from the equations
$DW=0$.  
Using this, one can argue that the continuous
flux approximation will well approximate
the total number of vacua in a region of radius $r$ satisfying
\begin{equation}\label{eq:appcond}
L > {K\over r^2} .
\end{equation}
Another rough model for the approximation might be a
Gaussian error model as in
\rfn{dNapprox}, with variance $\sigma \sim K/L$.
Finally, one can also understand the corrections to the large $L$
approximation as
a series in inverse fractional powers of $L$, using
mathematics discussed in \citeasnoun{Douglas:2005df}.

\subsubsection{General theory}
\label{sss:counttheory}

The result we just discussed is a particular case of a general
formula for the large $L$ limit of the index density
of supersymmetric flux vacua in IIb theory on an arbitrary
Calabi-Yau manifold $M$ \cite{Ashok:2003gk},
\begin{equation}\label{eq:adformula}
\sum_{L\le L_{max}} dI_{vac}(L)
 = {(2\pi L_{max})^{b_3}\over\pi^{b_3/2} b_3!}\ \det(-R-\omega) .
\end{equation}
We will explain what we mean by ``index density'' shortly; like
the vacuum counting distribution, it is a density on moduli space,
here  a $b_3/2$ complex dimensional space which is the product of
axion-dilaton and complex structure moduli spaces.
The prefactor depends on the tadpole number $L$
defined in Eq. (\ref{tadpole}), and on $b_3$, the third
Betti number of $M$.  Instead of the density for a single $L$, we
have added in all $L\le L_{max}$; in the large $L$ limit the
relation between these is the obvious one, but such a sum converges
to the limiting density far more quickly than results at fixed $L$.

The density $\det(-R-\omega\cdot 1)$ is entirely determined by the
metric on moduli space \rfn{metmod}; all the dependence on the other
data entering the flux superpotential Eq. (\ref{fluxsup}) cancels out of the
result.  It is a determinant of a $(b_3/2)\times (b_3/2)$ dimensional
matrix of two-forms, constructed from the K\"ahler form $\omega$ on
$\CM$, with the matrix valued curvature two-form $R$ constructed from
the metric on $\CM$.

Thus, while the volume form \rfn{volform} was a natural first guess
for the distribution of flux vacua, as we will see the actual
distribution can be rather different.  The agreement between 
\rfn{ttwolimdist} and \rfn{uhpmetric} in the example of 
\rfsss{rigidcount} was particular to this case, and
follows from $R\propto \omega$ for that moduli space.
Similar, though more complicated, explicit results can be obtained
for the actual vacuum counting distribution
\cite{Denef:2004ze,Douglas:2005df}, and 
distributions of nonsupersymmetric flux vacua \cite{Denef:2004cf}.

\begin{figure}
\label{fig:conifold}
\epsfbox{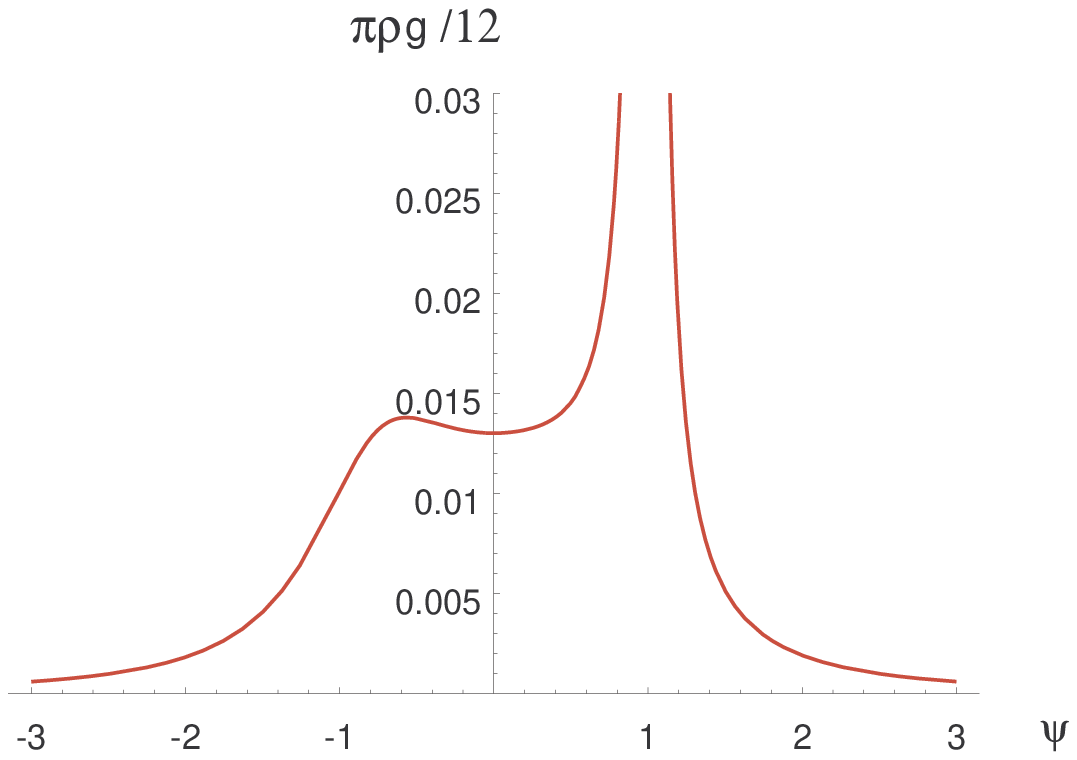}
\caption{The susy vacuum number density per unit
$\psi$ coordinate volume, on
the real $\psi$-axis, for the mirror quintic.
From  \citeasnoun{Denef:2004ze}.}
\end{figure}

Let us plot the number density in another example, compactification on
the mirror of the quintic CY \cite{Greene:1990ud,Candelas:1990rm}.  Here
$\CM_{C}(M)$ is one complex dimensional and thus the distribution
depends on two parameters; however it is a product distribution whose
dependence on the dilaton-axion is again \rfn{uhpmetric} for symmetry
reasons.  The dependence on the complex structure modulus is non-trivial;
if we plot it along a real slice, we get  \reffig{conifold}.

The striking enhancement as $\psi\rightarrow 1$ is because this limit
produces a conifold singularity as discussed in Sec. \ref{para:conifold}.
As discussed in \citeasnoun{Denef:2004ze}, near the conifold point 
\rfn{adformula} becomes\footnote{
While in general this formula is the index density
\rfn{dIdef}, it is not hard to
show that all vacua near the conifold point have $(-1)^F=+1$, so that
in this case it is also the number density.  More generally, 
globally supersymmetric vacua (which do not depend on the $(\partial K)W$
term in $DW$) always have $(-1)^F=+1$.  Conversely, the $(-1)^F=-1$ vacua are
in a sense ``K\"ahler stabilized.''}
\begin{equation} \label{eq:conifolddist}
 dN_{vac} \sim \frac{d^2\psi}{|\psi-1|^2 (\log |\psi-1|)^3}~. 
\end{equation}
As discussed in Sec. \ref{para:conifold}, under flux-gauge duality,
the parameter $\psi-1$ is dual to the dynamically induced scale (analogous
to $\Lambda_{QCD}$)
in the gauge theory, and thus dimensional transmutation
explains the leading $d^2\psi/|\psi-1|^2$ dependence here.  
However the $\log$ factors have to do with details of the sum over fluxes.

This distribution is (just barely) integrable; doing so over a disc,
the number of susy vacua with $L \leq L_*$ and $|\psi-1| \leq R$ is
\begin{equation} \label{eq:conifolddist2}
N_{vac} = \frac{\pi^4 L_*^4}{18 \ln
 \frac{1}{R^2}} .
\end{equation}
The logarithmic dependence on $R$ implies that a substantial
fraction of vacua are extremely close to the conifold point. For
example when $L_*=100$, there are still about one
million susy vacua with $|\psi-1|<10^{-100}$.  

Despite this enhancement, from the figure one sees that the majority
of vacua are not near the conifold point.  On the other hand,
in many parameter models, a sizable fraction of vacua can be
expected to contain conifold limits, by a simple probabilistic argument
we give in \rfsss{concentrate}.

Many of the other general results for flux vacuum distributions
which we called upon in \rfs{con} also follow from \rfn{adformula},
by inserting known behaviors of moduli space metrics, introducing
further constraints and so on.  For example,
\begin{itemize}
\item The fraction of flux vacua with string coupling 
$g_s \le \epsilon << 1$ goes
as $\epsilon$.  This follows from the expression \rfn{uhpmetric} for the
tree level metric on dilaton-axion moduli space.
\item The fraction of weakly coupled vacua 
with $e^K|W|^2 \le \epsilon$ goes as $\epsilon$.\footnote{
While we omit the derivation, the key step is \rfn{ddchangevar}.}
This is particular to IIb flux vacua, for reasons we discuss at the end
of this subsection.
\end{itemize}

\paragraph{Definition of the index density}
This is a sum over vacua, weighed by $\pm 1$ factors,
\begin{equation}\label{eq:dIdef}
dN_{vac}(T) = \sum_i \delta(T-T_i) (-1)^F_i .
\end{equation}
The factor $(-1)^F_i$ will be defined shortly; it is essentially the
sign of the determinant of the fermionic mass matrix.  

The primary reason to consider this quantity is that it leads to
much simpler explicit results than \rfn{dNdef}.  To explain why,
we recall the general
formula for the distribution of critical points of a random potential $V$.
As is well known in the theory of disordered systems, this is
\begin{equation}\label{eq:critdens}
dN_{vac}(z) = \vev{ \delta(V'(z)) |\det V''(z)| },
\end{equation}
where the expectation value is taken in the ensemble of random 
potentials; here the ensemble of flux potentials.
Formally, such a density is proportional to
the delta function $\delta(V'(z))$, however the integral of such a
delta function over field space is not $1$.  To get a normalized
density in which each vacuum has unit weight, we multiply by the
Jacobian factor.

Now, upon incorporating the sign factor in \rfn{dIdef}, this becomes
\begin{equation}\label{eq:inddens}
dI_{vac}(z) = \vev{ \delta(V'(z)) \det V''(z) },
\end{equation}
and the somewhat troublesome absolute value sign from the Jacobian 
is removed.  The virtue of this is that the index
turns out to be much simpler to compute than $dN_{vac}$,
yet provides a lower bound for the actual number of vacua.  There
is some evidence that the ratio of the index to the actual number
of vacua is of order $1/c^{b_3}$ for some order one $c$
\cite{Douglas:2004zu}.

We can use essentially the same formulae \rfn{critdens} and \rfn{inddens}
to count supersymmetric vacua, by replacing $V$ with a
flux superpotential $W(z)$, taking into account that it and
the chiral fields are complex.
Combining these ideas, and taking the continuous flux limit as in the
previous subsection, leads to the integral formula
\begin{eqnarray}\label{eq:adintform}
\lim_{L\rightarrow\infty} dI_{vac}(z;L) =& \\
 \int_{L=N\eta N} d^{2b_3}N\   &\delta^{(2n)}(DW(z))
 \det \left(\matrix{ D_i\bar D_j W& D_i D_j W\cr
\bar D_i\bar D_j \bar W& \bar D_i D_j \bar W}\right)
\nonumber
\end{eqnarray} 
where the tadpole constraint was schematically written
$L=N\eta N$ in terms of a known quadratic form $\eta$.

\paragraph{Computational techniques}
Without going into the details of the subsequent computations leading
to \rfn{adformula}, two general approaches have been used.  
In the original computation,
the integral over fluxes satisfying the tadpole
constraint was rewritten
as a Laplace transform of a Gaussian integral with weight
$\exp -N\eta N$.  In this way, one can think of the random superpotential
as defined by its two-point function,
$$
\vev{W(z_1)~{\bar W}({\bar z}_2)} = \exp -K(z_1,{\bar z}_2) ,
$$
where $K(z_1,{\bar z}_2)$ is the formal continuation of the K\"ahler
potential $K(z,\bar z)$ to independent holomorphic and antiholomorphic
variables.  In this sense, the flux superpotential is a Gaussian random
field, however a rather peculiar one as its correlations can grow with 
distance.  Still, one can proceed formally,
and then justify the final results.

The other approach \cite{Denef:2004ze} is to make a direct change of
variables from the original fluxes $F,H$ to the relevant derivatives
of the superpotential.  Since this provides more physical intuition
for the results, let us discuss it a bit.  

One of the main simplifications which allows obtaining explicit results for
a density such as \rfn{adformula}, is that its definition restricts attention
to a point $z$ in configuration space.
Because of this, we only need a finite amount of information,
namely the superpotential $W(z)$ and some finite number
of its derivatives at $z$, to compute it.

For example, to evaluate \rfn{adintform},
we only need $D_iW(z)$, $D_iD_jW(z)$, $D_i\bar D_jW(z)$,
and their complex conjugates.  Standard
results in supergravity (or, the fact that $W$ is a holomorphic section),
imply that $D_i\bar D_jW=g_{i\bar j}W$, so this is known in terms of $W$.
Thus, for a model with $n$ moduli,
we only need the joint distribution of
$1+n+n(n+1)/2=(n+1)(n+2)/2$ 
independent parameters derived from the potential
to compute the vacuum
counting index.  Let us give these names; in addition to $W\equiv W(z)$
we have\footnote{
Strictly speaking, one needs to include
the K\"ahler potential in these definitions, to get quantities which are
invariant under K\"ahler-Weyl transformations.  An alternate convention,
which saves a good deal of notation and which
we follow here, is to do a K\"ahler-Weyl transformation to set 
$K(z,\bar z)=0$ at the point $z$ under consideration, and use an orthonormal
frame for the tangent space to $z$; 
see \citeasnoun{Denef:2004ze} for more details.}
\begin{equation}\label{eq:FZUdef}
F_A = D_A W(z); \qquad Z_{AB} = D_A D_B W(z) .
\end{equation}
By substituting Eq. (\ref{fluxsup}) into these expressions and fixing
$z$, we get $F$ and $Z$ as functions of the fluxes $N$; in fact they
are {\it linear} in the $N$.

Now, we can rewrite \rfn{adintform} as
\begin{eqnarray}\label{eq:ddintform}
\lim_{L\rightarrow\infty} dI_{vac}(z;L) =& \\
 \int [d^2W~d^{2}F~d^2Z]_L\  \delta^{(2n)}(F_i)&
 \det \left(\matrix{ g_{i\bar j} W& Z_{ij}\cr
{\bar Z}_{\bar i\bar j}& g_{{\bar i}j} \bar W}\right) ,
\nonumber
\end{eqnarray} 
where the notation $[d^2W~d^2F~d^2Z]_L$ symbolizes the integral over
whatever subset of these variables corresponds to the original integral
over fluxes satisfying the tadpole condition.

What makes this rewriting very useful, is that the change of variables
$N\rightarrow (W,F,Z)$ turns out to be very simple, with
a constant Jacobian $\det\partial N/\partial(W,F,Z)$ (in
appropriate conventions, unity) \cite{Denef:2004ze}.
Let $n=b_3/2$, and denote the moduli as $t^i$ with
$i=0,\ldots,n-1$, where $i\ge 1$ label complex structure moduli and
$t^0$ is the dilaton-axion.  Then
\begin{equation}\label{eq:ddchangevar}
\int_{L=N\eta N} d^{2K}N \rightarrow
\int_{L=|W|^2-|F|^2+|Z|^2} d^2W d^2F_0 \prod_{i=1}^{n-1} d^{2}F_i~d^{2}Z_{0i}
\end{equation} 
The tadpole constraint is
also simple.  The other components $Z_{ij}$ are then
determined in terms of the $Z_{0i}$: $Z_{00}=0$ and
$$
Z_{ij} = {\cal F}_{ijk} g^{k\bar l} {\bar Z}_{\bar l} ; \qquad 
 1 \le i,j \le n-1,
$$
where ${\cal F}_{ijk}$ are the standard ``Yukawa couplings'' of
special geometry \cite{Candelas:1990rm}.  It is a short step from 
these formulae to \rfn{adformula} and its generalizations.

The rewriting \rfn{ddchangevar} is the simplest way to describe the
ensemble of IIb flux vacua, if one only needs to find distributions of
single vacua and their properties. 
On the other hand, the approach in which $W$ is a generalized Gaussian random
field, could also be used to compute distributions depending on the
properties of more than one vacuum, or on the effective potential away
from its critical points, for example average barrier heights between
vacua, or the average number of e-foldings of slow-roll inflationary
trajectories.  In fact, modelling the inflationary potential
as a Gaussian random field has been tried in cosmology
\cite{Tegmark:2004qd}; it would be interesting to
do the same with this more accurate description of the effective
potentials for flux vacua.

All of these precise results are in the continuous flux approximation.
As before, the general theory suggests that this should be good for
$L >> K$.  The results have been checked to some extent
by numerical study \cite{Giryavets:2004zr,Conlon:2004ds},
finding agreement with the distribution in $z$, and usually (though
not always) the predicted scaling with $L$.  It should be said
that numerous subtleties had to first be addressed in the works which 
eventually found agreement; such as the need to avoid double-counting
flux configurations related by duality, and the need to consider fairly
large values of the flux.

\paragraph{Other ensembles of flux vacua}
These can be treated by similar
methods, say by working out the analog to \rfn{ddchangevar}.  This was
done for $G_2$ compactifications in \citeasnoun{Acharya:2005ez}.
A useful
first picture can be formed by considering the ratio
\cite{DeWolfe:2005uu}
$$
\eta \equiv
 \frac{\rm number\ of\ independent\ fluxes}{\rm number\ of\ (real)\ moduli} ,
$$
as this determines the number of parameters $(W,F_i,Z_{ij},\ldots)$
which can be considered as roughly independent.  While for IIb flux
vacua $\eta = 2$, for all of the other well understood flux ensembles
(M theory, IIa,heterotic) $\eta=1$ as there is only one type of flux.

For $\eta=1$, one generally finds the uniform distribution
\rfn{volform}, and $|W|$ is of order the cutoff scale.  This is
because the conditions $D_iW=F_i=0$ already
set almost all of the fluxes, so there are too few fluxes to tune $W$ to
a small value.  This is perhaps the main reason why controlled small
volume compactifications are easier to discuss in the IIb theory.  Of
course, it may yet turn out that additional choices in the other
theories, less well understood at present, allow similar constructions
there.

\subsection{Scale of supersymmetry breaking}
\label{ss:susyscaletwo}

Let us now resume the discussion of \rfsss{susyscale}, combining results
from counting flux vacua with various general observations, to try to 
at least identify the important questions here.  We would like some
estimate of the number 
distribution of vacua described by spontaneously broken 
supergravity,\footnote{There are also vacua with no
such description, because supersymmetry is broken at the 
fundamental scale.  While these might further disfavor TeV scale
supersymmetry, at present it is hard to be quantitative about this.}
\begin{equation}\label{eq:susydist}
dN_{vac}[M_{susy},M_{EW},\Lambda]
\end{equation}
at the observed values of $\Lambda$ and $M_{EW}$,
with $M_{susy}$ as defined by \rfn{defMsusy}.
If this were approximately a power law,
$$
dN_{vac}[M_{susy},100 \GeV,0] \sim dM_{susy} M_{susy}^\alpha ,
$$
then for $\alpha<-1$ vacuum
statistics would favor low scale susy, while for $\alpha > -1$ it would 
not.

For purposes of comparison, let us begin with the prediction of
field theoretic naturalness. 
This is
\begin{equation}\label{eq:FTnatural}
dN_{vac}^{FT} 
 \sim 
\left( \frac{M_{EW}^2~M_{Pl}^2}{M_{susy}^4} \right)
\left( \frac{\Lambda}{M_{susy}^4} \right) f(M_{susy}) ,
\end{equation}
where the first factor follows from \rfn{gravmed}.
As for $f(M_{susy})$, if we grant that this is set by strong gauge
dyanmics, a reasonable ansatz might be $dM_{susy}/M_{susy}$, analogous to
\rfn{conifolddist}.  This would lead to $\alpha=-9$ and a clear 
(statistical) prediction.

Now, while we cannot say we have a rigorous disproof of \rfn{FTnatural},
the approach we are discussing gives us many reasons to disbelieve it,
based both on computation in toy models, and on simple intuitive
arguments.  Let us explain these in turn.

The simplest problem with \rfn{FTnatural}
is the factor $\Lambda/M_{susy}^4$.  Instead,
distributions of flux vacua generally go as
$\Lambda/M_{KK}^4$, $\Lambda/M_{Pl}^4$ or some other fundamental scale.
In other words,
{\it tuning the cosmological constant is not helped by supersymmetry}.

To see this, we start from \rfn{sugraV}, and the claim that $\Lambda$ is
the value of the potential at the minimum, so that
$\Lambda =  M_{susy}^4 - 3|W|^2/M_{Pl,4}^2$.
Intuitively, this formula expresses the cancellation between positive
energies due to supersymmetry breaking (the $F$ and $D$ terms), and a
negative ``compensating'' energy from the $-3|W|^2$ term.  However,
one should not fall into the trap of thinking that any of these terms
are going to ``adjust themselves'' to cancel the others.  Rather, there
is simply some complete set of vacua with some distribution of
$\Lambda$ values, out of which a $\Lambda\sim 0$ vacuum will be
selected by some {\it other} consideration (anthropic, cosmological,
or just fitting the data).  For the purpose of understanding this
distribution, it is best to forget about this later selection effect,
only bringing it in at the end.

On general grounds, since the cosmological constant is a sum of many
quasi-independent contributions, it is very plausible that it is 
roughly uniformly distributed out to some cutoff scale $M$, so that
the basic structure we are looking for in \rfn{susydist} is this scale.
Clearly by \rfn{sugraV} this is set by the cutoffs in
the $F$, $D$ and $W$ distributions; more specifically by the largest
of these.  

Let us now focus on the $W$ distribution, coming back to the $F$ and
$D$ distributions shortly.  According to the definition
Eq. (\ref{fluxsup}), the effective superpotential $W$ receives
contributions from all the fluxes, including those which preserve
supersymmetry.  Because of this, the distribution of $W$ values has
little to do with supersymmetry breaking; rather it is roughly uniform
(as a complex variable) out to a cutoff scale set by flux physics,
namely $M_{F}$ as defined in \rfn{defMF}.  Since
$$
d(|W|^2) = 2|W| d|W| = \frac{1}{\pi} d^2W ,
$$
this implies that $|W|^2$ is uniformly distributed out to this scale,
and thus that $\Lambda$ will be uniformly distributed at least out to this 
scale, leading to a tuning factor $\Lambda/M_F^4$.

To summarize, the distribution of the cosmological constant is not
directly tied to supersymmetry breaking, because it receives
contributions from supersymmetric sectors as well.  This correction
to \rfn{FTnatural} would result in $\alpha=-5$, still favoring low
scale supersymmetry, but rather less so.

Now, there is a clear loophole in this argument, namely that there
might be some reason for the supersymmetric contributions to $W$ to be
small.  In fact, one can get this by postulating an R symmetry, which
is only broken along with supersymmetry breaking.  However, within the
framework we are discussing, it is not enough just to say this to
resolve the problem.  Rather, one now has to count the vacua with the
proposed mechanism (here, R symmetry), and compare this to the total
number of vacua, to find the cost of assuming the mechanism.  Only if
this cost is outweighed by the gain (here a factor
$M_{F}^4/M_{susy}^4$) will the mechanism be relevant for the final
prediction.  We will come back and decide this shortly.

Before doing this, since the correction we just discussed would by
itself not change the prediction of low scale supersymmetry, we should
discuss the justification of the other factors in \rfn{FTnatural}.
First, we will grant the factor $M_{EW}^2/M_{susy}^4$, not because it
is beyond question -- after all this assumes some generic mechanism to
solve the $\mu$ problem -- but because the information we would need
about vacuum distributions has not yet been worked out.

On the other hand, the claim that the distribution of supersymmetry
breaking scales among string/M theory vacua is $dM_{susy}/M_{susy}$,
can also be questioned.  While this sounds like a reasonable
expectation for theories which break supersymmetry dynamically, one
has to ask whether there are other ways to break supersymmetry, what
distributions these lead to, and how many vacua realize these other
possibilities.

Given the definition of supersymmetry breaking vacuum we used in
\rfsss{susybr}, namely a metastable minimum of the effective potential
with $F$ or $D\ne 0$, one might well expect a generic effective
potential to contain many supersymmetry breaking vacua, not because
of any ``mechanism,'' but simply because generic functions have many
minima.  We discussed this idea in \ref{sp:fluxbreak}, and it
was shown to be generic for IIb flux vacua in 
\citeasnoun{Denef:2004cf}, leading to the distribution
\begin{equation}\label{eq:branchone}
dN_{vac}[M_{susy}] \sim \left( \frac{M_{susy}}{M_{F}} \right)^{12} .
\end{equation}
Although the high power $12$ may be surprising at first, it has a simple
explanation \cite{Dine:2005yq,Giudice:2006sn}.  
Let us consider a generic flux vacuum
with $M_{susy} << M_{F}$.
Since one needs a goldstino for spontaneous susy breaking, at least
one chiral superfield must have a low mass; call it $\phi$.
Generically, the flux potential gives order $M_{F}$ masses
to all the other chiral superfields, so they can be ignored, and
we can analyze the constraints in terms of an effective superpotential
reduced to depend on the single field $\phi$,
$$
W = W_0 + a \phi + b \phi^2 + c \phi^3 + \ldots .
$$
The form of the K\"ahler potential $K(\phi,\bar\phi)$ is also
important for this argument; however one can simplify this by
replacing $(a,b,c)$ by invariant variables generalizing \rfn{FZUdef},
$$
F \equiv D_\phi W;\ Z \equiv D_\phi D_\phi W;\ %
 U \equiv D_\phi\ D_\phi D_\phi W .
$$
In terms of these, the conditions for a metastable supersymmetric
vacuum are
$|F| = M_{susy}^2$ (by definition),
$|Z| = 2 |F|$ (this follows from the equation $V'=0$),
and finally
$|U| \sim |F|$ (as explained in \citeasnoun{Denef:2004cf} and many
previous discussions, this is necessary so that $V''>0$.
This also requires a
lower bound on the curvature of the moduli space metric).

Now, the distribution of the $(F,Z,U)$ parameters in flux superpotentials
can be worked out; we gave the result for $F$ and $Z$ in \rfn{ddintform},
and one can also find $U$ in terms of $(F,Z)$ and moduli space geometry.
A good zeroth order picture of the result is that $(F,Z,U)$ are
independent and uniformly distributed complex parameters, up to the 
flux potential cutoff scale $M_{F}$.  All three complex parameters
must be tuned to be
small in magnitude, leading directly to \rfn{branchone}.

The upshot is that ``generic'' supersymmetry breaking flux vacua
exist, but with a distribution heavily favoring the high scale, enough
to completely dominate the $1/M_{susy}^4$ benefit from solving
the hierarchy problem.  Indeed, this would be true for any set of
vacua arising from generic superpotentials constructed according to
the rules of traditional naturalness with a cutoff scale $M_{F}$.

The flaw in the naturalness argument in this case is very simple; one
needs to tune several parameters in the microscopic theory to
accomplish a single tuning at the low scale.  Of course, if the
underlying dynamics correlated these parameters, one could recover
natural low scale breaking.  This would be a reasonable expectation if
$W$ was entirely produced by dynamical effects, or perhaps in
some models in which it is a combination of dynamical and high scale
contributions.  Besides models based on gauge theory, it is entirely
possible that a more careful analysis of the distribution of flux
vacua on Calabi-Yau, going beyond the ``zeroth order picture'' we 
just described by taking into account more of the structure of the
actual moduli spaces, would predict such vacua as well.

Of course, even if such vacua exist, we must go on to decide how
numerous they are.  
Following \citeasnoun{Dine:2004is,Dine:2005yq,Dine:2004ct},
we can summarize the picture so far by dividing
the set of supersymmetry breaking vacua into ``three branches,''
\begin{enumerate}
\item Generic vacua; {\it i.e.} with all of the $F$, $D$ and $W$
distributions as predicted by the flux vacuum counting argument
we just discussed.
\item Vacua with dynamical supersymmetry breaking (DSB).  Here
we assume the distribution $dM_{susy}/M_{susy}$ for the breaking
parameters; however $W$ is uniformly distributed out to high scales.
\item Vacua with DSB and tree level R symmetry.  Besides the
$dM_{susy}/M_{susy}$ distribution, we also assume $W$ is produced
by the supersymmetry breaking physics.
\end{enumerate}
In option (1), TeV scale supersymmetry would seem very unlikely.
While both (2) and (3) lead to TeV scale supersymmetry, they can differ in 
their expectations for $|W|$ and thus the gravitino mass: in (3) this
should be low, while in (2) the prior distribution is neutral, so
the prediction depends on the details of mediation as
discussed in \rfsss{susybr}.

What can we say about which type of vacuum is more numerous in
string/M theory?  There is a simple argument against (3), and indeed
against most discrete symmetries in flux vacua \cite{Dine:2005gz}.
First, a discrete symmetry which acts on Calabi-Yau moduli space, will
have fixed points corresponding to particularly symmetric Calabi-Yau
manifolds; at one of these, it acts as a discrete symmetry of the
Calabi-Yau.  Such a symmetry of the Calabi-Yau will also act on the
fluxes, trivially on some and non-trivially others.  To get a flux
vacuum respecting the symmetry, one must turn on only invariant
fluxes.  Now, looking at examples, one finds that typically an order
one fraction of the fluxes transform non-trivially; for definiteness
let us say half of them.  Thus, applying \rfn{adformula} and putting
in some typical numbers for definiteness, we might estimate 
$$
\frac{N_{vac~symmetric}}{N_{vac~all}} \sim
 \frac{L^{K/2}}{L^{K}} \sim \frac{10^{100}}{10^{200}} .
$$
Thus, discrete symmetries of this type come with a huge penalty.
While one can imagine discrete symmetries with other origins for which
this argument might not apply, since $W$ receives flux contributions, 
it clearly applies to the R symmetry desired in branch (3), and
probably leads to
suppressions far outweighing the $(M_{F}/M_{susy})^4$ gain.

Thus, R symmetry appears to be heavily disfavored, with the exception
of R parity:
since the superpotential has R charge $2$, it is invariant under a
$\IZ_2$ R symmetry.  While crucial for other phenomenology, R parity does
not force small $W$.

What about branches (1) versus (2) ?  Among the many issues, we must
estimate what fraction of vacua realize
dynamical supersymmetry breaking.  Looking at the literature on this,
much of it adopts a very strong definition of supersymmetry breaking,
in which one requires that no supersymmetric vacua exist.  And,
although the situation is hardly clear, it appears that very few models
work according to this criterion.  This might be regarded as evidence
against (2).

However, this is a far stronger definition of
supersymmetry breaking than we used elsewhere in our review.  Rather,
the question we want to answer is the difficulty of realizing metastable
dynamical supersymmetry breaking vacua.  Recent work such as
\cite{Intriligator:2006dd,Dine:2006gm} 
suggests that this is not so difficult, but it is still a bit early to
evaluate this point.

Again, according to the point of view taken here, the goal is 
to show that metastable dynamical supersymmetry breaking vacua are
generic in a quantitative sense.  Doing this requires
having some knowledge about the distributions of gauge theories
among string/M theory vacua, to which we turn.

\subsection{Other distributions}

Understanding the total number and distribution of vacua requires
combining information from all sectors of the theory.  Here we discuss
some of the other sectors, while the problem of combining information
from different sectors is discussed in \citeasnoun{Douglas:2003um}.

\subsubsection{Gauge groups and matter content}

By now, the problem of trying to realize the Standard Model has been
studied in many classes of constructions.  Let us consider
type IIa orientifolds of a Calabi-Yau $M$, (see
\citeasnoun{Blumenhagen:2005mu} for a recent review).  
In the vast majority of such vacua which contain the SM, one finds
that the tadpole and other constraints force the inclusion of
``exotic matter,'' charged matter with unusual Standard Model quantum
numbers or with additional charges under other gauge groups.  One also
finds hidden sectors, analogous to the second $E_8$ of the original
CHSW models.  While less well studied, other constructions such as
more general heterotic vacua, F and M theory vacua, often contain
exotic matter as well.

All this might lead to striking predictions for new physics, if we
could form a clear picture of the possibilities, and which of them
were favored within string/M theory.  One is naturally led to
questions like: Should we expect to see such exotic matter at low
energies?  Could the extra matter be responsible for supersymmetry
breaking?  Could the hidden sectors be responsible for some or all of
the dark matter, or have other observable consequences?

A systematic base for addressing these questions would be to have a
list of all vacua, with their gauge groups and matter content, as well
as the other EFT data.  While this is a tall order, finding statistics of
large sets of vacua, such as the number of vacua with a given low energy gauge
group $G$ and matter representation $R$, is within current abilities
\cite{Blumenhagen:2004xx,Dijkstra:2004cc,%
Gmeiner:2006qw,Gmeiner:2005nh,Gmeiner:2005vz,%
Douglas:2006xy,Kumar:2005hf,Kumar:2004pv,Dienes:2006ut}.  
Besides providing a rough picture of the possibilities, such statistics
can guide a search for interesting vacua, or used to check that samples are
representative.

Thus, let us consider a vacuum counting distribution 
\begin{equation}\label{eq:GRcount}
N_{vac}(G,R) .
\end{equation}
To be precise, $G$ and $R$ should refer to
all matter with mass below some specified energy scale $\mu$.
The existing results count ${\cal N}=1$ supersymmetric
vacua and ignore quantum effects, considering gauge groups which remain
unbroken at all scales, and massless matter.

Most systematic surveys treat intersecting brane models (IBM's),
in which the possible gauge groups
$G$ are products of the classical groups $U(N)$, $SO(N)$ and $Sp(N)$,
while all charged matter transforms as two-index tensors: adjoint,
symmetric and anti-symmetric tensors, and bifundamentals.
In a theory with $r$ factors in the gauge group,
the charged matter content can be largely summarized in an $r\times r$
matrix $I_{ij}$, whose $(i,j)$ entry denotes the number of bifundamentals
in the $(N_i,{\bar N}_j)$, called the ``generalized intersection matrix.''
Thus, we can rewrite \rfn{GRcount} as
\begin{equation}\label{eq:GBcount}
N_{vac}(\{N_i\},I_{ij}) .
\end{equation}

Following the procedure outlined in \rfsss{branemodels}, one can make
lists of models, and compute \rfn{GBcount}.  The results from the studies so
far are rather intricate, so a basic question is to find some simple
approximate description of the result.  In particular, one would like to
know to what extent the data $(N_i,I_{ij})$ shows structure, such as 
preferred patterns of matter content, or other correlations, which might
lead to predictions.

Alternatively, one might propose a very simple model, such as that the
$(N_i,I_{ij})$ are (to some approximation) independent random
variables.\footnote{
While this cannot literally be true of the entire spectrum as this
must cancel anomalies, these constraints are relatively simple for
brane models, so the simplest model of the actual distribution is to
take a distribution of matter contents generated by taking these
parameters independent, and then keeping only anomaly free spectra.}
While one might be tempted to call this a ``negative''
claim, of course we should not be prejudiced about the outcome, and
methodologically it is useful to try to refute ``null hypotheses'' of
this sort.  Actually, since even in this case there would be preferred
distributions of the individual ranks and multiplicities, such a result
would carry important information.

The studies so far
\cite{Blumenhagen:2004xx,Douglas:2006xy,Dijkstra:2004cc} in general
are consistent with the null hypothesis, but suggest some places to
look for structure.  As yet they are rather exploratory and show only
partial agreement, even about distributions within the same model classes.

In \citeasnoun{Blumenhagen:2004xx}, the $T^6/\IZ_2\times\IZ_2$
orientifold (and simpler warm-up models) were studied, and all gauge
sectors enumerated.  Simple analytical models were proposed in which
\rfn{GBcount} is governed by the statistics of the number of ways of
partitioning the total tadpole among supersymmetric branes.  For
example, the total number of vacua with tadpole $L$ goes
roughly\footnote{There are $\log L$ corrections in the exponent.}  as
$\exp \sqrt{L}$, and the fraction containing an $SU(M)$ gauge group
goes as $\exp -M/\sqrt{L}$.  Computer surveys supported these claims,
and found evidence for a anticorrelation between total gauge group
rank and the signed number of chiral matter fields, and for a relative
suppression of three generation models.  However, it is not clear
whether these surveys used representative samples, for reasons
discussed in \citeasnoun{Douglas:2006xy}.

In \citeasnoun{Douglas:2006xy}, algorithms were developed to perform
complete enumerations of ``$k$-stack models,'' in other words the
distribution of $k$ of the gauge groups and associated matter.
These obey power law distributions such as $N_{vac} \sim L^n/N_i^\alpha$
with $\alpha$ depending on the types of branes involved.

The work \cite{Dijkstra:2004cc} enumerated orientifolds of
Gepner models, and restricted attention to the SM sector,
again finding that the majority of models had exotic matter, and multiple
Higgs doublets.

An element not fully discussed in any of these works is that to compute
\rfn{GRcount} as defined in \rfss{method}, one needs to stabilize all other
moduli, and incorporate multiplicities from these sectors.  One can try to
estimate these multiplicities in terms of the number of degrees of freedom
in the ``hidden'' (non-enumerated) sectors
by using generic results such as \rfn{adformula},
for example as is done in \citeasnoun{Kumar:2004pv}.

To state one conclusion on which all of these works agree, the
fraction of brane models containing the Standard Model gauge group and
matter representations is somewhere
around $10^{-10}$, as first suggested by heuristic arguments in
\citeasnoun{Douglas:2003um}.  In this sense, reproducing the SM is not the
hard part of model construction, and indeed has been done in
all model classes with
``sufficient complexity'' (for example, enough distinct homology classes)
which have been considered.  This is counting models with and without
exotic matter;
while it is clearly more difficult to get SM's without exotic matter,
we still await simple quantitative statements about
just how constraining this is, or how constraining it is to get
exotic matter which is consistent with current phenomenological constraints.

Among the many other open questions, it would be very interesting to know if
the heterotic constructions, which one might expect to favor GUTs and
thus work more generically, are in fact favored over the brane models.
The one existing survey \cite{Dienes:2006ut}, of non-supersymmetric models,
indeed finds GUT and SM gauge groups with far higher frequency.  However
a mere $10^{10}$ advantage here might well be swamped by multiplicities
from fluxes and other sectors.

\subsubsection{Yukawa couplings and other potential terms}

These have various sources in explicit constructions:
world-sheet instantons in IIa models; overlap between gauge theoretic
wave functions in IIb and heterotic models; all with additional
space-time instanton corrections.  While very interesting for
phenomenology, at this point none of this is understood in the
generality required to do statistical surveys.  

However one can suggest interesting pictures.  As an example,
let us suppose that in some large class of vacua,
quark and lepton masses were independent ``random'' variables, each with
distribution $d\mu(m)$.  Is there any $d\mu(m)$ with both plausible top-down
and bottom-up motivations?  In \citeasnoun{Donoghue:2003vs,Donoghue:2005cf},
the distribution $d\mu(m)\sim dm/m$ was proposed, both as a best fit
among power law distributions to the observed masses, and as naturally
arising from the combination of (1) uniform distributions of moduli $z$,
and (2) the general dependence of Yukawa couplings
$$
m \sim \lambda \sim \exp -z
$$
expected if they arise from world-sheet instantons.

\subsubsection{Calabi-Yau manifolds}
\label{sss:cystat}

All the explicit results we discussed assumed a choice of Calabi-Yau
manifold.  Now we do not know this choice {\it a priori}, so to count
all vacua we need to sum over it, and thus we need the distribution of
Calabi-Yau manifolds.  Of course, we might also use statistics to try
to decide {\it a priori} what is the most likely type of Calabi-Yau
to contain realistic models, or use this data in other ways.  In any
case it is very fundamental to this whole topic.

Unfortunately, we do not know this distribution.  The only large class
of Calabi-Yau manifolds which is understood in any detail at present
is the subset which can be realized as hypersurfaces in toric
varieties.  In more physical terms, these are the Calabi-Yau manifolds
which can be realized as linear sigma models with a superpotential of
the form $W=Pf(Z)$, leading to a single defining equation.
Mathematically, the toric varieties which can be used are in
one-to-one correspondence with reflexive polytopes in four dimensions.
Such a polytope encodes the geometry and determines the Betti numbers,
intersection forms, prepotential and flux superpotential, and
supersymmetric cycles; for examples of how this information is used in
explicitly constructing vacua see
\citeasnoun{Denef:2004dm,Denef:2005mm}.

We leave the definitions for the references, but the main
point for our present discussion is that this is a combinatorial
construction, so that the set of such polytopes can be shown to be
finite, and in principle listed.
In practice, the number of possibilities makes this rather
challenging.  Nevertheless, this was done by 
\citeasnoun{Kreuzer:2000qv,Kreuzer:2000xy,Kreuzer:2002uu}, 
who maintain databases and
software packages to work with this information.

\begin{figure}
\label{fig:toric}
\includegraphics[scale=0.5]{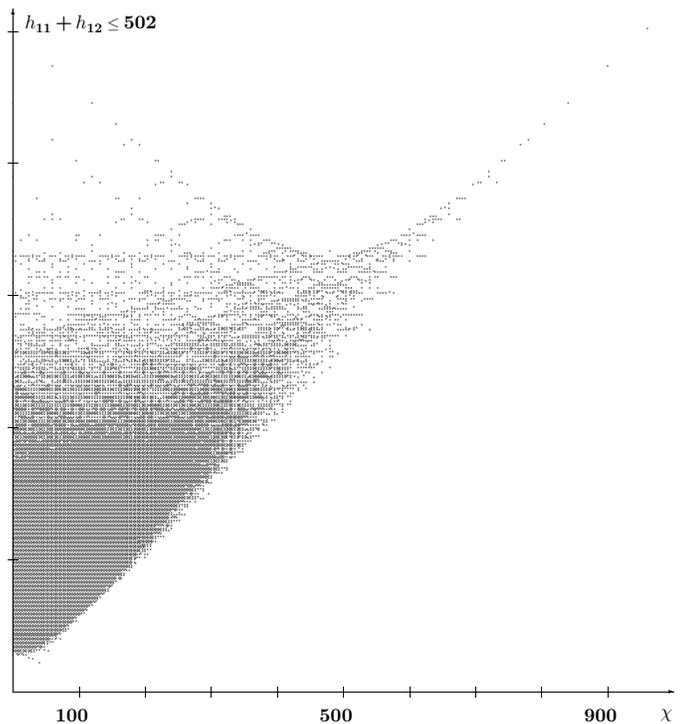}
\caption{The toric hypersurfaces with $\chi\ge 0$, from 
\citeasnoun{Kreuzer:2000xy}.
The vertical axis is $h^{1,1}+h^{2,1}$, while the horizontal axis
is $\chi=2(h^{1,1}-h^{2,1})$.  The full set also contains
the mirror manifolds obtained from these by taking $\chi\rightarrow -\chi$.}
\end{figure}

This data, as illustrated by 
\reffig{toric}, is the evidence for our earlier
assertion that ``most Calabi-Yau manifolds have $b\sim 20-300$,'' in
the range needed to solve the cosmological constant problem along the
lines of \citeasnoun{Bousso:2000xa}, but not leading to drastically
higher vacuum multiplicities.  

At present, the number of topologically distinct toric hypersurface
Calabi-Yau manifolds is not known.  While the 15122 points on this plot are
clearly distinct, one point can correspond to several polytopes;
furthermore the correspondence between polytopes and Calabi-Yau
manifolds is not one-to-one; thus one has only lower and upper bounds.
Furthermore, this set is known not to include all Calabi-Yau
manifolds.  One can at least hope that it is a representative subset;
most but not all mathematicians would agree that this is reasonable.

\subsubsection{Absolute numbers}

Combining the various sectors and multiplicities we discussed, leads
to rough estimates for numbers of vacua arising in different classes
of constructions.  The exploratory nature of much of the discussion,
combined with the theoretical uncertainties outlined in \rfs{eff},
make these estimates rather heuristic at present.  Let us quote a few
numbers anyways.  

To the extent that we can estimate numbers of other choices in
heterotic and IIa, they are subleading to numbers of IIb flux vacua.
One can get a lower bound on this from \rfn{adformula}, if one can
compute the integral over moduli space.  This has only been done in
one and two parameter examples, and for $T^6$ moduli space in
\citeasnoun{Ashok:2003gk}, and in these cases gave $\pi^{\dim \CM_C}$
times order one factors (one over the order of a discrete symmetry
group), and thus were subleading to the prefactor.  We will assume
this is generally true, but it would be worth checking, as it is not
inconceivable that CY moduli spaces have very large symmetry groups,
and this would drastically reduce the numbers.  See
\citeasnoun{Douglas:2005df,Zelditch:2006pi} for 
many more issues in making these estimates precise.

The number $L$ can be computed either by choosing a IIb
orientifolding, or using the relation to F theory on an elliptically
fibered fourfold $N$, for which $L=\chi(N)/24$.  While it would be
interesting to survey the expected number of flux vacua over all the
manifolds we discussed in \rfsss{cystat}, at present it is not
entirely clear that all of these allow stabilizing K\"ahler moduli.
For the three examples which were shown to do so in
\citeasnoun{Denef:2004dm}, one finds $10^{307}$, $10^{393}$, and $10^{506}$.
While there are further requirements, such as small $|W_0|$ and $g_s$,
metastability after supersymmetry breaking, and so on,
as discussed in \rfsss{counttheory} and elsewhere in the review
these cuts appear to lead to
comparatively small factors.  Thus, one can take $10^{500}$ as a
reasonable estimate at present, unless and until we can argue that
further conditions of the sort discussed in \rfs{eff} are required.

One might worry that this is an underestimate, as we have left out
many other known (and unknown) constructions.  The only handle we have
on this is the set of F theory compactifications, which are so similar
to IIb that the same formulas might be applied.  Since typical
fourfolds have $K \sim 1000$, this might drastically increase the
numbers, to say $10^{1000}$.  Or it might not,
both because one is typically 
not in the regime $L\ge K$ where these formulas are justified,
and because the additional moduli
(compared to IIb orientifolds) correspond to charged matter, leading
to additional  corrections to the superpotential.
More
generally, while one might expect that as more constructions come
under control the estimate will increase, this need not be, as 
new dualities between these constructions will also come into play.

\subsection{Model distributions and other arguments}

As we have seen, the computation of any distribution from microscopic
string theory considerations is a lot of work.  Since it is plausible
that many results will have simple explanations, having to do with
statistics and general features of the problem, it is tempting to try
to guess them in advance.

The simplest examples are the uniform distributions, such as
\rfn{volform}.  At first these may not look very interesting; for
example \rfn{uhpmetric} for the dilaton-axion prefers order one 
couplings.
Another well known example is a mass parameter in an EFT, such as a
boson mass $m^2\phi^2$.  The standard definition of naturalness
includes the idea that in a natural theory, this parameter will be
uniformly distributed up to the cutoff scale.  In some cases this
is a good model of the results, for example in
one parameter flux vacuum distributions away from singular points.

Even when individual parameters are simply distributed, on combining
many such parameters, one finds new structure, which can lead to
peaking and predictions.

\subsubsection{Central limit theorem}
As is very familiar, random variables which arise by combining
many different independent sources of randomness, tend to be Gaussian
(or normally) distributed.  This observation is made mathematically precise
by central limit theorems.
Thus, if we find that some observable in string theory is the sum (or
combination) of many moduli, or many independent choices in our definition
of vacuum, it becomes plausible that this observable will be normally
distributed as well.

One can design model field theory landscapes in which this postulate holds
\cite{Arkani-Hamed:2005yv,Dienes:2004pi,Distler:2005hi}.
A simple example is to take 
a large number $N$ of scalar fields $\phi_i$, with
scalar potential
\begin{equation}
V = \sum_i V_i(\phi_i)
\end{equation}
and where each $V_i$ is a quartic potential with two vacua, at 
$\phi_{i}^{\pm}$.  This kind of model would arise if the $N$ fields
are localized at distinct points in extra dimensions, for instance, so
their small wavefunction overlaps highly suppress cross terms in the potential. 
For simplicity, we will further take the quartics to be identical, though
our considerations would hold more generally.

It follows immediately from the central limit theorem that, despite the
fact that there are $2^N$ vacua, very few of them
have small cosmological constant!  
More concretely, let $V_{av}$ be the average of the energies of the
$\phi_{\pm}$ vacua, and $V_{diff}$ be the difference.  Then the 
distribution governing the vacuum energies of the vacua is
\begin{equation}
\rho(\Lambda) = {2^N \over \sqrt{2\pi N V_{diff}}} {\rm exp} \left( - 
{(\Lambda - N V_{av})^2 \over 2N V_{diff}^2}\right)~. 
\end{equation}
In a non-supersymmetric system with UV cutoff $M_*$, we would a priori
expect $V_{av} \sim M_*^4$, and therefore 
the distribution of vacua peaks at cosmological constant $N M_*^4$, with
a width of order $\sqrt{N} M_*^4$.  
Vacua around zero cosmological constant are ${\it not}$ scanned.
In some fraction of such ensembles of order $1/\sqrt{N}$, where for
some reason one fortuitously found $V_{av} \leq {1\over \sqrt{N}} M_*^4$,
one ${\it would}$ be able to scan around zero cosmological constant.
In a trivial supersymmetric generalization of this landscape, with an
unbroken R-symmetry (which guarantees that $V=0$ is special), 
again one ${\it would}$ be able to scan around zero
cosmological constant, while supersymmetric theories without R-symmetry
would not in general be expected to allow such scanning.

Suppose now that an observable coupling constant $g$ is controlled by
the sum of the vevs $\sum_i \phi_{i}$ characterizing a given vacuum.
The same logic would teach one that despite the vast landscape of
$2^N$ vacua, the coupling constant {\it does not} scan very much;
it fluctuates by ${\delta g / g} \sim {1 / \sqrt{N}}$ 
around its mean value.  More generally, there exist
landscapes with a large number of vacua, in which
many physical quantities can be predicted with $1/\sqrt{N}$ precision.
Since in nature only a few quantities seem plausibly to be environmentally
determined, while many others beg for explanations based on dynamics
and symmetries, one could hope that the cosmological term is one of 
a few variables that is scanned, while other quantities of interest 
do not scan \cite{Arkani-Hamed:2005yv}. 

To decide whether this is a good model for a particular parameter, one
must look at microscopic details.  As we mentioned, it is very
plausible that the cosmological constant works this way.  On the other
hand, there is no obvious sense in which a modulus is a sum of
independent random variables, and indeed the AD distribution
\rfn{adformula} does not look like this.  This would also be true for an
observable which was a simple function of one or a few moduli, for
example a gauge coupling in a brane model, proportional to the volume
of a cycle.  On the other hand, a hypothetical observable which was a
sum or combination of many moduli, might be well modelled in this way.
These observables would be the ones that are most clearly amenable to
prediction (or post-diction) using statistical techniques.

\subsubsection{Random matrix theory}
\label{sss:matrix}
Other universal distributions which appear very often in physics
are the random matrix ensembles, for example the 
gaussian unitary ensemble (GUE) \cite{Mehta:1991}.
In the large $N$ limit, these ``peak'' and
exhibit universal properties such as the semicircle law, level repulsion
and so on.

On general grounds, one might expect moduli masses to be modelled
by a random matrix distribution.  This was made more precise in
\citeasnoun{Denef:2004cf}, 
who observed that since the
matrix of fermion masses $D_iD_jW$ in supersymmetric
field theories is a complex symmetric matrix,
it can be modelled by the CI distribution of \citeasnoun{Altland}.
This leads to level repulsion between moduli masses $m_a$, characterized
by the distribution
\begin{equation}
 d\mu[m_a] = \prod_a
 d(m_a^2) \prod_{a<b} |m_a^2-m_b^2| .
\end{equation}
In particular, degenerate masses are non-generic.  This was 
important in the arguments for \rfn{branchone} as degenerate masses
would have led to an even larger exponent.

Another model for moduli masses was proposed in
\citeasnoun{Easther:2005zr}.  
They considered the large volume limit,
in which the superpotential is a sum of a flux term with
nonperturbative corrections, as in Eq. (\ref{kkltsup}).
In this limit, while most fields (complex structure moduli, dilaton and
others) obtain large masses, the axionic parts of the
K\"ahler moduli obtain small masses, depending on the expectation
values of the first set of fields.
Taking the number of K\"ahler moduli as $K$ and the number of the others
as $N$,
a reasonable model for the resulting mass matrix is
$$
(M^2)_{ij} = \sum_{\alpha\le K+N} H_{i\alpha} H^\dagger_{\alpha j} ,
$$
where $H$ is a
$K\times (K+N)$ matrix with randomly distributed entries.  For large
$K,N$, the limiting distribution for $M^2$ is very generally
the Marcenko-Pastur distribution, a simple distribution depending on 
the ratio $K/N$.

While these are interesting universal predictions, they apply to
moduli masses at scales $M_F$, and it is not completely obvious how
they would relate to observable physics.  In \citeasnoun{Easther:2005zr} it
was proposed that they favor ``N-flation,''
a mechanism for slow-roll inflation
\cite{Dimopoulos:2005ac}.

Similar ansatzes assuming less structure appear in 
\citeasnoun{Aazami:2005jf,Holman:2005ei,Kobakhidze:2004gm,%
Mersini-Houghton:2005im}.

\subsubsection{Other concentrations of measure}
\label{sss:concentrate}

This is the general term in mathematics for the ``large $N$ limits''
and other universal phenomena exhibited by integrals over high dimensional
spaces.

As a simple example, recall from \reffig{conifold} that in a one parameter
model, most flux vacua are not near a conifold point.  Suppose the
probability of a given modulus being {\it away} from a conifold point
is $1-\epsilon$, then the probability of $n$ moduli being away
from conifold points should be $(1-\epsilon)^n$, which for $n\epsilon>>1$
will be small.  In this sense, most vacua with many moduli will be near
some conifold point; some numbers are given in 
\citeasnoun{Hebecker:2006bn}.

Another example is that the vast bulk of an $n$-parameter CY moduli space
is at order one volume (of the CY itself); the fraction which sits at volume 
greater than ${\rm Vol}$ falls off as $({\rm Vol})^{-n/3}$
\cite{Denef:2004dm}.
This applies for example to the large volume regime we discussed
in \rfsss{matrix}; it is also relevant for IIb flux vacua in its 
mirror interpretation.

\subsubsection{Non-existence arguments}
Instead of doing statistics on explicit constructions, another
approach to characterizing the set of vacua is to find consistency
conditions or other {\it a priori} arguments that vacua with certain
properties cannot exist.

Perhaps the best known example is the statement that vacua of string
theory cannot have continuous global symmetries  \cite{Banks:1987cy}.  One
argument for this is based on general properties of theories of quantum
gravity, specifically the fact that absorption and radiation of
particles by black holes will violate these symmetries.  A very different
argument, from string world-sheet perturbation theory, is that such a
symmetry must correspond to a world-sheet conserved current, and such
a current can be used to construct a vertex operator for a vector boson
gauging the symmetry.

Recently, \citeasnoun{Arkani-Hamed:2006dz} have proposed
a quantitative extension of this result: 
in any theory of quantum gravity containing
a $U(1)$ gauge theory sector, there should
be a lower bound on the gauge coupling,
$$g > \frac{m}{M_P} ,
$$
where $m$ is the mass of the lightest charged particle.  Besides verifying
this in examples, they argue for this
by considering entropy bounds on the end states of charged black holes;
see also \citeasnoun{Banks:2006mm}.  It may be that such arguments, using only
general features of quantum gravity, can lead to further interesting
constraints \cite{Vafa:2005ui,Ooguri:2006in}.

\subsubsection{Finiteness arguments}
\label{sss:finiteness}

In counting vacua, one is implicitly assuming that the number of
quasi-realistic vacua of string/M theory is finite.  As it is easy to
write down effective potentials with an infinite number of local
minima, clearly this is a non-trivial hypothesis, which must be
checked.  Actually, if interpreted too literally, it is probably not true:
there are many well established infinite series of compactifications,
such as the original Freund-Rubin example of \rfsss{freundrubin}.
While the well established examples are not quasi-realistic, at first
one sees no obvious reason that such series cannot exist.

A basic reason to want a finite number of quasi-realistic vacua, under
some definition, is that if this is not true, one runs a real risk
that the theory can match any set of observations, and in this sense
will not be falsifiable.  Again, this may not be obvious at first, and
one can postulate hypothetical series which would not lead to a
problem, or even lead to more definite predictions.  Suppose for
example that the infinite series had an accumulation point, so that
almost all vacua made the same predictions; one might argue that
this accumulation point was the preferred prediction
\cite{Dvali:2003br}.

However, the problem which one will then face, is that any
general mechanism leading to infinite series of vacua in the
observable sector, would also be expected to lead to infinite sets of
choices in every other sector of the theory, including hidden sectors.
Now, while a hidden sector is not directly observable, still all
sectors are coupled (at least through gravity; in our considerations
through the structure of the moduli space as well), so choices made there do
have a small influence on observed physics.  For example, the precise
values of stabilized moduli in flux vacua, will depend on flux values in
the hidden sector.  Thus, an infinite-valued choice in this sector, would
be expected to lead to a set of vacua which densely populates even the
observable sector of theory space, eliminating any chance for statistical
predictions.  

This argument comes with loopholes of course; one of the most important
is that the measure factor can suppress infinite series.  Still, finiteness
is one of the most important questions about the distribution of vacua.

Let us consider the example of \rfsss{freundrubin}, in which the flux
$N$ can be an arbitrary positive integer.  Analogous infinite series
exist in its generalizations to the $G_2$ holonomy and IIa
examples of Sec. \ref{sec:IIaclass}, and so on.  In all of these series, the
compactification volume goes as a positive power of $N$.  Thus, if our
definition of ``quasi-realistic'' includes an upper bound on this
volume, these infinite series will not pose a problem.  Such a bound
follows from \rfn{kkscale} and a phenomenological lower bound on the 
fundamental scale, say $\MPd{D} > 1 \TeV$.

Various arguments have been given that the number of choices arising
from a particular part of the problem are finite in this sense: the
number of generations \cite{Douglas:2004yv}; the number of IIb flux vacua
\cite{Eguchi:2005eh,Douglas:2006zj}; the choice of compactification manifold
\cite{Acharya:2006zw}, and the choice of brane configuration
\cite{Douglas:2006xy}.  This rules out postulated infinite 
series such as that of
\citeasnoun{Dvali:2003br}, as well as others.
However at present there is no completely general
argument for finiteness, so this is an important point
to check in each new class of models.

\subsection{Interpretation}
\label{ss:interp}

We come finally to the question of how to use distributions such as
\rfn{dNdef} or \rfn{dmudef}.  One straightforward answer is that
they are useful in guiding the search for explicit vacua.  For example,
if it appears unlikely that a vacuum of some type exists, one should
probably not put a major effort into constructing it.

Going beyond this, distributions give us a useful shortcut to finding
explicit vacua with desired properties.  As one example, in the
explicit construction of \rfsss{quantumIIb}, we needed to assert that
IIb flux vacua exist with a specified small upper bound on $|W_0|$.
For many purposes, one does not need to know an explicit set of fluxes
with this property; a statistical argument that one exists would be
enough.
The cosmological constant itself is a very important example
because, for the reasons we discussed earlier, there is little hope in
this picture to find the actual vacua with small $\Lambda$.

Going further, it would be nice to know to what extent arguments
such as those in \rfsss{susyscale} and \rfss{susyscaletwo} could be
made precise, and what assumptions we would need to rely on.  At
first, one may think that such arguments require knowing the measure
factor, plunging us into the difficulties of \rfss{measure}.
However, if the absolute number of vacua is not too large, this is not
so; one could get strong predictions which are independent of this.
After all, if we make an observation $X$, and one has a convincing
argument that no vacuum reproducing $X$ exists, one has falsified the
theory, no matter what the probabilities of the other vacua might be.

These comments may seem a bit general, but when combined with the
formalism we just discussed, and under the hypothesis that there are
not too many vacua, could have force.  Let us return to the problem of
the scale of supersymmetry breaking.  According to the arguments of
\rfss{fluxsol} and the distribution results of \rfss{count}, tuning the
cosmological constant requires having $10^{120}$ vacua which, while
realizing a discretuum of cosmological constants, are otherwise
identical.  Now suppose we found only $10^{100}$ vacua with high scale
supersymmetry breaking; since finding the observed c.c. would require
an additional $10^{-20}$ tuning, we would have a good reason to
believe that high scale supersymmetry breaking is not just disfavored,
but {\it inconsistent} with string/M theory.  
While there would be a probabilistic aspect to this claim, it would
not be based on unknowns of cosmology or anthropic considerations, but
the theoretical approximations which were needed to get a definite
result.  If this were really the primary source of uncertainty in the
claim, one would have a clear path to improving it.

This argument is a simple justification for defining the ``stringy
naturalness'' of a property in terms of the number of string/M theory
vacua which realize it, or theoretical approximations to this number.
Of course, to the extent that one believes in a particular measure
factor or can bring in other considerations, one would prefer a
definition which takes these into account; however at present one
should probably stick to the simplest version of the idea.

The downside of this type of argument is, not having made additional
probabilistic assumptions, if there are ``too many'' vacua, so that
each alternative is represented by at least one vacuum, one gets no
predictions at all.  How many is ``not too many'' for this to have any
chance of succeeding?  A rough first estimate is, fewer than $10^{230}$.
This comes from multiplying together the observed accuracies of
dimensionless
couplings, the tuning factors of the dimensionful parameters, and the
estimated $10^{-10}$ difficulty of realizing the Standard Model
spectrum.
This produces roughly\footnote{
The exponent $70$ includes $\alpha_1$ (10), $\alpha_2$ (6), $\alpha_3$
(2), $m_{proton}$ (10), $m_n$ (10), and 14 less well measured SM parameters,
contributing say $32$.}
$10^{-70-120-30-10} \sim 10^{-230}$.
Neglecting all the further structure in the problem, one might say
that, if string/M theory has more than $10^{230}$ vacua, there is no
obvious barrier to reproducing the SM purely statistically, so
one should not be able to falsify the theory, on the basis of present data,
using statistical reasoning.
Conversely, if there are fewer vacua, in principle this might be possible.

The number $10^{230}$ is a lower bound; if the actual distribution of
vacua were highly peaked, or if we were interested in a rare property,
we could argue similarly with more vacua.  Let us illustrate this by
supposing that we find good evidence for a varying fine structure
constant.  As we discussed in \rfsss{moduliproblem}, fitting this
would require an effective potential which is almost independent of
$\alpha_{EM}$, and this is highly non-generic; in
\citeasnoun{Banks:2001qc}, it was argued that the first $8$
coefficients in the series expansion of $V(\alpha_{EM})$ would have to
be tuned away.  However, in a large enough landscape, even this
might happen statistically.  Taking the cutoff at a hypothetical
$M_{susy}\sim 10 \TeV$, this is a tuning factor of order $10^{-600}$,
so if string/M theory had fewer than $10^{800}$ or so vacua,
such an observation would rule it out
with some confidence, while if it had more, we would be less sure.

This is an instructive example, both because the point is clear, and
because the stated conclusions taken literally sound absurd.  If we
really thought the observations required a varying fine structure
constant, we would quickly proceed to the hypothesis that the
framework we are discussing based on the effective potential is wrong,
that there is some other mechanism for adjusting the c.c., or perhaps
some mechanism other than varying moduli for varying the apparent fine
structure constant.  Any such prediction depends on all of the
assumptions, including the basic ones, which should be suspected
first.  However, we can start to see how statistical and/or probabilistic
claims of this sort, might unavoidably enter the discussion.

But what if there are $10^{1000}$ vacua?  And what hope is there for
estimating the actual number of vacua?  All one can say about the
second question is that, while there are too many uncertainties to
make a convincing estimate at present, we have a fairly good record of
eventually answering well posed formal questions about string/M
theory.

Regarding the first question, in this case one probably needs to
introduce the measure factor, which will increase the predictivity.
This might be quantified by the standard concept
of the entropy of a probability distribution,
$$
S = \sum_i P_i~ \log (1/P_i).
$$
The smaller the entropy, the more concentrated the measure, and the
more predictive one expects the theory to be.  
To some extent, one can repeat the preceding discussion in this
context, by everywhere replacing the number of vacua with the
total statistical weight $e^S$.  However, justifying this
would require addressing the issues raised in \rfss{measure}.

There is another reason to call on the measure factor, namely
the infinite series of M theory and IIa vacua discussed in
\rfs{IIaclass}.  Since these run off to large volume, all but
a finite number are already ruled out, as discussed in
\rfsss{finiteness}.  However, since their number appears to grow
with volume, any sort of probabilistic reasoning is likely
to lead to the prediction that extra dimensions are just about to
be discovered, an optimistic but rather suspicious conclusion.

An alternate hypothesis \cite{Douglas-strings-2005} is that the
correct measure factor suppresses large extra dimensions, which would
be true for example if it had a factor $\exp -{\rm vol}(M)$.  Possible
origins for such a factor might be whatever dynamics selects
$3+1$ dimensions (some of the many suggestions include
\citeasnoun{Brandenberger:1988aj,Easther:2004sd}), or decoherence effects as
suggested in \citeasnoun{Firouzjahi:2004mx}.

One cannot go much further in the absence of more definite information
about the measure factor.  But an important hypothesis to confirm or
refute, is that its only important dependence is on the aspects of a
vacuum which are important in early cosmology, while for all other
aspects one can well approximate it by a uniform measure, in which the
probability that one of a set of $N$ similar vacua appears, is taken
to be $1/N$.

The former clearly include the scale of inflation and the size of the
extra dimensions, and may include other couplings which enter into the
physics of inflation and reheating.  However, since the physics of
inflation must take place at energy scales far above the scales of the
Standard Model, most features of the Standard Model, such as the
specific gauge group and matter content, the Yukawa couplings, and
perhaps the gauge couplings, are probably decorrelated from the
measure factor.  For such parameters, the uniform measure
$P(i)\sim 1/N$ should be a good approximation.  Regarding 
selection effects, we can try to bypass this discussion with the 
observation that we know that the Standard Model allows for our
existence, and we will not consider the question of whether in some
other vacuum the probability or number of observers might have been
larger.

It is not {\it a priori} obvious whether a measure factor will depend
on two particularly important parameters, the cosmological constant
and the supersymmetry breaking scale.  As we discussed in
\rfss{measure}, the cosmological constant does enter into some 
existing claims, but this leads to its own problems.  As for
supersymmetry breaking, one might argue that this should fall into the
category of physics below the scale of inflation and thus not enter
the measure factor, but clearly the importance of this point makes such
pat arguments unsatisfying.  See \citeasnoun{Kallosh:2004yh} for arguments
suggesting a link between these two scales.

Let us conclude by suggesting that, while an
understanding of the measure factor is clearly essential to put these
arguments on any firm footing, it might turn out that the actual
probabilities of vacua are essentially decorrelated from almost all
low energy observables, perhaps because they are
determined by the high scale physics of eternal inflation, perhaps
because they are controlled by the value of the c.c. which is
itself decorrelated from other observables, or perhaps for other
genericity reasons.  In any of these cases, decorrelation and the large
number of vacua would
justify using the uniform measure, and the style of probabilistic
reasoning we used in
\rfsss{susyscale} would turn out to be appropriate.

\section{CONCLUSIONS}
\label{sec:conclu}

The primary goal of superstring compactification is to find realistic
or quasi-realistic models.  Real world physics, both the Standard
Model and its various suggested extensions, is rather
complicated, and it should not be surprising that this goal is
taking time to achieve.

Already when the subject was introduced in the mid-1980's, good
plausibility arguments were given that the general framework of grand
unified theories and low energy supersymmetry could come out of string
theory.  While there were many gaps in the picture, and some of the
most interesting possibilities from a modern point of view were not
yet imagined, it seems fair to say that the framework we have
discussed in this review is the result of the accumulation of many
developments built on that original picture.

In this framework, we discussed how recent developments in flux
compactification and superstring duality, along with certain
additional assumptions such as the validity of the standard
interpretation of the effective potential, allow one to construct
models which solve more of the known problems of fundamental physics.
Most notably, this includes models with a small positive cosmological
constant, but also models of inflation and new models which solve the
hierarchy problem.

We emphasize that our discussion rests on assumptions which are by no
means beyond question.  We have done our best in this review to point
out these assumptions, so that they can be critically examined.  But
we would also say that they are not very radical or daring
assumptions, but rather follow general practice in the study of string
compactification, and 
in particle physics and other
areas.  Any of them might be false, but in our opinion that would in
itself be a significant discovery.

Even within the general framework we have discussed,
there are significant gaps in our knowledge of even the most basic
facts about the set of string vacua.  Our examples were largely based
on Calabi-Yau compactification of type II theories, where there are
tools coming from ${\cal N}=2$ supersymmetry that make the
calculations particularly tractable.  General ${\cal N}=1$ flux vacua
in these theories, which involve ``geometric flux'' (discretely
varying away from the Calabi-Yau metric) or even the non-geometric
compactifications of \rfss{mirrorsym},
are poorly
understood.  In the heterotic string, Calabi-Yau models do not admit a
sufficiently rich spectrum of fluxes to stabilize moduli in a regime
of control (though for work in this direction, see 
\cite{Gukov:2003cy,Becker:2004gw,Curio:2006dc}).
The more general non-K\"ahler compactifications, which
are dual to our type II constructions and should lead to
similar moduli potentials, are being intensely investigated as of this
writing \cite{Goldstein:2002pg,
LopesCardoso:2002hd,Becker:2003yv,Becker:2003gq,Becker:2003sh,
LopesCardoso:2003sp,Li:2004hx,Fu:2005sm,Fu:2006vj,Becker:2006et,
Kimura:2006af,Kim:2006qs}.
For work on moduli
potentials in $G_2$ compactifications of M-theory, which
also provide a promising home for SUSY GUTs; see 
\citeasnoun{Beasley:2002db,Acharya:2002kv,Acharya:2006ia}.

These investigations may still be of too limited a scope: in
a full survey, one should not require a strict
definition in terms of world-sheet conformal field theory.
For example, compactifications of non-critical strings
\cite{Myers:1987fv} should also be
explored.  There have been interesting investigations in this
direction \cite{Maloney:2002rr}, but as yet little is
yet known about the possible phenomenology of these models.

We think many readers will agree that what has emerged has at least
answered Pauli's famous criticism of a previous attempt at unification.
The picture is strange, perhaps strange enough to be true.
But is it true?  That is the question we now face.  

Let us briefly recap a few areas in which we might find testable
predictions of this framework, as outlined in \rfss{fluxphys}.
Perhaps the most straightforward application, at least conceptually,
is to inflation, as the physics we are discussing determines the
structure of the inflationary potential.  
There are by now many promising inflationary scenarios in string theory,
involving brane motion, moduli, or axions as inflatons.  In each scenario,
however, there are analogues of the infamous eta problem 
\cite{Copeland:1994vg}, where Planck-suppressed corrections to the
inflaton potential spoil flatness and require mild (1 part in 100)
tuning to achieve 60 e-foldings.  While this may be a small concern
relative to other hierarchies we have discussed, it has nevertheless
made it difficult to exhibit very explicit inflationary models in string
theory.  In addition to surmounting these obstacles through explicit
calculation in specific examples, it will be important
to develop some intuition for which classes of models are most generic;
this will involve sorting out the vexing issues of measure that were
discussed in III.E.  
Even lacking this top-down input, 
clear signatures for some classes of models have been
found, via cosmic string production \cite{Sarangi:2002yt,Copeland:2003bj} 
or non-Gaussianities of the perturbation
spectrum \cite{Alishahiha:2004eh}; perhaps our first clue will come from 
experiment.

Moduli could in principle lead to observable physics at later times,
such as a varying fine structure constant, or quintessence.  The first
is essentially ruled out, while the second appears even less natural than a
small cosmological term, with no comparable ``anthropic'' motivation.

Implicit in the word ``natural,'' is the fact that many predictions in
this framework are inherently statistical, referring to properties of
large sets but not all vacua.  The statistics of vacua provides
precise definitions of ``stringy naturalness,'' which take into
account not just values of couplings and the renormalization group,
but all of the choices involved in string compactification.  This
shares some features of ``traditional naturalness,'' but may differ
dramatically in others.

In particular, TeV scale supersymmetry is not an inevitable prediction
of string/M theory in this framework.  While we discussed many of the
ingredients which would go into making a well motivated string/M theory
prediction, we are not at present taking a position as to what the
eventual prediction might be.  Conceivably, after much further
theoretical development, we might find that TeV scale supersymmetry is
disfavored.  Of course, a successful prediction that Cern and Fermilab
will precisely confirm the Standard Model would be something of a
Pyrrhic victory.  As physicists, we would clearly be better off with
new data and new physics.

For the near term, the main goal here is not really prediction, but
rather to broaden the range of theories under discussion, as we will
need to keep an open mind in confronting the data.  The string
phenomenology literature contains many models with TeV scale
signatures; as examples inspired by this line of work, we can cite
\citeasnoun{Arkani-Hamed:2004fb,Giudice:2004tc,Arkani-Hamed:2005yv,%
Giudice:2006sn,Kane:2006yi}.  In the longer term, a statistical
approach may become an important element in bridging the large gap
between low energy data and fundamental theory.

We may stand at a crossroads; perhaps much more direct evidence for or
against string/M theory will be found before long, making statistical
predictions of secondary interest.  Or perhaps not; nature has hidden
her cards pretty well for the last twenty years, and perhaps we will have
to play the odds for some time to come.

\section*{Acknowledgements}

We would like to thank B. Acharya, 
N. Arkani-Hamed, S. Ashok, R. Bousso, M. Cvetic,
F. Denef, O. DeWolfe, S. Dimopoulos, M. Dine, R. Donagi, G. Dvali,
B. Florea, S. Giddings, A. Giryavets, G. Giudice, A. Grassi,
A. Guth, R. Kallosh, G. Kane, A. Kashani-Poor, P. Langacker,
A. Linde, J. Maldacena, L. McAllister, J. McGreevy, B. Ovrut, J. Polchinski, 
R. Rattazzi, N. Saulina, M. Schulz,  
N. Seiberg, S. Shenker, B. Shiffman, E. Silverstein, P. Steinhardt,
L. Susskind, 
W. Taylor, S. Thomas, A. Tomasiello, S. Trivedi, H. Verlinde,
A. Vilenkin, E. Witten and S. Zelditch for sharing
their understanding of these subjects with us over the years. 

The work of M.R.D. was supported by DOE grant DE-FG02-96ER40959;
he would also like to acknowledge the hospitality of the Isaac
Newton Institute, the KITP, the Banff Research Station,
and the Gordon Moore Distinguished Scholar program at Caltech.
The work of S.K. was supported in part by a David and Lucile Packard
Foundation Fellowship, the DOE under contract DE-AC02-76SF00515, and
the NSF under grant number 0244728.
He is grateful to KITP for hospitality during the completion
of this review.

\bibliographystyle{apsrmp}

\begin{thebibliography}{462}
\expandafter\ifx\csname natexlab\endcsname\relax\def\natexlab#1{#1}\fi
\expandafter\ifx\csname bibnamefont\endcsname\relax
  \def\bibnamefont#1{#1}\fi
\expandafter\ifx\csname bibfnamefont\endcsname\relax
  \def\bibfnamefont#1{#1}\fi
\expandafter\ifx\csname citenamefont\endcsname\relax
  \def\citenamefont#1{#1}\fi
\expandafter\ifx\csname url\endcsname\relax
  \def\url#1{\texttt{#1}}\fi
\expandafter\ifx\csname urlprefix\endcsname\relax\def\urlprefix{URL }\fi
\providecommand{\bibinfo}[2]{#2}
\providecommand{\eprint}[2][]{\url{#2}}

\bibitem[{\citenamefont{Aazami and Easther}(2006)}]{Aazami:2005jf}
\bibinfo{author}{\bibnamefont{Aazami}, \bibfnamefont{A.}}, and
  \bibinfo{author}{\bibfnamefont{R.}~\bibnamefont{Easther}},
  \bibinfo{year}{2006}, \bibinfo{journal}{JCAP}
  \textbf{\bibinfo{volume}{0603}}, \bibinfo{pages}{013}.

\bibitem[{\citenamefont{Abbott}(1985)}]{Abbott:1984qf}
\bibinfo{author}{\bibnamefont{Abbott}, \bibfnamefont{L.~F.}},
  \bibinfo{year}{1985}, \bibinfo{journal}{Phys. Lett.}
  \textbf{\bibinfo{volume}{B150}}, \bibinfo{pages}{427}.

\bibitem[{\citenamefont{Acharya}
  \emph{et~al.}(2006{\natexlab{a}})\citenamefont{Acharya, Bobkov, Kane, Kumar,
  and Vaman}}]{Acharya:2006ia}
\bibinfo{author}{\bibnamefont{Acharya}, \bibfnamefont{B.}},
  \bibinfo{author}{\bibfnamefont{K.}~\bibnamefont{Bobkov}},
  \bibinfo{author}{\bibfnamefont{G.}~\bibnamefont{Kane}},
  \bibinfo{author}{\bibfnamefont{P.}~\bibnamefont{Kumar}}, and
  \bibinfo{author}{\bibfnamefont{D.}~\bibnamefont{Vaman}},
  \bibinfo{year}{2006}{\natexlab{a}}, \eprint{hep-th/0606262}.

\bibitem[{\citenamefont{Acharya}(2002)}]{Acharya:2002kv}
\bibinfo{author}{\bibnamefont{Acharya}, \bibfnamefont{B.~S.}},
  \bibinfo{year}{2002}, \eprint{hep-th/0212294}.

\bibitem[{\citenamefont{Acharya}
  \emph{et~al.}(2006{\natexlab{b}})\citenamefont{Acharya, Benini, and
  Valandro}}]{Acharya:2006ne}
\bibinfo{author}{\bibnamefont{Acharya}, \bibfnamefont{B.~S.}},
  \bibinfo{author}{\bibfnamefont{F.}~\bibnamefont{Benini}}, and
  \bibinfo{author}{\bibfnamefont{R.}~\bibnamefont{Valandro}},
  \bibinfo{year}{2006}{\natexlab{b}}, \eprint{hep-th/0607223}.

\bibitem[{\citenamefont{Acharya} \emph{et~al.}(2003)\citenamefont{Acharya,
  Denef, Hofman, and Lambert}}]{Acharya:2003ii}
\bibinfo{author}{\bibnamefont{Acharya}, \bibfnamefont{B.~S.}},
  \bibinfo{author}{\bibfnamefont{F.}~\bibnamefont{Denef}},
  \bibinfo{author}{\bibfnamefont{C.}~\bibnamefont{Hofman}}, and
  \bibinfo{author}{\bibfnamefont{N.}~\bibnamefont{Lambert}},
  \bibinfo{year}{2003}, \eprint{hep-th/0308046}.

\bibitem[{\citenamefont{Acharya} \emph{et~al.}(2005)\citenamefont{Acharya,
  Denef, and Valandro}}]{Acharya:2005ez}
\bibinfo{author}{\bibnamefont{Acharya}, \bibfnamefont{B.~S.}},
  \bibinfo{author}{\bibfnamefont{F.}~\bibnamefont{Denef}}, and
  \bibinfo{author}{\bibfnamefont{R.}~\bibnamefont{Valandro}},
  \bibinfo{year}{2005}, \bibinfo{journal}{JHEP} \textbf{\bibinfo{volume}{06}},
  \bibinfo{pages}{056}.

\bibitem[{\citenamefont{Acharya and Douglas}(2006)}]{Acharya:2006zw}
\bibinfo{author}{\bibnamefont{Acharya}, \bibfnamefont{B.~S.}}, and
  \bibinfo{author}{\bibfnamefont{M.~R.} \bibnamefont{Douglas}},
  \bibinfo{year}{2006}, \eprint{hep-th/0606212}.

\bibitem[{\citenamefont{Adams} \emph{et~al.}(1993)\citenamefont{Adams, Bond,
  Freese, Frieman, and Olinto}}]{Adams:1992bn}
\bibinfo{author}{\bibnamefont{Adams}, \bibfnamefont{F.~C.}},
  \bibinfo{author}{\bibfnamefont{J.~R.} \bibnamefont{Bond}},
  \bibinfo{author}{\bibfnamefont{K.}~\bibnamefont{Freese}},
  \bibinfo{author}{\bibfnamefont{J.~A.} \bibnamefont{Frieman}}, and
  \bibinfo{author}{\bibfnamefont{A.~V.} \bibnamefont{Olinto}},
  \bibinfo{year}{1993}, \bibinfo{journal}{Phys. Rev.}
  \textbf{\bibinfo{volume}{D47}}, \bibinfo{pages}{426}.

\bibitem[{\citenamefont{Affleck} \emph{et~al.}(1984)\citenamefont{Affleck,
  Dine, and Seiberg}}]{Affleck:1983mk}
\bibinfo{author}{\bibnamefont{Affleck}, \bibfnamefont{I.}},
  \bibinfo{author}{\bibfnamefont{M.}~\bibnamefont{Dine}}, and
  \bibinfo{author}{\bibfnamefont{N.}~\bibnamefont{Seiberg}},
  \bibinfo{year}{1984}, \bibinfo{journal}{Nucl. Phys.}
  \textbf{\bibinfo{volume}{B241}}, \bibinfo{pages}{493}.

\bibitem[{\citenamefont{Affleck} \emph{et~al.}(1985)\citenamefont{Affleck,
  Dine, and Seiberg}}]{Affleck:1984xz}
\bibinfo{author}{\bibnamefont{Affleck}, \bibfnamefont{I.}},
  \bibinfo{author}{\bibfnamefont{M.}~\bibnamefont{Dine}}, and
  \bibinfo{author}{\bibfnamefont{N.}~\bibnamefont{Seiberg}},
  \bibinfo{year}{1985}, \bibinfo{journal}{Nucl. Phys.}
  \textbf{\bibinfo{volume}{B256}}, \bibinfo{pages}{557}.

\bibitem[{\citenamefont{Aharony} \emph{et~al.}(2000)\citenamefont{Aharony,
  Gubser, Maldacena, Ooguri, and Oz}}]{Aharony:1999ti}
\bibinfo{author}{\bibnamefont{Aharony}, \bibfnamefont{O.}},
  \bibinfo{author}{\bibfnamefont{S.~S.} \bibnamefont{Gubser}},
  \bibinfo{author}{\bibfnamefont{J.~M.} \bibnamefont{Maldacena}},
  \bibinfo{author}{\bibfnamefont{H.}~\bibnamefont{Ooguri}}, and
  \bibinfo{author}{\bibfnamefont{Y.}~\bibnamefont{Oz}}, \bibinfo{year}{2000},
  \bibinfo{journal}{Phys. Rept.} \textbf{\bibinfo{volume}{323}},
  \bibinfo{pages}{183}.

\bibitem[{\citenamefont{Aldazabal} \emph{et~al.}(2006)\citenamefont{Aldazabal,
  Camara, Font, and Ibanez}}]{Aldazabal:2006up}
\bibinfo{author}{\bibnamefont{Aldazabal}, \bibfnamefont{G.}},
  \bibinfo{author}{\bibfnamefont{P.~G.} \bibnamefont{Camara}},
  \bibinfo{author}{\bibfnamefont{A.}~\bibnamefont{Font}}, and
  \bibinfo{author}{\bibfnamefont{L.~E.} \bibnamefont{Ibanez}},
  \bibinfo{year}{2006}, \bibinfo{journal}{JHEP} \textbf{\bibinfo{volume}{05}},
  \bibinfo{pages}{070}.

\bibitem[{\citenamefont{Alexander}(2002)}]{Alexander:2001ks}
\bibinfo{author}{\bibnamefont{Alexander}, \bibfnamefont{S.~H.~S.}},
  \bibinfo{year}{2002}, \bibinfo{journal}{Phys. Rev.}
  \textbf{\bibinfo{volume}{D65}}, \bibinfo{pages}{023507}.

\bibitem[{\citenamefont{Alishahiha}
  \emph{et~al.}(2004)\citenamefont{Alishahiha, Silverstein, and
  Tong}}]{Alishahiha:2004eh}
\bibinfo{author}{\bibnamefont{Alishahiha}, \bibfnamefont{M.}},
  \bibinfo{author}{\bibfnamefont{E.}~\bibnamefont{Silverstein}}, and
  \bibinfo{author}{\bibfnamefont{D.}~\bibnamefont{Tong}}, \bibinfo{year}{2004},
  \bibinfo{journal}{Phys. Rev.} \textbf{\bibinfo{volume}{D70}},
  \bibinfo{pages}{123505}.

\bibitem[{\citenamefont{Allanach} \emph{et~al.}(2005)\citenamefont{Allanach,
  Brignole, and Ibanez}}]{Allanach:2005yq}
\bibinfo{author}{\bibnamefont{Allanach}, \bibfnamefont{B.~C.}},
  \bibinfo{author}{\bibfnamefont{A.}~\bibnamefont{Brignole}}, and
  \bibinfo{author}{\bibfnamefont{L.~E.} \bibnamefont{Ibanez}},
  \bibinfo{year}{2005}, \bibinfo{journal}{JHEP} \textbf{\bibinfo{volume}{05}},
  \bibinfo{pages}{030}.

\bibitem[{\citenamefont{Altland and Zirnbauer}(1997)}]{Altland}
\bibinfo{author}{\bibnamefont{Altland}, \bibfnamefont{A.}}, and
  \bibinfo{author}{\bibfnamefont{M.~R.} \bibnamefont{Zirnbauer}},
  \bibinfo{year}{1997}, \bibinfo{journal}{Phys. Rev.}
  \textbf{\bibinfo{volume}{B55}}, \bibinfo{pages}{1142}.

\bibitem[{\citenamefont{Alvarez-Gaume}
  \emph{et~al.}(1986)\citenamefont{Alvarez-Gaume, Ginsparg, Moore, and
  Vafa}}]{Alvarez-Gaume:1986jb}
\bibinfo{author}{\bibnamefont{Alvarez-Gaume}, \bibfnamefont{L.}},
  \bibinfo{author}{\bibfnamefont{P.~H.} \bibnamefont{Ginsparg}},
  \bibinfo{author}{\bibfnamefont{G.~W.} \bibnamefont{Moore}}, and
  \bibinfo{author}{\bibfnamefont{C.}~\bibnamefont{Vafa}}, \bibinfo{year}{1986},
  \bibinfo{journal}{Phys. Lett.} \textbf{\bibinfo{volume}{B171}},
  \bibinfo{pages}{155}.

\bibitem[{\citenamefont{de~Alwis}(2005)}]{deAlwis:2005tg}
\bibinfo{author}{\bibnamefont{de~Alwis}, \bibfnamefont{S.~P.}},
  \bibinfo{year}{2005}, \bibinfo{journal}{Phys. Lett.}
  \textbf{\bibinfo{volume}{B628}}, \bibinfo{pages}{183}.

\bibitem[{\citenamefont{Andrianopoli}
  \emph{et~al.}(2002{\natexlab{a}})\citenamefont{Andrianopoli, D'Auria, and
  Ferrara}}]{Andrianopoli:2001gm}
\bibinfo{author}{\bibnamefont{Andrianopoli}, \bibfnamefont{L.}},
  \bibinfo{author}{\bibfnamefont{R.}~\bibnamefont{D'Auria}}, and
  \bibinfo{author}{\bibfnamefont{S.}~\bibnamefont{Ferrara}},
  \bibinfo{year}{2002}{\natexlab{a}}, \bibinfo{journal}{Nucl. Phys.}
  \textbf{\bibinfo{volume}{B628}}, \bibinfo{pages}{387}.

\bibitem[{\citenamefont{Andrianopoli}
  \emph{et~al.}(2002{\natexlab{b}})\citenamefont{Andrianopoli, D'Auria,
  Ferrara, and Lledo}}]{Andrianopoli:2002mf}
\bibinfo{author}{\bibnamefont{Andrianopoli}, \bibfnamefont{L.}},
  \bibinfo{author}{\bibfnamefont{R.}~\bibnamefont{D'Auria}},
  \bibinfo{author}{\bibfnamefont{S.}~\bibnamefont{Ferrara}}, and
  \bibinfo{author}{\bibfnamefont{M.~A.} \bibnamefont{Lledo}},
  \bibinfo{year}{2002}{\natexlab{b}}, \bibinfo{journal}{JHEP}
  \textbf{\bibinfo{volume}{07}}, \bibinfo{pages}{010}.

\bibitem[{\citenamefont{Andrianopoli}
  \emph{et~al.}(2002{\natexlab{c}})\citenamefont{Andrianopoli, D'Auria,
  Ferrara, and Lledo}}]{Andrianopoli:2002rm}
\bibinfo{author}{\bibnamefont{Andrianopoli}, \bibfnamefont{L.}},
  \bibinfo{author}{\bibfnamefont{R.}~\bibnamefont{D'Auria}},
  \bibinfo{author}{\bibfnamefont{S.}~\bibnamefont{Ferrara}}, and
  \bibinfo{author}{\bibfnamefont{M.~A.} \bibnamefont{Lledo}},
  \bibinfo{year}{2002}{\natexlab{c}}, \bibinfo{journal}{Nucl. Phys.}
  \textbf{\bibinfo{volume}{B640}}, \bibinfo{pages}{46}.

\bibitem[{\citenamefont{Andrianopoli}
  \emph{et~al.}(2003{\natexlab{a}})\citenamefont{Andrianopoli, D'Auria,
  Ferrara, and Lledo}}]{Andrianopoli:2003jf}
\bibinfo{author}{\bibnamefont{Andrianopoli}, \bibfnamefont{L.}},
  \bibinfo{author}{\bibfnamefont{R.}~\bibnamefont{D'Auria}},
  \bibinfo{author}{\bibfnamefont{S.}~\bibnamefont{Ferrara}}, and
  \bibinfo{author}{\bibfnamefont{M.~A.} \bibnamefont{Lledo}},
  \bibinfo{year}{2003}{\natexlab{a}}, \bibinfo{journal}{JHEP}
  \textbf{\bibinfo{volume}{03}}, \bibinfo{pages}{044}.

\bibitem[{\citenamefont{Andrianopoli}
  \emph{et~al.}(2003{\natexlab{b}})\citenamefont{Andrianopoli, Ferrara, and
  Trigiante}}]{Andrianopoli:2003sa}
\bibinfo{author}{\bibnamefont{Andrianopoli}, \bibfnamefont{L.}},
  \bibinfo{author}{\bibfnamefont{S.}~\bibnamefont{Ferrara}}, and
  \bibinfo{author}{\bibfnamefont{M.}~\bibnamefont{Trigiante}},
  \bibinfo{year}{2003}{\natexlab{b}}, \eprint{hep-th/0307139}.

\bibitem[{\citenamefont{Angelantonj}
  \emph{et~al.}(2004)\citenamefont{Angelantonj, D'Auria, Ferrara, and
  Trigiante}}]{Angelantonj:2003zx}
\bibinfo{author}{\bibnamefont{Angelantonj}, \bibfnamefont{C.}},
  \bibinfo{author}{\bibfnamefont{R.}~\bibnamefont{D'Auria}},
  \bibinfo{author}{\bibfnamefont{S.}~\bibnamefont{Ferrara}}, and
  \bibinfo{author}{\bibfnamefont{M.}~\bibnamefont{Trigiante}},
  \bibinfo{year}{2004}, \bibinfo{journal}{Phys. Lett.}
  \textbf{\bibinfo{volume}{B583}}, \bibinfo{pages}{331}.

\bibitem[{\citenamefont{Angelantonj}
  \emph{et~al.}(2003)\citenamefont{Angelantonj, Ferrara, and
  Trigiante}}]{Angelantonj:2003rq}
\bibinfo{author}{\bibnamefont{Angelantonj}, \bibfnamefont{C.}},
  \bibinfo{author}{\bibfnamefont{S.}~\bibnamefont{Ferrara}}, and
  \bibinfo{author}{\bibfnamefont{M.}~\bibnamefont{Trigiante}},
  \bibinfo{year}{2003}, \bibinfo{journal}{JHEP} \textbf{\bibinfo{volume}{10}},
  \bibinfo{pages}{015}.

\bibitem[{\citenamefont{Antoniadis}
  \emph{et~al.}(1987)\citenamefont{Antoniadis, Bachas, and
  Kounnas}}]{Antoniadis:1986rn}
\bibinfo{author}{\bibnamefont{Antoniadis}, \bibfnamefont{I.}},
  \bibinfo{author}{\bibfnamefont{C.~P.} \bibnamefont{Bachas}}, and
  \bibinfo{author}{\bibfnamefont{C.}~\bibnamefont{Kounnas}},
  \bibinfo{year}{1987}, \bibinfo{journal}{Nucl. Phys.}
  \textbf{\bibinfo{volume}{B289}}, \bibinfo{pages}{87}.

\bibitem[{\citenamefont{Appelquist}
  \emph{et~al.}(1987)\citenamefont{Appelquist, Chodos, and
  Freund}}]{Appelquist:1987nr}
\bibinfo{author}{\bibnamefont{Appelquist}, \bibfnamefont{E.~., T.}},
  \bibinfo{author}{\bibfnamefont{E.~.} \bibnamefont{Chodos},
  \bibfnamefont{A.}}, and \bibinfo{author}{\bibfnamefont{E.~.}
  \bibnamefont{Freund}, \bibfnamefont{P.~G.~O.}}, \bibinfo{year}{1987},
  \bibinfo{note}{reading, USA: Addison-Wesley (1987) 619 P. (Frontiers in
  Physics, 65)}.

\bibitem[{\citenamefont{Arkani-Hamed}
  \emph{et~al.}(2003)\citenamefont{Arkani-Hamed, Cheng, Creminelli, and
  Randall}}]{Arkani-Hamed:2003wu}
\bibinfo{author}{\bibnamefont{Arkani-Hamed}, \bibfnamefont{N.}},
  \bibinfo{author}{\bibfnamefont{H.-C.} \bibnamefont{Cheng}},
  \bibinfo{author}{\bibfnamefont{P.}~\bibnamefont{Creminelli}}, and
  \bibinfo{author}{\bibfnamefont{L.}~\bibnamefont{Randall}},
  \bibinfo{year}{2003}, \bibinfo{journal}{Phys. Rev. Lett.}
  \textbf{\bibinfo{volume}{90}}, \bibinfo{pages}{221302}.

\bibitem[{\citenamefont{Arkani-Hamed and
  Dimopoulos}(2005)}]{Arkani-Hamed:2004fb}
\bibinfo{author}{\bibnamefont{Arkani-Hamed}, \bibfnamefont{N.}}, and
  \bibinfo{author}{\bibfnamefont{S.}~\bibnamefont{Dimopoulos}},
  \bibinfo{year}{2005}, \bibinfo{journal}{JHEP} \textbf{\bibinfo{volume}{06}},
  \bibinfo{pages}{073}.

\bibitem[{\citenamefont{Arkani-Hamed}
  \emph{et~al.}(1998)\citenamefont{Arkani-Hamed, Dimopoulos, and
  Dvali}}]{Arkani-Hamed:1998rs}
\bibinfo{author}{\bibnamefont{Arkani-Hamed}, \bibfnamefont{N.}},
  \bibinfo{author}{\bibfnamefont{S.}~\bibnamefont{Dimopoulos}}, and
  \bibinfo{author}{\bibfnamefont{G.~R.} \bibnamefont{Dvali}},
  \bibinfo{year}{1998}, \bibinfo{journal}{Phys. Lett.}
  \textbf{\bibinfo{volume}{B429}}, \bibinfo{pages}{263}.

\bibitem[{\citenamefont{Arkani-Hamed}
  \emph{et~al.}(2005{\natexlab{a}})\citenamefont{Arkani-Hamed, Dimopoulos,
  Giudice, and Romanino}}]{Arkani-Hamed:2004yi}
\bibinfo{author}{\bibnamefont{Arkani-Hamed}, \bibfnamefont{N.}},
  \bibinfo{author}{\bibfnamefont{S.}~\bibnamefont{Dimopoulos}},
  \bibinfo{author}{\bibfnamefont{G.~F.} \bibnamefont{Giudice}}, and
  \bibinfo{author}{\bibfnamefont{A.}~\bibnamefont{Romanino}},
  \bibinfo{year}{2005}{\natexlab{a}}, \bibinfo{journal}{Nucl. Phys.}
  \textbf{\bibinfo{volume}{B709}}, \bibinfo{pages}{3}.

\bibitem[{\citenamefont{Arkani-Hamed}
  \emph{et~al.}(2005{\natexlab{b}})\citenamefont{Arkani-Hamed, Dimopoulos, and
  Kachru}}]{Arkani-Hamed:2005yv}
\bibinfo{author}{\bibnamefont{Arkani-Hamed}, \bibfnamefont{N.}},
  \bibinfo{author}{\bibfnamefont{S.}~\bibnamefont{Dimopoulos}}, and
  \bibinfo{author}{\bibfnamefont{S.}~\bibnamefont{Kachru}},
  \bibinfo{year}{2005}{\natexlab{b}}, \eprint{hep-th/0501082}.

\bibitem[{\citenamefont{Arkani-Hamed}
  \emph{et~al.}(2006)\citenamefont{Arkani-Hamed, Motl, Nicolis, and
  Vafa}}]{Arkani-Hamed:2006dz}
\bibinfo{author}{\bibnamefont{Arkani-Hamed}, \bibfnamefont{N.}},
  \bibinfo{author}{\bibfnamefont{L.}~\bibnamefont{Motl}},
  \bibinfo{author}{\bibfnamefont{A.}~\bibnamefont{Nicolis}}, and
  \bibinfo{author}{\bibfnamefont{C.}~\bibnamefont{Vafa}}, \bibinfo{year}{2006},
  \eprint{hep-th/0601001}.

\bibitem[{\citenamefont{Ashok and Douglas}(2004)}]{Ashok:2003gk}
\bibinfo{author}{\bibnamefont{Ashok}, \bibfnamefont{S.}}, and
  \bibinfo{author}{\bibfnamefont{M.~R.} \bibnamefont{Douglas}},
  \bibinfo{year}{2004}, \bibinfo{journal}{JHEP} \textbf{\bibinfo{volume}{01}},
  \bibinfo{pages}{060}.

\bibitem[{\citenamefont{Aspinwall and Kallosh}(2005)}]{Aspinwall:2005ad}
\bibinfo{author}{\bibnamefont{Aspinwall}, \bibfnamefont{P.~S.}}, and
  \bibinfo{author}{\bibfnamefont{R.}~\bibnamefont{Kallosh}},
  \bibinfo{year}{2005}, \bibinfo{journal}{JHEP} \textbf{\bibinfo{volume}{10}},
  \bibinfo{pages}{001}.

\bibitem[{\citenamefont{Avram} \emph{et~al.}(1996)\citenamefont{Avram,
  Candelas, Jancic, and Mandelberg}}]{Avram:1995pu}
\bibinfo{author}{\bibnamefont{Avram}, \bibfnamefont{A.~C.}},
  \bibinfo{author}{\bibfnamefont{P.}~\bibnamefont{Candelas}},
  \bibinfo{author}{\bibfnamefont{D.}~\bibnamefont{Jancic}}, and
  \bibinfo{author}{\bibfnamefont{M.}~\bibnamefont{Mandelberg}},
  \bibinfo{year}{1996}, \bibinfo{journal}{Nucl. Phys.}
  \textbf{\bibinfo{volume}{B465}}, \bibinfo{pages}{458}.

\bibitem[{\citenamefont{Babich} \emph{et~al.}(2004)\citenamefont{Babich,
  Creminelli, and Zaldarriaga}}]{Babich:2004gb}
\bibinfo{author}{\bibnamefont{Babich}, \bibfnamefont{D.}},
  \bibinfo{author}{\bibfnamefont{P.}~\bibnamefont{Creminelli}}, and
  \bibinfo{author}{\bibfnamefont{M.}~\bibnamefont{Zaldarriaga}},
  \bibinfo{year}{2004}, \bibinfo{journal}{JCAP}
  \textbf{\bibinfo{volume}{0408}}, \bibinfo{pages}{009}.

\bibitem[{\citenamefont{Bachas} \emph{et~al.}(2002)\citenamefont{Bachas, Bilal,
  Douglas, Nekrasov, and David}}]{Bachas:2002mi}
\bibinfo{author}{\bibnamefont{Bachas}, \bibfnamefont{e.~., C.}},
  \bibinfo{author}{\bibfnamefont{e.~.} \bibnamefont{Bilal}, \bibfnamefont{A.}},
  \bibinfo{author}{\bibfnamefont{e.~.} \bibnamefont{Douglas},
  \bibfnamefont{M.}}, \bibinfo{author}{\bibfnamefont{e.~.}
  \bibnamefont{Nekrasov}, \bibfnamefont{N.}}, and
  \bibinfo{author}{\bibfnamefont{e.~.} \bibnamefont{David}, \bibfnamefont{F.}},
  \bibinfo{year}{2002}.

\bibitem[{\citenamefont{Balasubramanian and
  Berglund}(2004)}]{Balasubramanian:2004uy}
\bibinfo{author}{\bibnamefont{Balasubramanian}, \bibfnamefont{V.}}, and
  \bibinfo{author}{\bibfnamefont{P.}~\bibnamefont{Berglund}},
  \bibinfo{year}{2004}, \bibinfo{journal}{JHEP} \textbf{\bibinfo{volume}{11}},
  \bibinfo{pages}{085}.

\bibitem[{\citenamefont{Balasubramanian}
  \emph{et~al.}(2005)\citenamefont{Balasubramanian, Berglund, Conlon, and
  Quevedo}}]{Balasubramanian:2005zx}
\bibinfo{author}{\bibnamefont{Balasubramanian}, \bibfnamefont{V.}},
  \bibinfo{author}{\bibfnamefont{P.}~\bibnamefont{Berglund}},
  \bibinfo{author}{\bibfnamefont{J.~P.} \bibnamefont{Conlon}}, and
  \bibinfo{author}{\bibfnamefont{F.}~\bibnamefont{Quevedo}},
  \bibinfo{year}{2005}, \bibinfo{journal}{JHEP} \textbf{\bibinfo{volume}{03}},
  \bibinfo{pages}{007}.

\bibitem[{\citenamefont{Banks}(1984)}]{Banks:1984tw}
\bibinfo{author}{\bibnamefont{Banks}, \bibfnamefont{T.}}, \bibinfo{year}{1984},
  \bibinfo{journal}{Phys. Rev. Lett.} \textbf{\bibinfo{volume}{52}},
  \bibinfo{pages}{1461}.

\bibitem[{\citenamefont{Banks}(1985)}]{Banks:1984cw}
\bibinfo{author}{\bibnamefont{Banks}, \bibfnamefont{T.}}, \bibinfo{year}{1985},
  \bibinfo{journal}{Nucl. Phys.} \textbf{\bibinfo{volume}{B249}},
  \bibinfo{pages}{332}.

\bibitem[{\citenamefont{Banks}(1995{\natexlab{a}})}]{Banks:1995sw}
\bibinfo{author}{\bibnamefont{Banks}, \bibfnamefont{T.}},
  \bibinfo{year}{1995}{\natexlab{a}}, \bibinfo{note}{prepared for International
  School of Astrophysics, D. Chalonge: 4th Course: String Gravity and Physics
  at the Planck Energy scale (A NATO Advanced Study Institute), Erice, Italy,
  8-19 Sep 1995}.

\bibitem[{\citenamefont{Banks}(1995{\natexlab{b}})}]{Banks:1995uh}
\bibinfo{author}{\bibnamefont{Banks}, \bibfnamefont{T.}},
  \bibinfo{year}{1995}{\natexlab{b}}, \eprint{hep-th/9601151}.

\bibitem[{\citenamefont{Banks}(1999)}]{Banks:1999az}
\bibinfo{author}{\bibnamefont{Banks}, \bibfnamefont{T.}}, \bibinfo{year}{1999},
  \eprint{hep-th/9911068}.

\bibitem[{\citenamefont{Banks}(2000)}]{Banks:2000zy}
\bibinfo{author}{\bibnamefont{Banks}, \bibfnamefont{T.}}, \bibinfo{year}{2000},
  \eprint{hep-th/0011255}.

\bibitem[{\citenamefont{Banks}(2004)}]{Banks:2004xh}
\bibinfo{author}{\bibnamefont{Banks}, \bibfnamefont{T.}}, \bibinfo{year}{2004},
  \eprint{hep-th/0412129}.

\bibitem[{\citenamefont{Banks} \emph{et~al.}(2002)\citenamefont{Banks, Dine,
  and Douglas}}]{Banks:2001qc}
\bibinfo{author}{\bibnamefont{Banks}, \bibfnamefont{T.}},
  \bibinfo{author}{\bibfnamefont{M.}~\bibnamefont{Dine}}, and
  \bibinfo{author}{\bibfnamefont{M.~R.} \bibnamefont{Douglas}},
  \bibinfo{year}{2002}, \bibinfo{journal}{Phys. Rev. Lett.}
  \textbf{\bibinfo{volume}{88}}, \bibinfo{pages}{131301}.

\bibitem[{\citenamefont{Banks} \emph{et~al.}(2003)\citenamefont{Banks, Dine,
  Fox, and Gorbatov}}]{Banks:2003sx}
\bibinfo{author}{\bibnamefont{Banks}, \bibfnamefont{T.}},
  \bibinfo{author}{\bibfnamefont{M.}~\bibnamefont{Dine}},
  \bibinfo{author}{\bibfnamefont{P.~J.} \bibnamefont{Fox}}, and
  \bibinfo{author}{\bibfnamefont{E.}~\bibnamefont{Gorbatov}},
  \bibinfo{year}{2003}, \bibinfo{journal}{JCAP}
  \textbf{\bibinfo{volume}{0306}}, \bibinfo{pages}{001}.

\bibitem[{\citenamefont{Banks} \emph{et~al.}(2004)\citenamefont{Banks, Dine,
  and Gorbatov}}]{Banks:2003es}
\bibinfo{author}{\bibnamefont{Banks}, \bibfnamefont{T.}},
  \bibinfo{author}{\bibfnamefont{M.}~\bibnamefont{Dine}}, and
  \bibinfo{author}{\bibfnamefont{E.}~\bibnamefont{Gorbatov}},
  \bibinfo{year}{2004}, \bibinfo{journal}{JHEP} \textbf{\bibinfo{volume}{08}},
  \bibinfo{pages}{058}.

\bibitem[{\citenamefont{Banks} \emph{et~al.}(1991)\citenamefont{Banks, Dine,
  and Seiberg}}]{Banks:1991mb}
\bibinfo{author}{\bibnamefont{Banks}, \bibfnamefont{T.}},
  \bibinfo{author}{\bibfnamefont{M.}~\bibnamefont{Dine}}, and
  \bibinfo{author}{\bibfnamefont{N.}~\bibnamefont{Seiberg}},
  \bibinfo{year}{1991}, \bibinfo{journal}{Phys. Lett.}
  \textbf{\bibinfo{volume}{B273}}, \bibinfo{pages}{105}.

\bibitem[{\citenamefont{Banks} \emph{et~al.}(1988)\citenamefont{Banks, Dixon,
  Friedan, and Martinec}}]{Banks:1987cy}
\bibinfo{author}{\bibnamefont{Banks}, \bibfnamefont{T.}},
  \bibinfo{author}{\bibfnamefont{L.~J.} \bibnamefont{Dixon}},
  \bibinfo{author}{\bibfnamefont{D.}~\bibnamefont{Friedan}}, and
  \bibinfo{author}{\bibfnamefont{E.~J.} \bibnamefont{Martinec}},
  \bibinfo{year}{1988}, \bibinfo{journal}{Nucl. Phys.}
  \textbf{\bibinfo{volume}{B299}}, \bibinfo{pages}{613}.

\bibitem[{\citenamefont{Banks} \emph{et~al.}(2006)\citenamefont{Banks, Johnson,
  and Shomer}}]{Banks:2006mm}
\bibinfo{author}{\bibnamefont{Banks}, \bibfnamefont{T.}},
  \bibinfo{author}{\bibfnamefont{M.}~\bibnamefont{Johnson}}, and
  \bibinfo{author}{\bibfnamefont{A.}~\bibnamefont{Shomer}},
  \bibinfo{year}{2006}, \eprint{hep-th/0606277}.

\bibitem[{\citenamefont{Banks} \emph{et~al.}(1994)\citenamefont{Banks, Kaplan,
  and Nelson}}]{Banks:1993en}
\bibinfo{author}{\bibnamefont{Banks}, \bibfnamefont{T.}},
  \bibinfo{author}{\bibfnamefont{D.~B.} \bibnamefont{Kaplan}}, and
  \bibinfo{author}{\bibfnamefont{A.~E.} \bibnamefont{Nelson}},
  \bibinfo{year}{1994}, \bibinfo{journal}{Phys. Rev.}
  \textbf{\bibinfo{volume}{D49}}, \bibinfo{pages}{779}.

\bibitem[{\citenamefont{Barrow and Tipler}(1988)}]{Barrow}
\bibinfo{author}{\bibnamefont{Barrow}, \bibfnamefont{J.}}, and
  \bibinfo{author}{\bibfnamefont{F.}~\bibnamefont{Tipler}},
  \bibinfo{year}{1988}, \emph{\bibinfo{title}{The anthropic cosmological
  principle}} (\bibinfo{publisher}{Oxford University Press}).

\bibitem[{Baumann \emph{et~al.}(2006)\citenamefont{Baumann}
  \emph{et~al.}}]{Baumann:2006th}
\bibinfo{author}{\bibnamefont{Baumann}, \bibfnamefont{D.}}, \emph{et~al.},
  \bibinfo{year}{2006}, \eprint{hep-th/0607050}.

\bibitem[{\citenamefont{Beasley and Witten}(2002)}]{Beasley:2002db}
\bibinfo{author}{\bibnamefont{Beasley}, \bibfnamefont{C.}}, and
  \bibinfo{author}{\bibfnamefont{E.}~\bibnamefont{Witten}},
  \bibinfo{year}{2002}, \bibinfo{journal}{JHEP} \textbf{\bibinfo{volume}{07}},
  \bibinfo{pages}{46}.

\bibitem[{\citenamefont{Becker and Becker}(1996)}]{Becker:1996gj}
\bibinfo{author}{\bibnamefont{Becker}, \bibfnamefont{K.}}, and
  \bibinfo{author}{\bibfnamefont{M.}~\bibnamefont{Becker}},
  \bibinfo{year}{1996}, \bibinfo{journal}{Nucl. Phys.}
  \textbf{\bibinfo{volume}{B477}}, \bibinfo{pages}{155}.

\bibitem[{\citenamefont{Becker}
  \emph{et~al.}(2003{\natexlab{a}})\citenamefont{Becker, Becker, Dasgupta, and
  Green}}]{Becker:2003yv}
\bibinfo{author}{\bibnamefont{Becker}, \bibfnamefont{K.}},
  \bibinfo{author}{\bibfnamefont{M.}~\bibnamefont{Becker}},
  \bibinfo{author}{\bibfnamefont{K.}~\bibnamefont{Dasgupta}}, and
  \bibinfo{author}{\bibfnamefont{P.~S.} \bibnamefont{Green}},
  \bibinfo{year}{2003}{\natexlab{a}}, \bibinfo{journal}{JHEP}
  \textbf{\bibinfo{volume}{04}}, \bibinfo{pages}{007}.

\bibitem[{\citenamefont{Becker}
  \emph{et~al.}(2003{\natexlab{b}})\citenamefont{Becker, Becker, Dasgupta, and
  Prokushkin}}]{Becker:2003gq}
\bibinfo{author}{\bibnamefont{Becker}, \bibfnamefont{K.}},
  \bibinfo{author}{\bibfnamefont{M.}~\bibnamefont{Becker}},
  \bibinfo{author}{\bibfnamefont{K.}~\bibnamefont{Dasgupta}}, and
  \bibinfo{author}{\bibfnamefont{S.}~\bibnamefont{Prokushkin}},
  \bibinfo{year}{2003}{\natexlab{b}}, \bibinfo{journal}{Nucl. Phys.}
  \textbf{\bibinfo{volume}{B666}}, \bibinfo{pages}{144}.

\bibitem[{\citenamefont{Becker} \emph{et~al.}(2006)\citenamefont{Becker,
  Becker, Fu, Tseng, and Yau}}]{Becker:2006et}
\bibinfo{author}{\bibnamefont{Becker}, \bibfnamefont{K.}},
  \bibinfo{author}{\bibfnamefont{M.}~\bibnamefont{Becker}},
  \bibinfo{author}{\bibfnamefont{J.-X.} \bibnamefont{Fu}},
  \bibinfo{author}{\bibfnamefont{L.-S.} \bibnamefont{Tseng}}, and
  \bibinfo{author}{\bibfnamefont{S.-T.} \bibnamefont{Yau}},
  \bibinfo{year}{2006}, \bibinfo{journal}{Nucl. Phys.}
  \textbf{\bibinfo{volume}{B751}}, \bibinfo{pages}{108}.

\bibitem[{\citenamefont{Becker}
  \emph{et~al.}(2004{\natexlab{a}})\citenamefont{Becker, Becker, Green,
  Dasgupta, and Sharpe}}]{Becker:2003sh}
\bibinfo{author}{\bibnamefont{Becker}, \bibfnamefont{K.}},
  \bibinfo{author}{\bibfnamefont{M.}~\bibnamefont{Becker}},
  \bibinfo{author}{\bibfnamefont{P.~S.} \bibnamefont{Green}},
  \bibinfo{author}{\bibfnamefont{K.}~\bibnamefont{Dasgupta}}, and
  \bibinfo{author}{\bibfnamefont{E.}~\bibnamefont{Sharpe}},
  \bibinfo{year}{2004}{\natexlab{a}}, \bibinfo{journal}{Nucl. Phys.}
  \textbf{\bibinfo{volume}{B678}}, \bibinfo{pages}{19}.

\bibitem[{\citenamefont{Becker} \emph{et~al.}(2002)\citenamefont{Becker,
  Becker, Haack, and Louis}}]{Becker:2002nn}
\bibinfo{author}{\bibnamefont{Becker}, \bibfnamefont{K.}},
  \bibinfo{author}{\bibfnamefont{M.}~\bibnamefont{Becker}},
  \bibinfo{author}{\bibfnamefont{M.}~\bibnamefont{Haack}}, and
  \bibinfo{author}{\bibfnamefont{J.}~\bibnamefont{Louis}},
  \bibinfo{year}{2002}, \bibinfo{journal}{JHEP} \textbf{\bibinfo{volume}{06}},
  \bibinfo{pages}{060}.

\bibitem[{\citenamefont{Becker}
  \emph{et~al.}(2004{\natexlab{b}})\citenamefont{Becker, Curio, and
  Krause}}]{Becker:2004gw}
\bibinfo{author}{\bibnamefont{Becker}, \bibfnamefont{M.}},
  \bibinfo{author}{\bibfnamefont{G.}~\bibnamefont{Curio}}, and
  \bibinfo{author}{\bibfnamefont{A.}~\bibnamefont{Krause}},
  \bibinfo{year}{2004}{\natexlab{b}}, \bibinfo{journal}{Nucl. Phys.}
  \textbf{\bibinfo{volume}{B693}}, \bibinfo{pages}{223}.

\bibitem[{\citenamefont{Behrndt and
  Cvetic}(2005{\natexlab{a}})}]{Behrndt:2004km}
\bibinfo{author}{\bibnamefont{Behrndt}, \bibfnamefont{K.}}, and
  \bibinfo{author}{\bibfnamefont{M.}~\bibnamefont{Cvetic}},
  \bibinfo{year}{2005}{\natexlab{a}}, \bibinfo{journal}{Phys. Rev. Lett.}
  \textbf{\bibinfo{volume}{95}}, \bibinfo{pages}{021601}.

\bibitem[{\citenamefont{Behrndt and
  Cvetic}(2005{\natexlab{b}})}]{Behrndt:2004mj}
\bibinfo{author}{\bibnamefont{Behrndt}, \bibfnamefont{K.}}, and
  \bibinfo{author}{\bibfnamefont{M.}~\bibnamefont{Cvetic}},
  \bibinfo{year}{2005}{\natexlab{b}}, \bibinfo{journal}{Nucl. Phys.}
  \textbf{\bibinfo{volume}{B708}}, \bibinfo{pages}{45}.

\bibitem[{\citenamefont{Behrndt} \emph{et~al.}(2005)\citenamefont{Behrndt,
  Cvetic, and Gao}}]{Behrndt:2005bv}
\bibinfo{author}{\bibnamefont{Behrndt}, \bibfnamefont{K.}},
  \bibinfo{author}{\bibfnamefont{M.}~\bibnamefont{Cvetic}}, and
  \bibinfo{author}{\bibfnamefont{P.}~\bibnamefont{Gao}}, \bibinfo{year}{2005},
  \bibinfo{journal}{Nucl. Phys.} \textbf{\bibinfo{volume}{B721}},
  \bibinfo{pages}{287}.

\bibitem[{\citenamefont{Behrndt} \emph{et~al.}(2006)\citenamefont{Behrndt,
  Cvetic, and Liu}}]{Behrndt:2005im}
\bibinfo{author}{\bibnamefont{Behrndt}, \bibfnamefont{K.}},
  \bibinfo{author}{\bibfnamefont{M.}~\bibnamefont{Cvetic}}, and
  \bibinfo{author}{\bibfnamefont{T.}~\bibnamefont{Liu}}, \bibinfo{year}{2006},
  \bibinfo{journal}{Nucl. Phys.} \textbf{\bibinfo{volume}{B749}},
  \bibinfo{pages}{25}.

\bibitem[{\citenamefont{Benmachiche and Grimm}(2006)}]{Benmachiche:2006df}
\bibinfo{author}{\bibnamefont{Benmachiche}, \bibfnamefont{I.}}, and
  \bibinfo{author}{\bibfnamefont{T.~W.} \bibnamefont{Grimm}},
  \bibinfo{year}{2006}, \bibinfo{journal}{Nucl. Phys.}
  \textbf{\bibinfo{volume}{B748}}, \bibinfo{pages}{200}.

\bibitem[{\citenamefont{Berg} \emph{et~al.}(2005)\citenamefont{Berg, Haack, and
  Kors}}]{Berg:2005ja}
\bibinfo{author}{\bibnamefont{Berg}, \bibfnamefont{M.}},
  \bibinfo{author}{\bibfnamefont{M.}~\bibnamefont{Haack}}, and
  \bibinfo{author}{\bibfnamefont{B.}~\bibnamefont{Kors}}, \bibinfo{year}{2005},
  \bibinfo{journal}{JHEP} \textbf{\bibinfo{volume}{11}}, \bibinfo{pages}{030}.

\bibitem[{\citenamefont{Berg} \emph{et~al.}(2006)\citenamefont{Berg, Haack, and
  Kors}}]{Berg:2005yu}
\bibinfo{author}{\bibnamefont{Berg}, \bibfnamefont{M.}},
  \bibinfo{author}{\bibfnamefont{M.}~\bibnamefont{Haack}}, and
  \bibinfo{author}{\bibfnamefont{B.}~\bibnamefont{Kors}}, \bibinfo{year}{2006},
  \bibinfo{journal}{Phys. Rev. Lett.} \textbf{\bibinfo{volume}{96}},
  \bibinfo{pages}{021601}.

\bibitem[{\citenamefont{Berger}(1955)}]{Berger:1955}
\bibinfo{author}{\bibnamefont{Berger}, \bibfnamefont{M.}},
  \bibinfo{year}{1955}, \bibinfo{journal}{Bull. Soc. Math. de France}
  \textbf{\bibinfo{volume}{83}}, \bibinfo{pages}{279}.

\bibitem[{\citenamefont{Bergshoeff}
  \emph{et~al.}(1995)\citenamefont{Bergshoeff, Hull, and
  Ortin}}]{Bergshoeff:1995as}
\bibinfo{author}{\bibnamefont{Bergshoeff}, \bibfnamefont{E.}},
  \bibinfo{author}{\bibfnamefont{C.~M.} \bibnamefont{Hull}}, and
  \bibinfo{author}{\bibfnamefont{T.}~\bibnamefont{Ortin}},
  \bibinfo{year}{1995}, \bibinfo{journal}{Nucl. Phys.}
  \textbf{\bibinfo{volume}{B451}}, \bibinfo{pages}{547}.

\bibitem[{\citenamefont{Berkooz} \emph{et~al.}(1996)\citenamefont{Berkooz,
  Douglas, and Leigh}}]{Berkooz:1996km}
\bibinfo{author}{\bibnamefont{Berkooz}, \bibfnamefont{M.}},
  \bibinfo{author}{\bibfnamefont{M.~R.} \bibnamefont{Douglas}}, and
  \bibinfo{author}{\bibfnamefont{R.~G.} \bibnamefont{Leigh}},
  \bibinfo{year}{1996}, \bibinfo{journal}{Nucl. Phys.}
  \textbf{\bibinfo{volume}{B480}}, \bibinfo{pages}{265}.

\bibitem[{\citenamefont{Binetruy and Gaillard}(1986)}]{Binetruy:1986ss}
\bibinfo{author}{\bibnamefont{Binetruy}, \bibfnamefont{P.}}, and
  \bibinfo{author}{\bibfnamefont{M.~K.} \bibnamefont{Gaillard}},
  \bibinfo{year}{1986}, \bibinfo{journal}{Phys. Rev.}
  \textbf{\bibinfo{volume}{D34}}, \bibinfo{pages}{3069}.

\bibitem[{Blanco-Pillado \emph{et~al.}(2004)\citenamefont{Blanco-Pillado}
  \emph{et~al.}}]{Blanco-Pillado:2004ns}
\bibinfo{author}{\bibnamefont{Blanco-Pillado}, \bibfnamefont{J.~J.}},
  \emph{et~al.}, \bibinfo{year}{2004}, \bibinfo{journal}{JHEP}
  \textbf{\bibinfo{volume}{11}}, \bibinfo{pages}{063}.

\bibitem[{Blanco-Pillado \emph{et~al.}(2006)\citenamefont{Blanco-Pillado}
  \emph{et~al.}}]{Blanco-Pillado:2006he}
\bibinfo{author}{\bibnamefont{Blanco-Pillado}, \bibfnamefont{J.~J.}},
  \emph{et~al.}, \bibinfo{year}{2006}, \eprint{hep-th/0603129}.

\bibitem[{\citenamefont{Blumenhagen}
  \emph{et~al.}(2005{\natexlab{a}})\citenamefont{Blumenhagen, Cvetic,
  Langacker, and Shiu}}]{Blumenhagen:2005mu}
\bibinfo{author}{\bibnamefont{Blumenhagen}, \bibfnamefont{R.}},
  \bibinfo{author}{\bibfnamefont{M.}~\bibnamefont{Cvetic}},
  \bibinfo{author}{\bibfnamefont{P.}~\bibnamefont{Langacker}}, and
  \bibinfo{author}{\bibfnamefont{G.}~\bibnamefont{Shiu}},
  \bibinfo{year}{2005}{\natexlab{a}}, \bibinfo{journal}{Ann. Rev. Nucl. Part.
  Sci.} \textbf{\bibinfo{volume}{55}}, \bibinfo{pages}{71}.

\bibitem[{\citenamefont{Blumenhagen}
  \emph{et~al.}(2006)\citenamefont{Blumenhagen, Cvetic, and
  Weigand}}]{Blumenhagen:2006xt}
\bibinfo{author}{\bibnamefont{Blumenhagen}, \bibfnamefont{R.}},
  \bibinfo{author}{\bibfnamefont{M.}~\bibnamefont{Cvetic}}, and
  \bibinfo{author}{\bibfnamefont{T.}~\bibnamefont{Weigand}},
  \bibinfo{year}{2006}, \eprint{hep-th/0609191}.

\bibitem[{\citenamefont{Blumenhagen}
  \emph{et~al.}(2005{\natexlab{b}})\citenamefont{Blumenhagen, Gmeiner,
  Honecker, Lust, and Weigand}}]{Blumenhagen:2004xx}
\bibinfo{author}{\bibnamefont{Blumenhagen}, \bibfnamefont{R.}},
  \bibinfo{author}{\bibfnamefont{F.}~\bibnamefont{Gmeiner}},
  \bibinfo{author}{\bibfnamefont{G.}~\bibnamefont{Honecker}},
  \bibinfo{author}{\bibfnamefont{D.}~\bibnamefont{Lust}}, and
  \bibinfo{author}{\bibfnamefont{T.}~\bibnamefont{Weigand}},
  \bibinfo{year}{2005}{\natexlab{b}}, \bibinfo{journal}{Nucl. Phys.}
  \textbf{\bibinfo{volume}{B713}}, \bibinfo{pages}{83}.

\bibitem[{\citenamefont{Blumenhagen}
  \emph{et~al.}(2003)\citenamefont{Blumenhagen, Lust, and
  Taylor}}]{Blumenhagen:2003vr}
\bibinfo{author}{\bibnamefont{Blumenhagen}, \bibfnamefont{R.}},
  \bibinfo{author}{\bibfnamefont{D.}~\bibnamefont{Lust}}, and
  \bibinfo{author}{\bibfnamefont{T.~R.} \bibnamefont{Taylor}},
  \bibinfo{year}{2003}, \bibinfo{journal}{Nucl. Phys.}
  \textbf{\bibinfo{volume}{B663}}, \bibinfo{pages}{319}.

\bibitem[{\citenamefont{Bousso}(2006)}]{Bousso:2006ev}
\bibinfo{author}{\bibnamefont{Bousso}, \bibfnamefont{R.}},
  \bibinfo{year}{2006}, \eprint{hep-th/0605263}.

\bibitem[{\citenamefont{Bousso and Polchinski}(2000)}]{Bousso:2000xa}
\bibinfo{author}{\bibnamefont{Bousso}, \bibfnamefont{R.}}, and
  \bibinfo{author}{\bibfnamefont{J.}~\bibnamefont{Polchinski}},
  \bibinfo{year}{2000}, \bibinfo{journal}{JHEP} \textbf{\bibinfo{volume}{06}},
  \bibinfo{pages}{006}.

\bibitem[{\citenamefont{Bouwknegt} \emph{et~al.}(2004)\citenamefont{Bouwknegt,
  Evslin, and Mathai}}]{Bouwknegt:2003wp}
\bibinfo{author}{\bibnamefont{Bouwknegt}, \bibfnamefont{P.}},
  \bibinfo{author}{\bibfnamefont{J.}~\bibnamefont{Evslin}}, and
  \bibinfo{author}{\bibfnamefont{V.}~\bibnamefont{Mathai}},
  \bibinfo{year}{2004}, \bibinfo{journal}{Phys. Rev. Lett.}
  \textbf{\bibinfo{volume}{92}}, \bibinfo{pages}{181601}.

\bibitem[{\citenamefont{Bovy} \emph{et~al.}(2005)\citenamefont{Bovy, Lust, and
  Tsimpis}}]{Bovy:2005qq}
\bibinfo{author}{\bibnamefont{Bovy}, \bibfnamefont{J.}},
  \bibinfo{author}{\bibfnamefont{D.}~\bibnamefont{Lust}}, and
  \bibinfo{author}{\bibfnamefont{D.}~\bibnamefont{Tsimpis}},
  \bibinfo{year}{2005}, \bibinfo{journal}{JHEP} \textbf{\bibinfo{volume}{08}},
  \bibinfo{pages}{056}.

\bibitem[{\citenamefont{Brandenberger and Vafa}(1989)}]{Brandenberger:1988aj}
\bibinfo{author}{\bibnamefont{Brandenberger}, \bibfnamefont{R.~H.}}, and
  \bibinfo{author}{\bibfnamefont{C.}~\bibnamefont{Vafa}}, \bibinfo{year}{1989},
  \bibinfo{journal}{Nucl. Phys.} \textbf{\bibinfo{volume}{B316}},
  \bibinfo{pages}{391}.

\bibitem[{\citenamefont{Braun} \emph{et~al.}(2006)\citenamefont{Braun,
  Buchbinder, and Ovrut}}]{Braun:2006em}
\bibinfo{author}{\bibnamefont{Braun}, \bibfnamefont{V.}},
  \bibinfo{author}{\bibfnamefont{E.~I.} \bibnamefont{Buchbinder}}, and
  \bibinfo{author}{\bibfnamefont{B.~A.} \bibnamefont{Ovrut}},
  \bibinfo{year}{2006}, \bibinfo{journal}{Phys. Lett.}
  \textbf{\bibinfo{volume}{B639}}, \bibinfo{pages}{566}.

\bibitem[{\citenamefont{Brown and Teitelboim}(1987)}]{Brown:1987dd}
\bibinfo{author}{\bibnamefont{Brown}, \bibfnamefont{J.~D.}}, and
  \bibinfo{author}{\bibfnamefont{C.}~\bibnamefont{Teitelboim}},
  \bibinfo{year}{1987}, \bibinfo{journal}{Phys. Lett.}
  \textbf{\bibinfo{volume}{B195}}, \bibinfo{pages}{177}.

\bibitem[{\citenamefont{Brown and Teitelboim}(1988)}]{Brown:1988kg}
\bibinfo{author}{\bibnamefont{Brown}, \bibfnamefont{J.~D.}}, and
  \bibinfo{author}{\bibfnamefont{C.}~\bibnamefont{Teitelboim}},
  \bibinfo{year}{1988}, \bibinfo{journal}{Nucl. Phys.}
  \textbf{\bibinfo{volume}{B297}}, \bibinfo{pages}{787}.

\bibitem[{\citenamefont{Brummer} \emph{et~al.}(2006)\citenamefont{Brummer,
  Hebecker, and Trapletti}}]{Brummer:2006dg}
\bibinfo{author}{\bibnamefont{Brummer}, \bibfnamefont{F.}},
  \bibinfo{author}{\bibfnamefont{A.}~\bibnamefont{Hebecker}}, and
  \bibinfo{author}{\bibfnamefont{M.}~\bibnamefont{Trapletti}},
  \bibinfo{year}{2006}, \eprint{hep-th/0605232}.

\bibitem[{\citenamefont{Brustein} \emph{et~al.}(2003)\citenamefont{Brustein,
  De~Alwis, and Novak}}]{Brustein:2002mp}
\bibinfo{author}{\bibnamefont{Brustein}, \bibfnamefont{R.}},
  \bibinfo{author}{\bibfnamefont{S.~P.} \bibnamefont{De~Alwis}}, and
  \bibinfo{author}{\bibfnamefont{E.~G.} \bibnamefont{Novak}},
  \bibinfo{year}{2003}, \bibinfo{journal}{Phys. Rev.}
  \textbf{\bibinfo{volume}{D68}}, \bibinfo{pages}{023517}.

\bibitem[{\citenamefont{Burgess} \emph{et~al.}(2006)\citenamefont{Burgess,
  Escoda, and Quevedo}}]{Burgess:2005jx}
\bibinfo{author}{\bibnamefont{Burgess}, \bibfnamefont{C.~P.}},
  \bibinfo{author}{\bibfnamefont{C.}~\bibnamefont{Escoda}}, and
  \bibinfo{author}{\bibfnamefont{F.}~\bibnamefont{Quevedo}},
  \bibinfo{year}{2006}, \bibinfo{journal}{JHEP} \textbf{\bibinfo{volume}{06}},
  \bibinfo{pages}{044}.

\bibitem[{\citenamefont{Burgess} \emph{et~al.}(2003)\citenamefont{Burgess,
  Kallosh, and Quevedo}}]{Burgess:2003ic}
\bibinfo{author}{\bibnamefont{Burgess}, \bibfnamefont{C.~P.}},
  \bibinfo{author}{\bibfnamefont{R.}~\bibnamefont{Kallosh}}, and
  \bibinfo{author}{\bibfnamefont{F.}~\bibnamefont{Quevedo}},
  \bibinfo{year}{2003}, \bibinfo{journal}{JHEP} \textbf{\bibinfo{volume}{10}},
  \bibinfo{pages}{056}.

\bibitem[{\citenamefont{Burgess} \emph{et~al.}(2002)\citenamefont{Burgess,
  Martineau, Quevedo, Rajesh, and Zhang}}]{Burgess:2001vr}
\bibinfo{author}{\bibnamefont{Burgess}, \bibfnamefont{C.~P.}},
  \bibinfo{author}{\bibfnamefont{P.}~\bibnamefont{Martineau}},
  \bibinfo{author}{\bibfnamefont{F.}~\bibnamefont{Quevedo}},
  \bibinfo{author}{\bibfnamefont{G.}~\bibnamefont{Rajesh}}, and
  \bibinfo{author}{\bibfnamefont{R.~J.} \bibnamefont{Zhang}},
  \bibinfo{year}{2002}, \bibinfo{journal}{JHEP} \textbf{\bibinfo{volume}{03}},
  \bibinfo{pages}{052}.

\bibitem[{Burgess \emph{et~al.}(2001)\citenamefont{Burgess}
  \emph{et~al.}}]{Burgess:2001fx}
\bibinfo{author}{\bibnamefont{Burgess}, \bibfnamefont{C.~P.}}, \emph{et~al.},
  \bibinfo{year}{2001}, \bibinfo{journal}{JHEP} \textbf{\bibinfo{volume}{07}},
  \bibinfo{pages}{047}.

\bibitem[{\citenamefont{Buscher}(1987)}]{Buscher:1987sk}
\bibinfo{author}{\bibnamefont{Buscher}, \bibfnamefont{T.~H.}},
  \bibinfo{year}{1987}, \bibinfo{journal}{Phys. Lett.}
  \textbf{\bibinfo{volume}{B194}}, \bibinfo{pages}{59}.

\bibitem[{\citenamefont{Buscher}(1988)}]{Buscher:1987qj}
\bibinfo{author}{\bibnamefont{Buscher}, \bibfnamefont{T.~H.}},
  \bibinfo{year}{1988}, \bibinfo{journal}{Phys. Lett.}
  \textbf{\bibinfo{volume}{B201}}, \bibinfo{pages}{466}.

\bibitem[{\citenamefont{Callan and Coleman}(1977)}]{Callan:1977pt}
\bibinfo{author}{\bibnamefont{Callan}, \bibfnamefont{J., Curtis~G.}}, and
  \bibinfo{author}{\bibfnamefont{S.~R.} \bibnamefont{Coleman}},
  \bibinfo{year}{1977}, \bibinfo{journal}{Phys. Rev.}
  \textbf{\bibinfo{volume}{D16}}, \bibinfo{pages}{1762}.

\bibitem[{\citenamefont{Camara}
  \emph{et~al.}(2005{\natexlab{a}})\citenamefont{Camara, Font, and
  Ibanez}}]{Camara:2005dc}
\bibinfo{author}{\bibnamefont{Camara}, \bibfnamefont{P.~G.}},
  \bibinfo{author}{\bibfnamefont{A.}~\bibnamefont{Font}}, and
  \bibinfo{author}{\bibfnamefont{L.~E.} \bibnamefont{Ibanez}},
  \bibinfo{year}{2005}{\natexlab{a}}, \bibinfo{journal}{JHEP}
  \textbf{\bibinfo{volume}{09}}, \bibinfo{pages}{013}.

\bibitem[{\citenamefont{Camara} \emph{et~al.}(2004)\citenamefont{Camara,
  Ibanez, and Uranga}}]{Camara:2003ku}
\bibinfo{author}{\bibnamefont{Camara}, \bibfnamefont{P.~G.}},
  \bibinfo{author}{\bibfnamefont{L.~E.} \bibnamefont{Ibanez}}, and
  \bibinfo{author}{\bibfnamefont{A.~M.} \bibnamefont{Uranga}},
  \bibinfo{year}{2004}, \bibinfo{journal}{Nucl. Phys.}
  \textbf{\bibinfo{volume}{B689}}, \bibinfo{pages}{195}.

\bibitem[{\citenamefont{Camara}
  \emph{et~al.}(2005{\natexlab{b}})\citenamefont{Camara, Ibanez, and
  Uranga}}]{Camara:2004jj}
\bibinfo{author}{\bibnamefont{Camara}, \bibfnamefont{P.~G.}},
  \bibinfo{author}{\bibfnamefont{L.~E.} \bibnamefont{Ibanez}}, and
  \bibinfo{author}{\bibfnamefont{A.~M.} \bibnamefont{Uranga}},
  \bibinfo{year}{2005}{\natexlab{b}}, \bibinfo{journal}{Nucl. Phys.}
  \textbf{\bibinfo{volume}{B708}}, \bibinfo{pages}{268}.

\bibitem[{\citenamefont{Candelas} \emph{et~al.}(1991)\citenamefont{Candelas,
  De~La~Ossa, Green, and Parkes}}]{Candelas:1990rm}
\bibinfo{author}{\bibnamefont{Candelas}, \bibfnamefont{P.}},
  \bibinfo{author}{\bibfnamefont{X.~C.} \bibnamefont{De~La~Ossa}},
  \bibinfo{author}{\bibfnamefont{P.~S.} \bibnamefont{Green}}, and
  \bibinfo{author}{\bibfnamefont{L.}~\bibnamefont{Parkes}},
  \bibinfo{year}{1991}, \bibinfo{journal}{Nucl. Phys.}
  \textbf{\bibinfo{volume}{B359}}, \bibinfo{pages}{21}.

\bibitem[{\citenamefont{Candelas} \emph{et~al.}(1985)\citenamefont{Candelas,
  Horowitz, Strominger, and Witten}}]{Candelas:1985en}
\bibinfo{author}{\bibnamefont{Candelas}, \bibfnamefont{P.}},
  \bibinfo{author}{\bibfnamefont{G.~T.} \bibnamefont{Horowitz}},
  \bibinfo{author}{\bibfnamefont{A.}~\bibnamefont{Strominger}}, and
  \bibinfo{author}{\bibfnamefont{E.}~\bibnamefont{Witten}},
  \bibinfo{year}{1985}, \bibinfo{journal}{Nucl. Phys.}
  \textbf{\bibinfo{volume}{B258}}, \bibinfo{pages}{46}.

\bibitem[{\citenamefont{Candelas and de~la Ossa}(1991)}]{Candelas:1990pi}
\bibinfo{author}{\bibnamefont{Candelas}, \bibfnamefont{P.}}, and
  \bibinfo{author}{\bibfnamefont{X.}~\bibnamefont{de~la Ossa}},
  \bibinfo{year}{1991}, \bibinfo{journal}{Nucl. Phys.}
  \textbf{\bibinfo{volume}{B355}}, \bibinfo{pages}{455}.

\bibitem[{\citenamefont{Candelas} \emph{et~al.}(1995)\citenamefont{Candelas,
  de~la Ossa, and Katz}}]{Candelas:1994bu}
\bibinfo{author}{\bibnamefont{Candelas}, \bibfnamefont{P.}},
  \bibinfo{author}{\bibfnamefont{X.}~\bibnamefont{de~la Ossa}}, and
  \bibinfo{author}{\bibfnamefont{S.}~\bibnamefont{Katz}}, \bibinfo{year}{1995},
  \bibinfo{journal}{Nucl. Phys.} \textbf{\bibinfo{volume}{B450}},
  \bibinfo{pages}{267}.

\bibitem[{\citenamefont{Candelas and de~la Ossa}(1990)}]{Candelas:1989js}
\bibinfo{author}{\bibnamefont{Candelas}, \bibfnamefont{P.}}, and
  \bibinfo{author}{\bibfnamefont{X.~C.} \bibnamefont{de~la Ossa}},
  \bibinfo{year}{1990}, \bibinfo{journal}{Nucl. Phys.}
  \textbf{\bibinfo{volume}{B342}}, \bibinfo{pages}{246}.

\bibitem[{\citenamefont{de~Carlos} \emph{et~al.}(1993)\citenamefont{de~Carlos,
  Casas, Quevedo, and Roulet}}]{deCarlos:1993jw}
\bibinfo{author}{\bibnamefont{de~Carlos}, \bibfnamefont{B.}},
  \bibinfo{author}{\bibfnamefont{J.~A.} \bibnamefont{Casas}},
  \bibinfo{author}{\bibfnamefont{F.}~\bibnamefont{Quevedo}}, and
  \bibinfo{author}{\bibfnamefont{E.}~\bibnamefont{Roulet}},
  \bibinfo{year}{1993}, \bibinfo{journal}{Phys. Lett.}
  \textbf{\bibinfo{volume}{B318}}, \bibinfo{pages}{447}.

\bibitem[{\citenamefont{Carroll}(2001)}]{Carroll:2000fy}
\bibinfo{author}{\bibnamefont{Carroll}, \bibfnamefont{S.~M.}},
  \bibinfo{year}{2001}, \bibinfo{journal}{Living Rev. Rel.}
  \textbf{\bibinfo{volume}{4}}, \bibinfo{pages}{1}.

\bibitem[{\citenamefont{Cascales} \emph{et~al.}(2004)\citenamefont{Cascales,
  Garcia~del Moral, Quevedo, and Uranga}}]{Cascales:2003wn}
\bibinfo{author}{\bibnamefont{Cascales}, \bibfnamefont{J.~F.~G.}},
  \bibinfo{author}{\bibfnamefont{M.~P.} \bibnamefont{Garcia~del Moral}},
  \bibinfo{author}{\bibfnamefont{F.}~\bibnamefont{Quevedo}}, and
  \bibinfo{author}{\bibfnamefont{A.~M.} \bibnamefont{Uranga}},
  \bibinfo{year}{2004}, \bibinfo{journal}{JHEP} \textbf{\bibinfo{volume}{02}},
  \bibinfo{pages}{031}.

\bibitem[{\citenamefont{Cascales} \emph{et~al.}(2005)\citenamefont{Cascales,
  Saad, and Uranga}}]{Cascales:2005rj}
\bibinfo{author}{\bibnamefont{Cascales}, \bibfnamefont{J.~F.~G.}},
  \bibinfo{author}{\bibfnamefont{F.}~\bibnamefont{Saad}}, and
  \bibinfo{author}{\bibfnamefont{A.~M.} \bibnamefont{Uranga}},
  \bibinfo{year}{2005}, \bibinfo{journal}{JHEP} \textbf{\bibinfo{volume}{11}},
  \bibinfo{pages}{047}.

\bibitem[{\citenamefont{Cascales and
  Uranga}(2003{\natexlab{a}})}]{Cascales:2003zp}
\bibinfo{author}{\bibnamefont{Cascales}, \bibfnamefont{J.~F.~G.}}, and
  \bibinfo{author}{\bibfnamefont{A.~M.} \bibnamefont{Uranga}},
  \bibinfo{year}{2003}{\natexlab{a}}, \bibinfo{journal}{JHEP}
  \textbf{\bibinfo{volume}{05}}, \bibinfo{pages}{011}.

\bibitem[{\citenamefont{Cascales and
  Uranga}(2003{\natexlab{b}})}]{Cascales:2003pt}
\bibinfo{author}{\bibnamefont{Cascales}, \bibfnamefont{J.~F.~G.}}, and
  \bibinfo{author}{\bibfnamefont{A.~M.} \bibnamefont{Uranga}},
  \bibinfo{year}{2003}{\natexlab{b}}, \eprint{hep-th/0311250}.

\bibitem[{\citenamefont{Ceresole} \emph{et~al.}(2006)\citenamefont{Ceresole,
  Dall'Agata, Giryavets, Kallosh, and Linde}}]{Ceresole:2006iq}
\bibinfo{author}{\bibnamefont{Ceresole}, \bibfnamefont{A.}},
  \bibinfo{author}{\bibfnamefont{G.}~\bibnamefont{Dall'Agata}},
  \bibinfo{author}{\bibfnamefont{A.}~\bibnamefont{Giryavets}},
  \bibinfo{author}{\bibfnamefont{R.}~\bibnamefont{Kallosh}}, and
  \bibinfo{author}{\bibfnamefont{A.}~\bibnamefont{Linde}},
  \bibinfo{year}{2006}, \eprint{hep-th/0605266}.

\bibitem[{\citenamefont{Chan} \emph{et~al.}(2000)\citenamefont{Chan, Paul, and
  Verlinde}}]{Chan:2000ms}
\bibinfo{author}{\bibnamefont{Chan}, \bibfnamefont{C.~S.}},
  \bibinfo{author}{\bibfnamefont{P.~L.} \bibnamefont{Paul}}, and
  \bibinfo{author}{\bibfnamefont{H.~L.} \bibnamefont{Verlinde}},
  \bibinfo{year}{2000}, \bibinfo{journal}{Nucl. Phys.}
  \textbf{\bibinfo{volume}{B581}}, \bibinfo{pages}{156}.

\bibitem[{\citenamefont{Chen}(2005{\natexlab{a}})}]{Chen:2005ad}
\bibinfo{author}{\bibnamefont{Chen}, \bibfnamefont{X.}},
  \bibinfo{year}{2005}{\natexlab{a}}, \bibinfo{journal}{JHEP}
  \textbf{\bibinfo{volume}{08}}, \bibinfo{pages}{045}.

\bibitem[{\citenamefont{Chen}(2005{\natexlab{b}})}]{Chen:2004gc}
\bibinfo{author}{\bibnamefont{Chen}, \bibfnamefont{X.}},
  \bibinfo{year}{2005}{\natexlab{b}}, \bibinfo{journal}{Phys. Rev.}
  \textbf{\bibinfo{volume}{D71}}, \bibinfo{pages}{063506}.

\bibitem[{\citenamefont{Chen} \emph{et~al.}(2006)\citenamefont{Chen, Huang,
  Kachru, and Shiu}}]{Chen:2006nt}
\bibinfo{author}{\bibnamefont{Chen}, \bibfnamefont{X.}},
  \bibinfo{author}{\bibfnamefont{M.-x.} \bibnamefont{Huang}},
  \bibinfo{author}{\bibfnamefont{S.}~\bibnamefont{Kachru}}, and
  \bibinfo{author}{\bibfnamefont{G.}~\bibnamefont{Shiu}}, \bibinfo{year}{2006},
  \eprint{hep-th/0605045}.

\bibitem[{\citenamefont{Chiantese} \emph{et~al.}(2006)\citenamefont{Chiantese,
  Gmeiner, and Jeschek}}]{Chiantese:2004tx}
\bibinfo{author}{\bibnamefont{Chiantese}, \bibfnamefont{S.}},
  \bibinfo{author}{\bibfnamefont{F.}~\bibnamefont{Gmeiner}}, and
  \bibinfo{author}{\bibfnamefont{C.}~\bibnamefont{Jeschek}},
  \bibinfo{year}{2006}, \bibinfo{journal}{Int. J. Mod. Phys.}
  \textbf{\bibinfo{volume}{A21}}, \bibinfo{pages}{2377}.

\bibitem[{\citenamefont{Chiossi and Salamon}(2002)}]{Chiossi:2002tw}
\bibinfo{author}{\bibnamefont{Chiossi}, \bibfnamefont{S.}}, and
  \bibinfo{author}{\bibfnamefont{S.}~\bibnamefont{Salamon}},
  \bibinfo{year}{2002}, \eprint{math.dg/0202282}.

\bibitem[{\citenamefont{Choi}
  \emph{et~al.}(2005{\natexlab{a}})\citenamefont{Choi, Falkowski, Nilles, and
  Olechowski}}]{Choi:2005ge}
\bibinfo{author}{\bibnamefont{Choi}, \bibfnamefont{K.}},
  \bibinfo{author}{\bibfnamefont{A.}~\bibnamefont{Falkowski}},
  \bibinfo{author}{\bibfnamefont{H.~P.} \bibnamefont{Nilles}}, and
  \bibinfo{author}{\bibfnamefont{M.}~\bibnamefont{Olechowski}},
  \bibinfo{year}{2005}{\natexlab{a}}, \bibinfo{journal}{Nucl. Phys.}
  \textbf{\bibinfo{volume}{B718}}, \bibinfo{pages}{113}.

\bibitem[{\citenamefont{Choi} \emph{et~al.}(2004)\citenamefont{Choi, Falkowski,
  Nilles, Olechowski, and Pokorski}}]{Choi:2004sx}
\bibinfo{author}{\bibnamefont{Choi}, \bibfnamefont{K.}},
  \bibinfo{author}{\bibfnamefont{A.}~\bibnamefont{Falkowski}},
  \bibinfo{author}{\bibfnamefont{H.~P.} \bibnamefont{Nilles}},
  \bibinfo{author}{\bibfnamefont{M.}~\bibnamefont{Olechowski}}, and
  \bibinfo{author}{\bibfnamefont{S.}~\bibnamefont{Pokorski}},
  \bibinfo{year}{2004}, \bibinfo{journal}{JHEP} \textbf{\bibinfo{volume}{11}},
  \bibinfo{pages}{076}.

\bibitem[{\citenamefont{Choi} \emph{et~al.}(2006)\citenamefont{Choi, Jeong,
  Kobayashi, and Okumura}}]{Choi:2005hd}
\bibinfo{author}{\bibnamefont{Choi}, \bibfnamefont{K.}},
  \bibinfo{author}{\bibfnamefont{K.~S.} \bibnamefont{Jeong}},
  \bibinfo{author}{\bibfnamefont{T.}~\bibnamefont{Kobayashi}}, and
  \bibinfo{author}{\bibfnamefont{K.-i.} \bibnamefont{Okumura}},
  \bibinfo{year}{2006}, \bibinfo{journal}{Phys. Lett.}
  \textbf{\bibinfo{volume}{B633}}, \bibinfo{pages}{355}.

\bibitem[{\citenamefont{Choi}
  \emph{et~al.}(2005{\natexlab{b}})\citenamefont{Choi, Jeong, and
  Okumura}}]{Choi:2005uz}
\bibinfo{author}{\bibnamefont{Choi}, \bibfnamefont{K.}},
  \bibinfo{author}{\bibfnamefont{K.~S.} \bibnamefont{Jeong}}, and
  \bibinfo{author}{\bibfnamefont{K.-i.} \bibnamefont{Okumura}},
  \bibinfo{year}{2005}{\natexlab{b}}, \bibinfo{journal}{JHEP}
  \textbf{\bibinfo{volume}{09}}, \bibinfo{pages}{039}.

\bibitem[{\citenamefont{Chuang} \emph{et~al.}(2005)\citenamefont{Chuang,
  Kachru, and Tomasiello}}]{Chuang:2005qd}
\bibinfo{author}{\bibnamefont{Chuang}, \bibfnamefont{W.-y.}},
  \bibinfo{author}{\bibfnamefont{S.}~\bibnamefont{Kachru}}, and
  \bibinfo{author}{\bibfnamefont{A.}~\bibnamefont{Tomasiello}},
  \bibinfo{year}{2005}, \eprint{hep-th/0510042}.

\bibitem[{\citenamefont{Coleman}(1977)}]{Coleman:1977py}
\bibinfo{author}{\bibnamefont{Coleman}, \bibfnamefont{S.~R.}},
  \bibinfo{year}{1977}, \bibinfo{journal}{Phys. Rev.}
  \textbf{\bibinfo{volume}{D15}}, \bibinfo{pages}{2929}.

\bibitem[{\citenamefont{Coleman}(1988)}]{Coleman:1988tj}
\bibinfo{author}{\bibnamefont{Coleman}, \bibfnamefont{S.~R.}},
  \bibinfo{year}{1988}, \bibinfo{journal}{Nucl. Phys.}
  \textbf{\bibinfo{volume}{B310}}, \bibinfo{pages}{643}.

\bibitem[{\citenamefont{Coleman and De~Luccia}(1980)}]{Coleman:1980aw}
\bibinfo{author}{\bibnamefont{Coleman}, \bibfnamefont{S.~R.}}, and
  \bibinfo{author}{\bibfnamefont{F.}~\bibnamefont{De~Luccia}},
  \bibinfo{year}{1980}, \bibinfo{journal}{Phys. Rev.}
  \textbf{\bibinfo{volume}{D21}}, \bibinfo{pages}{3305}.

\bibitem[{\citenamefont{Conlon and Quevedo}(2004)}]{Conlon:2004ds}
\bibinfo{author}{\bibnamefont{Conlon}, \bibfnamefont{J.~P.}}, and
  \bibinfo{author}{\bibfnamefont{F.}~\bibnamefont{Quevedo}},
  \bibinfo{year}{2004}, \bibinfo{journal}{JHEP} \textbf{\bibinfo{volume}{10}},
  \bibinfo{pages}{039}.

\bibitem[{\citenamefont{Conlon and Quevedo}(2006)}]{Conlon:2006us}
\bibinfo{author}{\bibnamefont{Conlon}, \bibfnamefont{J.~P.}}, and
  \bibinfo{author}{\bibfnamefont{F.}~\bibnamefont{Quevedo}},
  \bibinfo{year}{2006}, \bibinfo{journal}{JHEP} \textbf{\bibinfo{volume}{06}},
  \bibinfo{pages}{029}.

\bibitem[{\citenamefont{Conlon} \emph{et~al.}(2005)\citenamefont{Conlon,
  Quevedo, and Suruliz}}]{Conlon:2005ki}
\bibinfo{author}{\bibnamefont{Conlon}, \bibfnamefont{J.~P.}},
  \bibinfo{author}{\bibfnamefont{F.}~\bibnamefont{Quevedo}}, and
  \bibinfo{author}{\bibfnamefont{K.}~\bibnamefont{Suruliz}},
  \bibinfo{year}{2005}, \bibinfo{journal}{JHEP} \textbf{\bibinfo{volume}{08}},
  \bibinfo{pages}{007}.

\bibitem[{\citenamefont{Copeland} \emph{et~al.}(1994)\citenamefont{Copeland,
  Liddle, Lyth, Stewart, and Wands}}]{Copeland:1994vg}
\bibinfo{author}{\bibnamefont{Copeland}, \bibfnamefont{E.~J.}},
  \bibinfo{author}{\bibfnamefont{A.~R.} \bibnamefont{Liddle}},
  \bibinfo{author}{\bibfnamefont{D.~H.} \bibnamefont{Lyth}},
  \bibinfo{author}{\bibfnamefont{E.~D.} \bibnamefont{Stewart}}, and
  \bibinfo{author}{\bibfnamefont{D.}~\bibnamefont{Wands}},
  \bibinfo{year}{1994}, \bibinfo{journal}{Phys. Rev.}
  \textbf{\bibinfo{volume}{D49}}, \bibinfo{pages}{6410}.

\bibitem[{\citenamefont{Copeland} \emph{et~al.}(2004)\citenamefont{Copeland,
  Myers, and Polchinski}}]{Copeland:2003bj}
\bibinfo{author}{\bibnamefont{Copeland}, \bibfnamefont{E.~J.}},
  \bibinfo{author}{\bibfnamefont{R.~C.} \bibnamefont{Myers}}, and
  \bibinfo{author}{\bibfnamefont{J.}~\bibnamefont{Polchinski}},
  \bibinfo{year}{2004}, \bibinfo{journal}{JHEP} \textbf{\bibinfo{volume}{06}},
  \bibinfo{pages}{013}.

\bibitem[{\citenamefont{Copeland} \emph{et~al.}(2006)\citenamefont{Copeland,
  Sami, and Tsujikawa}}]{Copeland:2006wr}
\bibinfo{author}{\bibnamefont{Copeland}, \bibfnamefont{E.~J.}},
  \bibinfo{author}{\bibfnamefont{M.}~\bibnamefont{Sami}}, and
  \bibinfo{author}{\bibfnamefont{S.}~\bibnamefont{Tsujikawa}},
  \bibinfo{year}{2006}, \eprint{hep-th/0603057}.

\bibitem[{\citenamefont{Cremmer} \emph{et~al.}(1983)\citenamefont{Cremmer,
  Ferrara, Kounnas, and Nanopoulos}}]{Cremmer:1983bf}
\bibinfo{author}{\bibnamefont{Cremmer}, \bibfnamefont{E.}},
  \bibinfo{author}{\bibfnamefont{S.}~\bibnamefont{Ferrara}},
  \bibinfo{author}{\bibfnamefont{C.}~\bibnamefont{Kounnas}}, and
  \bibinfo{author}{\bibfnamefont{D.~V.} \bibnamefont{Nanopoulos}},
  \bibinfo{year}{1983}, \bibinfo{journal}{Phys. Lett.}
  \textbf{\bibinfo{volume}{B133}}, \bibinfo{pages}{61}.

\bibitem[{\citenamefont{Cremmer and Scherk}(1976)}]{Cremmer:1976ir}
\bibinfo{author}{\bibnamefont{Cremmer}, \bibfnamefont{E.}}, and
  \bibinfo{author}{\bibfnamefont{J.}~\bibnamefont{Scherk}},
  \bibinfo{year}{1976}, \bibinfo{journal}{Nucl. Phys.}
  \textbf{\bibinfo{volume}{B108}}, \bibinfo{pages}{409}.

\bibitem[{\citenamefont{Curio} \emph{et~al.}(2002)\citenamefont{Curio, Klemm,
  Kors, and Lust}}]{Curio:2001ae}
\bibinfo{author}{\bibnamefont{Curio}, \bibfnamefont{G.}},
  \bibinfo{author}{\bibfnamefont{A.}~\bibnamefont{Klemm}},
  \bibinfo{author}{\bibfnamefont{B.}~\bibnamefont{Kors}}, and
  \bibinfo{author}{\bibfnamefont{D.}~\bibnamefont{Lust}}, \bibinfo{year}{2002},
  \bibinfo{journal}{Nucl. Phys.} \textbf{\bibinfo{volume}{B620}},
  \bibinfo{pages}{237}.

\bibitem[{\citenamefont{Curio} \emph{et~al.}(2001)\citenamefont{Curio, Klemm,
  Lust, and Theisen}}]{Curio:2000sc}
\bibinfo{author}{\bibnamefont{Curio}, \bibfnamefont{G.}},
  \bibinfo{author}{\bibfnamefont{A.}~\bibnamefont{Klemm}},
  \bibinfo{author}{\bibfnamefont{D.}~\bibnamefont{Lust}}, and
  \bibinfo{author}{\bibfnamefont{S.}~\bibnamefont{Theisen}},
  \bibinfo{year}{2001}, \bibinfo{journal}{Nucl. Phys.}
  \textbf{\bibinfo{volume}{B609}}, \bibinfo{pages}{3}.

\bibitem[{\citenamefont{Curio and Krause}(2006)}]{Curio:2006dc}
\bibinfo{author}{\bibnamefont{Curio}, \bibfnamefont{G.}}, and
  \bibinfo{author}{\bibfnamefont{A.}~\bibnamefont{Krause}},
  \bibinfo{year}{2006}, \eprint{hep-th/0606243}.

\bibitem[{\citenamefont{Cvetic} \emph{et~al.}(2005)\citenamefont{Cvetic, Li,
  and Liu}}]{Cvetic:2005bn}
\bibinfo{author}{\bibnamefont{Cvetic}, \bibfnamefont{M.}},
  \bibinfo{author}{\bibfnamefont{T.}~\bibnamefont{Li}}, and
  \bibinfo{author}{\bibfnamefont{T.}~\bibnamefont{Liu}}, \bibinfo{year}{2005},
  \bibinfo{journal}{Phys. Rev.} \textbf{\bibinfo{volume}{D71}},
  \bibinfo{pages}{106008}.

\bibitem[{\citenamefont{Cvetic and Liu}(2005)}]{Cvetic:2004xx}
\bibinfo{author}{\bibnamefont{Cvetic}, \bibfnamefont{M.}}, and
  \bibinfo{author}{\bibfnamefont{T.}~\bibnamefont{Liu}}, \bibinfo{year}{2005},
  \bibinfo{journal}{Phys. Lett.} \textbf{\bibinfo{volume}{B610}},
  \bibinfo{pages}{122}.

\bibitem[{\citenamefont{Dabholkar and Hull}(2003)}]{Dabholkar:2002sy}
\bibinfo{author}{\bibnamefont{Dabholkar}, \bibfnamefont{A.}}, and
  \bibinfo{author}{\bibfnamefont{C.}~\bibnamefont{Hull}}, \bibinfo{year}{2003},
  \bibinfo{journal}{JHEP} \textbf{\bibinfo{volume}{09}}, \bibinfo{pages}{054}.

\bibitem[{\citenamefont{Dabholkar and Hull}(2006)}]{Dabholkar:2005ve}
\bibinfo{author}{\bibnamefont{Dabholkar}, \bibfnamefont{A.}}, and
  \bibinfo{author}{\bibfnamefont{C.}~\bibnamefont{Hull}}, \bibinfo{year}{2006},
  \bibinfo{journal}{JHEP} \textbf{\bibinfo{volume}{05}}, \bibinfo{pages}{009}.

\bibitem[{\citenamefont{Dall'Agata}(2001)}]{Dall'Agata:2001zh}
\bibinfo{author}{\bibnamefont{Dall'Agata}, \bibfnamefont{G.}},
  \bibinfo{year}{2001}, \bibinfo{journal}{JHEP} \textbf{\bibinfo{volume}{11}},
  \bibinfo{pages}{005}.

\bibitem[{\citenamefont{Dall'Agata}(2004{\natexlab{a}})}]{Dall'Agata:2004dk}
\bibinfo{author}{\bibnamefont{Dall'Agata}, \bibfnamefont{G.}},
  \bibinfo{year}{2004}{\natexlab{a}}, \bibinfo{journal}{Nucl. Phys.}
  \textbf{\bibinfo{volume}{B695}}, \bibinfo{pages}{243}.

\bibitem[{\citenamefont{Dall'Agata}(2004{\natexlab{b}})}]{Dall'Agata:2004nw}
\bibinfo{author}{\bibnamefont{Dall'Agata}, \bibfnamefont{G.}},
  \bibinfo{year}{2004}{\natexlab{b}}, \bibinfo{journal}{Class. Quant. Grav.}
  \textbf{\bibinfo{volume}{21}}, \bibinfo{pages}{S1479}.

\bibitem[{\citenamefont{Dasgupta} \emph{et~al.}(2002)\citenamefont{Dasgupta,
  Herdeiro, Hirano, and Kallosh}}]{Dasgupta:2002ew}
\bibinfo{author}{\bibnamefont{Dasgupta}, \bibfnamefont{K.}},
  \bibinfo{author}{\bibfnamefont{C.}~\bibnamefont{Herdeiro}},
  \bibinfo{author}{\bibfnamefont{S.}~\bibnamefont{Hirano}}, and
  \bibinfo{author}{\bibfnamefont{R.}~\bibnamefont{Kallosh}},
  \bibinfo{year}{2002}, \bibinfo{journal}{Phys. Rev.}
  \textbf{\bibinfo{volume}{D65}}, \bibinfo{pages}{126002}.

\bibitem[{\citenamefont{Dasgupta} \emph{et~al.}(1999)\citenamefont{Dasgupta,
  Rajesh, and Sethi}}]{Dasgupta:1999ss}
\bibinfo{author}{\bibnamefont{Dasgupta}, \bibfnamefont{K.}},
  \bibinfo{author}{\bibfnamefont{G.}~\bibnamefont{Rajesh}}, and
  \bibinfo{author}{\bibfnamefont{S.}~\bibnamefont{Sethi}},
  \bibinfo{year}{1999}, \bibinfo{journal}{JHEP} \textbf{\bibinfo{volume}{08}},
  \bibinfo{pages}{023}.

\bibitem[{\citenamefont{D'Auria}
  \emph{et~al.}(2003{\natexlab{a}})\citenamefont{D'Auria, Ferrara, Gargiulo,
  Trigiante, and Vaula}}]{D'Auria:2003jk}
\bibinfo{author}{\bibnamefont{D'Auria}, \bibfnamefont{R.}},
  \bibinfo{author}{\bibfnamefont{S.}~\bibnamefont{Ferrara}},
  \bibinfo{author}{\bibfnamefont{F.}~\bibnamefont{Gargiulo}},
  \bibinfo{author}{\bibfnamefont{M.}~\bibnamefont{Trigiante}}, and
  \bibinfo{author}{\bibfnamefont{S.}~\bibnamefont{Vaula}},
  \bibinfo{year}{2003}{\natexlab{a}}, \bibinfo{journal}{JHEP}
  \textbf{\bibinfo{volume}{06}}, \bibinfo{pages}{045}.

\bibitem[{\citenamefont{D'Auria}
  \emph{et~al.}(2003{\natexlab{b}})\citenamefont{D'Auria, Ferrara, Lledo, and
  Vaula}}]{D'Auria:2002th}
\bibinfo{author}{\bibnamefont{D'Auria}, \bibfnamefont{R.}},
  \bibinfo{author}{\bibfnamefont{S.}~\bibnamefont{Ferrara}},
  \bibinfo{author}{\bibfnamefont{M.~A.} \bibnamefont{Lledo}}, and
  \bibinfo{author}{\bibfnamefont{S.}~\bibnamefont{Vaula}},
  \bibinfo{year}{2003}{\natexlab{b}}, \bibinfo{journal}{Phys. Lett.}
  \textbf{\bibinfo{volume}{B557}}, \bibinfo{pages}{278}.

\bibitem[{\citenamefont{D'Auria} \emph{et~al.}(2002)\citenamefont{D'Auria,
  Ferrara, and Vaula}}]{D'Auria:2002tc}
\bibinfo{author}{\bibnamefont{D'Auria}, \bibfnamefont{R.}},
  \bibinfo{author}{\bibfnamefont{S.}~\bibnamefont{Ferrara}}, and
  \bibinfo{author}{\bibfnamefont{S.}~\bibnamefont{Vaula}},
  \bibinfo{year}{2002}, \bibinfo{journal}{New J. Phys.}
  \textbf{\bibinfo{volume}{4}}, \bibinfo{pages}{71}.

\bibitem[{\citenamefont{Denef and Douglas}(2004)}]{Denef:2004ze}
\bibinfo{author}{\bibnamefont{Denef}, \bibfnamefont{F.}}, and
  \bibinfo{author}{\bibfnamefont{M.~R.} \bibnamefont{Douglas}},
  \bibinfo{year}{2004}, \bibinfo{journal}{JHEP} \textbf{\bibinfo{volume}{05}},
  \bibinfo{pages}{072}.

\bibitem[{\citenamefont{Denef and Douglas}(2005)}]{Denef:2004cf}
\bibinfo{author}{\bibnamefont{Denef}, \bibfnamefont{F.}}, and
  \bibinfo{author}{\bibfnamefont{M.~R.} \bibnamefont{Douglas}},
  \bibinfo{year}{2005}, \bibinfo{journal}{JHEP} \textbf{\bibinfo{volume}{03}},
  \bibinfo{pages}{061}.

\bibitem[{\citenamefont{Denef and Douglas}(2006)}]{Denef:2006ad}
\bibinfo{author}{\bibnamefont{Denef}, \bibfnamefont{F.}}, and
  \bibinfo{author}{\bibfnamefont{M.~R.} \bibnamefont{Douglas}},
  \bibinfo{year}{2006}, \eprint{hep-th/0602072}.

\bibitem[{\citenamefont{Denef} \emph{et~al.}(2004)\citenamefont{Denef, Douglas,
  and Florea}}]{Denef:2004dm}
\bibinfo{author}{\bibnamefont{Denef}, \bibfnamefont{F.}},
  \bibinfo{author}{\bibfnamefont{M.~R.} \bibnamefont{Douglas}}, and
  \bibinfo{author}{\bibfnamefont{B.}~\bibnamefont{Florea}},
  \bibinfo{year}{2004}, \bibinfo{journal}{JHEP} \textbf{\bibinfo{volume}{06}},
  \bibinfo{pages}{034}.

\bibitem[{\citenamefont{Denef} \emph{et~al.}(2005)\citenamefont{Denef, Douglas,
  Florea, Grassi, and Kachru}}]{Denef:2005mm}
\bibinfo{author}{\bibnamefont{Denef}, \bibfnamefont{F.}},
  \bibinfo{author}{\bibfnamefont{M.~R.} \bibnamefont{Douglas}},
  \bibinfo{author}{\bibfnamefont{B.}~\bibnamefont{Florea}},
  \bibinfo{author}{\bibfnamefont{A.}~\bibnamefont{Grassi}}, and
  \bibinfo{author}{\bibfnamefont{S.}~\bibnamefont{Kachru}},
  \bibinfo{year}{2005}, \eprint{hep-th/0503124}.

\bibitem[{\citenamefont{Derendinger}
  \emph{et~al.}(2005{\natexlab{a}})\citenamefont{Derendinger, Kounnas,
  Petropoulos, and Zwirner}}]{Derendinger:2005ph}
\bibinfo{author}{\bibnamefont{Derendinger}, \bibfnamefont{J.~P.}},
  \bibinfo{author}{\bibfnamefont{C.}~\bibnamefont{Kounnas}},
  \bibinfo{author}{\bibfnamefont{P.~M.} \bibnamefont{Petropoulos}}, and
  \bibinfo{author}{\bibfnamefont{F.}~\bibnamefont{Zwirner}},
  \bibinfo{year}{2005}{\natexlab{a}}, \bibinfo{journal}{Fortsch. Phys.}
  \textbf{\bibinfo{volume}{53}}, \bibinfo{pages}{926}.

\bibitem[{\citenamefont{Derendinger}
  \emph{et~al.}(2005{\natexlab{b}})\citenamefont{Derendinger, Kounnas,
  Petropoulos, and Zwirner}}]{Derendinger:2004jn}
\bibinfo{author}{\bibnamefont{Derendinger}, \bibfnamefont{J.-P.}},
  \bibinfo{author}{\bibfnamefont{C.}~\bibnamefont{Kounnas}},
  \bibinfo{author}{\bibfnamefont{P.~M.} \bibnamefont{Petropoulos}}, and
  \bibinfo{author}{\bibfnamefont{F.}~\bibnamefont{Zwirner}},
  \bibinfo{year}{2005}{\natexlab{b}}, \bibinfo{journal}{Nucl. Phys.}
  \textbf{\bibinfo{volume}{B715}}, \bibinfo{pages}{211}.

\bibitem[{\citenamefont{DeWolfe}(2005)}]{DeWolfe:2005gy}
\bibinfo{author}{\bibnamefont{DeWolfe}, \bibfnamefont{O.}},
  \bibinfo{year}{2005}, \bibinfo{journal}{JHEP} \textbf{\bibinfo{volume}{10}},
  \bibinfo{pages}{066}.

\bibitem[{\citenamefont{DeWolfe and Giddings}(2003)}]{DeWolfe:2002nn}
\bibinfo{author}{\bibnamefont{DeWolfe}, \bibfnamefont{O.}}, and
  \bibinfo{author}{\bibfnamefont{S.~B.} \bibnamefont{Giddings}},
  \bibinfo{year}{2003}, \bibinfo{journal}{Phys. Rev.}
  \textbf{\bibinfo{volume}{D67}}, \bibinfo{pages}{066008}.

\bibitem[{\citenamefont{DeWolfe}
  \emph{et~al.}(2005{\natexlab{a}})\citenamefont{DeWolfe, Giryavets, Kachru,
  and Taylor}}]{DeWolfe:2004ns}
\bibinfo{author}{\bibnamefont{DeWolfe}, \bibfnamefont{O.}},
  \bibinfo{author}{\bibfnamefont{A.}~\bibnamefont{Giryavets}},
  \bibinfo{author}{\bibfnamefont{S.}~\bibnamefont{Kachru}}, and
  \bibinfo{author}{\bibfnamefont{W.}~\bibnamefont{Taylor}},
  \bibinfo{year}{2005}{\natexlab{a}}, \bibinfo{journal}{JHEP}
  \textbf{\bibinfo{volume}{02}}, \bibinfo{pages}{037}.

\bibitem[{\citenamefont{DeWolfe}
  \emph{et~al.}(2005{\natexlab{b}})\citenamefont{DeWolfe, Giryavets, Kachru,
  and Taylor}}]{DeWolfe:2005uu}
\bibinfo{author}{\bibnamefont{DeWolfe}, \bibfnamefont{O.}},
  \bibinfo{author}{\bibfnamefont{A.}~\bibnamefont{Giryavets}},
  \bibinfo{author}{\bibfnamefont{S.}~\bibnamefont{Kachru}}, and
  \bibinfo{author}{\bibfnamefont{W.}~\bibnamefont{Taylor}},
  \bibinfo{year}{2005}{\natexlab{b}}, \bibinfo{journal}{JHEP}
  \textbf{\bibinfo{volume}{07}}, \bibinfo{pages}{066}.

\bibitem[{\citenamefont{DeWolfe} \emph{et~al.}(2004)\citenamefont{DeWolfe,
  Kachru, and Verlinde}}]{DeWolfe:2004qx}
\bibinfo{author}{\bibnamefont{DeWolfe}, \bibfnamefont{O.}},
  \bibinfo{author}{\bibfnamefont{S.}~\bibnamefont{Kachru}}, and
  \bibinfo{author}{\bibfnamefont{H.~L.} \bibnamefont{Verlinde}},
  \bibinfo{year}{2004}, \bibinfo{journal}{JHEP} \textbf{\bibinfo{volume}{05}},
  \bibinfo{pages}{017}.

\bibitem[{\citenamefont{Diaconescu}
  \emph{et~al.}(2006)\citenamefont{Diaconescu, Florea, Kachru, and
  Svrcek}}]{Diaconescu:2005pc}
\bibinfo{author}{\bibnamefont{Diaconescu}, \bibfnamefont{D.-E.}},
  \bibinfo{author}{\bibfnamefont{B.}~\bibnamefont{Florea}},
  \bibinfo{author}{\bibfnamefont{S.}~\bibnamefont{Kachru}}, and
  \bibinfo{author}{\bibfnamefont{P.}~\bibnamefont{Svrcek}},
  \bibinfo{year}{2006}, \bibinfo{journal}{JHEP} \textbf{\bibinfo{volume}{02}},
  \bibinfo{pages}{020}.

\bibitem[{\citenamefont{Dienes}(2006)}]{Dienes:2006ut}
\bibinfo{author}{\bibnamefont{Dienes}, \bibfnamefont{K.~R.}},
  \bibinfo{year}{2006}, \bibinfo{journal}{Phys. Rev.}
  \textbf{\bibinfo{volume}{D73}}, \bibinfo{pages}{106010}.

\bibitem[{\citenamefont{Dienes} \emph{et~al.}(2005)\citenamefont{Dienes, Dudas,
  and Gherghetta}}]{Dienes:2004pi}
\bibinfo{author}{\bibnamefont{Dienes}, \bibfnamefont{K.~R.}},
  \bibinfo{author}{\bibfnamefont{E.}~\bibnamefont{Dudas}}, and
  \bibinfo{author}{\bibfnamefont{T.}~\bibnamefont{Gherghetta}},
  \bibinfo{year}{2005}, \bibinfo{journal}{Phys. Rev.}
  \textbf{\bibinfo{volume}{D72}}, \bibinfo{pages}{026005}.

\bibitem[{\citenamefont{Dijkstra} \emph{et~al.}(2005)\citenamefont{Dijkstra,
  Huiszoon, and Schellekens}}]{Dijkstra:2004cc}
\bibinfo{author}{\bibnamefont{Dijkstra}, \bibfnamefont{T.~P.~T.}},
  \bibinfo{author}{\bibfnamefont{L.~R.} \bibnamefont{Huiszoon}}, and
  \bibinfo{author}{\bibfnamefont{A.~N.} \bibnamefont{Schellekens}},
  \bibinfo{year}{2005}, \bibinfo{journal}{Nucl. Phys.}
  \textbf{\bibinfo{volume}{B710}}, \bibinfo{pages}{3}.

\bibitem[{\citenamefont{Dimopoulos and Georgi}(1981)}]{Dimopoulos:1981zb}
\bibinfo{author}{\bibnamefont{Dimopoulos}, \bibfnamefont{S.}}, and
  \bibinfo{author}{\bibfnamefont{H.}~\bibnamefont{Georgi}},
  \bibinfo{year}{1981}, \bibinfo{journal}{Nucl. Phys.}
  \textbf{\bibinfo{volume}{B193}}, \bibinfo{pages}{150}.

\bibitem[{\citenamefont{Dimopoulos and Giudice}(1996)}]{Dimopoulos:1996kp}
\bibinfo{author}{\bibnamefont{Dimopoulos}, \bibfnamefont{S.}}, and
  \bibinfo{author}{\bibfnamefont{G.~F.} \bibnamefont{Giudice}},
  \bibinfo{year}{1996}, \bibinfo{journal}{Phys. Lett.}
  \textbf{\bibinfo{volume}{B379}}, \bibinfo{pages}{105}.

\bibitem[{\citenamefont{Dimopoulos}
  \emph{et~al.}(2005)\citenamefont{Dimopoulos, Kachru, McGreevy, and
  Wacker}}]{Dimopoulos:2005ac}
\bibinfo{author}{\bibnamefont{Dimopoulos}, \bibfnamefont{S.}},
  \bibinfo{author}{\bibfnamefont{S.}~\bibnamefont{Kachru}},
  \bibinfo{author}{\bibfnamefont{J.}~\bibnamefont{McGreevy}}, and
  \bibinfo{author}{\bibfnamefont{J.~G.} \bibnamefont{Wacker}},
  \bibinfo{year}{2005}, \eprint{hep-th/0507205}.

\bibitem[{\citenamefont{Dine}(2004{\natexlab{a}})}]{Dine:2004fw}
\bibinfo{author}{\bibnamefont{Dine}, \bibfnamefont{M.}},
  \bibinfo{year}{2004}{\natexlab{a}}, \eprint{hep-th/0402101}.

\bibitem[{\citenamefont{Dine}(2004{\natexlab{b}})}]{Dine:2004ct}
\bibinfo{author}{\bibnamefont{Dine}, \bibfnamefont{M.}},
  \bibinfo{year}{2004}{\natexlab{b}}, \eprint{hep-th/0410201}.

\bibitem[{\citenamefont{Dine} \emph{et~al.}(2006)\citenamefont{Dine, Feng, and
  Silverstein}}]{Dine:2006gm}
\bibinfo{author}{\bibnamefont{Dine}, \bibfnamefont{M.}},
  \bibinfo{author}{\bibfnamefont{J.~L.} \bibnamefont{Feng}}, and
  \bibinfo{author}{\bibfnamefont{E.}~\bibnamefont{Silverstein}},
  \bibinfo{year}{2006}, \eprint{hep-th/0608159}.

\bibitem[{\citenamefont{Dine} \emph{et~al.}(2004)\citenamefont{Dine, Gorbatov,
  and Thomas}}]{Dine:2004is}
\bibinfo{author}{\bibnamefont{Dine}, \bibfnamefont{M.}},
  \bibinfo{author}{\bibfnamefont{E.}~\bibnamefont{Gorbatov}}, and
  \bibinfo{author}{\bibfnamefont{S.~D.} \bibnamefont{Thomas}},
  \bibinfo{year}{2004}, \eprint{hep-th/0407043}.

\bibitem[{\citenamefont{Dine} \emph{et~al.}(2005)\citenamefont{Dine, O'Neil,
  and Sun}}]{Dine:2005yq}
\bibinfo{author}{\bibnamefont{Dine}, \bibfnamefont{M.}},
  \bibinfo{author}{\bibfnamefont{D.}~\bibnamefont{O'Neil}}, and
  \bibinfo{author}{\bibfnamefont{Z.}~\bibnamefont{Sun}}, \bibinfo{year}{2005},
  \bibinfo{journal}{JHEP} \textbf{\bibinfo{volume}{07}}, \bibinfo{pages}{014}.

\bibitem[{\citenamefont{Dine and Sun}(2006)}]{Dine:2005gz}
\bibinfo{author}{\bibnamefont{Dine}, \bibfnamefont{M.}}, and
  \bibinfo{author}{\bibfnamefont{Z.}~\bibnamefont{Sun}}, \bibinfo{year}{2006},
  \bibinfo{journal}{JHEP} \textbf{\bibinfo{volume}{01}}, \bibinfo{pages}{129}.

\bibitem[{\citenamefont{Distler and Varadarajan}(2005)}]{Distler:2005hi}
\bibinfo{author}{\bibnamefont{Distler}, \bibfnamefont{J.}}, and
  \bibinfo{author}{\bibfnamefont{U.}~\bibnamefont{Varadarajan}},
  \bibinfo{year}{2005}, \eprint{hep-th/0507090}.

\bibitem[{\citenamefont{Dixon and Harvey}(1986)}]{Dixon:1986iz}
\bibinfo{author}{\bibnamefont{Dixon}, \bibfnamefont{L.~J.}}, and
  \bibinfo{author}{\bibfnamefont{J.~A.} \bibnamefont{Harvey}},
  \bibinfo{year}{1986}, \bibinfo{journal}{Nucl. Phys.}
  \textbf{\bibinfo{volume}{B274}}, \bibinfo{pages}{93}.

\bibitem[{\citenamefont{Dixon} \emph{et~al.}(1990)\citenamefont{Dixon,
  Kaplunovsky, and Louis}}]{Dixon:1989fj}
\bibinfo{author}{\bibnamefont{Dixon}, \bibfnamefont{L.~J.}},
  \bibinfo{author}{\bibfnamefont{V.}~\bibnamefont{Kaplunovsky}}, and
  \bibinfo{author}{\bibfnamefont{J.}~\bibnamefont{Louis}},
  \bibinfo{year}{1990}, \bibinfo{journal}{Nucl. Phys.}
  \textbf{\bibinfo{volume}{B329}}, \bibinfo{pages}{27}.

\bibitem[{\citenamefont{Dixon} \emph{et~al.}(1987)\citenamefont{Dixon,
  Kaplunovsky, and Vafa}}]{Dixon:1987yp}
\bibinfo{author}{\bibnamefont{Dixon}, \bibfnamefont{L.~J.}},
  \bibinfo{author}{\bibfnamefont{V.}~\bibnamefont{Kaplunovsky}}, and
  \bibinfo{author}{\bibfnamefont{C.}~\bibnamefont{Vafa}}, \bibinfo{year}{1987},
  \bibinfo{journal}{Nucl. Phys.} \textbf{\bibinfo{volume}{B294}},
  \bibinfo{pages}{43}.

\bibitem[{\citenamefont{Donagi} \emph{et~al.}(2005)\citenamefont{Donagi, He,
  Ovrut, and Reinbacher}}]{Donagi:2004ub}
\bibinfo{author}{\bibnamefont{Donagi}, \bibfnamefont{R.}},
  \bibinfo{author}{\bibfnamefont{Y.-H.} \bibnamefont{He}},
  \bibinfo{author}{\bibfnamefont{B.~A.} \bibnamefont{Ovrut}}, and
  \bibinfo{author}{\bibfnamefont{R.}~\bibnamefont{Reinbacher}},
  \bibinfo{year}{2005}, \bibinfo{journal}{JHEP} \textbf{\bibinfo{volume}{06}},
  \bibinfo{pages}{070}.

\bibitem[{\citenamefont{Donoghue}(2004)}]{Donoghue:2003vs}
\bibinfo{author}{\bibnamefont{Donoghue}, \bibfnamefont{J.~F.}},
  \bibinfo{year}{2004}, \bibinfo{journal}{Phys. Rev.}
  \textbf{\bibinfo{volume}{D69}}, \bibinfo{pages}{106012}.

\bibitem[{\citenamefont{Donoghue} \emph{et~al.}(2006)\citenamefont{Donoghue,
  Dutta, and Ross}}]{Donoghue:2005cf}
\bibinfo{author}{\bibnamefont{Donoghue}, \bibfnamefont{J.~F.}},
  \bibinfo{author}{\bibfnamefont{K.}~\bibnamefont{Dutta}}, and
  \bibinfo{author}{\bibfnamefont{A.}~\bibnamefont{Ross}}, \bibinfo{year}{2006},
  \bibinfo{journal}{Phys. Rev.} \textbf{\bibinfo{volume}{D73}},
  \bibinfo{pages}{113002}.

\bibitem[{\citenamefont{Douglas and Lu}(2006)}]{Douglas:2006zj}
\bibinfo{author}{\bibnamefont{Douglas}, \bibfnamefont{M.}}, and
  \bibinfo{author}{\bibfnamefont{Z.}~\bibnamefont{Lu}}, \bibinfo{year}{2006},
  \eprint{math.dg/0603414}.

\bibitem[{\citenamefont{Douglas}(2003)}]{Douglas:2003um}
\bibinfo{author}{\bibnamefont{Douglas}, \bibfnamefont{M.~R.}},
  \bibinfo{year}{2003}, \bibinfo{journal}{JHEP} \textbf{\bibinfo{volume}{05}},
  \bibinfo{pages}{046}.

\bibitem[{\citenamefont{Douglas}(2004{\natexlab{a}})}]{Douglas:2004zg}
\bibinfo{author}{\bibnamefont{Douglas}, \bibfnamefont{M.~R.}},
  \bibinfo{year}{2004}{\natexlab{a}}, \bibinfo{journal}{Comptes Rendus
  Physique} \textbf{\bibinfo{volume}{5}}, \bibinfo{pages}{965}.

\bibitem[{\citenamefont{Douglas}(2004{\natexlab{b}})}]{Douglas:2004qg}
\bibinfo{author}{\bibnamefont{Douglas}, \bibfnamefont{M.~R.}},
  \bibinfo{year}{2004}{\natexlab{b}}, \eprint{hep-th/0405279}.

\bibitem[{\citenamefont{Douglas}(2004{\natexlab{c}})}]{Douglas:2004kp}
\bibinfo{author}{\bibnamefont{Douglas}, \bibfnamefont{M.~R.}},
  \bibinfo{year}{2004}{\natexlab{c}}, \eprint{hep-ph/0401004}.

\bibitem[{\citenamefont{Douglas}(2005)}]{Douglas-strings-2005}
\bibinfo{author}{\bibnamefont{Douglas}, \bibfnamefont{M.~R.}},
  \bibinfo{year}{2005}.

\bibitem[{\citenamefont{Douglas}
  \emph{et~al.}(2006{\natexlab{a}})\citenamefont{Douglas, Karp, Lukic, and
  Reinbacher}}]{Douglas:2006hz}
\bibinfo{author}{\bibnamefont{Douglas}, \bibfnamefont{M.~R.}},
  \bibinfo{author}{\bibfnamefont{R.~L.} \bibnamefont{Karp}},
  \bibinfo{author}{\bibfnamefont{S.}~\bibnamefont{Lukic}}, and
  \bibinfo{author}{\bibfnamefont{R.}~\bibnamefont{Reinbacher}},
  \bibinfo{year}{2006}{\natexlab{a}}, \eprint{hep-th/0606261}.

\bibitem[{\citenamefont{Douglas} \emph{et~al.}(2004)\citenamefont{Douglas,
  Shiffman, and Zelditch}}]{Douglas:2004zu}
\bibinfo{author}{\bibnamefont{Douglas}, \bibfnamefont{M.~R.}},
  \bibinfo{author}{\bibfnamefont{B.}~\bibnamefont{Shiffman}}, and
  \bibinfo{author}{\bibfnamefont{S.}~\bibnamefont{Zelditch}},
  \bibinfo{year}{2004}, \bibinfo{journal}{Commun. Math. Phys.}
  \textbf{\bibinfo{volume}{252}}, \bibinfo{pages}{325}.

\bibitem[{\citenamefont{Douglas}
  \emph{et~al.}(2006{\natexlab{b}})\citenamefont{Douglas, Shiffman, and
  Zelditch}}]{Douglas:2005df}
\bibinfo{author}{\bibnamefont{Douglas}, \bibfnamefont{M.~R.}},
  \bibinfo{author}{\bibfnamefont{B.}~\bibnamefont{Shiffman}}, and
  \bibinfo{author}{\bibfnamefont{S.}~\bibnamefont{Zelditch}},
  \bibinfo{year}{2006}{\natexlab{b}}, \bibinfo{journal}{Commun. Math. Phys.}
  \textbf{\bibinfo{volume}{265}}, \bibinfo{pages}{617}.

\bibitem[{\citenamefont{Douglas and Taylor}(2006)}]{Douglas:2006xy}
\bibinfo{author}{\bibnamefont{Douglas}, \bibfnamefont{M.~R.}}, and
  \bibinfo{author}{\bibfnamefont{W.}~\bibnamefont{Taylor}},
  \bibinfo{year}{2006}, \eprint{hep-th/0606109}.

\bibitem[{\citenamefont{Douglas and Zhou}(2004)}]{Douglas:2004yv}
\bibinfo{author}{\bibnamefont{Douglas}, \bibfnamefont{M.~R.}}, and
  \bibinfo{author}{\bibfnamefont{C.-g.} \bibnamefont{Zhou}},
  \bibinfo{year}{2004}, \bibinfo{journal}{JHEP} \textbf{\bibinfo{volume}{06}},
  \bibinfo{pages}{014}.

\bibitem[{\citenamefont{Duff}(1996)}]{Duff:1996aw}
\bibinfo{author}{\bibnamefont{Duff}, \bibfnamefont{M.~J.}},
  \bibinfo{year}{1996}, \bibinfo{journal}{Int. J. Mod. Phys.}
  \textbf{\bibinfo{volume}{A11}}, \bibinfo{pages}{5623}.

\bibitem[{\citenamefont{Dvali and Vilenkin}(2004{\natexlab{a}})}]{Dvali:2003br}
\bibinfo{author}{\bibnamefont{Dvali}, \bibfnamefont{G.}}, and
  \bibinfo{author}{\bibfnamefont{A.}~\bibnamefont{Vilenkin}},
  \bibinfo{year}{2004}{\natexlab{a}}, \bibinfo{journal}{Phys. Rev.}
  \textbf{\bibinfo{volume}{D70}}, \bibinfo{pages}{063501}.

\bibitem[{\citenamefont{Dvali and Vilenkin}(2004{\natexlab{b}})}]{Dvali:2003zj}
\bibinfo{author}{\bibnamefont{Dvali}, \bibfnamefont{G.}}, and
  \bibinfo{author}{\bibfnamefont{A.}~\bibnamefont{Vilenkin}},
  \bibinfo{year}{2004}{\natexlab{b}}, \bibinfo{journal}{JCAP}
  \textbf{\bibinfo{volume}{0403}}, \bibinfo{pages}{010}.

\bibitem[{\citenamefont{Dvali}(1995)}]{Dvali:1995mj}
\bibinfo{author}{\bibnamefont{Dvali}, \bibfnamefont{G.~R.}},
  \bibinfo{year}{1995}, \eprint{hep-ph/9503259}.

\bibitem[{\citenamefont{Dvali} \emph{et~al.}(2001)\citenamefont{Dvali, Shafi,
  and Solganik}}]{Dvali:2001fw}
\bibinfo{author}{\bibnamefont{Dvali}, \bibfnamefont{G.~R.}},
  \bibinfo{author}{\bibfnamefont{Q.}~\bibnamefont{Shafi}}, and
  \bibinfo{author}{\bibfnamefont{S.}~\bibnamefont{Solganik}},
  \bibinfo{year}{2001}, \eprint{hep-th/0105203}.

\bibitem[{\citenamefont{Dvali and Tye}(1999)}]{Dvali:1998pa}
\bibinfo{author}{\bibnamefont{Dvali}, \bibfnamefont{G.~R.}}, and
  \bibinfo{author}{\bibfnamefont{S.~H.~H.} \bibnamefont{Tye}},
  \bibinfo{year}{1999}, \bibinfo{journal}{Phys. Lett.}
  \textbf{\bibinfo{volume}{B450}}, \bibinfo{pages}{72}.

\bibitem[{\citenamefont{Easther} \emph{et~al.}(2005)\citenamefont{Easther,
  Greene, Jackson, and Kabat}}]{Easther:2004sd}
\bibinfo{author}{\bibnamefont{Easther}, \bibfnamefont{R.}},
  \bibinfo{author}{\bibfnamefont{B.~R.} \bibnamefont{Greene}},
  \bibinfo{author}{\bibfnamefont{M.~G.} \bibnamefont{Jackson}}, and
  \bibinfo{author}{\bibfnamefont{D.}~\bibnamefont{Kabat}},
  \bibinfo{year}{2005}, \bibinfo{journal}{JCAP}
  \textbf{\bibinfo{volume}{0502}}, \bibinfo{pages}{009}.

\bibitem[{\citenamefont{Easther and McAllister}(2006)}]{Easther:2005zr}
\bibinfo{author}{\bibnamefont{Easther}, \bibfnamefont{R.}}, and
  \bibinfo{author}{\bibfnamefont{L.}~\bibnamefont{McAllister}},
  \bibinfo{year}{2006}, \bibinfo{journal}{JCAP}
  \textbf{\bibinfo{volume}{0605}}, \bibinfo{pages}{018}.

\bibitem[{\citenamefont{Eguchi and Tachikawa}(2006)}]{Eguchi:2005eh}
\bibinfo{author}{\bibnamefont{Eguchi}, \bibfnamefont{T.}}, and
  \bibinfo{author}{\bibfnamefont{Y.}~\bibnamefont{Tachikawa}},
  \bibinfo{year}{2006}, \bibinfo{journal}{JHEP} \textbf{\bibinfo{volume}{01}},
  \bibinfo{pages}{100}.

\bibitem[{\citenamefont{Ellis} \emph{et~al.}(1984)\citenamefont{Ellis, Lahanas,
  Nanopoulos, and Tamvakis}}]{Ellis:1983sf}
\bibinfo{author}{\bibnamefont{Ellis}, \bibfnamefont{J.~R.}},
  \bibinfo{author}{\bibfnamefont{A.~B.} \bibnamefont{Lahanas}},
  \bibinfo{author}{\bibfnamefont{D.~V.} \bibnamefont{Nanopoulos}}, and
  \bibinfo{author}{\bibfnamefont{K.}~\bibnamefont{Tamvakis}},
  \bibinfo{year}{1984}, \bibinfo{journal}{Phys. Lett.}
  \textbf{\bibinfo{volume}{B134}}, \bibinfo{pages}{429}.

\bibitem[{\citenamefont{Farhi} \emph{et~al.}(1990)\citenamefont{Farhi, Guth,
  and Guven}}]{Farhi:1989yr}
\bibinfo{author}{\bibnamefont{Farhi}, \bibfnamefont{E.}},
  \bibinfo{author}{\bibfnamefont{A.~H.} \bibnamefont{Guth}}, and
  \bibinfo{author}{\bibfnamefont{J.}~\bibnamefont{Guven}},
  \bibinfo{year}{1990}, \bibinfo{journal}{Nucl. Phys.}
  \textbf{\bibinfo{volume}{B339}}, \bibinfo{pages}{417}.

\bibitem[{\citenamefont{Feng} \emph{et~al.}(2001)\citenamefont{Feng,
  March-Russell, Sethi, and Wilczek}}]{Feng:2000if}
\bibinfo{author}{\bibnamefont{Feng}, \bibfnamefont{J.~L.}},
  \bibinfo{author}{\bibfnamefont{J.}~\bibnamefont{March-Russell}},
  \bibinfo{author}{\bibfnamefont{S.}~\bibnamefont{Sethi}}, and
  \bibinfo{author}{\bibfnamefont{F.}~\bibnamefont{Wilczek}},
  \bibinfo{year}{2001}, \bibinfo{journal}{Nucl. Phys.}
  \textbf{\bibinfo{volume}{B602}}, \bibinfo{pages}{307}.

\bibitem[{\citenamefont{Ferrara}(2002)}]{Ferrara:2002hb}
\bibinfo{author}{\bibnamefont{Ferrara}, \bibfnamefont{S.}},
  \bibinfo{year}{2002}, \eprint{hep-th/0211116}.

\bibitem[{\citenamefont{Ferrara and Porrati}(2002)}]{Ferrara:2002bt}
\bibinfo{author}{\bibnamefont{Ferrara}, \bibfnamefont{S.}}, and
  \bibinfo{author}{\bibfnamefont{M.}~\bibnamefont{Porrati}},
  \bibinfo{year}{2002}, \bibinfo{journal}{Phys. Lett.}
  \textbf{\bibinfo{volume}{B545}}, \bibinfo{pages}{411}.

\bibitem[{\citenamefont{Fidanza} \emph{et~al.}(2004)\citenamefont{Fidanza,
  Minasian, and Tomasiello}}]{Fidanza:2004wa}
\bibinfo{author}{\bibnamefont{Fidanza}, \bibfnamefont{S.}},
  \bibinfo{author}{\bibfnamefont{R.}~\bibnamefont{Minasian}}, and
  \bibinfo{author}{\bibfnamefont{A.}~\bibnamefont{Tomasiello}},
  \bibinfo{year}{2004}, \bibinfo{journal}{Fortsch. Phys.}
  \textbf{\bibinfo{volume}{52}}, \bibinfo{pages}{618}.

\bibitem[{\citenamefont{Fidanza} \emph{et~al.}(2005)\citenamefont{Fidanza,
  Minasian, and Tomasiello}}]{Fidanza:2003zi}
\bibinfo{author}{\bibnamefont{Fidanza}, \bibfnamefont{S.}},
  \bibinfo{author}{\bibfnamefont{R.}~\bibnamefont{Minasian}}, and
  \bibinfo{author}{\bibfnamefont{A.}~\bibnamefont{Tomasiello}},
  \bibinfo{year}{2005}, \bibinfo{journal}{Commun. Math. Phys.}
  \textbf{\bibinfo{volume}{254}}, \bibinfo{pages}{401}.

\bibitem[{\citenamefont{Firouzjahi}
  \emph{et~al.}(2004)\citenamefont{Firouzjahi, Sarangi, and
  Tye}}]{Firouzjahi:2004mx}
\bibinfo{author}{\bibnamefont{Firouzjahi}, \bibfnamefont{H.}},
  \bibinfo{author}{\bibfnamefont{S.}~\bibnamefont{Sarangi}}, and
  \bibinfo{author}{\bibfnamefont{S.~H.~H.} \bibnamefont{Tye}},
  \bibinfo{year}{2004}, \bibinfo{journal}{JHEP} \textbf{\bibinfo{volume}{09}},
  \bibinfo{pages}{060}.

\bibitem[{\citenamefont{Fischler} \emph{et~al.}(2001)\citenamefont{Fischler,
  Kashani-Poor, McNees, and Paban}}]{Fischler:2001yj}
\bibinfo{author}{\bibnamefont{Fischler}, \bibfnamefont{W.}},
  \bibinfo{author}{\bibfnamefont{A.}~\bibnamefont{Kashani-Poor}},
  \bibinfo{author}{\bibfnamefont{R.}~\bibnamefont{McNees}}, and
  \bibinfo{author}{\bibfnamefont{S.}~\bibnamefont{Paban}},
  \bibinfo{year}{2001}, \bibinfo{journal}{JHEP} \textbf{\bibinfo{volume}{07}},
  \bibinfo{pages}{003}.

\bibitem[{\citenamefont{Florea} \emph{et~al.}(2006)\citenamefont{Florea,
  Kachru, McGreevy, and Saulina}}]{Florea:2006si}
\bibinfo{author}{\bibnamefont{Florea}, \bibfnamefont{B.}},
  \bibinfo{author}{\bibfnamefont{S.}~\bibnamefont{Kachru}},
  \bibinfo{author}{\bibfnamefont{J.}~\bibnamefont{McGreevy}}, and
  \bibinfo{author}{\bibfnamefont{N.}~\bibnamefont{Saulina}},
  \bibinfo{year}{2006}, \eprint{hep-th/0610003}.

\bibitem[{\citenamefont{Flournoy} \emph{et~al.}(2005)\citenamefont{Flournoy,
  Wecht, and Williams}}]{Flournoy:2004vn}
\bibinfo{author}{\bibnamefont{Flournoy}, \bibfnamefont{A.}},
  \bibinfo{author}{\bibfnamefont{B.}~\bibnamefont{Wecht}}, and
  \bibinfo{author}{\bibfnamefont{B.}~\bibnamefont{Williams}},
  \bibinfo{year}{2005}, \bibinfo{journal}{Nucl. Phys.}
  \textbf{\bibinfo{volume}{B706}}, \bibinfo{pages}{127}.

\bibitem[{\citenamefont{Flournoy and Williams}(2006)}]{Flournoy:2005xe}
\bibinfo{author}{\bibnamefont{Flournoy}, \bibfnamefont{A.}}, and
  \bibinfo{author}{\bibfnamefont{B.}~\bibnamefont{Williams}},
  \bibinfo{year}{2006}, \bibinfo{journal}{JHEP} \textbf{\bibinfo{volume}{01}},
  \bibinfo{pages}{166}.

\bibitem[{\citenamefont{Font and Ibanez}(2005)}]{Font:2004cx}
\bibinfo{author}{\bibnamefont{Font}, \bibfnamefont{A.}}, and
  \bibinfo{author}{\bibfnamefont{L.~E.} \bibnamefont{Ibanez}},
  \bibinfo{year}{2005}, \bibinfo{journal}{JHEP} \textbf{\bibinfo{volume}{03}},
  \bibinfo{pages}{040}.

\bibitem[{\citenamefont{Franco} \emph{et~al.}(2005)\citenamefont{Franco,
  Hanany, and Uranga}}]{Franco:2005fd}
\bibinfo{author}{\bibnamefont{Franco}, \bibfnamefont{S.}},
  \bibinfo{author}{\bibfnamefont{A.}~\bibnamefont{Hanany}}, and
  \bibinfo{author}{\bibfnamefont{A.~M.} \bibnamefont{Uranga}},
  \bibinfo{year}{2005}, \bibinfo{journal}{JHEP} \textbf{\bibinfo{volume}{09}},
  \bibinfo{pages}{028}.

\bibitem[{\citenamefont{Franco and Uranga}(2006)}]{Franco:2006es}
\bibinfo{author}{\bibnamefont{Franco}, \bibfnamefont{S.}}, and
  \bibinfo{author}{\bibfnamefont{A.~M.~.} \bibnamefont{Uranga}},
  \bibinfo{year}{2006}, \bibinfo{journal}{JHEP} \textbf{\bibinfo{volume}{06}},
  \bibinfo{pages}{031}.

\bibitem[{\citenamefont{Freedman} \emph{et~al.}(1976)\citenamefont{Freedman,
  van Nieuwenhuizen, and Ferrara}}]{Freedman:1976xh}
\bibinfo{author}{\bibnamefont{Freedman}, \bibfnamefont{D.~Z.}},
  \bibinfo{author}{\bibfnamefont{P.}~\bibnamefont{van Nieuwenhuizen}}, and
  \bibinfo{author}{\bibfnamefont{S.}~\bibnamefont{Ferrara}},
  \bibinfo{year}{1976}, \bibinfo{journal}{Phys. Rev.}
  \textbf{\bibinfo{volume}{D13}}, \bibinfo{pages}{3214}.

\bibitem[{\citenamefont{Freese} \emph{et~al.}(1990)\citenamefont{Freese,
  Frieman, and Olinto}}]{Freese:1990rb}
\bibinfo{author}{\bibnamefont{Freese}, \bibfnamefont{K.}},
  \bibinfo{author}{\bibfnamefont{J.~A.} \bibnamefont{Frieman}}, and
  \bibinfo{author}{\bibfnamefont{A.~V.} \bibnamefont{Olinto}},
  \bibinfo{year}{1990}, \bibinfo{journal}{Phys. Rev. Lett.}
  \textbf{\bibinfo{volume}{65}}, \bibinfo{pages}{3233}.

\bibitem[{\citenamefont{Freund and Rubin}(1980)}]{Freund:1980xh}
\bibinfo{author}{\bibnamefont{Freund}, \bibfnamefont{P.~G.~O.}}, and
  \bibinfo{author}{\bibfnamefont{M.~A.} \bibnamefont{Rubin}},
  \bibinfo{year}{1980}, \bibinfo{journal}{Phys. Lett.}
  \textbf{\bibinfo{volume}{B97}}, \bibinfo{pages}{233}.

\bibitem[{\citenamefont{Frey}(2003)}]{Frey:2003tf}
\bibinfo{author}{\bibnamefont{Frey}, \bibfnamefont{A.~R.}},
  \bibinfo{year}{2003}, \eprint{hep-th/0308156}.

\bibitem[{\citenamefont{Frey} \emph{et~al.}(2003)\citenamefont{Frey, Lippert,
  and Williams}}]{Frey:2003dm}
\bibinfo{author}{\bibnamefont{Frey}, \bibfnamefont{A.~R.}},
  \bibinfo{author}{\bibfnamefont{M.}~\bibnamefont{Lippert}}, and
  \bibinfo{author}{\bibfnamefont{B.}~\bibnamefont{Williams}},
  \bibinfo{year}{2003}, \bibinfo{journal}{Phys. Rev.}
  \textbf{\bibinfo{volume}{D68}}, \bibinfo{pages}{046008}.

\bibitem[{\citenamefont{Frey and Maharana}(2006)}]{Frey:2006wv}
\bibinfo{author}{\bibnamefont{Frey}, \bibfnamefont{A.~R.}}, and
  \bibinfo{author}{\bibfnamefont{A.}~\bibnamefont{Maharana}},
  \bibinfo{year}{2006}, \bibinfo{journal}{JHEP} \textbf{\bibinfo{volume}{08}},
  \bibinfo{pages}{021}.

\bibitem[{\citenamefont{Frey and Polchinski}(2002)}]{Frey:2002hf}
\bibinfo{author}{\bibnamefont{Frey}, \bibfnamefont{A.~R.}}, and
  \bibinfo{author}{\bibfnamefont{J.}~\bibnamefont{Polchinski}},
  \bibinfo{year}{2002}, \bibinfo{journal}{Phys. Rev.}
  \textbf{\bibinfo{volume}{D65}}, \bibinfo{pages}{126009}.

\bibitem[{\citenamefont{Fu and Yau}(2005)}]{Fu:2005sm}
\bibinfo{author}{\bibnamefont{Fu}, \bibfnamefont{J.-X.}}, and
  \bibinfo{author}{\bibfnamefont{S.-T.} \bibnamefont{Yau}},
  \bibinfo{year}{2005}, \eprint{hep-th/0509028}.

\bibitem[{\citenamefont{Fu and Yau}(2006)}]{Fu:2006vj}
\bibinfo{author}{\bibnamefont{Fu}, \bibfnamefont{J.-X.}}, and
  \bibinfo{author}{\bibfnamefont{S.-T.} \bibnamefont{Yau}},
  \bibinfo{year}{2006}, \eprint{hep-th/0604063}.

\bibitem[{\citenamefont{Garcia-Bellido}
  \emph{et~al.}(2002)\citenamefont{Garcia-Bellido, Rabadan, and
  Zamora}}]{Garcia-Bellido:2001ky}
\bibinfo{author}{\bibnamefont{Garcia-Bellido}, \bibfnamefont{J.}},
  \bibinfo{author}{\bibfnamefont{R.}~\bibnamefont{Rabadan}}, and
  \bibinfo{author}{\bibfnamefont{F.}~\bibnamefont{Zamora}},
  \bibinfo{year}{2002}, \bibinfo{journal}{JHEP} \textbf{\bibinfo{volume}{01}},
  \bibinfo{pages}{036}.

\bibitem[{\citenamefont{Garcia-Etxebarria}
  \emph{et~al.}(2006)\citenamefont{Garcia-Etxebarria, Saad, and
  Uranga}}]{Garcia-Etxebarria:2006rw}
\bibinfo{author}{\bibnamefont{Garcia-Etxebarria}, \bibfnamefont{I.}},
  \bibinfo{author}{\bibfnamefont{F.}~\bibnamefont{Saad}}, and
  \bibinfo{author}{\bibfnamefont{A.~M.} \bibnamefont{Uranga}},
  \bibinfo{year}{2006}, \bibinfo{journal}{JHEP} \textbf{\bibinfo{volume}{08}},
  \bibinfo{pages}{069}.

\bibitem[{\citenamefont{Gepner}(1987)}]{Gepner:1987vz}
\bibinfo{author}{\bibnamefont{Gepner}, \bibfnamefont{D.}},
  \bibinfo{year}{1987}, \bibinfo{journal}{Phys. Lett.}
  \textbf{\bibinfo{volume}{B199}}, \bibinfo{pages}{380}.

\bibitem[{\citenamefont{von Gersdorff and
  Hebecker}(2005)}]{vonGersdorff:2005bf}
\bibinfo{author}{\bibnamefont{von Gersdorff}, \bibfnamefont{G.}}, and
  \bibinfo{author}{\bibfnamefont{A.}~\bibnamefont{Hebecker}},
  \bibinfo{year}{2005}, \bibinfo{journal}{Phys. Lett.}
  \textbf{\bibinfo{volume}{B624}}, \bibinfo{pages}{270}.

\bibitem[{\citenamefont{Gherghetta and Giedt}(2006)}]{Gherghetta:2006yq}
\bibinfo{author}{\bibnamefont{Gherghetta}, \bibfnamefont{T.}}, and
  \bibinfo{author}{\bibfnamefont{J.}~\bibnamefont{Giedt}},
  \bibinfo{year}{2006}, \bibinfo{journal}{Phys. Rev.}
  \textbf{\bibinfo{volume}{D74}}, \bibinfo{pages}{066007}.

\bibitem[{\citenamefont{Gibbons}(1984)}]{Gibbons:1984kp}
\bibinfo{author}{\bibnamefont{Gibbons}, \bibfnamefont{G.~W.}},
  \bibinfo{year}{1984}, \bibinfo{note}{three lectures given at GIFT Seminar on
  Theoretical Physics, San Feliu de Guixols, Spain, Jun 4-11, 1984}.

\bibitem[{\citenamefont{Gibbons and Turok}(2006)}]{Gibbons:2006pa}
\bibinfo{author}{\bibnamefont{Gibbons}, \bibfnamefont{G.~W.}}, and
  \bibinfo{author}{\bibfnamefont{N.}~\bibnamefont{Turok}},
  \bibinfo{year}{2006}, \eprint{hep-th/0609095}.

\bibitem[{\citenamefont{Giddings}(2003)}]{Giddings:2003zw}
\bibinfo{author}{\bibnamefont{Giddings}, \bibfnamefont{S.~B.}},
  \bibinfo{year}{2003}, \bibinfo{journal}{Phys. Rev.}
  \textbf{\bibinfo{volume}{D68}}, \bibinfo{pages}{026006}.

\bibitem[{\citenamefont{Giddings}(2006)}]{Giddings:2006sj}
\bibinfo{author}{\bibnamefont{Giddings}, \bibfnamefont{S.~B.}},
  \bibinfo{year}{2006}, \eprint{hep-th/0605196}.

\bibitem[{\citenamefont{Giddings} \emph{et~al.}(2002)\citenamefont{Giddings,
  Kachru, and Polchinski}}]{Giddings:2001yu}
\bibinfo{author}{\bibnamefont{Giddings}, \bibfnamefont{S.~B.}},
  \bibinfo{author}{\bibfnamefont{S.}~\bibnamefont{Kachru}}, and
  \bibinfo{author}{\bibfnamefont{J.}~\bibnamefont{Polchinski}},
  \bibinfo{year}{2002}, \bibinfo{journal}{Phys. Rev.}
  \textbf{\bibinfo{volume}{D66}}, \bibinfo{pages}{106006}.

\bibitem[{\citenamefont{Giddings and Maharana}(2006)}]{Giddings:2005ff}
\bibinfo{author}{\bibnamefont{Giddings}, \bibfnamefont{S.~B.}}, and
  \bibinfo{author}{\bibfnamefont{A.}~\bibnamefont{Maharana}},
  \bibinfo{year}{2006}, \bibinfo{journal}{Phys. Rev.}
  \textbf{\bibinfo{volume}{D73}}, \bibinfo{pages}{126003}.

\bibitem[{\citenamefont{Gimon and Polchinski}(1996)}]{Gimon:1996rq}
\bibinfo{author}{\bibnamefont{Gimon}, \bibfnamefont{E.~G.}}, and
  \bibinfo{author}{\bibfnamefont{J.}~\bibnamefont{Polchinski}},
  \bibinfo{year}{1996}, \bibinfo{journal}{Phys. Rev.}
  \textbf{\bibinfo{volume}{D54}}, \bibinfo{pages}{1667}.

\bibitem[{\citenamefont{Giryavets}
  \emph{et~al.}(2004{\natexlab{a}})\citenamefont{Giryavets, Kachru, and
  Tripathy}}]{Giryavets:2004zr}
\bibinfo{author}{\bibnamefont{Giryavets}, \bibfnamefont{A.}},
  \bibinfo{author}{\bibfnamefont{S.}~\bibnamefont{Kachru}}, and
  \bibinfo{author}{\bibfnamefont{P.~K.} \bibnamefont{Tripathy}},
  \bibinfo{year}{2004}{\natexlab{a}}, \bibinfo{journal}{JHEP}
  \textbf{\bibinfo{volume}{08}}, \bibinfo{pages}{002}.

\bibitem[{\citenamefont{Giryavets}
  \emph{et~al.}(2004{\natexlab{b}})\citenamefont{Giryavets, Kachru, Tripathy,
  and Trivedi}}]{Giryavets:2003vd}
\bibinfo{author}{\bibnamefont{Giryavets}, \bibfnamefont{A.}},
  \bibinfo{author}{\bibfnamefont{S.}~\bibnamefont{Kachru}},
  \bibinfo{author}{\bibfnamefont{P.~K.} \bibnamefont{Tripathy}}, and
  \bibinfo{author}{\bibfnamefont{S.~P.} \bibnamefont{Trivedi}},
  \bibinfo{year}{2004}{\natexlab{b}}, \bibinfo{journal}{JHEP}
  \textbf{\bibinfo{volume}{04}}, \bibinfo{pages}{003}.

\bibitem[{\citenamefont{Giudice and Rattazzi}(1998)}]{Giudice:1998ic}
\bibinfo{author}{\bibnamefont{Giudice}, \bibfnamefont{G.}}, and
  \bibinfo{author}{\bibfnamefont{R.}~\bibnamefont{Rattazzi}},
  \bibinfo{year}{1998}, \bibinfo{note}{in *Kane, G.L. (ed.): Perspectives on
  supersymmetry* 355- 377}.

\bibitem[{\citenamefont{Giudice} \emph{et~al.}(1998)\citenamefont{Giudice,
  Luty, Murayama, and Rattazzi}}]{Giudice:1998xp}
\bibinfo{author}{\bibnamefont{Giudice}, \bibfnamefont{G.~F.}},
  \bibinfo{author}{\bibfnamefont{M.~A.} \bibnamefont{Luty}},
  \bibinfo{author}{\bibfnamefont{H.}~\bibnamefont{Murayama}}, and
  \bibinfo{author}{\bibfnamefont{R.}~\bibnamefont{Rattazzi}},
  \bibinfo{year}{1998}, \bibinfo{journal}{JHEP} \textbf{\bibinfo{volume}{12}},
  \bibinfo{pages}{027}.

\bibitem[{\citenamefont{Giudice and Rattazzi}(1999)}]{Giudice:1998bp}
\bibinfo{author}{\bibnamefont{Giudice}, \bibfnamefont{G.~F.}}, and
  \bibinfo{author}{\bibfnamefont{R.}~\bibnamefont{Rattazzi}},
  \bibinfo{year}{1999}, \bibinfo{journal}{Phys. Rept.}
  \textbf{\bibinfo{volume}{322}}, \bibinfo{pages}{419}.

\bibitem[{\citenamefont{Giudice and Rattazzi}(2006)}]{Giudice:2006sn}
\bibinfo{author}{\bibnamefont{Giudice}, \bibfnamefont{G.~F.}}, and
  \bibinfo{author}{\bibfnamefont{R.}~\bibnamefont{Rattazzi}},
  \bibinfo{year}{2006}, \eprint{hep-ph/0606105}.

\bibitem[{\citenamefont{Giudice and Romanino}(2004)}]{Giudice:2004tc}
\bibinfo{author}{\bibnamefont{Giudice}, \bibfnamefont{G.~F.}}, and
  \bibinfo{author}{\bibfnamefont{A.}~\bibnamefont{Romanino}},
  \bibinfo{year}{2004}, \bibinfo{journal}{Nucl. Phys.}
  \textbf{\bibinfo{volume}{B699}}, \bibinfo{pages}{65}.

\bibitem[{\citenamefont{Gmeiner}(2006{\natexlab{a}})}]{Gmeiner:2006qw}
\bibinfo{author}{\bibnamefont{Gmeiner}, \bibfnamefont{F.}},
  \bibinfo{year}{2006}{\natexlab{a}}, \eprint{hep-th/0608227}.

\bibitem[{\citenamefont{Gmeiner}(2006{\natexlab{b}})}]{Gmeiner:2005nh}
\bibinfo{author}{\bibnamefont{Gmeiner}, \bibfnamefont{F.}},
  \bibinfo{year}{2006}{\natexlab{b}}, \bibinfo{journal}{Fortsch. Phys.}
  \textbf{\bibinfo{volume}{54}}, \bibinfo{pages}{391}.

\bibitem[{\citenamefont{Gmeiner} \emph{et~al.}(2006)\citenamefont{Gmeiner,
  Blumenhagen, Honecker, Lust, and Weigand}}]{Gmeiner:2005vz}
\bibinfo{author}{\bibnamefont{Gmeiner}, \bibfnamefont{F.}},
  \bibinfo{author}{\bibfnamefont{R.}~\bibnamefont{Blumenhagen}},
  \bibinfo{author}{\bibfnamefont{G.}~\bibnamefont{Honecker}},
  \bibinfo{author}{\bibfnamefont{D.}~\bibnamefont{Lust}}, and
  \bibinfo{author}{\bibfnamefont{T.}~\bibnamefont{Weigand}},
  \bibinfo{year}{2006}, \bibinfo{journal}{JHEP} \textbf{\bibinfo{volume}{01}},
  \bibinfo{pages}{004}.

\bibitem[{\citenamefont{Goldstein and Prokushkin}(2004)}]{Goldstein:2002pg}
\bibinfo{author}{\bibnamefont{Goldstein}, \bibfnamefont{E.}}, and
  \bibinfo{author}{\bibfnamefont{S.}~\bibnamefont{Prokushkin}},
  \bibinfo{year}{2004}, \bibinfo{journal}{Commun. Math. Phys.}
  \textbf{\bibinfo{volume}{251}}, \bibinfo{pages}{65}.

\bibitem[{\citenamefont{Gomez-Reino and Zavala}(2002)}]{Gomez-Reino:2002fs}
\bibinfo{author}{\bibnamefont{Gomez-Reino}, \bibfnamefont{M.}}, and
  \bibinfo{author}{\bibfnamefont{I.}~\bibnamefont{Zavala}},
  \bibinfo{year}{2002}, \bibinfo{journal}{JHEP} \textbf{\bibinfo{volume}{09}},
  \bibinfo{pages}{020}.

\bibitem[{\citenamefont{Gomis} \emph{et~al.}(2005)\citenamefont{Gomis,
  Marchesano, and Mateos}}]{Gomis:2005wc}
\bibinfo{author}{\bibnamefont{Gomis}, \bibfnamefont{J.}},
  \bibinfo{author}{\bibfnamefont{F.}~\bibnamefont{Marchesano}}, and
  \bibinfo{author}{\bibfnamefont{D.}~\bibnamefont{Mateos}},
  \bibinfo{year}{2005}, \bibinfo{journal}{JHEP} \textbf{\bibinfo{volume}{11}},
  \bibinfo{pages}{021}.

\bibitem[{\citenamefont{Gorlich} \emph{et~al.}(2004)\citenamefont{Gorlich,
  Kachru, Tripathy, and Trivedi}}]{Gorlich:2004qm}
\bibinfo{author}{\bibnamefont{Gorlich}, \bibfnamefont{L.}},
  \bibinfo{author}{\bibfnamefont{S.}~\bibnamefont{Kachru}},
  \bibinfo{author}{\bibfnamefont{P.~K.} \bibnamefont{Tripathy}}, and
  \bibinfo{author}{\bibfnamefont{S.~P.} \bibnamefont{Trivedi}},
  \bibinfo{year}{2004}, \bibinfo{journal}{JHEP} \textbf{\bibinfo{volume}{12}},
  \bibinfo{pages}{074}.

\bibitem[{\citenamefont{Grana}(2006)}]{Grana:2005jc}
\bibinfo{author}{\bibnamefont{Grana}, \bibfnamefont{M.}}, \bibinfo{year}{2006},
  \bibinfo{journal}{Phys. Rept.} \textbf{\bibinfo{volume}{423}},
  \bibinfo{pages}{91}.

\bibitem[{\citenamefont{Grana}
  \emph{et~al.}(2006{\natexlab{a}})\citenamefont{Grana, Louis, and
  Waldram}}]{Grana:2005ny}
\bibinfo{author}{\bibnamefont{Grana}, \bibfnamefont{M.}},
  \bibinfo{author}{\bibfnamefont{J.}~\bibnamefont{Louis}}, and
  \bibinfo{author}{\bibfnamefont{D.}~\bibnamefont{Waldram}},
  \bibinfo{year}{2006}{\natexlab{a}}, \bibinfo{journal}{JHEP}
  \textbf{\bibinfo{volume}{01}}, \bibinfo{pages}{008}.

\bibitem[{\citenamefont{Grana}
  \emph{et~al.}(2006{\natexlab{b}})\citenamefont{Grana, Minasian, Petrini, and
  Tomasiello}}]{Grana:2006kf}
\bibinfo{author}{\bibnamefont{Grana}, \bibfnamefont{M.}},
  \bibinfo{author}{\bibfnamefont{R.}~\bibnamefont{Minasian}},
  \bibinfo{author}{\bibfnamefont{M.}~\bibnamefont{Petrini}}, and
  \bibinfo{author}{\bibfnamefont{A.}~\bibnamefont{Tomasiello}},
  \bibinfo{year}{2006}{\natexlab{b}}, \eprint{hep-th/0609124}.

\bibitem[{\citenamefont{Green}
  \emph{et~al.}(1987{\natexlab{a}})\citenamefont{Green, Schwarz, and
  Witten}}]{Green:1987sp}
\bibinfo{author}{\bibnamefont{Green}, \bibfnamefont{M.~B.}},
  \bibinfo{author}{\bibfnamefont{J.~H.} \bibnamefont{Schwarz}}, and
  \bibinfo{author}{\bibfnamefont{E.}~\bibnamefont{Witten}},
  \bibinfo{year}{1987}{\natexlab{a}}, \bibinfo{note}{cambridge, Uk: Univ. Pr. (
  1987) 469 P. ( Cambridge Monographs On Mathematical Physics)}.

\bibitem[{\citenamefont{Green}
  \emph{et~al.}(1987{\natexlab{b}})\citenamefont{Green, Schwarz, and
  Witten}}]{Green:1987mn}
\bibinfo{author}{\bibnamefont{Green}, \bibfnamefont{M.~B.}},
  \bibinfo{author}{\bibfnamefont{J.~H.} \bibnamefont{Schwarz}}, and
  \bibinfo{author}{\bibfnamefont{E.}~\bibnamefont{Witten}},
  \bibinfo{year}{1987}{\natexlab{b}}, \bibinfo{note}{cambridge, Uk: Univ. Pr. (
  1987) 596 P. ( Cambridge Monographs On Mathematical Physics)}.

\bibitem[{\citenamefont{Greene and Weltman}(2006)}]{Greene:2005rn}
\bibinfo{author}{\bibnamefont{Greene}, \bibfnamefont{B.}}, and
  \bibinfo{author}{\bibfnamefont{A.}~\bibnamefont{Weltman}},
  \bibinfo{year}{2006}, \bibinfo{journal}{JHEP} \textbf{\bibinfo{volume}{03}},
  \bibinfo{pages}{035}.

\bibitem[{\citenamefont{Greene} \emph{et~al.}(1986)\citenamefont{Greene,
  Kirklin, Miron, and Ross}}]{Greene:1986bm}
\bibinfo{author}{\bibnamefont{Greene}, \bibfnamefont{B.~R.}},
  \bibinfo{author}{\bibfnamefont{K.~H.} \bibnamefont{Kirklin}},
  \bibinfo{author}{\bibfnamefont{P.~J.} \bibnamefont{Miron}}, and
  \bibinfo{author}{\bibfnamefont{G.~G.} \bibnamefont{Ross}},
  \bibinfo{year}{1986}, \bibinfo{journal}{Nucl. Phys.}
  \textbf{\bibinfo{volume}{B278}}, \bibinfo{pages}{667}.

\bibitem[{\citenamefont{Greene} \emph{et~al.}(1995)\citenamefont{Greene,
  Morrison, and Strominger}}]{Greene:1995hu}
\bibinfo{author}{\bibnamefont{Greene}, \bibfnamefont{B.~R.}},
  \bibinfo{author}{\bibfnamefont{D.~R.} \bibnamefont{Morrison}}, and
  \bibinfo{author}{\bibfnamefont{A.}~\bibnamefont{Strominger}},
  \bibinfo{year}{1995}, \bibinfo{journal}{Nucl. Phys.}
  \textbf{\bibinfo{volume}{B451}}, \bibinfo{pages}{109}.

\bibitem[{\citenamefont{Greene and Plesser}(1990)}]{Greene:1990ud}
\bibinfo{author}{\bibnamefont{Greene}, \bibfnamefont{B.~R.}}, and
  \bibinfo{author}{\bibfnamefont{M.~R.} \bibnamefont{Plesser}},
  \bibinfo{year}{1990}, \bibinfo{journal}{Nucl. Phys.}
  \textbf{\bibinfo{volume}{B338}}, \bibinfo{pages}{15}.

\bibitem[{\citenamefont{Greene} \emph{et~al.}(2000)\citenamefont{Greene,
  Schalm, and Shiu}}]{Greene:2000gh}
\bibinfo{author}{\bibnamefont{Greene}, \bibfnamefont{B.~R.}},
  \bibinfo{author}{\bibfnamefont{K.}~\bibnamefont{Schalm}}, and
  \bibinfo{author}{\bibfnamefont{G.}~\bibnamefont{Shiu}}, \bibinfo{year}{2000},
  \bibinfo{journal}{Nucl. Phys.} \textbf{\bibinfo{volume}{B584}},
  \bibinfo{pages}{480}.

\bibitem[{\citenamefont{Grimm and Louis}(2004)}]{Grimm:2004uq}
\bibinfo{author}{\bibnamefont{Grimm}, \bibfnamefont{T.~W.}}, and
  \bibinfo{author}{\bibfnamefont{J.}~\bibnamefont{Louis}},
  \bibinfo{year}{2004}, \bibinfo{journal}{Nucl. Phys.}
  \textbf{\bibinfo{volume}{B699}}, \bibinfo{pages}{387}.

\bibitem[{\citenamefont{Grimm and Louis}(2005)}]{Grimm:2004ua}
\bibinfo{author}{\bibnamefont{Grimm}, \bibfnamefont{T.~W.}}, and
  \bibinfo{author}{\bibfnamefont{J.}~\bibnamefont{Louis}},
  \bibinfo{year}{2005}, \bibinfo{journal}{Nucl. Phys.}
  \textbf{\bibinfo{volume}{B718}}, \bibinfo{pages}{153}.

\bibitem[{\citenamefont{Gukov} \emph{et~al.}(2004)\citenamefont{Gukov, Kachru,
  Liu, and McAllister}}]{Gukov:2003cy}
\bibinfo{author}{\bibnamefont{Gukov}, \bibfnamefont{S.}},
  \bibinfo{author}{\bibfnamefont{S.}~\bibnamefont{Kachru}},
  \bibinfo{author}{\bibfnamefont{X.}~\bibnamefont{Liu}}, and
  \bibinfo{author}{\bibfnamefont{L.}~\bibnamefont{McAllister}},
  \bibinfo{year}{2004}, \bibinfo{journal}{Phys. Rev.}
  \textbf{\bibinfo{volume}{D69}}, \bibinfo{pages}{086008}.

\bibitem[{\citenamefont{Gukov} \emph{et~al.}(2000)\citenamefont{Gukov, Vafa,
  and Witten}}]{Gukov:1999ya}
\bibinfo{author}{\bibnamefont{Gukov}, \bibfnamefont{S.}},
  \bibinfo{author}{\bibfnamefont{C.}~\bibnamefont{Vafa}}, and
  \bibinfo{author}{\bibfnamefont{E.}~\bibnamefont{Witten}},
  \bibinfo{year}{2000}, \bibinfo{journal}{Nucl. Phys.}
  \textbf{\bibinfo{volume}{B584}}, \bibinfo{pages}{69}.

\bibitem[{\citenamefont{Gurrieri} \emph{et~al.}(2003)\citenamefont{Gurrieri,
  Louis, Micu, and Waldram}}]{Gurrieri:2002wz}
\bibinfo{author}{\bibnamefont{Gurrieri}, \bibfnamefont{S.}},
  \bibinfo{author}{\bibfnamefont{J.}~\bibnamefont{Louis}},
  \bibinfo{author}{\bibfnamefont{A.}~\bibnamefont{Micu}}, and
  \bibinfo{author}{\bibfnamefont{D.}~\bibnamefont{Waldram}},
  \bibinfo{year}{2003}, \bibinfo{journal}{Nucl. Phys.}
  \textbf{\bibinfo{volume}{B654}}, \bibinfo{pages}{61}.

\bibitem[{\citenamefont{Guth}(2000)}]{Guth:2000ka}
\bibinfo{author}{\bibnamefont{Guth}, \bibfnamefont{A.~H.}},
  \bibinfo{year}{2000}, \bibinfo{journal}{Phys. Rept.}
  \textbf{\bibinfo{volume}{333}}, \bibinfo{pages}{555}.

\bibitem[{\citenamefont{Haack} \emph{et~al.}(2006)\citenamefont{Haack, Krefl,
  Lust, Van~Proeyen, and Zagermann}}]{Haack:2006cy}
\bibinfo{author}{\bibnamefont{Haack}, \bibfnamefont{M.}},
  \bibinfo{author}{\bibfnamefont{D.}~\bibnamefont{Krefl}},
  \bibinfo{author}{\bibfnamefont{D.}~\bibnamefont{Lust}},
  \bibinfo{author}{\bibfnamefont{A.}~\bibnamefont{Van~Proeyen}}, and
  \bibinfo{author}{\bibfnamefont{M.}~\bibnamefont{Zagermann}},
  \bibinfo{year}{2006}, \eprint{hep-th/0609211}.

\bibitem[{\citenamefont{Hartle and Hawking}(1983)}]{Hartle:1983ai}
\bibinfo{author}{\bibnamefont{Hartle}, \bibfnamefont{J.~B.}}, and
  \bibinfo{author}{\bibfnamefont{S.~W.} \bibnamefont{Hawking}},
  \bibinfo{year}{1983}, \bibinfo{journal}{Phys. Rev.}
  \textbf{\bibinfo{volume}{D28}}, \bibinfo{pages}{2960}.

\bibitem[{\citenamefont{Hassan}(2000)}]{Hassan:1999bv}
\bibinfo{author}{\bibnamefont{Hassan}, \bibfnamefont{S.~F.}},
  \bibinfo{year}{2000}, \bibinfo{journal}{Nucl. Phys.}
  \textbf{\bibinfo{volume}{B568}}, \bibinfo{pages}{145}.

\bibitem[{\citenamefont{Hawking}(1984)}]{Hawking:1984hk}
\bibinfo{author}{\bibnamefont{Hawking}, \bibfnamefont{S.~W.}},
  \bibinfo{year}{1984}, \bibinfo{journal}{Phys. Lett.}
  \textbf{\bibinfo{volume}{B134}}, \bibinfo{pages}{403}.

\bibitem[{\citenamefont{Hawking and Moss}(1982)}]{Hawking:1981fz}
\bibinfo{author}{\bibnamefont{Hawking}, \bibfnamefont{S.~W.}}, and
  \bibinfo{author}{\bibfnamefont{I.~G.} \bibnamefont{Moss}},
  \bibinfo{year}{1982}, \bibinfo{journal}{Phys. Lett.}
  \textbf{\bibinfo{volume}{B110}}, \bibinfo{pages}{35}.

\bibitem[{\citenamefont{Headrick and Wiseman}(2005)}]{Headrick:2005ch}
\bibinfo{author}{\bibnamefont{Headrick}, \bibfnamefont{M.}}, and
  \bibinfo{author}{\bibfnamefont{T.}~\bibnamefont{Wiseman}},
  \bibinfo{year}{2005}, \bibinfo{journal}{Class. Quant. Grav.}
  \textbf{\bibinfo{volume}{22}}, \bibinfo{pages}{4931}.

\bibitem[{\citenamefont{Hebecker and March-Russell}(2006)}]{Hebecker:2006bn}
\bibinfo{author}{\bibnamefont{Hebecker}, \bibfnamefont{A.}}, and
  \bibinfo{author}{\bibfnamefont{J.}~\bibnamefont{March-Russell}},
  \bibinfo{year}{2006}, \eprint{hep-th/0607120}.

\bibitem[{\citenamefont{Hellerman} \emph{et~al.}(2001)\citenamefont{Hellerman,
  Kaloper, and Susskind}}]{Hellerman:2001yi}
\bibinfo{author}{\bibnamefont{Hellerman}, \bibfnamefont{S.}},
  \bibinfo{author}{\bibfnamefont{N.}~\bibnamefont{Kaloper}}, and
  \bibinfo{author}{\bibfnamefont{L.}~\bibnamefont{Susskind}},
  \bibinfo{year}{2001}, \bibinfo{journal}{JHEP} \textbf{\bibinfo{volume}{06}},
  \bibinfo{pages}{003}.

\bibitem[{\citenamefont{Hellerman} \emph{et~al.}(2004)\citenamefont{Hellerman,
  McGreevy, and Williams}}]{Hellerman:2002ax}
\bibinfo{author}{\bibnamefont{Hellerman}, \bibfnamefont{S.}},
  \bibinfo{author}{\bibfnamefont{J.}~\bibnamefont{McGreevy}}, and
  \bibinfo{author}{\bibfnamefont{B.}~\bibnamefont{Williams}},
  \bibinfo{year}{2004}, \bibinfo{journal}{JHEP} \textbf{\bibinfo{volume}{01}},
  \bibinfo{pages}{024}.

\bibitem[{\citenamefont{Herdeiro} \emph{et~al.}(2001)\citenamefont{Herdeiro,
  Hirano, and Kallosh}}]{Herdeiro:2001zb}
\bibinfo{author}{\bibnamefont{Herdeiro}, \bibfnamefont{C.}},
  \bibinfo{author}{\bibfnamefont{S.}~\bibnamefont{Hirano}}, and
  \bibinfo{author}{\bibfnamefont{R.}~\bibnamefont{Kallosh}},
  \bibinfo{year}{2001}, \bibinfo{journal}{JHEP} \textbf{\bibinfo{volume}{12}},
  \bibinfo{pages}{027}.

\bibitem[{\citenamefont{Holman and Mersini-Houghton}(2005)}]{Holman:2005ei}
\bibinfo{author}{\bibnamefont{Holman}, \bibfnamefont{R.}}, and
  \bibinfo{author}{\bibfnamefont{L.}~\bibnamefont{Mersini-Houghton}},
  \bibinfo{year}{2005}, \eprint{hep-th/0511102}.

\bibitem[{\citenamefont{Horava and Witten}(1996)}]{Horava:1995qa}
\bibinfo{author}{\bibnamefont{Horava}, \bibfnamefont{P.}}, and
  \bibinfo{author}{\bibfnamefont{E.}~\bibnamefont{Witten}},
  \bibinfo{year}{1996}, \bibinfo{journal}{Nucl. Phys.}
  \textbf{\bibinfo{volume}{B460}}, \bibinfo{pages}{506}.

\bibitem[{\citenamefont{Horne and Moore}(1994)}]{Horne:1994mi}
\bibinfo{author}{\bibnamefont{Horne}, \bibfnamefont{J.~H.}}, and
  \bibinfo{author}{\bibfnamefont{G.~W.} \bibnamefont{Moore}},
  \bibinfo{year}{1994}, \bibinfo{journal}{Nucl. Phys.}
  \textbf{\bibinfo{volume}{B432}}, \bibinfo{pages}{109}.

\bibitem[{\citenamefont{House and Palti}(2005)}]{House:2005yc}
\bibinfo{author}{\bibnamefont{House}, \bibfnamefont{T.}}, and
  \bibinfo{author}{\bibfnamefont{E.}~\bibnamefont{Palti}},
  \bibinfo{year}{2005}, \bibinfo{journal}{Phys. Rev.}
  \textbf{\bibinfo{volume}{D72}}, \bibinfo{pages}{026004}.

\bibitem[{\citenamefont{Hull}(2005)}]{Hull:2004in}
\bibinfo{author}{\bibnamefont{Hull}, \bibfnamefont{C.~M.}},
  \bibinfo{year}{2005}, \bibinfo{journal}{JHEP} \textbf{\bibinfo{volume}{10}},
  \bibinfo{pages}{065}.

\bibitem[{\citenamefont{Hull}(2006{\natexlab{a}})}]{Hull:2006va}
\bibinfo{author}{\bibnamefont{Hull}, \bibfnamefont{C.~M.}},
  \bibinfo{year}{2006}{\natexlab{a}}, \eprint{hep-th/0605149}.

\bibitem[{\citenamefont{Hull}(2006{\natexlab{b}})}]{Hull:2006qs}
\bibinfo{author}{\bibnamefont{Hull}, \bibfnamefont{C.~M.}},
  \bibinfo{year}{2006}{\natexlab{b}}, \eprint{hep-th/0604178}.

\bibitem[{\citenamefont{Hull and Reid-Edwards}(2005)}]{Hull:2005hk}
\bibinfo{author}{\bibnamefont{Hull}, \bibfnamefont{C.~M.}}, and
  \bibinfo{author}{\bibfnamefont{R.~A.} \bibnamefont{Reid-Edwards}},
  \bibinfo{year}{2005}, \eprint{hep-th/0503114}.

\bibitem[{\citenamefont{Ibanez}(2005)}]{Ibanez:2004iv}
\bibinfo{author}{\bibnamefont{Ibanez}, \bibfnamefont{L.~E.}},
  \bibinfo{year}{2005}, \bibinfo{journal}{Phys. Rev.}
  \textbf{\bibinfo{volume}{D71}}, \bibinfo{pages}{055005}.

\bibitem[{\citenamefont{Ibanez and Uranga}(2006)}]{Ibanez:2006da}
\bibinfo{author}{\bibnamefont{Ibanez}, \bibfnamefont{L.~E.}}, and
  \bibinfo{author}{\bibfnamefont{A.~M.} \bibnamefont{Uranga}},
  \bibinfo{year}{2006}, \eprint{hep-th/0609213}.

\bibitem[{\citenamefont{Ihl and Wrase}(2006)}]{Ihl:2006pp}
\bibinfo{author}{\bibnamefont{Ihl}, \bibfnamefont{M.}}, and
  \bibinfo{author}{\bibfnamefont{T.}~\bibnamefont{Wrase}},
  \bibinfo{year}{2006}, \bibinfo{journal}{JHEP} \textbf{\bibinfo{volume}{07}},
  \bibinfo{pages}{027}.

\bibitem[{\citenamefont{Intriligator}
  \emph{et~al.}(2006)\citenamefont{Intriligator, Seiberg, and
  Shih}}]{Intriligator:2006dd}
\bibinfo{author}{\bibnamefont{Intriligator}, \bibfnamefont{K.}},
  \bibinfo{author}{\bibfnamefont{N.}~\bibnamefont{Seiberg}}, and
  \bibinfo{author}{\bibfnamefont{D.}~\bibnamefont{Shih}}, \bibinfo{year}{2006},
  \bibinfo{journal}{JHEP} \textbf{\bibinfo{volume}{04}}, \bibinfo{pages}{021}.

\bibitem[{\citenamefont{Intriligator}
  \emph{et~al.}(1994)\citenamefont{Intriligator, Leigh, and
  Seiberg}}]{Intriligator:1994jr}
\bibinfo{author}{\bibnamefont{Intriligator}, \bibfnamefont{K.~A.}},
  \bibinfo{author}{\bibfnamefont{R.~G.} \bibnamefont{Leigh}}, and
  \bibinfo{author}{\bibfnamefont{N.}~\bibnamefont{Seiberg}},
  \bibinfo{year}{1994}, \bibinfo{journal}{Phys. Rev.}
  \textbf{\bibinfo{volume}{D50}}, \bibinfo{pages}{1092}.

\bibitem[{\citenamefont{Itzhaki}(2006)}]{Itzhaki:2006re}
\bibinfo{author}{\bibnamefont{Itzhaki}, \bibfnamefont{N.}},
  \bibinfo{year}{2006}, \bibinfo{journal}{JHEP} \textbf{\bibinfo{volume}{08}},
  \bibinfo{pages}{020}.

\bibitem[{\citenamefont{Jatkar} \emph{et~al.}(2002)\citenamefont{Jatkar,
  Mandal, Wadia, and Yogendran}}]{Jatkar:2001uh}
\bibinfo{author}{\bibnamefont{Jatkar}, \bibfnamefont{D.~P.}},
  \bibinfo{author}{\bibfnamefont{G.}~\bibnamefont{Mandal}},
  \bibinfo{author}{\bibfnamefont{S.~R.} \bibnamefont{Wadia}}, and
  \bibinfo{author}{\bibfnamefont{K.~P.} \bibnamefont{Yogendran}},
  \bibinfo{year}{2002}, \bibinfo{journal}{JHEP} \textbf{\bibinfo{volume}{01}},
  \bibinfo{pages}{039}.

\bibitem[{\citenamefont{Jockers and
  Louis}(2005{\natexlab{a}})}]{Jockers:2005zy}
\bibinfo{author}{\bibnamefont{Jockers}, \bibfnamefont{H.}}, and
  \bibinfo{author}{\bibfnamefont{J.}~\bibnamefont{Louis}},
  \bibinfo{year}{2005}{\natexlab{a}}, \bibinfo{journal}{Nucl. Phys.}
  \textbf{\bibinfo{volume}{B718}}, \bibinfo{pages}{203}.

\bibitem[{\citenamefont{Jockers and
  Louis}(2005{\natexlab{b}})}]{Jockers:2004yj}
\bibinfo{author}{\bibnamefont{Jockers}, \bibfnamefont{H.}}, and
  \bibinfo{author}{\bibfnamefont{J.}~\bibnamefont{Louis}},
  \bibinfo{year}{2005}{\natexlab{b}}, \bibinfo{journal}{Nucl. Phys.}
  \textbf{\bibinfo{volume}{B705}}, \bibinfo{pages}{167}.

\bibitem[{\citenamefont{Johnson}(2003)}]{Johnson:2003gi}
\bibinfo{author}{\bibnamefont{Johnson}, \bibfnamefont{C.}},
  \bibinfo{year}{2003}, \emph{\bibinfo{title}{D-branes}}
  (\bibinfo{publisher}{Cambridge Univ. Press}).

\bibitem[{\citenamefont{Jones} \emph{et~al.}(2002)\citenamefont{Jones, Stoica,
  and Tye}}]{Jones:2002cv}
\bibinfo{author}{\bibnamefont{Jones}, \bibfnamefont{N.~T.}},
  \bibinfo{author}{\bibfnamefont{H.}~\bibnamefont{Stoica}}, and
  \bibinfo{author}{\bibfnamefont{S.~H.~H.} \bibnamefont{Tye}},
  \bibinfo{year}{2002}, \bibinfo{journal}{JHEP} \textbf{\bibinfo{volume}{07}},
  \bibinfo{pages}{051}.

\bibitem[{\citenamefont{Jones} \emph{et~al.}(2003)\citenamefont{Jones, Stoica,
  and Tye}}]{Jones:2003da}
\bibinfo{author}{\bibnamefont{Jones}, \bibfnamefont{N.~T.}},
  \bibinfo{author}{\bibfnamefont{H.}~\bibnamefont{Stoica}}, and
  \bibinfo{author}{\bibfnamefont{S.~H.~H.} \bibnamefont{Tye}},
  \bibinfo{year}{2003}, \bibinfo{journal}{Phys. Lett.}
  \textbf{\bibinfo{volume}{B563}}, \bibinfo{pages}{6}.

\bibitem[{\citenamefont{Kachru}
  \emph{et~al.}(2003{\natexlab{a}})\citenamefont{Kachru, Kallosh, Linde, and
  Trivedi}}]{Kachru:2003aw}
\bibinfo{author}{\bibnamefont{Kachru}, \bibfnamefont{S.}},
  \bibinfo{author}{\bibfnamefont{R.}~\bibnamefont{Kallosh}},
  \bibinfo{author}{\bibfnamefont{A.}~\bibnamefont{Linde}}, and
  \bibinfo{author}{\bibfnamefont{S.~P.} \bibnamefont{Trivedi}},
  \bibinfo{year}{2003}{\natexlab{a}}, \bibinfo{journal}{Phys. Rev.}
  \textbf{\bibinfo{volume}{D68}}, \bibinfo{pages}{046005}.

\bibitem[{\citenamefont{Kachru and Kashani-Poor}(2005)}]{Kachru:2004jr}
\bibinfo{author}{\bibnamefont{Kachru}, \bibfnamefont{S.}}, and
  \bibinfo{author}{\bibfnamefont{A.-K.} \bibnamefont{Kashani-Poor}},
  \bibinfo{year}{2005}, \bibinfo{journal}{JHEP} \textbf{\bibinfo{volume}{03}},
  \bibinfo{pages}{066}.

\bibitem[{\citenamefont{Kachru} \emph{et~al.}(1996)\citenamefont{Kachru, Klemm,
  Lerche, Mayr, and Vafa}}]{Kachru:1995fv}
\bibinfo{author}{\bibnamefont{Kachru}, \bibfnamefont{S.}},
  \bibinfo{author}{\bibfnamefont{A.}~\bibnamefont{Klemm}},
  \bibinfo{author}{\bibfnamefont{W.}~\bibnamefont{Lerche}},
  \bibinfo{author}{\bibfnamefont{P.}~\bibnamefont{Mayr}}, and
  \bibinfo{author}{\bibfnamefont{C.}~\bibnamefont{Vafa}}, \bibinfo{year}{1996},
  \bibinfo{journal}{Nucl. Phys.} \textbf{\bibinfo{volume}{B459}},
  \bibinfo{pages}{537}.

\bibitem[{\citenamefont{Kachru} \emph{et~al.}(2006)\citenamefont{Kachru,
  McGreevy, and Svrcek}}]{Kachru:2006em}
\bibinfo{author}{\bibnamefont{Kachru}, \bibfnamefont{S.}},
  \bibinfo{author}{\bibfnamefont{J.}~\bibnamefont{McGreevy}}, and
  \bibinfo{author}{\bibfnamefont{P.}~\bibnamefont{Svrcek}},
  \bibinfo{year}{2006}, \bibinfo{journal}{JHEP} \textbf{\bibinfo{volume}{04}},
  \bibinfo{pages}{023}.

\bibitem[{\citenamefont{Kachru} \emph{et~al.}(2002)\citenamefont{Kachru,
  Pearson, and Verlinde}}]{Kachru:2002gs}
\bibinfo{author}{\bibnamefont{Kachru}, \bibfnamefont{S.}},
  \bibinfo{author}{\bibfnamefont{J.}~\bibnamefont{Pearson}}, and
  \bibinfo{author}{\bibfnamefont{H.~L.} \bibnamefont{Verlinde}},
  \bibinfo{year}{2002}, \bibinfo{journal}{JHEP} \textbf{\bibinfo{volume}{06}},
  \bibinfo{pages}{021}.

\bibitem[{\citenamefont{Kachru}
  \emph{et~al.}(2003{\natexlab{b}})\citenamefont{Kachru, Schulz, Tripathy, and
  Trivedi}}]{Kachru:2002sk}
\bibinfo{author}{\bibnamefont{Kachru}, \bibfnamefont{S.}},
  \bibinfo{author}{\bibfnamefont{M.~B.} \bibnamefont{Schulz}},
  \bibinfo{author}{\bibfnamefont{P.~K.} \bibnamefont{Tripathy}}, and
  \bibinfo{author}{\bibfnamefont{S.~P.} \bibnamefont{Trivedi}},
  \bibinfo{year}{2003}{\natexlab{b}}, \bibinfo{journal}{JHEP}
  \textbf{\bibinfo{volume}{03}}, \bibinfo{pages}{061}.

\bibitem[{\citenamefont{Kachru}
  \emph{et~al.}(2003{\natexlab{c}})\citenamefont{Kachru, Schulz, and
  Trivedi}}]{Kachru:2002he}
\bibinfo{author}{\bibnamefont{Kachru}, \bibfnamefont{S.}},
  \bibinfo{author}{\bibfnamefont{M.~B.} \bibnamefont{Schulz}}, and
  \bibinfo{author}{\bibfnamefont{S.}~\bibnamefont{Trivedi}},
  \bibinfo{year}{2003}{\natexlab{c}}, \bibinfo{journal}{JHEP}
  \textbf{\bibinfo{volume}{10}}, \bibinfo{pages}{007}.

\bibitem[{\citenamefont{Kachru and Vafa}(1995)}]{Kachru:1995wm}
\bibinfo{author}{\bibnamefont{Kachru}, \bibfnamefont{S.}}, and
  \bibinfo{author}{\bibfnamefont{C.}~\bibnamefont{Vafa}}, \bibinfo{year}{1995},
  \bibinfo{journal}{Nucl. Phys.} \textbf{\bibinfo{volume}{B450}},
  \bibinfo{pages}{69}.

\bibitem[{Kachru \emph{et~al.}(2003)\citenamefont{Kachru}
  \emph{et~al.}}]{Kachru:2003sx}
\bibinfo{author}{\bibnamefont{Kachru}, \bibfnamefont{S.}}, \emph{et~al.},
  \bibinfo{year}{2003}, \bibinfo{journal}{JCAP}
  \textbf{\bibinfo{volume}{0310}}, \bibinfo{pages}{013}.

\bibitem[{\citenamefont{Kallosh} \emph{et~al.}(2005)\citenamefont{Kallosh,
  Kashani-Poor, and Tomasiello}}]{Kallosh:2005gs}
\bibinfo{author}{\bibnamefont{Kallosh}, \bibfnamefont{R.}},
  \bibinfo{author}{\bibfnamefont{A.-K.} \bibnamefont{Kashani-Poor}}, and
  \bibinfo{author}{\bibfnamefont{A.}~\bibnamefont{Tomasiello}},
  \bibinfo{year}{2005}, \bibinfo{journal}{JHEP} \textbf{\bibinfo{volume}{06}},
  \bibinfo{pages}{069}.

\bibitem[{\citenamefont{Kallosh and Linde}(2004)}]{Kallosh:2004yh}
\bibinfo{author}{\bibnamefont{Kallosh}, \bibfnamefont{R.}}, and
  \bibinfo{author}{\bibfnamefont{A.}~\bibnamefont{Linde}},
  \bibinfo{year}{2004}, \bibinfo{journal}{JHEP} \textbf{\bibinfo{volume}{12}},
  \bibinfo{pages}{004}.

\bibitem[{\citenamefont{Kaluza}(1921)}]{Kaluza:1921tu}
\bibinfo{author}{\bibnamefont{Kaluza}, \bibfnamefont{T.}},
  \bibinfo{year}{1921}, \bibinfo{journal}{Sitzungsber. Preuss. Akad. Wiss.
  Berlin (Math. Phys. )} \textbf{\bibinfo{volume}{1921}}, \bibinfo{pages}{966}.

\bibitem[{\citenamefont{Kane} \emph{et~al.}(2006)\citenamefont{Kane, Kumar, and
  Shao}}]{Kane:2006yi}
\bibinfo{author}{\bibnamefont{Kane}, \bibfnamefont{G.~L.}},
  \bibinfo{author}{\bibfnamefont{P.}~\bibnamefont{Kumar}}, and
  \bibinfo{author}{\bibfnamefont{J.}~\bibnamefont{Shao}}, \bibinfo{year}{2006},
  \eprint{hep-ph/0610038}.

\bibitem[{\citenamefont{Kaplunovsky}(1988)}]{Kaplunovsky:1987rp}
\bibinfo{author}{\bibnamefont{Kaplunovsky}, \bibfnamefont{V.~S.}},
  \bibinfo{year}{1988}, \bibinfo{journal}{Nucl. Phys.}
  \textbf{\bibinfo{volume}{B307}}, \bibinfo{pages}{145}.

\bibitem[{\citenamefont{Katz} \emph{et~al.}(1997)\citenamefont{Katz, Klemm, and
  Vafa}}]{Katz:1996fh}
\bibinfo{author}{\bibnamefont{Katz}, \bibfnamefont{S.}},
  \bibinfo{author}{\bibfnamefont{A.}~\bibnamefont{Klemm}}, and
  \bibinfo{author}{\bibfnamefont{C.}~\bibnamefont{Vafa}}, \bibinfo{year}{1997},
  \bibinfo{journal}{Nucl. Phys.} \textbf{\bibinfo{volume}{B497}},
  \bibinfo{pages}{173}.

\bibitem[{\citenamefont{Katz and Vafa}(1997)}]{Katz:1996th}
\bibinfo{author}{\bibnamefont{Katz}, \bibfnamefont{S.}}, and
  \bibinfo{author}{\bibfnamefont{C.}~\bibnamefont{Vafa}}, \bibinfo{year}{1997},
  \bibinfo{journal}{Nucl. Phys.} \textbf{\bibinfo{volume}{B497}},
  \bibinfo{pages}{196}.

\bibitem[{\citenamefont{Kawai} \emph{et~al.}(1986)\citenamefont{Kawai,
  Lewellen, and Tye}}]{Kawai:1986va}
\bibinfo{author}{\bibnamefont{Kawai}, \bibfnamefont{H.}},
  \bibinfo{author}{\bibfnamefont{D.~C.} \bibnamefont{Lewellen}}, and
  \bibinfo{author}{\bibfnamefont{S.~H.~H.} \bibnamefont{Tye}},
  \bibinfo{year}{1986}, \bibinfo{journal}{Phys. Rev. Lett.}
  \textbf{\bibinfo{volume}{57}}, \bibinfo{pages}{1832}.

\bibitem[{\citenamefont{Kawai} \emph{et~al.}(1987)\citenamefont{Kawai,
  Lewellen, and Tye}}]{Kawai:1986ah}
\bibinfo{author}{\bibnamefont{Kawai}, \bibfnamefont{H.}},
  \bibinfo{author}{\bibfnamefont{D.~C.} \bibnamefont{Lewellen}}, and
  \bibinfo{author}{\bibfnamefont{S.~H.~H.} \bibnamefont{Tye}},
  \bibinfo{year}{1987}, \bibinfo{journal}{Nucl. Phys.}
  \textbf{\bibinfo{volume}{B288}}, \bibinfo{pages}{1}.

\bibitem[{\citenamefont{Kim and Yi}(2006)}]{Kim:2006qs}
\bibinfo{author}{\bibnamefont{Kim}, \bibfnamefont{S.}}, and
  \bibinfo{author}{\bibfnamefont{P.}~\bibnamefont{Yi}}, \bibinfo{year}{2006},
  \eprint{hep-th/0607091}.

\bibitem[{\citenamefont{Kimura and Yi}(2006)}]{Kimura:2006af}
\bibinfo{author}{\bibnamefont{Kimura}, \bibfnamefont{T.}}, and
  \bibinfo{author}{\bibfnamefont{P.}~\bibnamefont{Yi}}, \bibinfo{year}{2006},
  \bibinfo{journal}{JHEP} \textbf{\bibinfo{volume}{07}}, \bibinfo{pages}{030}.

\bibitem[{\citenamefont{Kitano and Nomura}(2005)}]{Kitano:2005wc}
\bibinfo{author}{\bibnamefont{Kitano}, \bibfnamefont{R.}}, and
  \bibinfo{author}{\bibfnamefont{Y.}~\bibnamefont{Nomura}},
  \bibinfo{year}{2005}, \bibinfo{journal}{Phys. Lett.}
  \textbf{\bibinfo{volume}{B631}}, \bibinfo{pages}{58}.

\bibitem[{\citenamefont{Klebanov and Strassler}(2000)}]{Klebanov:2000hb}
\bibinfo{author}{\bibnamefont{Klebanov}, \bibfnamefont{I.~R.}}, and
  \bibinfo{author}{\bibfnamefont{M.~J.} \bibnamefont{Strassler}},
  \bibinfo{year}{2000}, \bibinfo{journal}{JHEP} \textbf{\bibinfo{volume}{08}},
  \bibinfo{pages}{052}.

\bibitem[{\citenamefont{Klein}(1926)}]{Klein:1926tv}
\bibinfo{author}{\bibnamefont{Klein}, \bibfnamefont{O.}}, \bibinfo{year}{1926},
  \bibinfo{journal}{Z. Phys.} \textbf{\bibinfo{volume}{37}},
  \bibinfo{pages}{895}.

\bibitem[{\citenamefont{Kobakhidze and
  Mersini-Houghton}(2004)}]{Kobakhidze:2004gm}
\bibinfo{author}{\bibnamefont{Kobakhidze}, \bibfnamefont{A.}}, and
  \bibinfo{author}{\bibfnamefont{L.}~\bibnamefont{Mersini-Houghton}},
  \bibinfo{year}{2004}, \eprint{hep-th/0410213}.

\bibitem[{Kofman \emph{et~al.}(2004)\citenamefont{Kofman}
  \emph{et~al.}}]{Kofman:2004yc}
\bibinfo{author}{\bibnamefont{Kofman}, \bibfnamefont{L.}}, \emph{et~al.},
  \bibinfo{year}{2004}, \bibinfo{journal}{JHEP} \textbf{\bibinfo{volume}{05}},
  \bibinfo{pages}{030}.

\bibitem[{\citenamefont{Kreuzer} \emph{et~al.}(1992)\citenamefont{Kreuzer,
  Schimmrigk, and Skarke}}]{Kreuzer:1991gf}
\bibinfo{author}{\bibnamefont{Kreuzer}, \bibfnamefont{M.}},
  \bibinfo{author}{\bibfnamefont{R.}~\bibnamefont{Schimmrigk}}, and
  \bibinfo{author}{\bibfnamefont{H.}~\bibnamefont{Skarke}},
  \bibinfo{year}{1992}, \bibinfo{journal}{Nucl. Phys.}
  \textbf{\bibinfo{volume}{B372}}, \bibinfo{pages}{61}.

\bibitem[{\citenamefont{Kreuzer and
  Skarke}(2002{\natexlab{a}})}]{Kreuzer:2000xy}
\bibinfo{author}{\bibnamefont{Kreuzer}, \bibfnamefont{M.}}, and
  \bibinfo{author}{\bibfnamefont{H.}~\bibnamefont{Skarke}},
  \bibinfo{year}{2002}{\natexlab{a}}, \bibinfo{journal}{Adv. Theor. Math.
  Phys.} \textbf{\bibinfo{volume}{4}}, \bibinfo{pages}{1209}.

\bibitem[{\citenamefont{Kreuzer and
  Skarke}(2002{\natexlab{b}})}]{Kreuzer:2000qv}
\bibinfo{author}{\bibnamefont{Kreuzer}, \bibfnamefont{M.}}, and
  \bibinfo{author}{\bibfnamefont{H.}~\bibnamefont{Skarke}},
  \bibinfo{year}{2002}{\natexlab{b}}, \bibinfo{journal}{Rev. Math. Phys.}
  \textbf{\bibinfo{volume}{14}}, \bibinfo{pages}{343}.

\bibitem[{\citenamefont{Kreuzer and Skarke}(2004)}]{Kreuzer:2002uu}
\bibinfo{author}{\bibnamefont{Kreuzer}, \bibfnamefont{M.}}, and
  \bibinfo{author}{\bibfnamefont{H.}~\bibnamefont{Skarke}},
  \bibinfo{year}{2004}, \bibinfo{journal}{Comput. Phys. Commun.}
  \textbf{\bibinfo{volume}{157}}, \bibinfo{pages}{87}.

\bibitem[{\citenamefont{Kumar}(2006)}]{Kumar:2006tn}
\bibinfo{author}{\bibnamefont{Kumar}, \bibfnamefont{J.}}, \bibinfo{year}{2006},
  \bibinfo{journal}{Int. J. Mod. Phys.} \textbf{\bibinfo{volume}{A21}},
  \bibinfo{pages}{3441}.

\bibitem[{\citenamefont{Kumar and Wells}(2005{\natexlab{a}})}]{Kumar:2004pv}
\bibinfo{author}{\bibnamefont{Kumar}, \bibfnamefont{J.}}, and
  \bibinfo{author}{\bibfnamefont{J.~D.} \bibnamefont{Wells}},
  \bibinfo{year}{2005}{\natexlab{a}}, \bibinfo{journal}{Phys. Rev.}
  \textbf{\bibinfo{volume}{D71}}, \bibinfo{pages}{026009}.

\bibitem[{\citenamefont{Kumar and Wells}(2005{\natexlab{b}})}]{Kumar:2005hf}
\bibinfo{author}{\bibnamefont{Kumar}, \bibfnamefont{J.}}, and
  \bibinfo{author}{\bibfnamefont{J.~D.} \bibnamefont{Wells}},
  \bibinfo{year}{2005}{\natexlab{b}}, \bibinfo{journal}{JHEP}
  \textbf{\bibinfo{volume}{09}}, \bibinfo{pages}{067}.

\bibitem[{\citenamefont{Landsberg}(2006)}]{Landsberg:2006mm}
\bibinfo{author}{\bibnamefont{Landsberg}, \bibfnamefont{G.}},
  \bibinfo{year}{2006}, \bibinfo{journal}{J. Phys.}
  \textbf{\bibinfo{volume}{G32}}, \bibinfo{pages}{R337}.

\bibitem[{\citenamefont{Lawrence and
  McGreevy}(2004{\natexlab{a}})}]{Lawrence:2004zk}
\bibinfo{author}{\bibnamefont{Lawrence}, \bibfnamefont{A.}}, and
  \bibinfo{author}{\bibfnamefont{J.}~\bibnamefont{McGreevy}},
  \bibinfo{year}{2004}{\natexlab{a}}, \bibinfo{journal}{JHEP}
  \textbf{\bibinfo{volume}{06}}, \bibinfo{pages}{007}.

\bibitem[{\citenamefont{Lawrence and
  McGreevy}(2004{\natexlab{b}})}]{Lawrence:2004kj}
\bibinfo{author}{\bibnamefont{Lawrence}, \bibfnamefont{A.}}, and
  \bibinfo{author}{\bibfnamefont{J.}~\bibnamefont{McGreevy}},
  \bibinfo{year}{2004}{\natexlab{b}}, \eprint{hep-th/0401233}.

\bibitem[{\citenamefont{Lawrence} \emph{et~al.}(2006)\citenamefont{Lawrence,
  Schulz, and Wecht}}]{Lawrence:2006ma}
\bibinfo{author}{\bibnamefont{Lawrence}, \bibfnamefont{A.}},
  \bibinfo{author}{\bibfnamefont{M.~B.} \bibnamefont{Schulz}}, and
  \bibinfo{author}{\bibfnamefont{B.}~\bibnamefont{Wecht}},
  \bibinfo{year}{2006}, \bibinfo{journal}{JHEP} \textbf{\bibinfo{volume}{07}},
  \bibinfo{pages}{038}.

\bibitem[{\citenamefont{Lerche} \emph{et~al.}(1987)\citenamefont{Lerche, Lust,
  and Schellekens}}]{Lerche:1986cx}
\bibinfo{author}{\bibnamefont{Lerche}, \bibfnamefont{W.}},
  \bibinfo{author}{\bibfnamefont{D.}~\bibnamefont{Lust}}, and
  \bibinfo{author}{\bibfnamefont{A.~N.} \bibnamefont{Schellekens}},
  \bibinfo{year}{1987}, \bibinfo{journal}{Nucl. Phys.}
  \textbf{\bibinfo{volume}{B287}}, \bibinfo{pages}{477}.

\bibitem[{\citenamefont{Li and Yau}(2004)}]{Li:2004hx}
\bibinfo{author}{\bibnamefont{Li}, \bibfnamefont{J.}}, and
  \bibinfo{author}{\bibfnamefont{S.-T.} \bibnamefont{Yau}},
  \bibinfo{year}{2004}, \eprint{hep-th/0411136}.

\bibitem[{\citenamefont{Linde}(1984)}]{Linde:1984ir}
\bibinfo{author}{\bibnamefont{Linde}, \bibfnamefont{A.~D.}},
  \bibinfo{year}{1984}, \bibinfo{journal}{Rept. Prog. Phys.}
  \textbf{\bibinfo{volume}{47}}, \bibinfo{pages}{925}.

\bibitem[{\citenamefont{Linde}(1986{\natexlab{a}})}]{Linde:1986fc}
\bibinfo{author}{\bibnamefont{Linde}, \bibfnamefont{A.~D.}},
  \bibinfo{year}{1986}{\natexlab{a}}, \bibinfo{journal}{Mod. Phys. Lett.}
  \textbf{\bibinfo{volume}{A1}}, \bibinfo{pages}{81}.

\bibitem[{\citenamefont{Linde}(1986{\natexlab{b}})}]{Linde:1986fd}
\bibinfo{author}{\bibnamefont{Linde}, \bibfnamefont{A.~D.}},
  \bibinfo{year}{1986}{\natexlab{b}}, \bibinfo{journal}{Phys. Lett.}
  \textbf{\bibinfo{volume}{B175}}, \bibinfo{pages}{395}.

\bibitem[{\citenamefont{Linde}(1994)}]{Linde:1993cn}
\bibinfo{author}{\bibnamefont{Linde}, \bibfnamefont{A.~D.}},
  \bibinfo{year}{1994}, \bibinfo{journal}{Phys. Rev.}
  \textbf{\bibinfo{volume}{D49}}, \bibinfo{pages}{748}.

\bibitem[{\citenamefont{Linde}(2005)}]{Linde:2005ht}
\bibinfo{author}{\bibnamefont{Linde}, \bibfnamefont{A.~D.}},
  \bibinfo{year}{2005}, \bibinfo{journal}{Contemp. Concepts Phys.}
  \textbf{\bibinfo{volume}{5}}, \bibinfo{pages}{1}.

\bibitem[{\citenamefont{Loeb}(2006)}]{Loeb:2006en}
\bibinfo{author}{\bibnamefont{Loeb}, \bibfnamefont{A.}}, \bibinfo{year}{2006},
  \bibinfo{journal}{JCAP} \textbf{\bibinfo{volume}{0605}},
  \bibinfo{pages}{009}.

\bibitem[{\citenamefont{Lopes~Cardoso}
  \emph{et~al.}(2004)\citenamefont{Lopes~Cardoso, Curio, Dall'Agata, and
  Lust}}]{LopesCardoso:2003sp}
\bibinfo{author}{\bibnamefont{Lopes~Cardoso}, \bibfnamefont{G.}},
  \bibinfo{author}{\bibfnamefont{G.}~\bibnamefont{Curio}},
  \bibinfo{author}{\bibfnamefont{G.}~\bibnamefont{Dall'Agata}}, and
  \bibinfo{author}{\bibfnamefont{D.}~\bibnamefont{Lust}}, \bibinfo{year}{2004},
  \bibinfo{journal}{Fortsch. Phys.} \textbf{\bibinfo{volume}{52}},
  \bibinfo{pages}{483}.

\bibitem[{Lopes~Cardoso \emph{et~al.}(2003)\citenamefont{Lopes~Cardoso}
  \emph{et~al.}}]{LopesCardoso:2002hd}
\bibinfo{author}{\bibnamefont{Lopes~Cardoso}, \bibfnamefont{G.}},
  \emph{et~al.}, \bibinfo{year}{2003}, \bibinfo{journal}{Nucl. Phys.}
  \textbf{\bibinfo{volume}{B652}}, \bibinfo{pages}{5}.

\bibitem[{\citenamefont{Luciani}(1978)}]{Luciani:1977zv}
\bibinfo{author}{\bibnamefont{Luciani}, \bibfnamefont{J.~F.}},
  \bibinfo{year}{1978}, \bibinfo{journal}{Nucl. Phys.}
  \textbf{\bibinfo{volume}{B135}}, \bibinfo{pages}{111}.

\bibitem[{\citenamefont{Lust}
  \emph{et~al.}(2006{\natexlab{a}})\citenamefont{Lust, Mayr, Reffert, and
  Stieberger}}]{Lust:2005bd}
\bibinfo{author}{\bibnamefont{Lust}, \bibfnamefont{D.}},
  \bibinfo{author}{\bibfnamefont{P.}~\bibnamefont{Mayr}},
  \bibinfo{author}{\bibfnamefont{S.}~\bibnamefont{Reffert}}, and
  \bibinfo{author}{\bibfnamefont{S.}~\bibnamefont{Stieberger}},
  \bibinfo{year}{2006}{\natexlab{a}}, \bibinfo{journal}{Nucl. Phys.}
  \textbf{\bibinfo{volume}{B732}}, \bibinfo{pages}{243}.

\bibitem[{\citenamefont{Lust}
  \emph{et~al.}(2006{\natexlab{b}})\citenamefont{Lust, Reffert, Scheidegger,
  Schulgin, and Stieberger}}]{Lust:2006zg}
\bibinfo{author}{\bibnamefont{Lust}, \bibfnamefont{D.}},
  \bibinfo{author}{\bibfnamefont{S.}~\bibnamefont{Reffert}},
  \bibinfo{author}{\bibfnamefont{E.}~\bibnamefont{Scheidegger}},
  \bibinfo{author}{\bibfnamefont{W.}~\bibnamefont{Schulgin}}, and
  \bibinfo{author}{\bibfnamefont{S.}~\bibnamefont{Stieberger}},
  \bibinfo{year}{2006}{\natexlab{b}}, \eprint{hep-th/0609013}.

\bibitem[{\citenamefont{Lust}
  \emph{et~al.}(2005{\natexlab{a}})\citenamefont{Lust, Reffert, Schulgin, and
  Stieberger}}]{Lust:2005dy}
\bibinfo{author}{\bibnamefont{Lust}, \bibfnamefont{D.}},
  \bibinfo{author}{\bibfnamefont{S.}~\bibnamefont{Reffert}},
  \bibinfo{author}{\bibfnamefont{W.}~\bibnamefont{Schulgin}}, and
  \bibinfo{author}{\bibfnamefont{S.}~\bibnamefont{Stieberger}},
  \bibinfo{year}{2005}{\natexlab{a}}, \eprint{hep-th/0506090}.

\bibitem[{\citenamefont{Lust}
  \emph{et~al.}(2006{\natexlab{c}})\citenamefont{Lust, Reffert, Schulgin, and
  Tripathy}}]{Lust:2005cu}
\bibinfo{author}{\bibnamefont{Lust}, \bibfnamefont{D.}},
  \bibinfo{author}{\bibfnamefont{S.}~\bibnamefont{Reffert}},
  \bibinfo{author}{\bibfnamefont{W.}~\bibnamefont{Schulgin}}, and
  \bibinfo{author}{\bibfnamefont{P.~K.} \bibnamefont{Tripathy}},
  \bibinfo{year}{2006}{\natexlab{c}}, \bibinfo{journal}{JHEP}
  \textbf{\bibinfo{volume}{08}}, \bibinfo{pages}{071}.

\bibitem[{\citenamefont{Lust}
  \emph{et~al.}(2005{\natexlab{b}})\citenamefont{Lust, Reffert, and
  Stieberger}}]{Lust:2004fi}
\bibinfo{author}{\bibnamefont{Lust}, \bibfnamefont{D.}},
  \bibinfo{author}{\bibfnamefont{S.}~\bibnamefont{Reffert}}, and
  \bibinfo{author}{\bibfnamefont{S.}~\bibnamefont{Stieberger}},
  \bibinfo{year}{2005}{\natexlab{b}}, \bibinfo{journal}{Nucl. Phys.}
  \textbf{\bibinfo{volume}{B706}}, \bibinfo{pages}{3}.

\bibitem[{\citenamefont{Lust}
  \emph{et~al.}(2005{\natexlab{c}})\citenamefont{Lust, Reffert, and
  Stieberger}}]{Lust:2004dn}
\bibinfo{author}{\bibnamefont{Lust}, \bibfnamefont{D.}},
  \bibinfo{author}{\bibfnamefont{S.}~\bibnamefont{Reffert}}, and
  \bibinfo{author}{\bibfnamefont{S.}~\bibnamefont{Stieberger}},
  \bibinfo{year}{2005}{\natexlab{c}}, \bibinfo{journal}{Nucl. Phys.}
  \textbf{\bibinfo{volume}{B727}}, \bibinfo{pages}{264}.

\bibitem[{\citenamefont{Lust and Tsimpis}(2005)}]{Lust:2004ig}
\bibinfo{author}{\bibnamefont{Lust}, \bibfnamefont{D.}}, and
  \bibinfo{author}{\bibfnamefont{D.}~\bibnamefont{Tsimpis}},
  \bibinfo{year}{2005}, \bibinfo{journal}{JHEP} \textbf{\bibinfo{volume}{02}},
  \bibinfo{pages}{027}.

\bibitem[{\citenamefont{Luty}(2005)}]{Luty:2005sn}
\bibinfo{author}{\bibnamefont{Luty}, \bibfnamefont{M.~A.}},
  \bibinfo{year}{2005}, \eprint{hep-th/0509029}.

\bibitem[{\citenamefont{Lyth and Riotto}(1999)}]{Lyth:1998xn}
\bibinfo{author}{\bibnamefont{Lyth}, \bibfnamefont{D.~H.}}, and
  \bibinfo{author}{\bibfnamefont{A.}~\bibnamefont{Riotto}},
  \bibinfo{year}{1999}, \bibinfo{journal}{Phys. Rept.}
  \textbf{\bibinfo{volume}{314}}, \bibinfo{pages}{1}.

\bibitem[{\citenamefont{MacKay}(2003)}]{MacKay:2003}
\bibinfo{author}{\bibnamefont{MacKay}, \bibfnamefont{D.~J.~C.}},
  \bibinfo{year}{2003}, \emph{\bibinfo{title}{Information Theory, Inference and
  Learning Algorithms}} (\bibinfo{publisher}{Cambridge Univ. Press}).

\bibitem[{\citenamefont{Maldacena}(1998)}]{Maldacena:1997re}
\bibinfo{author}{\bibnamefont{Maldacena}, \bibfnamefont{J.~M.}},
  \bibinfo{year}{1998}, \bibinfo{journal}{Adv. Theor. Math. Phys.}
  \textbf{\bibinfo{volume}{2}}, \bibinfo{pages}{231}.

\bibitem[{\citenamefont{Maldacena and Nunez}(2001)}]{Maldacena:2000mw}
\bibinfo{author}{\bibnamefont{Maldacena}, \bibfnamefont{J.~M.}}, and
  \bibinfo{author}{\bibfnamefont{C.}~\bibnamefont{Nunez}},
  \bibinfo{year}{2001}, \bibinfo{journal}{Int. J. Mod. Phys.}
  \textbf{\bibinfo{volume}{A16}}, \bibinfo{pages}{822}.

\bibitem[{\citenamefont{Maloney} \emph{et~al.}(2002)\citenamefont{Maloney,
  Silverstein, and Strominger}}]{Maloney:2002rr}
\bibinfo{author}{\bibnamefont{Maloney}, \bibfnamefont{A.}},
  \bibinfo{author}{\bibfnamefont{E.}~\bibnamefont{Silverstein}}, and
  \bibinfo{author}{\bibfnamefont{A.}~\bibnamefont{Strominger}},
  \bibinfo{year}{2002}, \eprint{hep-th/0205316}.

\bibitem[{\citenamefont{Marchesano and Shiu}(2004)}]{Marchesano:2004xz}
\bibinfo{author}{\bibnamefont{Marchesano}, \bibfnamefont{F.}}, and
  \bibinfo{author}{\bibfnamefont{G.}~\bibnamefont{Shiu}}, \bibinfo{year}{2004},
  \bibinfo{journal}{JHEP} \textbf{\bibinfo{volume}{11}}, \bibinfo{pages}{041}.

\bibitem[{\citenamefont{Marchesano and Shiu}(2005)}]{Marchesano:2004yq}
\bibinfo{author}{\bibnamefont{Marchesano}, \bibfnamefont{F.}}, and
  \bibinfo{author}{\bibfnamefont{G.}~\bibnamefont{Shiu}}, \bibinfo{year}{2005},
  \bibinfo{journal}{Phys. Rev.} \textbf{\bibinfo{volume}{D71}},
  \bibinfo{pages}{011701}.

\bibitem[{\citenamefont{Marchesano}
  \emph{et~al.}(2005)\citenamefont{Marchesano, Shiu, and
  Wang}}]{Marchesano:2004yn}
\bibinfo{author}{\bibnamefont{Marchesano}, \bibfnamefont{F.}},
  \bibinfo{author}{\bibfnamefont{G.}~\bibnamefont{Shiu}}, and
  \bibinfo{author}{\bibfnamefont{L.-T.} \bibnamefont{Wang}},
  \bibinfo{year}{2005}, \bibinfo{journal}{Nucl. Phys.}
  \textbf{\bibinfo{volume}{B712}}, \bibinfo{pages}{20}.

\bibitem[{\citenamefont{Martin}(1997)}]{Martin:1997ns}
\bibinfo{author}{\bibnamefont{Martin}, \bibfnamefont{S.~P.}},
  \bibinfo{year}{1997}, \eprint{hep-ph/9709356}.

\bibitem[{\citenamefont{Mayr}(2001)}]{Mayr:2000hh}
\bibinfo{author}{\bibnamefont{Mayr}, \bibfnamefont{P.}}, \bibinfo{year}{2001},
  \bibinfo{journal}{Nucl. Phys.} \textbf{\bibinfo{volume}{B593}},
  \bibinfo{pages}{99}.

\bibitem[{\citenamefont{Mehta}(1991)}]{Mehta:1991}
\bibinfo{author}{\bibnamefont{Mehta}, \bibfnamefont{M.~L.}},
  \bibinfo{year}{1991}, \emph{\bibinfo{title}{Random matrices (2nd ed)}}
  (\bibinfo{publisher}{Academic Press}).

\bibitem[{\citenamefont{Mersini-Houghton}(2005)}]{Mersini-Houghton:2005im}
\bibinfo{author}{\bibnamefont{Mersini-Houghton}, \bibfnamefont{L.}},
  \bibinfo{year}{2005}, \bibinfo{journal}{Class. Quant. Grav.}
  \textbf{\bibinfo{volume}{22}}, \bibinfo{pages}{3481}.

\bibitem[{\citenamefont{Michelson}(1997)}]{Michelson:1996pn}
\bibinfo{author}{\bibnamefont{Michelson}, \bibfnamefont{J.}},
  \bibinfo{year}{1997}, \bibinfo{journal}{Nucl. Phys.}
  \textbf{\bibinfo{volume}{B495}}, \bibinfo{pages}{127}.

\bibitem[{\citenamefont{Moore}(2003)}]{Moore:2003vf}
\bibinfo{author}{\bibnamefont{Moore}, \bibfnamefont{G.~W.}},
  \bibinfo{year}{2003}, \eprint{hep-th/0304018}.

\bibitem[{\citenamefont{Garcia~del Moral}(2006)}]{GarciadelMoral:2005js}
\bibinfo{author}{\bibnamefont{Garcia~del Moral}, \bibfnamefont{M.~P.}},
  \bibinfo{year}{2006}, \bibinfo{journal}{JHEP} \textbf{\bibinfo{volume}{04}},
  \bibinfo{pages}{022}.

\bibitem[{\citenamefont{Murphy} \emph{et~al.}(2003)\citenamefont{Murphy, Webb,
  and Flambaum}}]{Murphy:2003hw}
\bibinfo{author}{\bibnamefont{Murphy}, \bibfnamefont{M.~T.}},
  \bibinfo{author}{\bibfnamefont{J.~K.} \bibnamefont{Webb}}, and
  \bibinfo{author}{\bibfnamefont{V.~V.} \bibnamefont{Flambaum}},
  \bibinfo{year}{2003}, \bibinfo{journal}{Mon. Not. Roy. Astron. Soc.}
  \textbf{\bibinfo{volume}{345}}, \bibinfo{pages}{609}.

\bibitem[{\citenamefont{Myers}(1987)}]{Myers:1987fv}
\bibinfo{author}{\bibnamefont{Myers}, \bibfnamefont{R.~C.}},
  \bibinfo{year}{1987}, \bibinfo{journal}{Phys. Lett.}
  \textbf{\bibinfo{volume}{B199}}, \bibinfo{pages}{371}.

\bibitem[{\citenamefont{Myers}(1999)}]{Myers:1999ps}
\bibinfo{author}{\bibnamefont{Myers}, \bibfnamefont{R.~C.}},
  \bibinfo{year}{1999}, \bibinfo{journal}{JHEP} \textbf{\bibinfo{volume}{12}},
  \bibinfo{pages}{022}.

\bibitem[{\citenamefont{Narain} \emph{et~al.}(1987)\citenamefont{Narain,
  Sarmadi, and Vafa}}]{Narain:1986qm}
\bibinfo{author}{\bibnamefont{Narain}, \bibfnamefont{K.~S.}},
  \bibinfo{author}{\bibfnamefont{M.~H.} \bibnamefont{Sarmadi}}, and
  \bibinfo{author}{\bibfnamefont{C.}~\bibnamefont{Vafa}}, \bibinfo{year}{1987},
  \bibinfo{journal}{Nucl. Phys.} \textbf{\bibinfo{volume}{B288}},
  \bibinfo{pages}{551}.

\bibitem[{\citenamefont{Nepomechie}(1985)}]{Nepomechie:1984wu}
\bibinfo{author}{\bibnamefont{Nepomechie}, \bibfnamefont{R.~I.}},
  \bibinfo{year}{1985}, \bibinfo{journal}{Phys. Rev.}
  \textbf{\bibinfo{volume}{D31}}, \bibinfo{pages}{1921}.

\bibitem[{\citenamefont{Nobbenhuis}(2004)}]{Nobbenhuis:2004wn}
\bibinfo{author}{\bibnamefont{Nobbenhuis}, \bibfnamefont{S.}},
  \bibinfo{year}{2004}, \eprint{gr-qc/0411093}.

\bibitem[{\citenamefont{Ooguri and Vafa}(2006)}]{Ooguri:2006in}
\bibinfo{author}{\bibnamefont{Ooguri}, \bibfnamefont{H.}}, and
  \bibinfo{author}{\bibfnamefont{C.}~\bibnamefont{Vafa}}, \bibinfo{year}{2006},
  \eprint{hep-th/0605264}.

\bibitem[{\citenamefont{Ooguri} \emph{et~al.}(2005)\citenamefont{Ooguri, Vafa,
  and Verlinde}}]{Ooguri:2005vr}
\bibinfo{author}{\bibnamefont{Ooguri}, \bibfnamefont{H.}},
  \bibinfo{author}{\bibfnamefont{C.}~\bibnamefont{Vafa}}, and
  \bibinfo{author}{\bibfnamefont{E.~P.} \bibnamefont{Verlinde}},
  \bibinfo{year}{2005}, \bibinfo{journal}{Lett. Math. Phys.}
  \textbf{\bibinfo{volume}{74}}, \bibinfo{pages}{311}.

\bibitem[{\citenamefont{Padmanabhan}(2003)}]{Padmanabhan:2002ji}
\bibinfo{author}{\bibnamefont{Padmanabhan}, \bibfnamefont{T.}},
  \bibinfo{year}{2003}, \bibinfo{journal}{Phys. Rept.}
  \textbf{\bibinfo{volume}{380}}, \bibinfo{pages}{235}.

\bibitem[{\citenamefont{Peebles and Ratra}(1988)}]{Peebles:1987ek}
\bibinfo{author}{\bibnamefont{Peebles}, \bibfnamefont{P.~J.~E.}}, and
  \bibinfo{author}{\bibfnamefont{B.}~\bibnamefont{Ratra}},
  \bibinfo{year}{1988}, \bibinfo{journal}{Astrophys. J.}
  \textbf{\bibinfo{volume}{325}}, \bibinfo{pages}{L17}.

\bibitem[{\citenamefont{Polchinski}(1995)}]{Polchinski:1995mt}
\bibinfo{author}{\bibnamefont{Polchinski}, \bibfnamefont{J.}},
  \bibinfo{year}{1995}, \bibinfo{journal}{Phys. Rev. Lett.}
  \textbf{\bibinfo{volume}{75}}, \bibinfo{pages}{4724}.

\bibitem[{\citenamefont{Polchinski}(1996)}]{Polchinski:1996nb}
\bibinfo{author}{\bibnamefont{Polchinski}, \bibfnamefont{J.}},
  \bibinfo{year}{1996}, \bibinfo{journal}{Rev. Mod. Phys.}
  \textbf{\bibinfo{volume}{68}}, \bibinfo{pages}{1245}.

\bibitem[{\citenamefont{Polchinski}(1998{\natexlab{a}})}]{Polchinski:1998rr}
\bibinfo{author}{\bibnamefont{Polchinski}, \bibfnamefont{J.}},
  \bibinfo{year}{1998}{\natexlab{a}}, \emph{\bibinfo{title}{String theory. Vol.
  2: Superstring theory and beyond}} (\bibinfo{publisher}{Cambridge Univ.
  Press}).

\bibitem[{\citenamefont{Polchinski}(1998{\natexlab{b}})}]{Polchinski:1998rq}
\bibinfo{author}{\bibnamefont{Polchinski}, \bibfnamefont{J.}},
  \bibinfo{year}{1998}{\natexlab{b}}, \emph{\bibinfo{title}{String theory, Vol.
  I: An introduction to the bosonic string}} (\bibinfo{publisher}{Cambridge
  Univ. Press}).

\bibitem[{\citenamefont{Polchinski}(2006)}]{Polchinski:2006gy}
\bibinfo{author}{\bibnamefont{Polchinski}, \bibfnamefont{J.}},
  \bibinfo{year}{2006}, \eprint{hep-th/0603249}.

\bibitem[{\citenamefont{Polchinski and Strominger}(1996)}]{Polchinski:1995sm}
\bibinfo{author}{\bibnamefont{Polchinski}, \bibfnamefont{J.}}, and
  \bibinfo{author}{\bibfnamefont{A.}~\bibnamefont{Strominger}},
  \bibinfo{year}{1996}, \bibinfo{journal}{Phys. Lett.}
  \textbf{\bibinfo{volume}{B388}}, \bibinfo{pages}{736}.

\bibitem[{\citenamefont{Poppitz and Trivedi}(1998)}]{Poppitz:1998vd}
\bibinfo{author}{\bibnamefont{Poppitz}, \bibfnamefont{E.}}, and
  \bibinfo{author}{\bibfnamefont{S.~P.} \bibnamefont{Trivedi}},
  \bibinfo{year}{1998}, \bibinfo{journal}{Ann. Rev. Nucl. Part. Sci.}
  \textbf{\bibinfo{volume}{48}}, \bibinfo{pages}{307}.

\bibitem[{\citenamefont{Quevedo}(2002)}]{Quevedo:2002xw}
\bibinfo{author}{\bibnamefont{Quevedo}, \bibfnamefont{F.}},
  \bibinfo{year}{2002}, \bibinfo{journal}{Class. Quant. Grav.}
  \textbf{\bibinfo{volume}{19}}, \bibinfo{pages}{5721}.

\bibitem[{\citenamefont{Randall and
  Sundrum}(1999{\natexlab{a}})}]{Randall:1999ee}
\bibinfo{author}{\bibnamefont{Randall}, \bibfnamefont{L.}}, and
  \bibinfo{author}{\bibfnamefont{R.}~\bibnamefont{Sundrum}},
  \bibinfo{year}{1999}{\natexlab{a}}, \bibinfo{journal}{Phys. Rev. Lett.}
  \textbf{\bibinfo{volume}{83}}, \bibinfo{pages}{3370}.

\bibitem[{\citenamefont{Randall and
  Sundrum}(1999{\natexlab{b}})}]{Randall:1998uk}
\bibinfo{author}{\bibnamefont{Randall}, \bibfnamefont{L.}}, and
  \bibinfo{author}{\bibfnamefont{R.}~\bibnamefont{Sundrum}},
  \bibinfo{year}{1999}{\natexlab{b}}, \bibinfo{journal}{Nucl. Phys.}
  \textbf{\bibinfo{volume}{B557}}, \bibinfo{pages}{79}.

\bibitem[{\citenamefont{Randall and Thomas}(1995)}]{Randall:1994fr}
\bibinfo{author}{\bibnamefont{Randall}, \bibfnamefont{L.}}, and
  \bibinfo{author}{\bibfnamefont{S.~D.} \bibnamefont{Thomas}},
  \bibinfo{year}{1995}, \bibinfo{journal}{Nucl. Phys.}
  \textbf{\bibinfo{volume}{B449}}, \bibinfo{pages}{229}.

\bibitem[{\citenamefont{Rubakov}(2000)}]{Rubakov:1999aq}
\bibinfo{author}{\bibnamefont{Rubakov}, \bibfnamefont{V.}},
  \bibinfo{year}{2000}, \bibinfo{journal}{Phys. Rev.}
  \textbf{\bibinfo{volume}{D61}}, \bibinfo{pages}{061501}.

\bibitem[{\citenamefont{Sakharov}(1984)}]{Sakharov:1984ir}
\bibinfo{author}{\bibnamefont{Sakharov}, \bibfnamefont{A.~D.}},
  \bibinfo{year}{1984}, \bibinfo{journal}{Sov. Phys. JETP}
  \textbf{\bibinfo{volume}{60}}, \bibinfo{pages}{214}.

\bibitem[{\citenamefont{Saltman and Silverstein}(2004)}]{Saltman:2004sn}
\bibinfo{author}{\bibnamefont{Saltman}, \bibfnamefont{A.}}, and
  \bibinfo{author}{\bibfnamefont{E.}~\bibnamefont{Silverstein}},
  \bibinfo{year}{2004}, \bibinfo{journal}{JHEP} \textbf{\bibinfo{volume}{11}},
  \bibinfo{pages}{066}.

\bibitem[{\citenamefont{Saltman and Silverstein}(2006)}]{Saltman:2004jh}
\bibinfo{author}{\bibnamefont{Saltman}, \bibfnamefont{A.}}, and
  \bibinfo{author}{\bibfnamefont{E.}~\bibnamefont{Silverstein}},
  \bibinfo{year}{2006}, \bibinfo{journal}{JHEP} \textbf{\bibinfo{volume}{01}},
  \bibinfo{pages}{139}.

\bibitem[{\citenamefont{Sarangi and Tye}(2002)}]{Sarangi:2002yt}
\bibinfo{author}{\bibnamefont{Sarangi}, \bibfnamefont{S.}}, and
  \bibinfo{author}{\bibfnamefont{S.~H.~H.} \bibnamefont{Tye}},
  \bibinfo{year}{2002}, \bibinfo{journal}{Phys. Lett.}
  \textbf{\bibinfo{volume}{B536}}, \bibinfo{pages}{185}.

\bibitem[{\citenamefont{Sarangi and Tye}(2006)}]{Sarangi:2006eb}
\bibinfo{author}{\bibnamefont{Sarangi}, \bibfnamefont{S.}}, and
  \bibinfo{author}{\bibfnamefont{S.~H.~H.} \bibnamefont{Tye}},
  \bibinfo{year}{2006}, \eprint{hep-th/0603237}.

\bibitem[{\citenamefont{Saueressig}
  \emph{et~al.}(2006)\citenamefont{Saueressig, Theis, and
  Vandoren}}]{Saueressig:2005es}
\bibinfo{author}{\bibnamefont{Saueressig}, \bibfnamefont{F.}},
  \bibinfo{author}{\bibfnamefont{U.}~\bibnamefont{Theis}}, and
  \bibinfo{author}{\bibfnamefont{S.}~\bibnamefont{Vandoren}},
  \bibinfo{year}{2006}, \bibinfo{journal}{Phys. Lett.}
  \textbf{\bibinfo{volume}{B633}}, \bibinfo{pages}{125}.

\bibitem[{\citenamefont{Saulina}(2005)}]{Saulina:2005ve}
\bibinfo{author}{\bibnamefont{Saulina}, \bibfnamefont{N.}},
  \bibinfo{year}{2005}, \bibinfo{journal}{Nucl. Phys.}
  \textbf{\bibinfo{volume}{B720}}, \bibinfo{pages}{203}.

\bibitem[{\citenamefont{Schellekens}(1998)}]{Schellekens:2006}
\bibinfo{author}{\bibnamefont{Schellekens}, \bibfnamefont{A.~N.}},
  \bibinfo{year}{1998}, \bibinfo{title}{The landscape ``avant la lettre''},
  \urlprefix\url{http://arXiv.org/physics/0604134}.

\bibitem[{\citenamefont{Scherk and Schwarz}(1979)}]{Scherk:1979zr}
\bibinfo{author}{\bibnamefont{Scherk}, \bibfnamefont{J.}}, and
  \bibinfo{author}{\bibfnamefont{J.~H.} \bibnamefont{Schwarz}},
  \bibinfo{year}{1979}, \bibinfo{journal}{Nucl. Phys.}
  \textbf{\bibinfo{volume}{B153}}, \bibinfo{pages}{61}.

\bibitem[{\citenamefont{Schnabl}(2005)}]{Schnabl:2005gv}
\bibinfo{author}{\bibnamefont{Schnabl}, \bibfnamefont{M.}},
  \bibinfo{year}{2005}, \eprint{hep-th/0511286}.

\bibitem[{\citenamefont{Schulz}(2004)}]{Schulz:2004ub}
\bibinfo{author}{\bibnamefont{Schulz}, \bibfnamefont{M.~B.}},
  \bibinfo{year}{2004}, \bibinfo{journal}{Fortsch. Phys.}
  \textbf{\bibinfo{volume}{52}}, \bibinfo{pages}{963}.

\bibitem[{\citenamefont{Schulz}(2006)}]{Schulz:2004tt}
\bibinfo{author}{\bibnamefont{Schulz}, \bibfnamefont{M.~B.}},
  \bibinfo{year}{2006}, \bibinfo{journal}{JHEP} \textbf{\bibinfo{volume}{05}},
  \bibinfo{pages}{023}.

\bibitem[{\citenamefont{Seiberg}(1988)}]{Seiberg:1988pf}
\bibinfo{author}{\bibnamefont{Seiberg}, \bibfnamefont{N.}},
  \bibinfo{year}{1988}, \bibinfo{journal}{Nucl. Phys.}
  \textbf{\bibinfo{volume}{B303}}, \bibinfo{pages}{286}.

\bibitem[{\citenamefont{Seiberg and Witten}(1986)}]{Seiberg:1986by}
\bibinfo{author}{\bibnamefont{Seiberg}, \bibfnamefont{N.}}, and
  \bibinfo{author}{\bibfnamefont{E.}~\bibnamefont{Witten}},
  \bibinfo{year}{1986}, \bibinfo{journal}{Nucl. Phys.}
  \textbf{\bibinfo{volume}{B276}}, \bibinfo{pages}{272}.

\bibitem[{\citenamefont{Seiberg and Witten}(1994)}]{Seiberg:1994rs}
\bibinfo{author}{\bibnamefont{Seiberg}, \bibfnamefont{N.}}, and
  \bibinfo{author}{\bibfnamefont{E.}~\bibnamefont{Witten}},
  \bibinfo{year}{1994}, \bibinfo{journal}{Nucl. Phys.}
  \textbf{\bibinfo{volume}{B426}}, \bibinfo{pages}{19}.

\bibitem[{\citenamefont{Sen}(1997)}]{Sen:1997gv}
\bibinfo{author}{\bibnamefont{Sen}, \bibfnamefont{A.}}, \bibinfo{year}{1997},
  \bibinfo{journal}{Phys. Rev.} \textbf{\bibinfo{volume}{D55}},
  \bibinfo{pages}{7345}.

\bibitem[{\citenamefont{Sethi} \emph{et~al.}(1996)\citenamefont{Sethi, Vafa,
  and Witten}}]{Sethi:1996es}
\bibinfo{author}{\bibnamefont{Sethi}, \bibfnamefont{S.}},
  \bibinfo{author}{\bibfnamefont{C.}~\bibnamefont{Vafa}}, and
  \bibinfo{author}{\bibfnamefont{E.}~\bibnamefont{Witten}},
  \bibinfo{year}{1996}, \bibinfo{journal}{Nucl. Phys.}
  \textbf{\bibinfo{volume}{B480}}, \bibinfo{pages}{213}.

\bibitem[{\citenamefont{Shadmi and Shirman}(2000)}]{Shadmi:1999jy}
\bibinfo{author}{\bibnamefont{Shadmi}, \bibfnamefont{Y.}}, and
  \bibinfo{author}{\bibfnamefont{Y.}~\bibnamefont{Shirman}},
  \bibinfo{year}{2000}, \bibinfo{journal}{Rev. Mod. Phys.}
  \textbf{\bibinfo{volume}{72}}, \bibinfo{pages}{25}.

\bibitem[{\citenamefont{Shelton} \emph{et~al.}(2005)\citenamefont{Shelton,
  Taylor, and Wecht}}]{Shelton:2005cf}
\bibinfo{author}{\bibnamefont{Shelton}, \bibfnamefont{J.}},
  \bibinfo{author}{\bibfnamefont{W.}~\bibnamefont{Taylor}}, and
  \bibinfo{author}{\bibfnamefont{B.}~\bibnamefont{Wecht}},
  \bibinfo{year}{2005}, \bibinfo{journal}{JHEP} \textbf{\bibinfo{volume}{10}},
  \bibinfo{pages}{085}.

\bibitem[{\citenamefont{Shenker}(1990)}]{Shenker:1990uf}
\bibinfo{author}{\bibnamefont{Shenker}, \bibfnamefont{S.~H.}},
  \bibinfo{year}{1990}, \bibinfo{note}{presented at the Cargese Workshop on
  Random Surfaces, Quantum Gravity and Strings, Cargese, France, May 28 - Jun
  1, 1990}.

\bibitem[{\citenamefont{Shifman and Vainshtein}(1991)}]{Shifman:1991dz}
\bibinfo{author}{\bibnamefont{Shifman}, \bibfnamefont{M.~A.}}, and
  \bibinfo{author}{\bibfnamefont{A.~I.} \bibnamefont{Vainshtein}},
  \bibinfo{year}{1991}, \bibinfo{journal}{Nucl. Phys.}
  \textbf{\bibinfo{volume}{B359}}, \bibinfo{pages}{571}.

\bibitem[{\citenamefont{Shiu and Tye}(2001)}]{Shiu:2001sy}
\bibinfo{author}{\bibnamefont{Shiu}, \bibfnamefont{G.}}, and
  \bibinfo{author}{\bibfnamefont{S.~H.~H.} \bibnamefont{Tye}},
  \bibinfo{year}{2001}, \bibinfo{journal}{Phys. Lett.}
  \textbf{\bibinfo{volume}{B516}}, \bibinfo{pages}{421}.

\bibitem[{\citenamefont{Silverstein}(2001)}]{Silverstein:2001xn}
\bibinfo{author}{\bibnamefont{Silverstein}, \bibfnamefont{E.}},
  \bibinfo{year}{2001}, \eprint{hep-th/0106209}.

\bibitem[{\citenamefont{Silverstein}(2003)}]{Silverstein:2003jp}
\bibinfo{author}{\bibnamefont{Silverstein}, \bibfnamefont{E.}},
  \bibinfo{year}{2003}, \eprint{hep-th/0308175}.

\bibitem[{\citenamefont{Silverstein}(2004{\natexlab{a}})}]{Silverstein:2004sh}
\bibinfo{author}{\bibnamefont{Silverstein}, \bibfnamefont{E.}},
  \bibinfo{year}{2004}{\natexlab{a}}, \eprint{hep-th/0407202}.

\bibitem[{\citenamefont{Silverstein}(2004{\natexlab{b}})}]{Silverstein:2004id}
\bibinfo{author}{\bibnamefont{Silverstein}, \bibfnamefont{E.}},
  \bibinfo{year}{2004}{\natexlab{b}}, \eprint{hep-th/0405068}.

\bibitem[{\citenamefont{Silverstein and Tong}(2004)}]{Silverstein:2003hf}
\bibinfo{author}{\bibnamefont{Silverstein}, \bibfnamefont{E.}}, and
  \bibinfo{author}{\bibfnamefont{D.}~\bibnamefont{Tong}}, \bibinfo{year}{2004},
  \bibinfo{journal}{Phys. Rev.} \textbf{\bibinfo{volume}{D70}},
  \bibinfo{pages}{103505}.

\bibitem[{\citenamefont{Smolin}(1997)}]{Smolin:1997}
\bibinfo{author}{\bibnamefont{Smolin}, \bibfnamefont{L.}},
  \bibinfo{year}{1997}, \emph{\bibinfo{title}{The life of the cosmos}}
  (\bibinfo{publisher}{Oxford University Press}).

\bibitem[{Spergel \emph{et~al.}(2006)\citenamefont{Spergel}
  \emph{et~al.}}]{Spergel:2006hy}
\bibinfo{author}{\bibnamefont{Spergel}, \bibfnamefont{D.~N.}}, \emph{et~al.},
  \bibinfo{year}{2006}, \eprint{astro-ph/0603449}.

\bibitem[{\citenamefont{Steinhardt and Turok}(2006)}]{Steinhardt:2006bf}
\bibinfo{author}{\bibnamefont{Steinhardt}, \bibfnamefont{P.~J.}}, and
  \bibinfo{author}{\bibfnamefont{N.}~\bibnamefont{Turok}},
  \bibinfo{year}{2006}, \bibinfo{journal}{Science}
  \textbf{\bibinfo{volume}{312}}, \bibinfo{pages}{1180}.

\bibitem[{\citenamefont{Strominger}(1986)}]{Strominger:1986uh}
\bibinfo{author}{\bibnamefont{Strominger}, \bibfnamefont{A.}},
  \bibinfo{year}{1986}, \bibinfo{journal}{Nucl. Phys.}
  \textbf{\bibinfo{volume}{B274}}, \bibinfo{pages}{253}.

\bibitem[{\citenamefont{Strominger}(1990)}]{Strominger:1990pd}
\bibinfo{author}{\bibnamefont{Strominger}, \bibfnamefont{A.}},
  \bibinfo{year}{1990}, \bibinfo{journal}{Commun. Math. Phys.}
  \textbf{\bibinfo{volume}{133}}, \bibinfo{pages}{163}.

\bibitem[{\citenamefont{Strominger}
  \emph{et~al.}(1996)\citenamefont{Strominger, Yau, and
  Zaslow}}]{Strominger:1996it}
\bibinfo{author}{\bibnamefont{Strominger}, \bibfnamefont{A.}},
  \bibinfo{author}{\bibfnamefont{S.-T.} \bibnamefont{Yau}}, and
  \bibinfo{author}{\bibfnamefont{E.}~\bibnamefont{Zaslow}},
  \bibinfo{year}{1996}, \bibinfo{journal}{Nucl. Phys.}
  \textbf{\bibinfo{volume}{B479}}, \bibinfo{pages}{243}.

\bibitem[{\citenamefont{Susskind}(2003)}]{Susskind:2003kw}
\bibinfo{author}{\bibnamefont{Susskind}, \bibfnamefont{L.}},
  \bibinfo{year}{2003}, \eprint{hep-th/0302219}.

\bibitem[{\citenamefont{Susskind}(2004)}]{Susskind:2004uv}
\bibinfo{author}{\bibnamefont{Susskind}, \bibfnamefont{L.}},
  \bibinfo{year}{2004}, \eprint{hep-th/0405189}.

\bibitem[{\citenamefont{Susskind}(2005)}]{Susskind:2005bd}
\bibinfo{author}{\bibnamefont{Susskind}, \bibfnamefont{L.}},
  \bibinfo{year}{2005}, \emph{\bibinfo{title}{The cosmic landscape: String
  theory and the illusion of intelligent design}} (\bibinfo{publisher}{Little,
  Brown}).

\bibitem[{\citenamefont{Svrcek}(2006)}]{Svrcek:2006hf}
\bibinfo{author}{\bibnamefont{Svrcek}, \bibfnamefont{P.}},
  \bibinfo{year}{2006}, \eprint{hep-th/0607086}.

\bibitem[{\citenamefont{Taylor and Vafa}(2000)}]{Taylor:1999ii}
\bibinfo{author}{\bibnamefont{Taylor}, \bibfnamefont{T.~R.}}, and
  \bibinfo{author}{\bibfnamefont{C.}~\bibnamefont{Vafa}}, \bibinfo{year}{2000},
  \bibinfo{journal}{Phys. Lett.} \textbf{\bibinfo{volume}{B474}},
  \bibinfo{pages}{130}.

\bibitem[{\citenamefont{Tegmark}(2005)}]{Tegmark:2004qd}
\bibinfo{author}{\bibnamefont{Tegmark}, \bibfnamefont{M.}},
  \bibinfo{year}{2005}, \bibinfo{journal}{JCAP}
  \textbf{\bibinfo{volume}{0504}}, \bibinfo{pages}{001}.

\bibitem[{\citenamefont{Teitelboim}(1986)}]{Teitelboim:1985yc}
\bibinfo{author}{\bibnamefont{Teitelboim}, \bibfnamefont{C.}},
  \bibinfo{year}{1986}, \bibinfo{journal}{Phys. Lett.}
  \textbf{\bibinfo{volume}{B167}}, \bibinfo{pages}{69}.

\bibitem[{\citenamefont{Tian and Yau}(1986)}]{Tian:1986ic}
\bibinfo{author}{\bibnamefont{Tian}, \bibfnamefont{G.}}, and
  \bibinfo{author}{\bibfnamefont{S.-T.} \bibnamefont{Yau}},
  \bibinfo{year}{1986}, \bibinfo{note}{in San Diego 1986 proceedings,
  Mathematical aspects of string theory, 543-559.}

\bibitem[{\citenamefont{Tomasiello}(2005)}]{Tomasiello:2005bp}
\bibinfo{author}{\bibnamefont{Tomasiello}, \bibfnamefont{A.}},
  \bibinfo{year}{2005}, \bibinfo{journal}{JHEP} \textbf{\bibinfo{volume}{06}},
  \bibinfo{pages}{067}.

\bibitem[{\citenamefont{Tripathy and Trivedi}(2003)}]{Tripathy:2002qw}
\bibinfo{author}{\bibnamefont{Tripathy}, \bibfnamefont{P.~K.}}, and
  \bibinfo{author}{\bibfnamefont{S.~P.} \bibnamefont{Trivedi}},
  \bibinfo{year}{2003}, \bibinfo{journal}{JHEP} \textbf{\bibinfo{volume}{03}},
  \bibinfo{pages}{028}.

\bibitem[{\citenamefont{Uzan}(2003)}]{Uzan:2002vq}
\bibinfo{author}{\bibnamefont{Uzan}, \bibfnamefont{J.-P.}},
  \bibinfo{year}{2003}, \bibinfo{journal}{Rev. Mod. Phys.}
  \textbf{\bibinfo{volume}{75}}, \bibinfo{pages}{403}.

\bibitem[{\citenamefont{Uzan}(2005)}]{Uzan:2004qr}
\bibinfo{author}{\bibnamefont{Uzan}, \bibfnamefont{J.-P.}},
  \bibinfo{year}{2005}, \bibinfo{journal}{AIP Conf. Proc.}
  \textbf{\bibinfo{volume}{736}}, \bibinfo{pages}{3}.

\bibitem[{\citenamefont{Vafa}(1996)}]{Vafa:1996xn}
\bibinfo{author}{\bibnamefont{Vafa}, \bibfnamefont{C.}}, \bibinfo{year}{1996},
  \bibinfo{journal}{Nucl. Phys.} \textbf{\bibinfo{volume}{B469}},
  \bibinfo{pages}{403}.

\bibitem[{\citenamefont{Vafa}(2001)}]{Vafa:2000wi}
\bibinfo{author}{\bibnamefont{Vafa}, \bibfnamefont{C.}}, \bibinfo{year}{2001},
  \bibinfo{journal}{J. Math. Phys.} \textbf{\bibinfo{volume}{42}},
  \bibinfo{pages}{2798}.

\bibitem[{\citenamefont{Vafa}(2005)}]{Vafa:2005ui}
\bibinfo{author}{\bibnamefont{Vafa}, \bibfnamefont{C.}}, \bibinfo{year}{2005},
  \eprint{hep-th/0509212}.

\bibitem[{\citenamefont{Vanchurin and Vilenkin}(2006)}]{Vanchurin:2006qp}
\bibinfo{author}{\bibnamefont{Vanchurin}, \bibfnamefont{V.}}, and
  \bibinfo{author}{\bibfnamefont{A.}~\bibnamefont{Vilenkin}},
  \bibinfo{year}{2006}, \bibinfo{journal}{Phys. Rev.}
  \textbf{\bibinfo{volume}{D74}}, \bibinfo{pages}{043520}.

\bibitem[{\citenamefont{Veneziano and Yankielowicz}(1982)}]{Veneziano:1982ah}
\bibinfo{author}{\bibnamefont{Veneziano}, \bibfnamefont{G.}}, and
  \bibinfo{author}{\bibfnamefont{S.}~\bibnamefont{Yankielowicz}},
  \bibinfo{year}{1982}, \bibinfo{journal}{Phys. Lett.}
  \textbf{\bibinfo{volume}{B113}}, \bibinfo{pages}{231}.

\bibitem[{\citenamefont{Verlinde}(2000)}]{Verlinde:1999fy}
\bibinfo{author}{\bibnamefont{Verlinde}, \bibfnamefont{H.~L.}},
  \bibinfo{year}{2000}, \bibinfo{journal}{Nucl. Phys.}
  \textbf{\bibinfo{volume}{B580}}, \bibinfo{pages}{264}.

\bibitem[{\citenamefont{Vilenkin}(1983)}]{Vilenkin:1983xq}
\bibinfo{author}{\bibnamefont{Vilenkin}, \bibfnamefont{A.}},
  \bibinfo{year}{1983}, \bibinfo{journal}{Phys. Rev.}
  \textbf{\bibinfo{volume}{D27}}, \bibinfo{pages}{2848}.

\bibitem[{\citenamefont{Vilenkin}(2006)}]{Vilenkin:2006xv}
\bibinfo{author}{\bibnamefont{Vilenkin}, \bibfnamefont{A.}},
  \bibinfo{year}{2006}, \eprint{hep-th/0609193}.

\bibitem[{\citenamefont{Villadoro and Zwirner}(2005)}]{Villadoro:2005cu}
\bibinfo{author}{\bibnamefont{Villadoro}, \bibfnamefont{G.}}, and
  \bibinfo{author}{\bibfnamefont{F.}~\bibnamefont{Zwirner}},
  \bibinfo{year}{2005}, \bibinfo{journal}{JHEP} \textbf{\bibinfo{volume}{06}},
  \bibinfo{pages}{047}.

\bibitem[{\citenamefont{Wales}(2003)}]{Wales:2003}
\bibinfo{author}{\bibnamefont{Wales}, \bibfnamefont{D.~J.}},
  \bibinfo{year}{2003}, \emph{\bibinfo{title}{Energy Landscapes}}
  (\bibinfo{publisher}{Cambridge Univ. Press}).

\bibitem[{\citenamefont{Weinberg}(1987)}]{Weinberg:1987dv}
\bibinfo{author}{\bibnamefont{Weinberg}, \bibfnamefont{S.}},
  \bibinfo{year}{1987}, \bibinfo{journal}{Phys. Rev. Lett.}
  \textbf{\bibinfo{volume}{59}}, \bibinfo{pages}{2607}.

\bibitem[{\citenamefont{Weinberg}(1989)}]{Weinberg:1988cp}
\bibinfo{author}{\bibnamefont{Weinberg}, \bibfnamefont{S.}},
  \bibinfo{year}{1989}, \bibinfo{journal}{Rev. Mod. Phys.}
  \textbf{\bibinfo{volume}{61}}, \bibinfo{pages}{1}.

\bibitem[{\citenamefont{Wess and Bagger}(1992)}]{Wess:1992cp}
\bibinfo{author}{\bibnamefont{Wess}, \bibfnamefont{J.}}, and
  \bibinfo{author}{\bibfnamefont{J.}~\bibnamefont{Bagger}},
  \bibinfo{year}{1992}, \emph{\bibinfo{title}{Supersymmetry and supergravity}}
  (\bibinfo{publisher}{Princeton Univ. Press}).

\bibitem[{\citenamefont{Wilczek}(2005)}]{Wilczek:2005aj}
\bibinfo{author}{\bibnamefont{Wilczek}, \bibfnamefont{F.}},
  \bibinfo{year}{2005}, \eprint{hep-ph/0512187}.

\bibitem[{\citenamefont{de~Wit} \emph{et~al.}(1987)\citenamefont{de~Wit, Smit,
  and Hari~Dass}}]{deWit:1986xg}
\bibinfo{author}{\bibnamefont{de~Wit}, \bibfnamefont{B.}},
  \bibinfo{author}{\bibfnamefont{D.~J.} \bibnamefont{Smit}}, and
  \bibinfo{author}{\bibfnamefont{N.~D.} \bibnamefont{Hari~Dass}},
  \bibinfo{year}{1987}, \bibinfo{journal}{Nucl. Phys.}
  \textbf{\bibinfo{volume}{B283}}, \bibinfo{pages}{165}.

\bibitem[{\citenamefont{Witten}(1981{\natexlab{a}})}]{Witten:1981nf}
\bibinfo{author}{\bibnamefont{Witten}, \bibfnamefont{E.}},
  \bibinfo{year}{1981}{\natexlab{a}}, \bibinfo{journal}{Nucl. Phys.}
  \textbf{\bibinfo{volume}{B188}}, \bibinfo{pages}{513}.

\bibitem[{\citenamefont{Witten}(1981{\natexlab{b}})}]{Witten:1981me}
\bibinfo{author}{\bibnamefont{Witten}, \bibfnamefont{E.}},
  \bibinfo{year}{1981}{\natexlab{b}}, \bibinfo{journal}{Nucl. Phys.}
  \textbf{\bibinfo{volume}{B186}}, \bibinfo{pages}{412}.

\bibitem[{\citenamefont{Witten}(1996{\natexlab{a}})}]{Witten:1995im}
\bibinfo{author}{\bibnamefont{Witten}, \bibfnamefont{E.}},
  \bibinfo{year}{1996}{\natexlab{a}}, \bibinfo{journal}{Nucl. Phys.}
  \textbf{\bibinfo{volume}{B460}}, \bibinfo{pages}{335}.

\bibitem[{\citenamefont{Witten}(1996{\natexlab{b}})}]{Witten:1996bn}
\bibinfo{author}{\bibnamefont{Witten}, \bibfnamefont{E.}},
  \bibinfo{year}{1996}{\natexlab{b}}, \bibinfo{journal}{Nucl. Phys.}
  \textbf{\bibinfo{volume}{B474}}, \bibinfo{pages}{343}.

\bibitem[{\citenamefont{Witten}(1996{\natexlab{c}})}]{Witten:1996mz}
\bibinfo{author}{\bibnamefont{Witten}, \bibfnamefont{E.}},
  \bibinfo{year}{1996}{\natexlab{c}}, \bibinfo{journal}{Nucl. Phys.}
  \textbf{\bibinfo{volume}{B471}}, \bibinfo{pages}{135}.

\bibitem[{\citenamefont{Witten and Witten}(1987)}]{Witten:1986kg}
\bibinfo{author}{\bibnamefont{Witten}, \bibfnamefont{L.}}, and
  \bibinfo{author}{\bibfnamefont{E.}~\bibnamefont{Witten}},
  \bibinfo{year}{1987}, \bibinfo{journal}{Nucl. Phys.}
  \textbf{\bibinfo{volume}{B281}}, \bibinfo{pages}{109}.

\bibitem[{\citenamefont{Wright}(1932)}]{Wright:1932}
\bibinfo{author}{\bibnamefont{Wright}, \bibfnamefont{S.}},
  \bibinfo{year}{1932}, \bibinfo{journal}{Proceedings of the Sixth
  International Congress on Genetics} \textbf{\bibinfo{volume}{1}},
  \bibinfo{pages}{356}.

\bibitem[{\citenamefont{Yau}(1977)}]{Yau:1977ms}
\bibinfo{author}{\bibnamefont{Yau}, \bibfnamefont{S.-T.}},
  \bibinfo{year}{1977}, \bibinfo{journal}{Proc. Nat. Acad. Sci.}
  \textbf{\bibinfo{volume}{74}}, \bibinfo{pages}{1798}.

\bibitem[{\citenamefont{Zelditch}(2006)}]{Zelditch:2006pi}
\bibinfo{author}{\bibnamefont{Zelditch}, \bibfnamefont{S.}},
  \bibinfo{year}{2006}, \eprint{math-ph/0603066}.

\bibitem[{\citenamefont{Zwiebach}(1993)}]{Zwiebach:1993cs}
\bibinfo{author}{\bibnamefont{Zwiebach}, \bibfnamefont{B.}},
  \bibinfo{year}{1993}, \eprint{hep-th/9305026}.

\bibitem[{\citenamefont{Zwiebach}(2004)}]{Zwiebach:2004tj}
\bibinfo{author}{\bibnamefont{Zwiebach}, \bibfnamefont{B.}},
  \bibinfo{year}{2004}, \emph{\bibinfo{title}{A first course in string theory}}
  (\bibinfo{publisher}{Cambridge Univ. Press}).

\end{thebibliography}

\end{document}